\documentclass[twocolumn]{aastex631}

\usepackage{graphicx}
\usepackage{CJK}
\usepackage{enumitem}
\usepackage{hyperref}
\usepackage{amsmath}
\usepackage{amssymb}
\usepackage{footmisc}
\usepackage{booktabs}
\usepackage{listings}

\hypersetup{linkcolor=red,citecolor=green,filecolor=cyan,urlcolor=magenta}

\submitjournal{\apj}

\shorttitle{Evolved Massive Stars at low-Z \uppercase\expandafter{\romannumeral7}. the Lower Mass Limit of RSGs in the LMC}
\shortauthors{Yang et al.}    
\graphicspath{{./}{figures/}}      

\begin{document}

\begin{CJK*}{UTF8}{gbsn}

\title{Evolved Massive Stars at Low-metallicity \uppercase\expandafter{\romannumeral7}.\\ the Lower Mass Limit of Red Supergiant Population in the Large Magellanic Cloud}

\correspondingauthor{Ming Yang (杨明)}
\email{myang@nao.cas.cn}

\author[0000-0001-8247-4936]{Ming Yang (杨明)}
\affiliation{Key Laboratory of Space Astronomy and Technology, National Astronomical Observatories, Chinese Academy of Sciences, Beijing 100101, People's Republic of China}

\author[0000-0002-6434-7201]{Bo Zhang (章博)}
\affiliation{Key Laboratory of Space Astronomy and Technology, National Astronomical Observatories, Chinese Academy of Sciences, Beijing 100101, People's Republic of China}

\author[0000-0003-3168-2617]{Biwei Jiang (姜碧沩)}
\affiliation{Institute for Frontiers in Astronomy and Astrophysics, Beijing Normal University, Beijing 102206, People's Republic of China}
\affiliation{Department of Astronomy, Beijing Normal University, Beijing 100875, People's Republic of China} 

\author[0000-0003-4195-0195]{Jian Gao (高健)}
\affiliation{Institute for Frontiers in Astronomy and Astrophysics, Beijing Normal University, Beijing 102206, People's Republic of China}
\affiliation{Department of Astronomy, Beijing Normal University, Beijing 100875, People's Republic of China} 

\author[0000-0003-1218-8699]{Yi Ren (任逸)}
\affiliation{College of Physics and Electronic Engineering, Qilu Normal University, Jinan 250200, People's Republic of China}

\author[0000-0003-4489-9794]{Shu Wang (王舒)}
\affiliation{Key Laboratory of Optical Astronomy, National Astronomical Observatories, Chinese Academy of Sciences, Beijing 100101, People's Republic of China}

\author{Man I Lam (林敏仪)}
\affiliation{Key Laboratory of Space Astronomy and Technology, National Astronomical Observatories, Chinese Academy of Sciences, Beijing 100101, People's Republic of China}

\author[0000-0003-3347-7596]{Hao Tian (田浩)}
\affiliation{Key Laboratory of Space Astronomy and Technology, National Astronomical Observatories, Chinese Academy of Sciences, Beijing 100101, People's Republic of China}
\affiliation{Institute for Frontiers in Astronomy and Astrophysics, Beijing Normal University, Beijing, 102206, People's Republic of China}

\author{Changqing Luo (罗常青)}
\affiliation{Key Laboratory of Space Astronomy and Technology, National Astronomical Observatories, Chinese Academy of Sciences, Beijing 100101, People's Republic of China}

\author{Bingqiu Chen (陈丙秋)}
\affiliation{South-Western Institute for Astronomy Research, Yunnan University, Kunming 650500, People's Republic of China}

\author{Jing Wen (文静)}
\affiliation{Institute for Frontiers in Astronomy and Astrophysics, Beijing Normal University, Beijing 102206, People's Republic of China}
\affiliation{Department of Astronomy, Beijing Normal University, Beijing 100875, People's Republic of China}

\begin{abstract}
The precise definition of the lower mass limit of red supergiant stars (RSGs) is an open question in astrophysics and does not attract too much attention. Here we assemble a spectroscopic evolved cool star sample with 6,602 targets, including RSGs, asymptotic giant branch stars, and red giant branch stars, in the Large Magellanic Cloud based on \textit{Gaia} DR3 and SDSS-IV/APOGEE-2. The reference spectrum of each stellar population is built according to the quantile range of relative intensity ($1\%\sim99\%$). Five different methods, e.g., chi-square ($\chi^2$), cosine similarity (CS), machine learning (ML), equivalent width (EW), and line ratio (LR), are used in order to separate different stellar populations. The ML and $\chi^2$ provide the best and relatively consistent prediction of certain population. The derived lower limit of the RSG population is able to reach to the $\rm K_S$-band tip of red giant branch ($\rm K_S~$$\approx12.0$ mag), indicating a luminosity as low as about $10^{3.5}~L_{\sun}$, which corresponds to a stellar radius only about $100~R_{\sun}$. Given the mass-luminosity relation of $L/L_\sun =f(M/M_\sun)^3$ with $f\approx15.5\pm3$ and taking into account of the mass loss of faint RSGs up to now, the minimal initial mass of the RSG population would be about $6.1\pm0.4~M_\sun$, which is much lower than the traditional threshold of $8~M_\sun$ for the massive stars. This is the first spectroscopic evidence, indicating that the lower mass limit of RSG population is around $6~M_\sun$. However, the destinies of such faint RSGs are still elusive and may have large impact on the stellar evolutionary and supernova models.

\end{abstract}

\keywords{Red supergiant stars (1375) --- Infrared spectroscopy (2285) --- Stellar populations (1622) --- Stellar classification (1589)}

\section{Introduction} \label{sec:intro}

As one of the massive and extreme stellar populations on the Hertzsprung-Russell (H-R) diagram, red supergiant stars (RSGs) have unique properties of moderately high initial masses ($\sim8-40~M_\sun$), high luminosities ($\sim4000-400000~L_\sun$), low effective temperatures ($T_{\rm eff} \sim3500-4500~K$), large radii ($\sim200-1500~R_\sun$), and young ages ($\rm 8-20~Myr$) \citep{Humphreys1979, Massey1998, Massey2003, Levesque2010, Davies2013, Gonzalez2015, Neugent2020}. These distinctive properties make them an important role in the formation and evolution of both stars and galaxies. 

However, when talking about RSGs, the definition of both upper and lower mass limits seem a bit vague. For the upper limit, it is model-dependent and is typically referred to as ``Humphreys-Davidson limit'', which delineates the empirical maximum luminosity of cool supergiant stars \citep{Humphreys1979}. It is often explained as the consequence of strong stellar winds or episodic mass-loss that stripping off the stellar envelopes, which is still under debate \citep{Higgins2020, Gilkis2021, Agrawal2022, McDonald2022}. 

On the other hand, the lower mass limit of RSGs, or to say, the the minimal initial mass ($M_{min}$), seems less important and pedestrian. From the theoretical point of view, stars below $M_{min}$ shall develop a degenerate core without carbon burning, meaning that they will probably \textit{NOT} to form a collapsing core and progress to supernovae or black holes. Meanwhile, the exact value of $M_{min}$ depends upon many parameters, like metallicity, mixing length, convection, overshooting, and so on \citep{Woosley1986, Woosley1995, Woosley2002}, for which $8~M_\sun$ is commonly adopted \citep{Iben1983}. However, during the main sequence, a large overshoot beyond the convective core can significantly reduce this mass, e.g., down to $5-6~M_\sun$ \citep{Bressan1993}. From the observational point of view, the ``traditional'' RSG population is the \textit{upper part} of the red helium burning sequence (RHeBS; stars with initial masses $>2~M_\sun$ and that evolved off the main sequence with core helium burning; \citealt{DohmPalmer1997, Cole1999, Weisz2008, McQuinn2011, Dalcanton2012, Schombert2015}), where the RHeBS is a continuous uniform sequence ranging from the red clump stars to the RSGs on the color-magnitude diagram (CMD; see good examples of Figure 6 of \citealt{RadburnSmith2011} or Figure 2 of \citealt{McQuinn2011}). Our previous studies also indicated that, there was a distinct branch extending continuously from the top of the cool luminous region (RSG population) towards the relatively faint warm area on the CMD, reaching approximately the tip of red giant branch (TRGB) without blending into the asymptotic giant branch (AGB) population \citep{Yang2019, Yang2020, Ren2021a, Ren2021b, Yang2021a, Yang2021b, Ren2022, Yang2023}. Meanwhile, the criteria for defining a star as a RSG are relatively subjective and observational, e.g., simple cutoffs in magnitude and color, which marks the RSGs as bright and red (presumably massive) stars in the upper right corner of the CMD. In that sense, the faint end (as well as the definition) of RSG population is a bit ambiguous, since it is hard to put a cut-off on a continuous uniform sequence. Even for our previous paper, one key question from the anonymous referee was, whether targets with luminosities as low as $L\approx10^{3.5}~L_{\sun}$ should be considered as RSGs, since they were far too faint. Thus, from our point of view, the exact lower mass limit of the RSG population is still on debate.

Based on our previous works and recent released large-scale spectroscopic data, we re-studied the RSGs in the Large Magellnic Cloud (LMC), in order to investigate the lower mass limit of the RSG population. Here a true distance modulus of $18.493\pm0.055$ mag for the LMC was adopted in this work \citep{Pietrzynski2013}. The sample selection and building reference spectra are presented in \textsection2 and \textsection3, respectively. The classification of different stellar populations is described in \textsection4. The discussion and summary are given in \textsection5 and \textsection6, respectively .

\section{Cool Evolved Star Samples in the Large Magellanic Clouds} \label{sec:sample}

Initially, the sample used in this work was taken from the Apache Point Observatory Galactic Evolution Experiment 2 (APOGEE-2; \citealt{Majewski2017, Jonsson2020}) in the seventeenth data release (DR17; \citealt{Abdurrouf2022}) of the fourth phase of Sloan Digital Sky Surveys (SDSS-IV; \citealt{Blanton2017}). APOGEE-2 is one of the major programs of SDSS-IV, which has collected infrared spectra (H-band; 1.51-1.70 $\mu$m) for over 730,000 stars in the Milky Way and extragalactic regions with high resolution (R$\sim$22,500) and high signal-to-noise ratio ($>$100). 

\begin{figure}
\center
\includegraphics[scale=0.45]{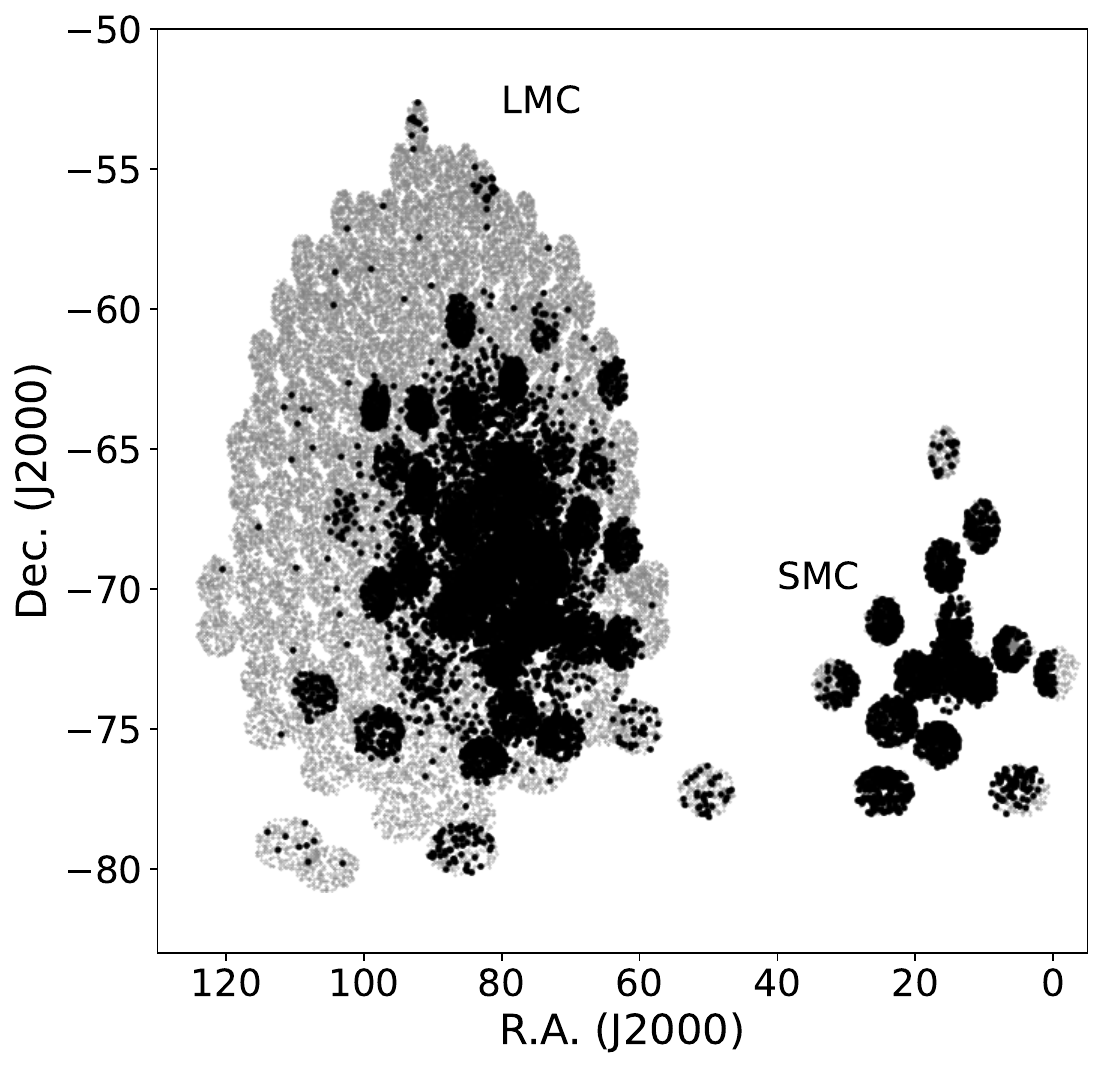}
\caption{The spatial distribution of initial sample (grey dots; same below) and MCs sample (black dots; same below) from the APOGEE-2/SDSS-IV.
\label{fig:spatial}}
\end{figure}

For our study, we first selected targets within the region of the Magellanic Clouds (MCs; Figure~\ref{fig:spatial}, grey dots) from the APOGEE-2/SDSS-IV summative catalog\footnote{https://data.sdss.org/sas/dr17/apogee/spectro/aspcap/\\dr17/synspec\_rev1/allStar-dr17-synspec\_rev1.fits}, which resulted in 67,289 records. Then the duplications (multiple observations for the same target) were removed, which left 60,338 targets in the sample. Notice that, for each target, there is no neighbour within 3'', so that there is basically no \textit{serious} blending issue. The sample was crossmatched with \textit{Gaia} DR3 \citep{Gaia2016, Gaia2022} with a search radius of 1'', for which the sample size was shrank to 57,366 targets. 

\begin{figure}
\center
\includegraphics[scale=0.45]{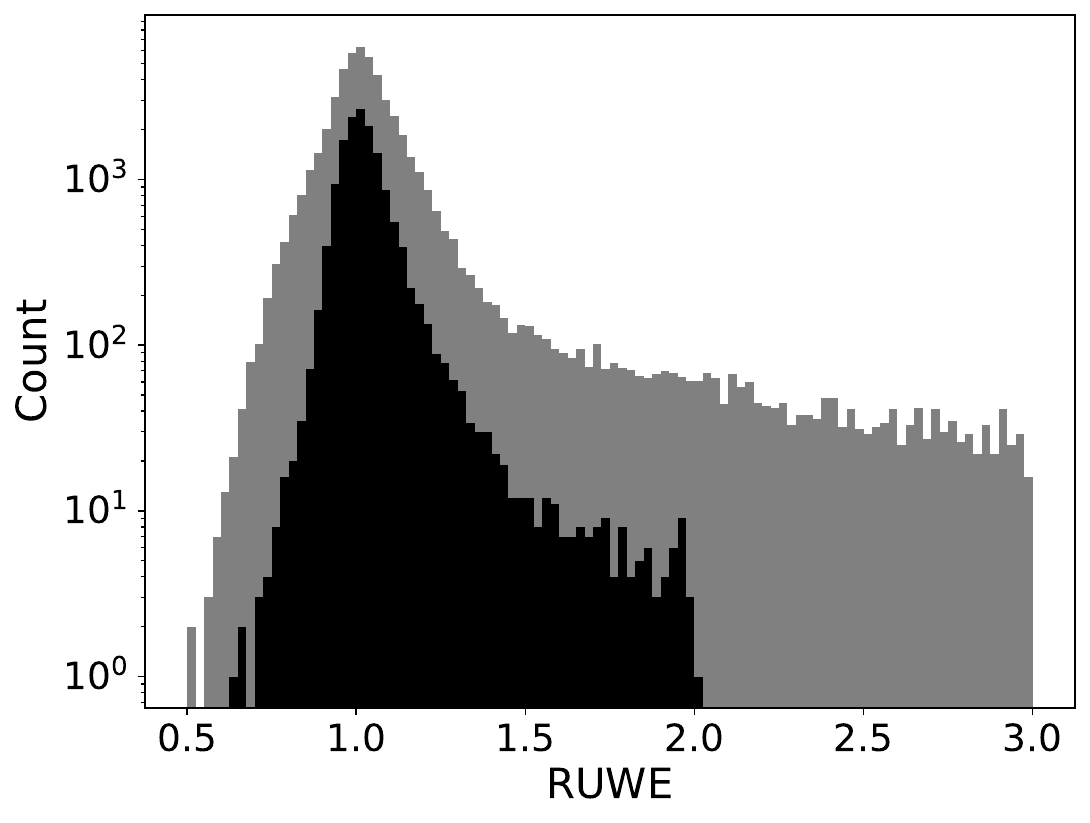}
\caption{Histogram of the RUWE from \textit{Gaia} DR3. The limit of our sample was set to be less than 2.0.
\label{fig:ruwe}}
\end{figure}

\begin{figure}
\center
\includegraphics[scale=0.45]{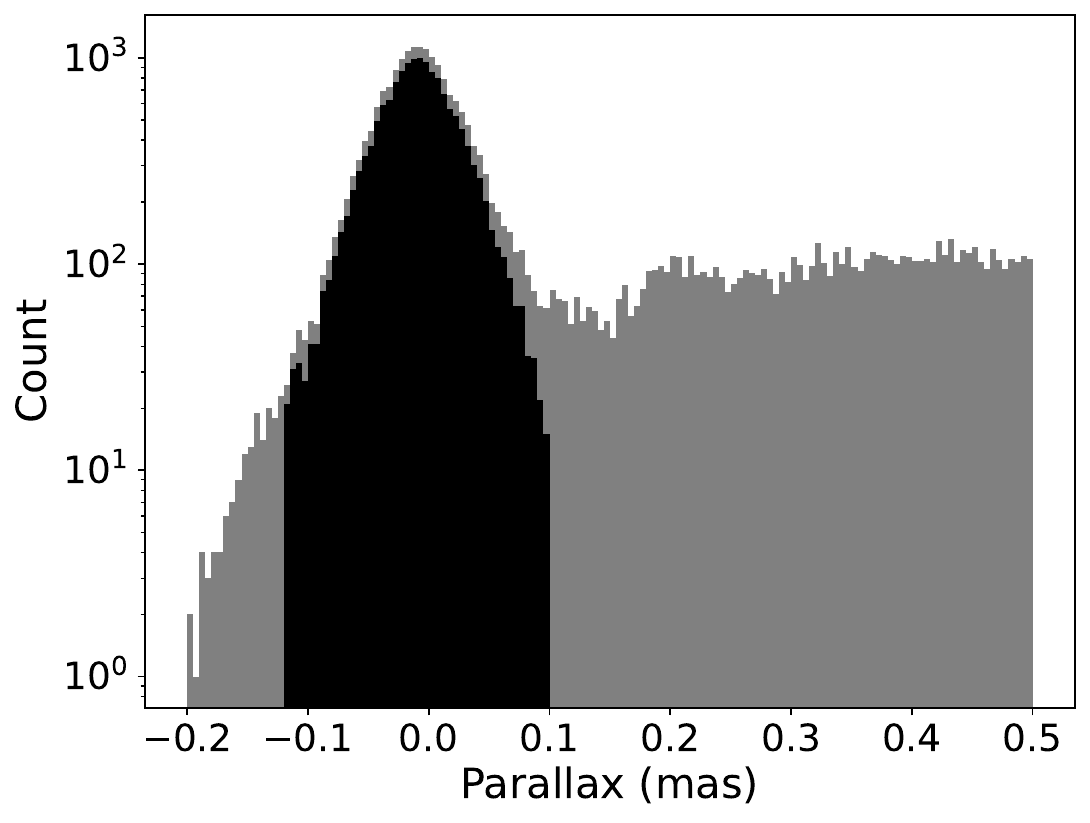}
\caption{Histogram of the parallax from \textit{Gaia} DR3, which was constrained to be in the range between -0.12 and 0.1 mas.
\label{fig:plx}}
\end{figure}

\begin{figure}
\center
\includegraphics[scale=0.45]{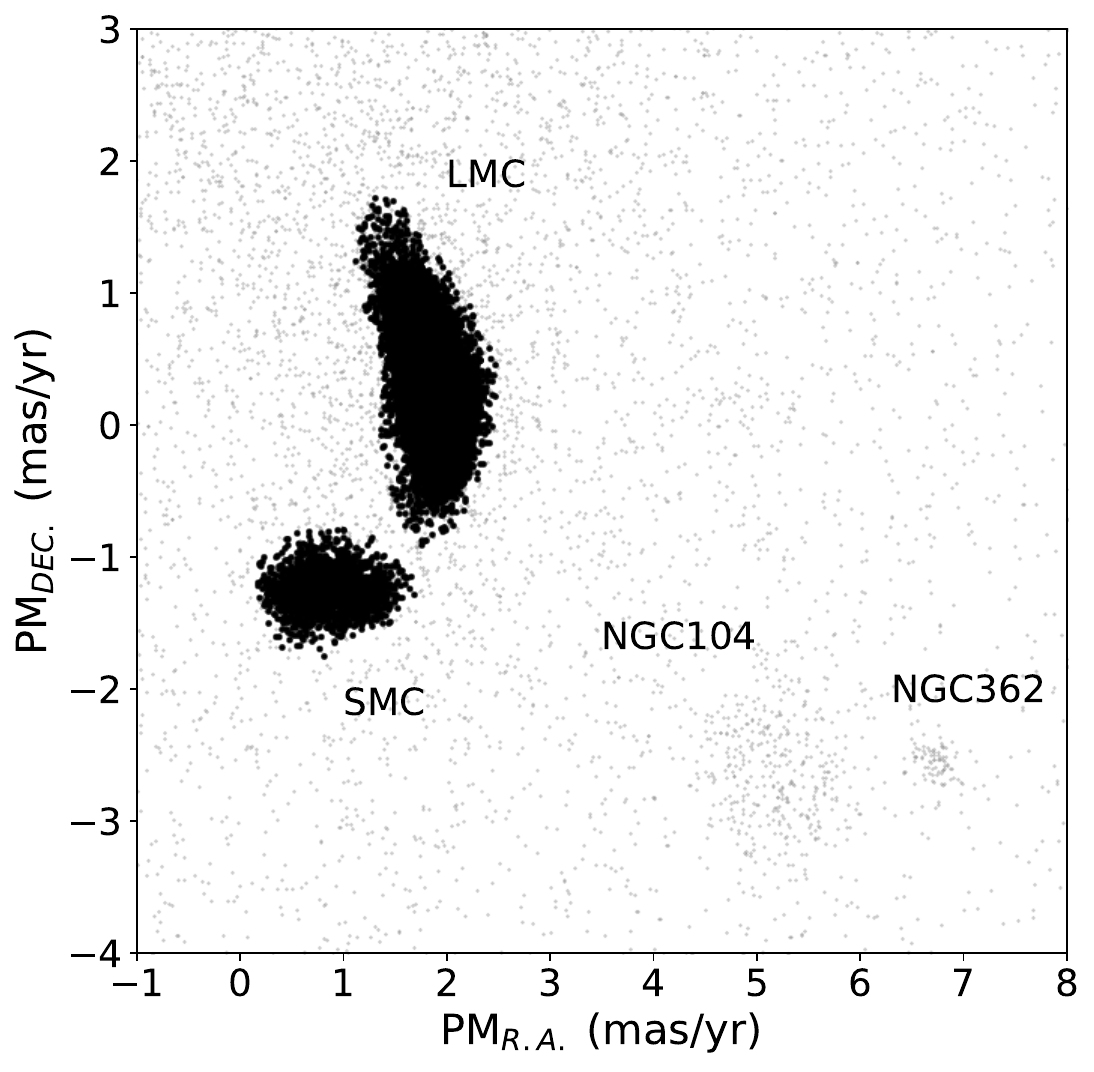}
\caption{PM in right ascension versus PM in declination from \textit{Gaia} DR3. MCs members are easily observed. NGC 104 and NGC 362 are also marked on the diagram.
\label{fig:pm}}
\end{figure}

\begin{figure*}
\center
\includegraphics[scale=0.43]{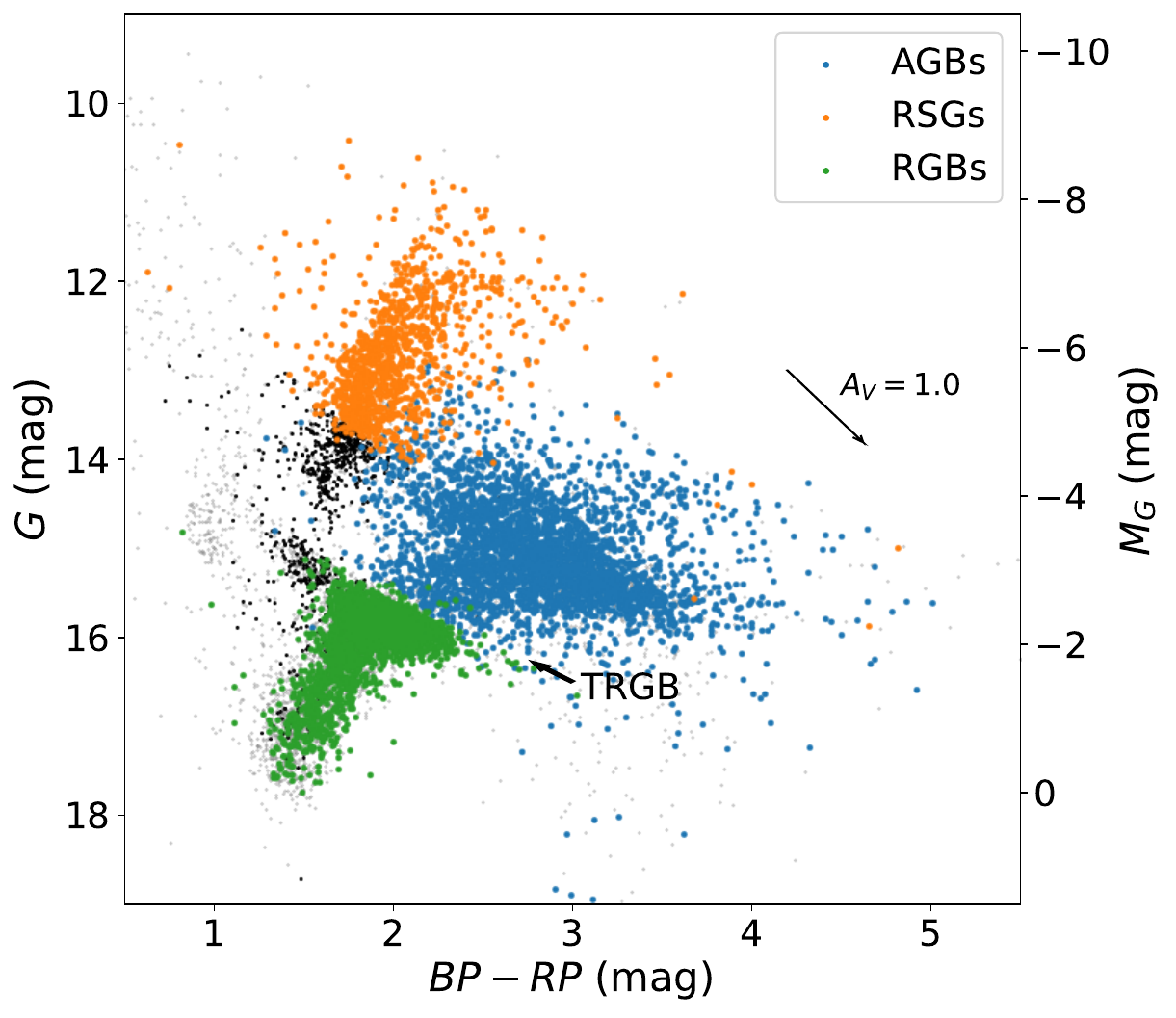}
\includegraphics[scale=0.43]{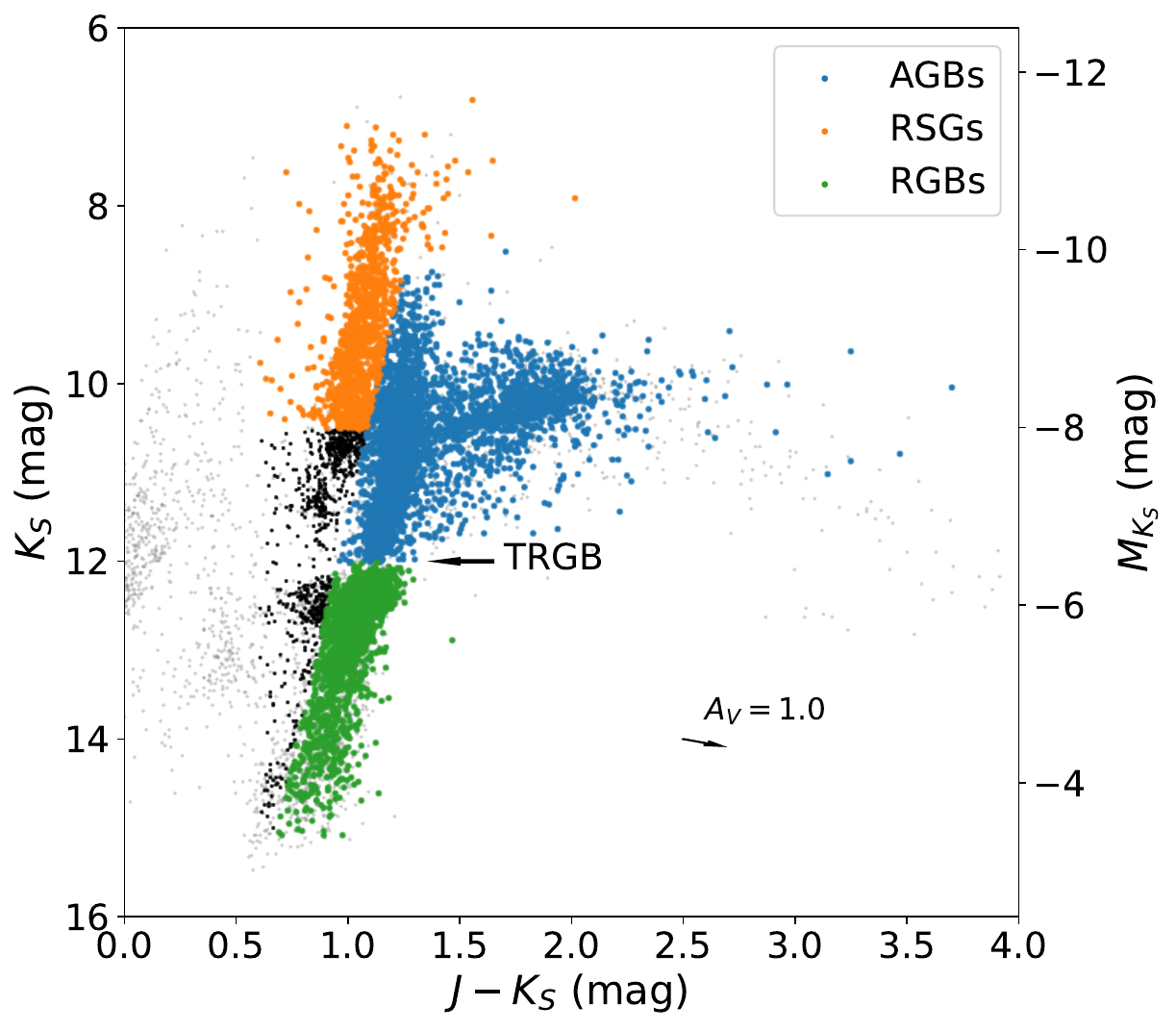}
\caption{The \textit{Gaia} and 2MASS color-magnitude diagrams for the LMC sample, where RSG, AGB, and RGB populations, as well as the TRGB, are marked on the diagram. The gray and black points indicate the LMC member and unclassified cool evolved stars, respectively. A reddening vector of $A_{\rm V}=1.0$ mag is shown as a reference in each panel (same below). 
\label{fig:cmd_lmc}}
\end{figure*}

The astrometric parameters, namely the proper motions (PMs), parallax, and their errors (including the renormalised unit weight error; RUWE), from \textit{Gaia} DR3 were then used to determine the membership of the MCs. We constrained the errors of PMs (mas/yr) and parallax (mas) to be no larger than 0.5. The RUWE was set to be less than 2.0 as shown in Figure~\ref{fig:ruwe}, which was justified by the RUWE of WOH G64 ($\rm RUWE\approx1.9$; one of the most famous RSGs; \citealt{Ohnaka2008, Levesque2009, Yang2018}) in the LMC. The parallax was limited to be in the range between -0.12 and 0.1 mas as shown in Figure~\ref{fig:plx} \citep{Gaia2018, Yang2019, Gaia2021, Yang2021a}. The clustering of LMC and SMC targets on the PMs diagram was then easily observed in Figure~\ref{fig:pm}, for which we selected the members of the MCs by eye since their distributions were asymmetric. In total, there are 17,081 targets in the MCs with 13,603 ($\sim80\%$) in the LMC and 3,478 ($\sim20\%$) in the SMC (the boundary between the LMC and SMC is set to be at $\rm R.A.=40.0~deg$), respectively. Due to the relatively small number of targets in the SMC, we decided to continue the investigation based only on the LMC data. Furthermore, we focused our study on the cool evolved stars with $\rm J-K_{\rm S}\geq0.6$, that 11,682 targets were left in the sample. Finally, the ``STARFLAG'', which identify issues associated with spectral processing, radial velocity measurement, and spectral combination, was set to 0 in order to avoid any data reduction issues. This resulted in 7,938 targets in the LMC sample. Notice that, we did not constrain the ``ASPCAPFLAG'', since the models from the Apogee Stellar Parameter and Chemical Abundances Pipeline (ASPCAP; \citealt{Garcia2016}) are not well converged in this region of $T_{\rm eff}-\log{g}$ plane (e.g., see Figure 2 of \citealt{Jonsson2020}). For example, out of 7,938 targets in the sample, there are 4,917 targets brighter than the $\rm K_S$-band TRGB ($\rm K_S$-TRGB$~\approx12.0$ mag; see also Figure~\ref{fig:cmd_lmc}). However, among them, there are only 175 targets ($\sim$3.6\%) with $\rm ASPCAPFLAG=0$ (no problem at all) and 1,946 targets ($\sim$39.6\%) with $\rm ASPCAPFLAG=4$ (warning on the microturbulent velocity). Thus, one thing has to be kept in mind that, some of the derived stellar parameters and chemical abundances from ASPCAP are not quite reliable for the bright cool evolved stars (e.g., RSGs and AGBs).

Next, we followed previous studies to build the reference sample by separating the cool evolved stars into RSG, AGB, and RGB populations on the Two Micron All Sky Survey (2MASS; \citealt{Skrutskie2006}) CMD \citep{Cioni2006, Boyer2011, Yang2018, Yang2019, Neugent2020, Yang2020, Massey2021, Yang2021a, Yang2021b}, since the effect of extinction and variability are much less at near-infrared bands compared to the optical bands (e.g., the V-band). As the foreground extinction was about only $A_V\approx0.1$ mag at the line of sight of the LMC (corresponding to about $A_J = 0.025$ mag and $A_{K_S}=0.008$ mag, respectively; \citealt{Wang2019, Wang2023}), if $E(B-V) \approx 0.06$ mag with the Galactic average value of $R_V = 3.1$ was adopted \citep{Oestreicher1995, Dobashi2008, Gao2013}, we did not correct the foreground extinction. The major goal of our study is to understand the lower mass limit of the RSG population. In that sense, we calculated the bolometric luminosities of the sample based on different bolometric corrections (BCs), e.g., from \citet{Josselin2000}, \citet{Davies2013}, \citet{Neugent2020}, or \citet{Yang2023}. Generally, the resultant luminosities from these BCs agreed within $\pm0.1~dex$. Thus, the threshold of the ``genuine'' RSGs was set to be brighter than $K_{\rm S}=10.5$ mag ($L\gtrsim10^{4.0}~L_{\sun}$) by comparing different BCs. This ensured that in any case we selected the true RSGs without any major confusion (inevitably, we might still have small contamination from AGBs at the red end of the RSG sample). Table~\ref{tbl:constable} shows the criteria for the selection of different stellar populations, while Figure~\ref{fig:cmd_lmc} shows the resulted \textit{Gaia} and 2MASS CMDs. Unfortunately, as can be seen from the diagrams, due to the target selection strategy of APOGEE-2, there is an obvious gap at the faint end of the RSG branch \citep{Santana2021}, which will be discussed later. On the other hand, still, hundreds of targets remain in the region that can be used for our study (black points in Figure~\ref{fig:cmd_lmc}). 

\begin{figure*}
\center
\includegraphics[scale=0.27]{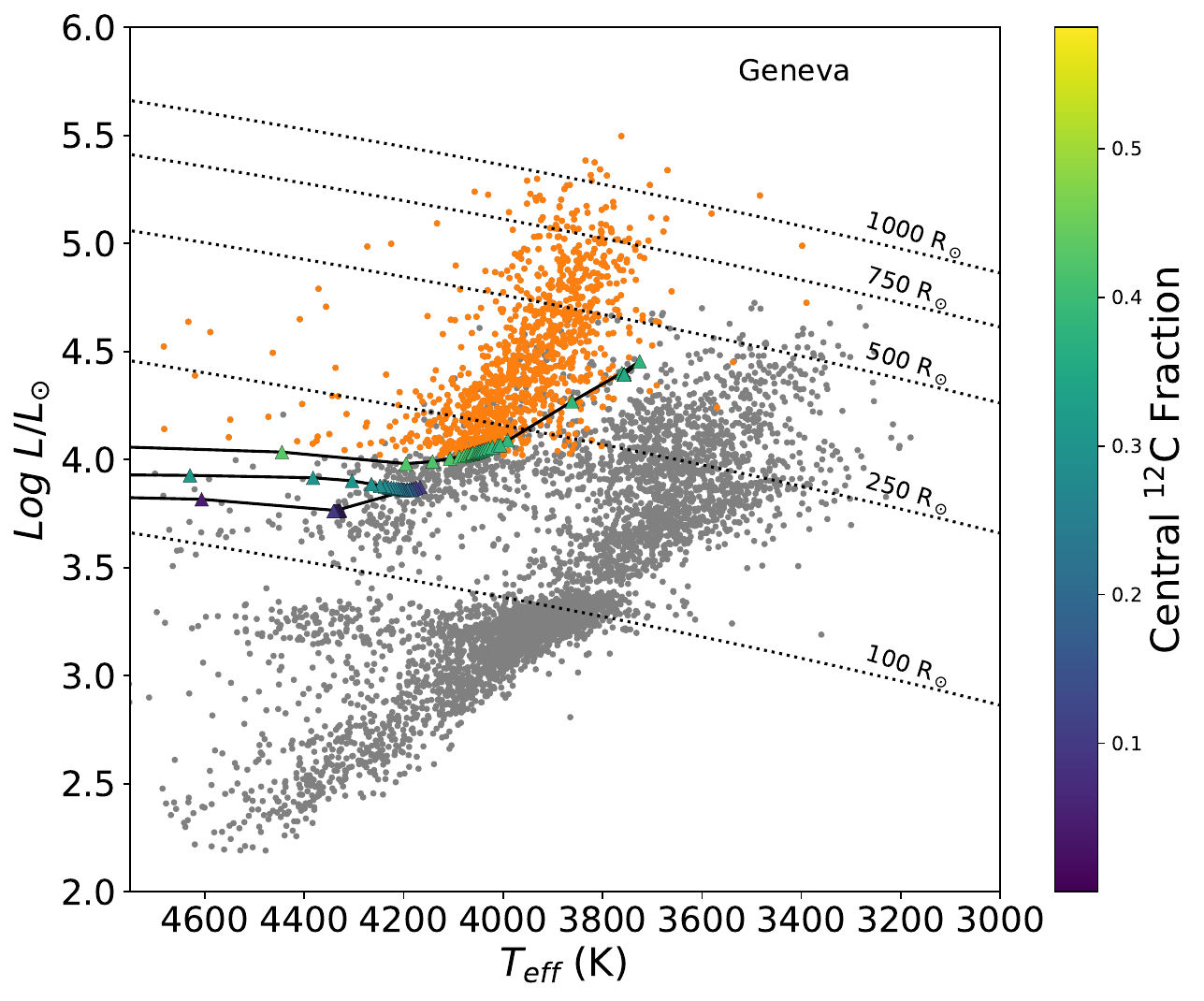}
\includegraphics[scale=0.27]{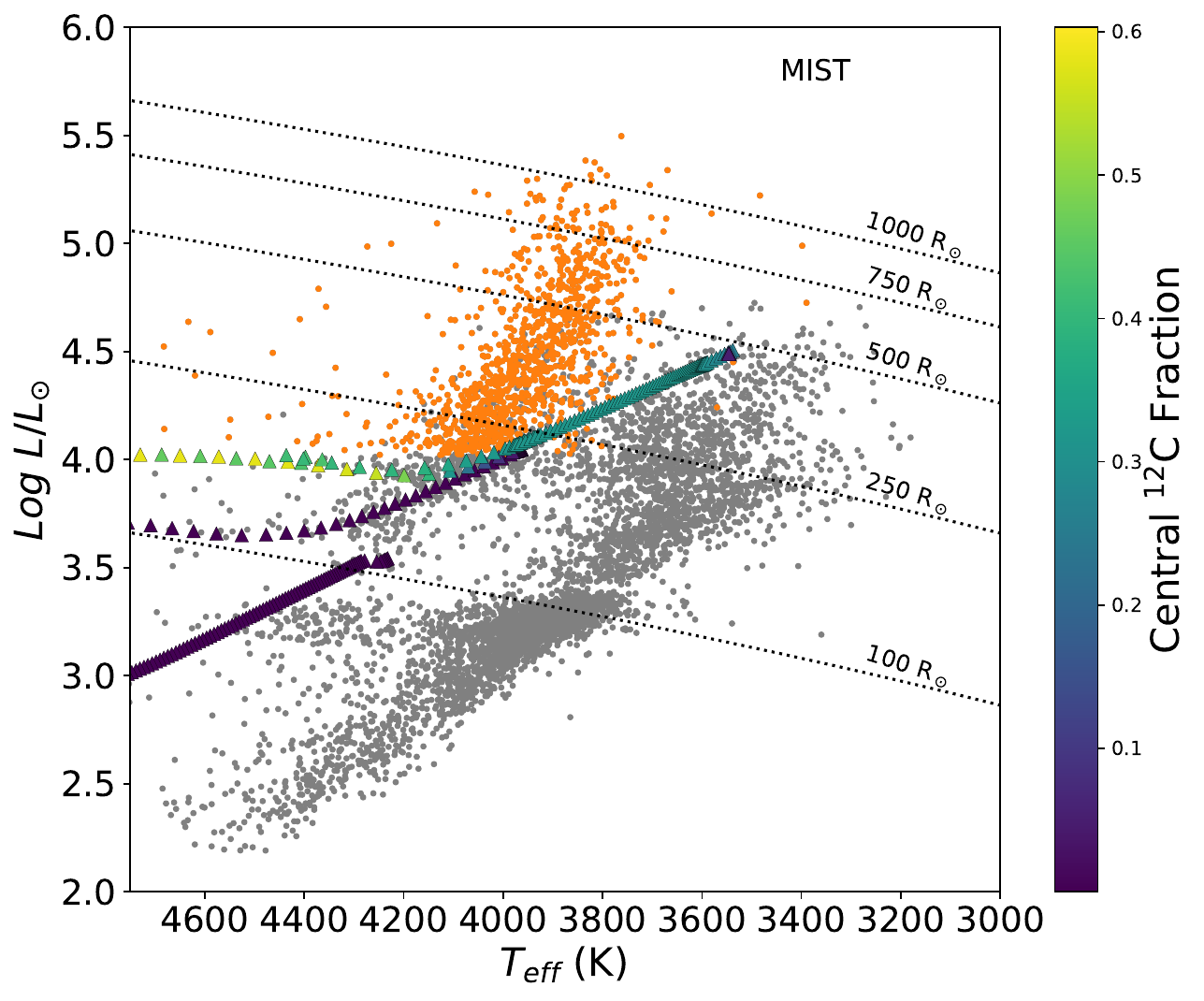}
\includegraphics[scale=0.27]{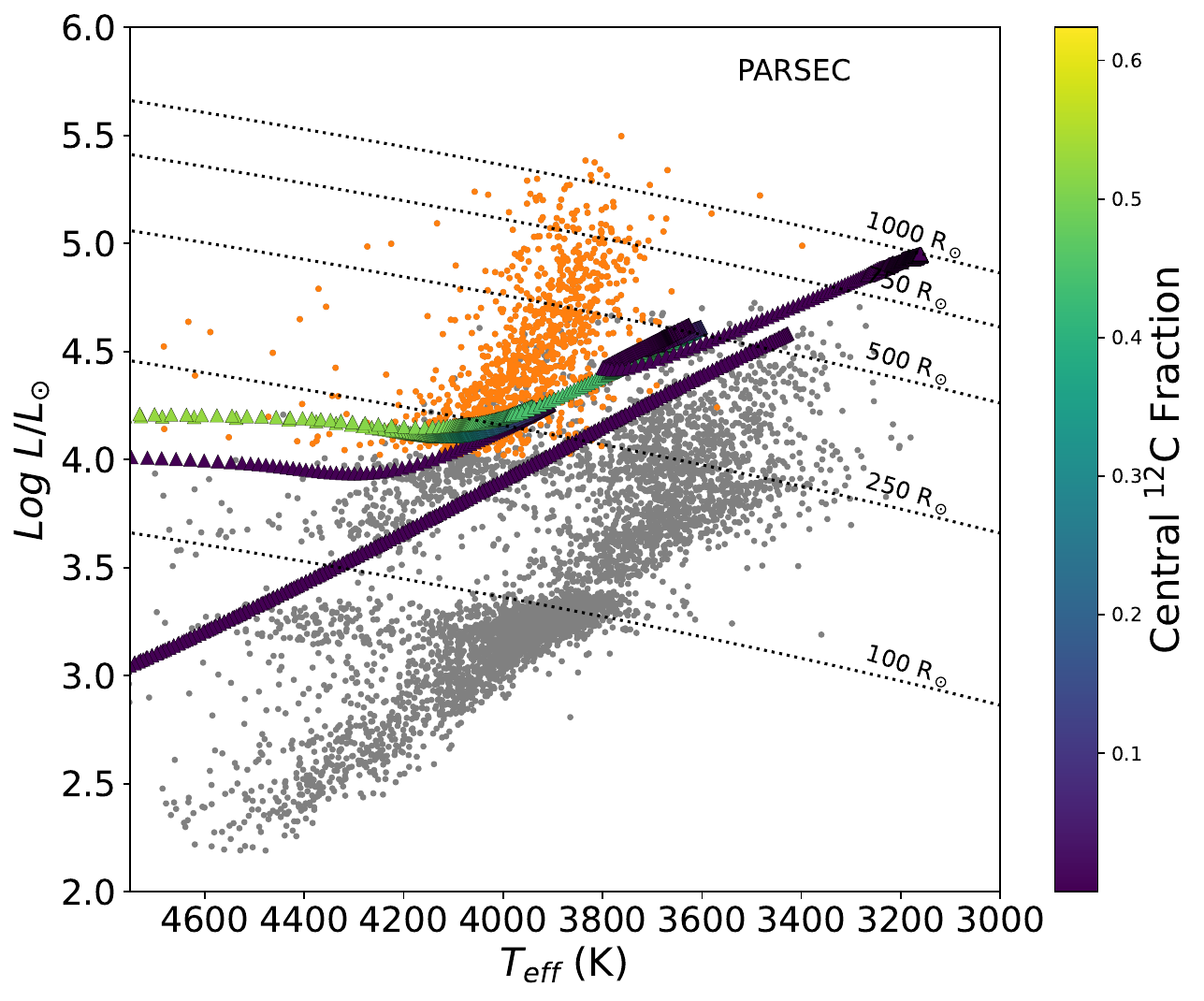}
\caption{Hertzsprung-Russell diagrams for the LMC sample. The evolutionary tracks of 8 $M_\sun$ from Geneva (left), MIST (middle), and PARSEC (right) are shown in the diagrams, where the selected RSG population is in orange color. The color map of evolutionary tracks represents the central $^{12}$C mass fraction.
\label{fig:hrd}}
\end{figure*}

Figure~\ref{fig:hrd} shows the H-R diagrams, where the luminosities were derived using $BC_K=2.69$ from \citet{Davies2013} (the difference between $K$ and $K_S$ bands can be ignored; the luminosities of AGBs and RGBs are only approximate values by using the same BC), and the $T_{\rm eff}$ were adopted from APOGEE\footnote{As mentioned before, warning on the microturbulent velocity might indicate unreliable stellar parameters from ASPCAP. However, we compared the $T_{\rm eff}$ derived from both APOGEE and other methods, e.g., $J-K_S$ color, and found that they were consistent with each other within the uncertainty.}. The evolutionary tracks of 8 $M_\sun$ from Geneva stellar evolution models \citep{Ekstrom2012, Georgy2013, Groh2019, Eggenberger2021, Murphy2021, Yusof2022}, MESA Isochrones \& Stellar Tracks (MIST; \citealt{Paxton2011, Paxton2013, Paxton2015, Choi2016, Dotter2016, Paxton2018, Paxton2019}), and PAdova and TRieste Stellar Evolution Code (PARSEC; \citealt{Bressan2012, Tang2014, Chen2015, Fu2018, Nguyen2022}) were plotted in the diagrams, indicating that our threshold of RSG population was appropriate for the traditional lower mass limit of massive stars. Notice that, the available metallicity was 0.008 for both MIST and PARSEC, while it was 0.006 for Geneva. The rotation rate of $V/V_{crit}$ was 0.4 for both Geneva and MIST, while $\Omega/\Omega_{crit}$ were 0.568, 0.6, and 0.4 for Geneva, PARSEC, and MIST, respectively ($V_{crit}$ and $\Omega_{crit}$ are the critical surface linear and angular velocities, respectively). The 8 $M_\sun$ track of Geneva models was an interpolated model, which might be differ from a real computed one, especially in phases of instability.

\begin{deluxetable*}{ccc}
\tablecaption{Constraints for the Cool Evolved Stars on the 2MASS Color-Magnitude Diagram \label{tbl:constable}}
\tablewidth{0pt}
\tabletypesize{\scriptsize}
\tablehead{
\colhead{Type} & \colhead{Constraints} & \colhead{Numbers}
} 
\startdata
AGBs         & $((J - K_S > -0.09\times(K_S - 12) + 0.95)$ and $K_S\leq12$ and $K_S - 18.493 > -10)$     & 3,410 \\
             & or $(J - K_S > 1.5$ and $K_S - 18.493 > -10)$                                             &      \\
RSGs         & $((J - K_S < -0.09\times(K_S - 12) + 0.95)$ and $K_S\leq12$) or ($K_S - 18.493 \leq -10$) & 1,098 \\
             & and $(K_S\leq10.5)$                                                                       &      \\
RGBs         & $(J - K_S > -0.09\times(K_S - 12) + 0.95)$ and $(K_S > 12)$ and $(J - K_S \leq 1.5)$      & 2,766 \\
Unclassified &                                                                                           & 664  \\
Total        &                                                                                           & 7,938 \\
\enddata
\tablecomments{$J-K_S\geq0.6$ and $STARFLAG=0$ are applied for all types of cool evolved stars in the LMC.}
\end{deluxetable*}

\section{Reference Spectra of Cool Evolved Stars from APOGEE-2/SDSS-IV} \label{sec:spectra}

The pseudo-continuum normalized and best fit spectra from APOGEE-2 were then used for our analysis, for which the spectra were velocity-corrected (to vacuum wavelengths) and resampled on to a common logarithmically-spaced wavelength scale by combining individual visits spectra. These spectra also contain a synthetic spectrum computed for the best fit parameters from the ASPCAP. For our study, we also converted vacuum wavelengths into air wavelengths at standard temperature and pressure (S.T.P.) following the instruction on the SDSS website\footnote{https://www.sdss4.org/dr17/irspec/spectra/}. 

To better facilitate the study, we also applied a post-processing for the spectra, e.g., removing the bad data points with values less than 0.001, or data points with errors larger than 1.0, or data points with error larger than the 99\% quantile when errors less than 1.0. As there are small gaps between three different detectors (``chips'') of APOGEE-2, which results in gaps in the wavelength coverage, the pseudo-continuum normalization is imperfect in different chips. For each target, we then simply re-adjusted the pseudo-continuum level by subtracting the difference between 1.0 and 90\% quantile of the post-processed spectrum in each chip.

\begin{figure*}
\center
\includegraphics[scale=0.18]{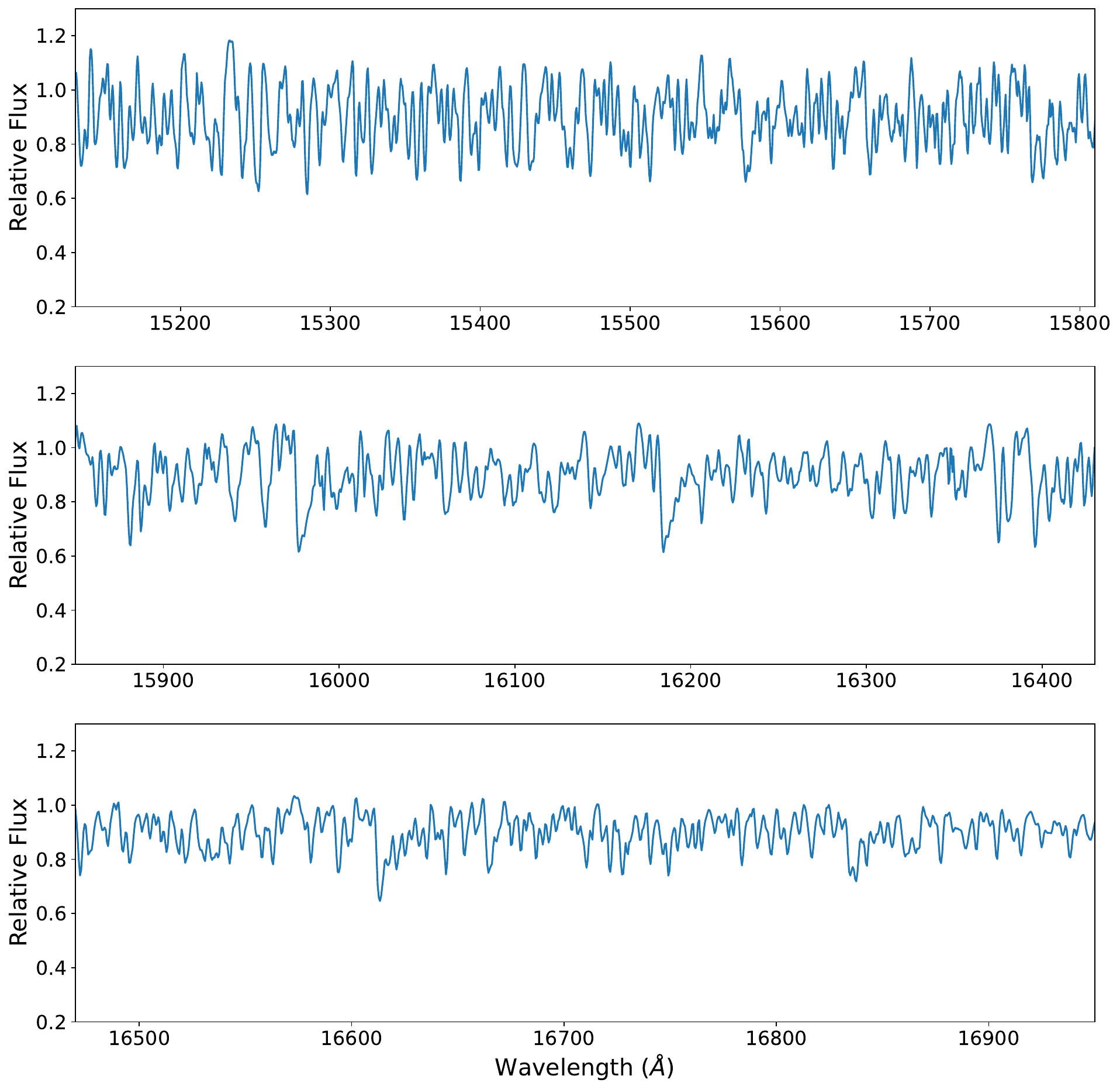}
\includegraphics[scale=0.18]{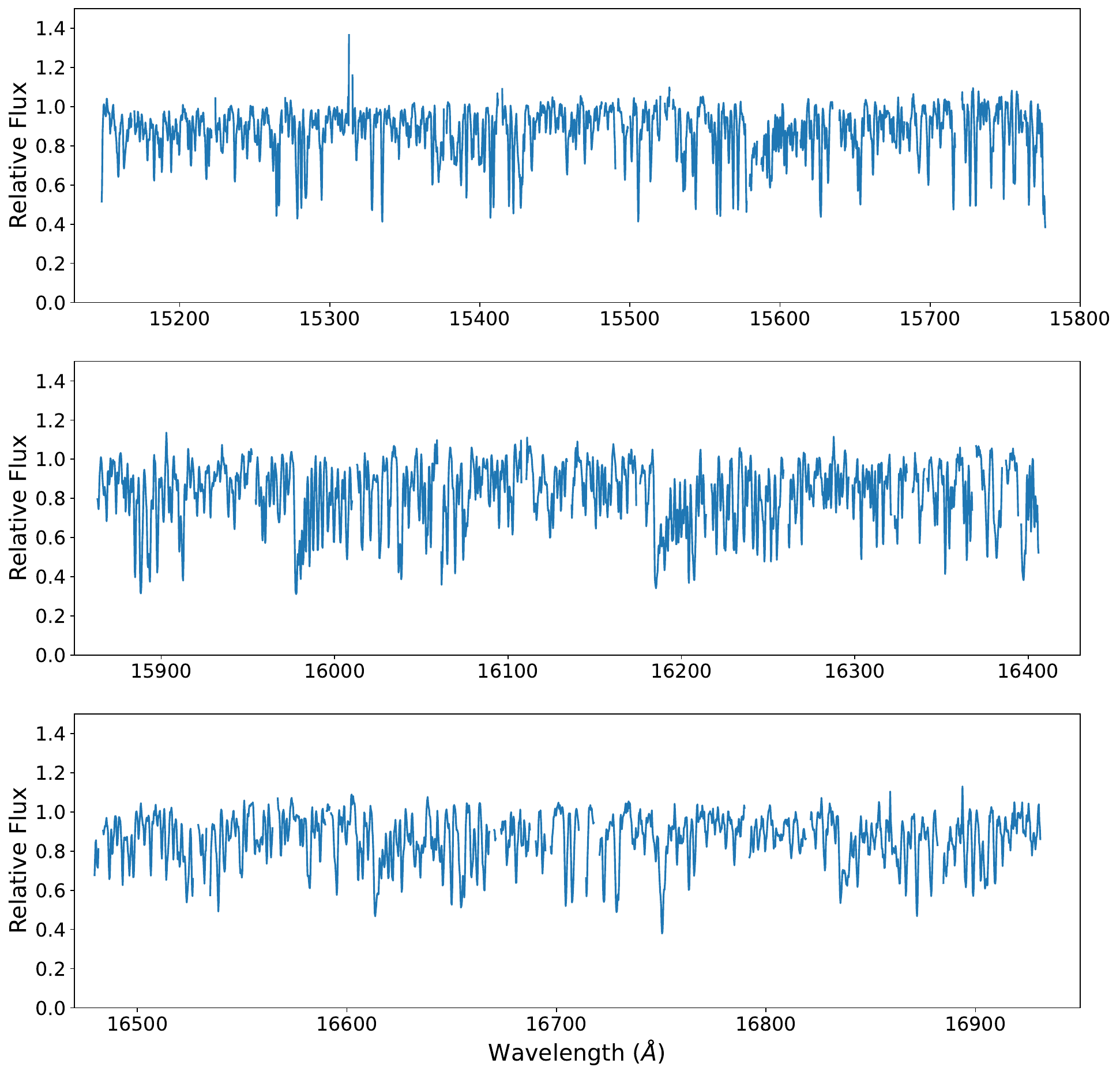}
\includegraphics[scale=0.18]{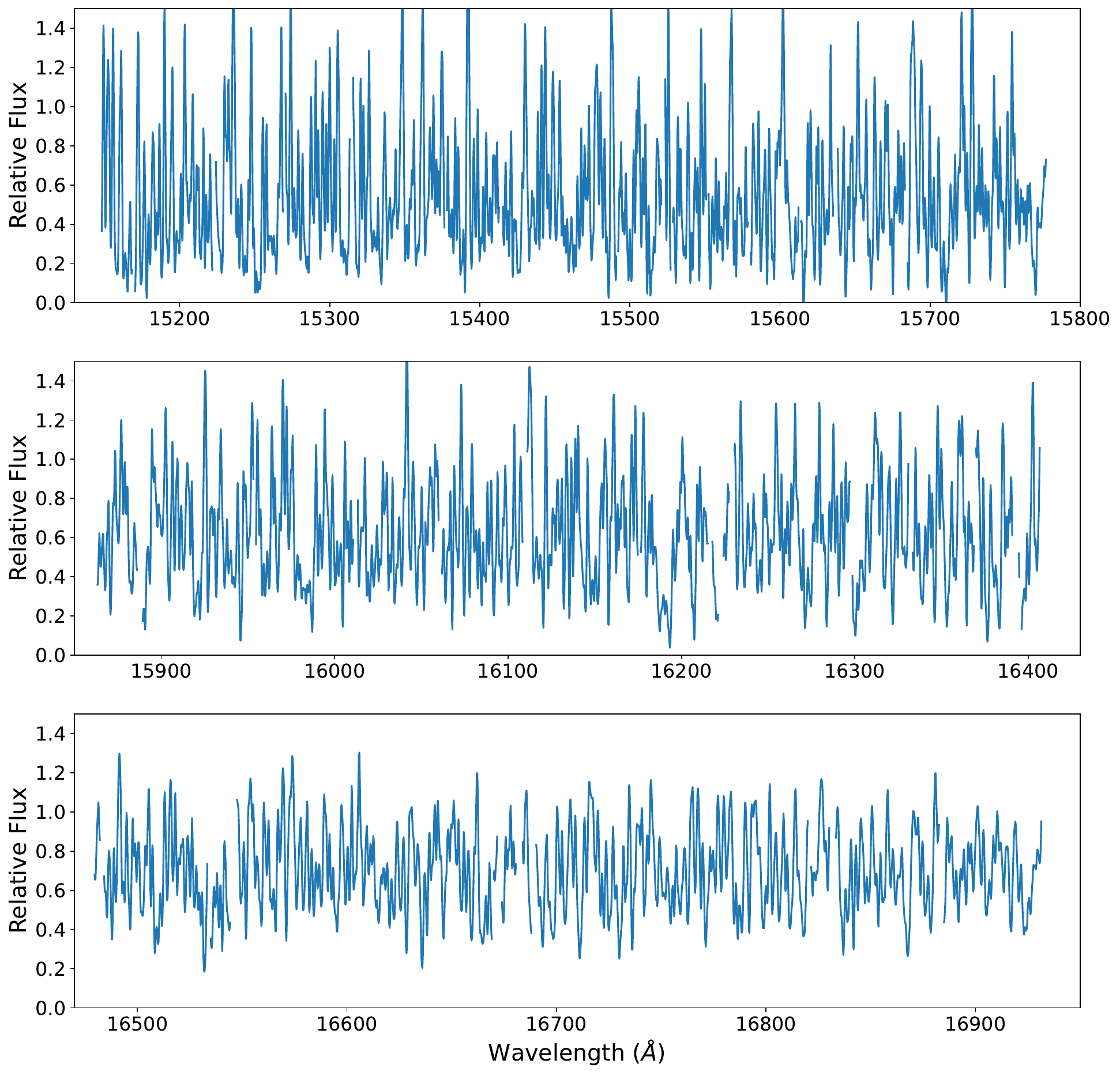}
\caption{Examples of normal spectra of C-AGBs from both XSL (SHV 0518161-683543; left) and APOGEE-2 (2M05413193-7110110; middle), and abnormal spectrum of C-AGBs from APOGEE-2 (2M06260466-6931120; right), respectively.
\label{fig:spectra_examples_poor_cagb}}
\end{figure*}

\begin{figure}
\center
\includegraphics[scale=0.45]{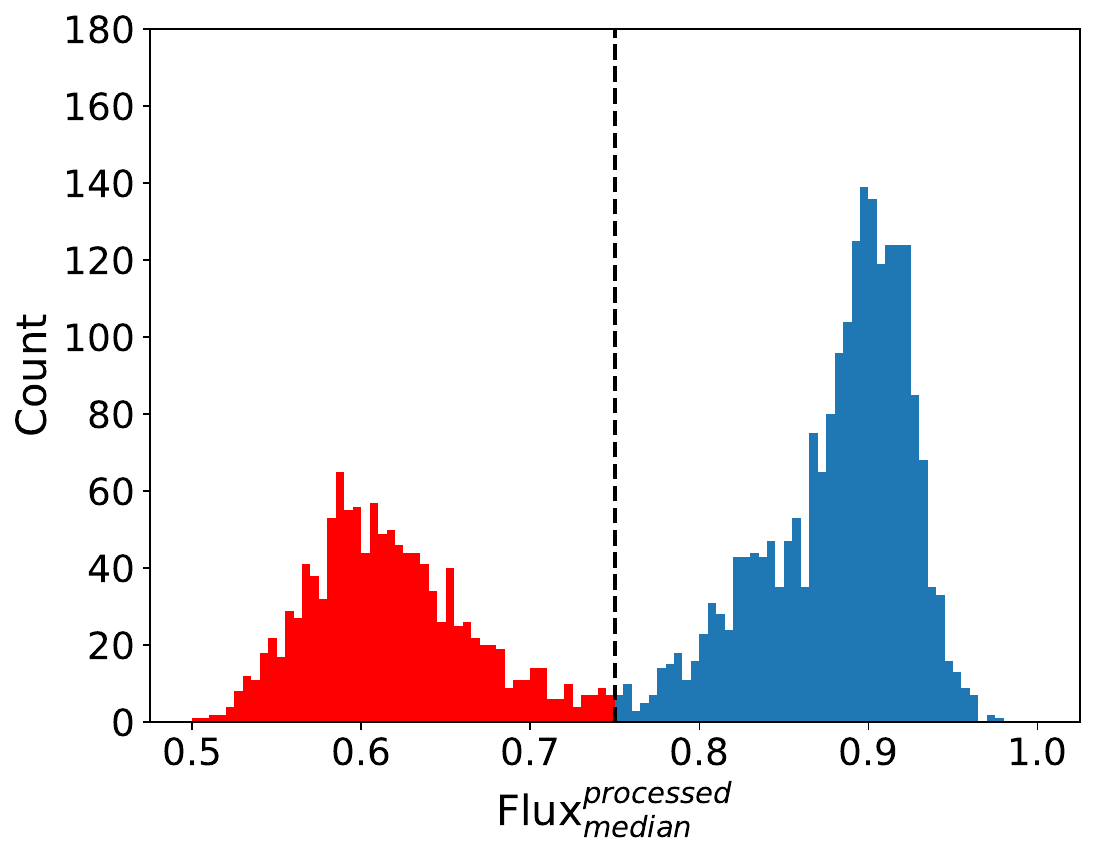}
\caption{Histogram of the post-processing median flux of AGBs. Red color indicates the targets with low spectral qualities (median flux $<0.75$).
\label{fig:cool_stars_flux}}
\end{figure}

\begin{figure}
\center
\includegraphics[scale=0.41]{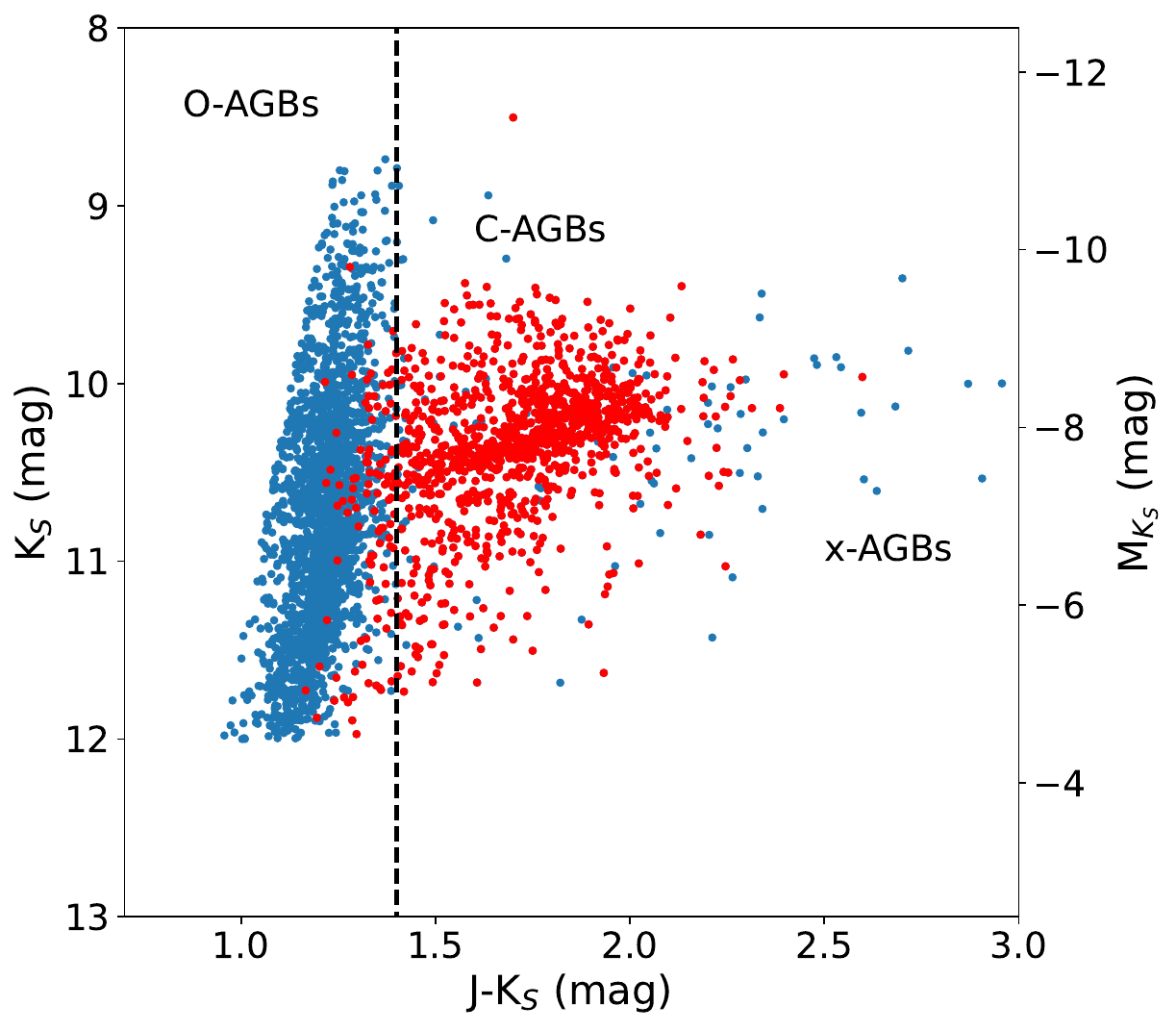}
\caption{Color-magnitude diagram of AGBs in the LMC. Red color represents the targets with abnormal spectra. The vertical dashed line indicates $\rm J-K_{\rm S}>1.4$.
\label{fig:cmd_poor}}
\end{figure}

\begin{figure*}
\center
\includegraphics[scale=0.43]{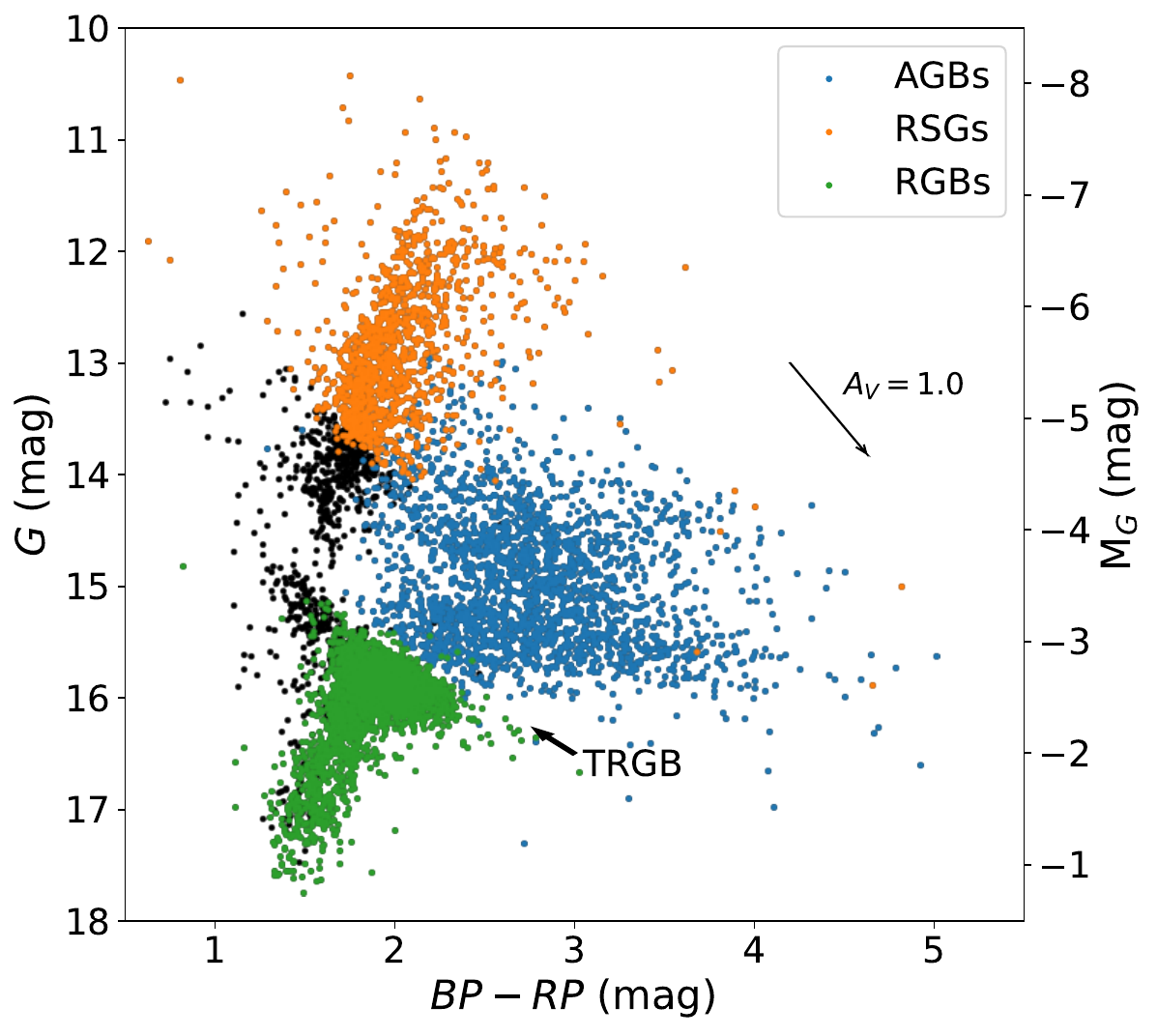}
\includegraphics[scale=0.43]{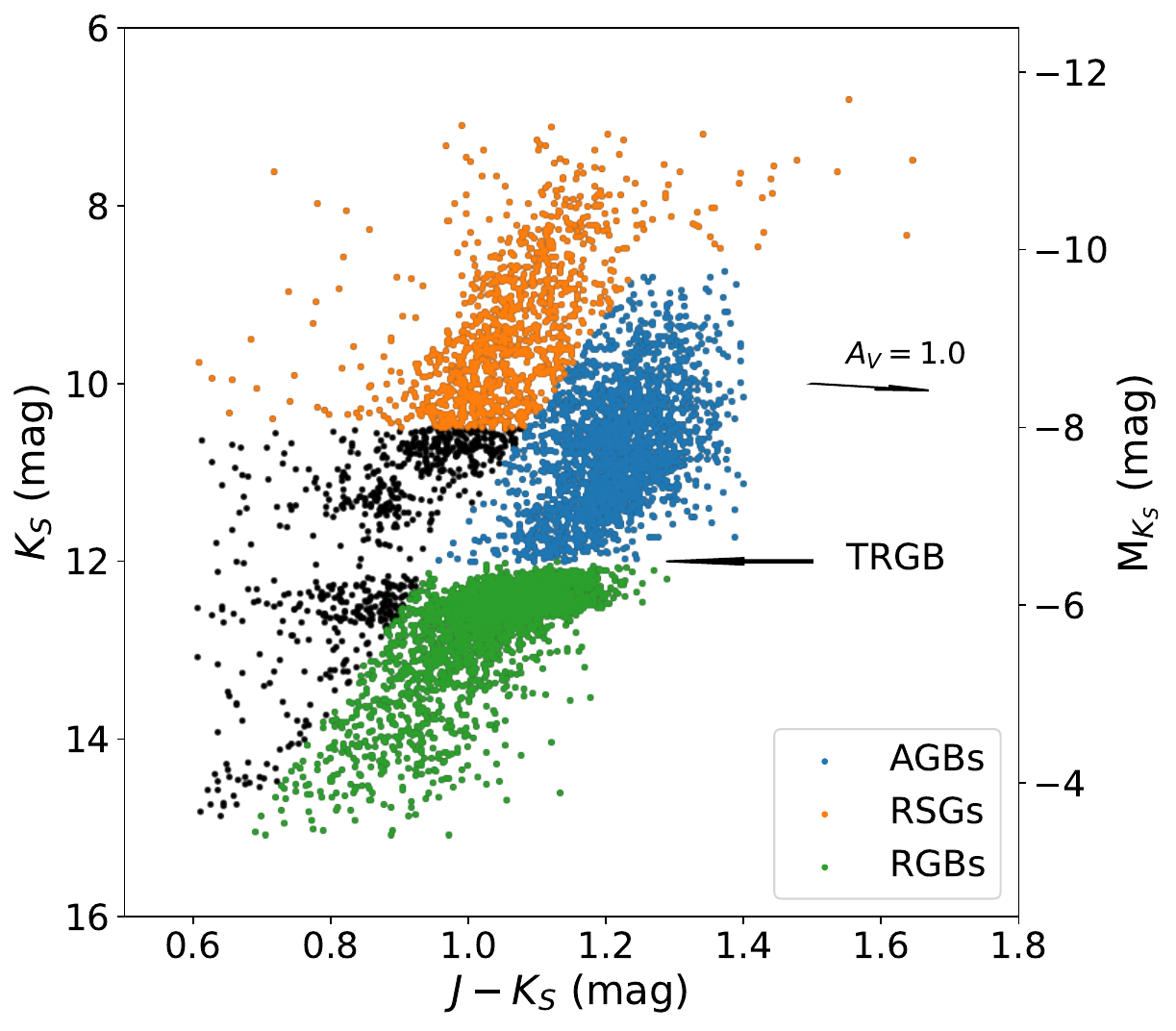}
\caption{Color-magnitude diagram of the final sample of 6,602 targets in the LMC. The color convention is the same as Figure~\ref{fig:cmd_lmc}.
\label{fig:cmd_lmc_con}}
\end{figure*}

However, during the following analysis, we found quite some unexpected results, which led to recheck the initial datasets. As a result, we identified a bunch of stars with (presumably) abnormal spectra\footnote{This problem has been reported to SDSS, but we believe that the exact reason is still not very clear. A response from SDSS indicated that such jagged spectra might be due to many molecular lines involving carbon in C-AGBs.}, for which an example is shown in the right panel of Figure~\ref{fig:spectra_examples_poor_cagb}. For comparison, it also shows the examples of normal spectra of C-AGBs from both X-Shooter Spectral Library (XSL; \citealt{Chen2014, Gonneau2016, Gonneau2020, Verro2022}) and APOGEE-2, where the wedge-shaped carbon monoxide (CO) bandheads are clearly visible. Typically, such abnormal spectra are very noisy that even strong spectral features (e.g., several CO bandheads) are barely visible. Fortunately, these spectra could be empirically separated from the normal spectra by calculating the median flux ($<0.75$) of the post-processed spectra (or similarly, $<0.95$ for the original spectra) as shown in Figure~\ref{fig:cool_stars_flux}. The vast majority of these spectra were carbon-rich AGBs (C-AGBs) as shown in Figure~\ref{fig:cmd_poor}. Moreover, to be on the safe side, we also visually inspected targets with median flux larger than 0.75 and $\rm J-K_{\rm S}>1.4$ (99 targets), for which many of them were extreme-AGBs (x-AGBs). It turned out that only a few of those targets showed normal spectra (less than 1/3). Therefore, we further constrained our sample with median flux larger than 0.75, $\rm J-K_{\rm S}<1.4$, or $M_{K_{\rm S}} > -10$ mag, that is to say, mostly RSGs, oxygen-rich AGBs (O-AGBs), and RGBs, as shown in the Figure~\ref{fig:cmd_lmc_con}. This again posed additional constraint on our sample size, which resulted in 6,602 targets in the final sample, including 1,096 RSGs, 2,083 AGBs, 2,762 RGBs, and 661 unclassified targets.

\begin{figure*}
\center
\includegraphics[scale=0.56]{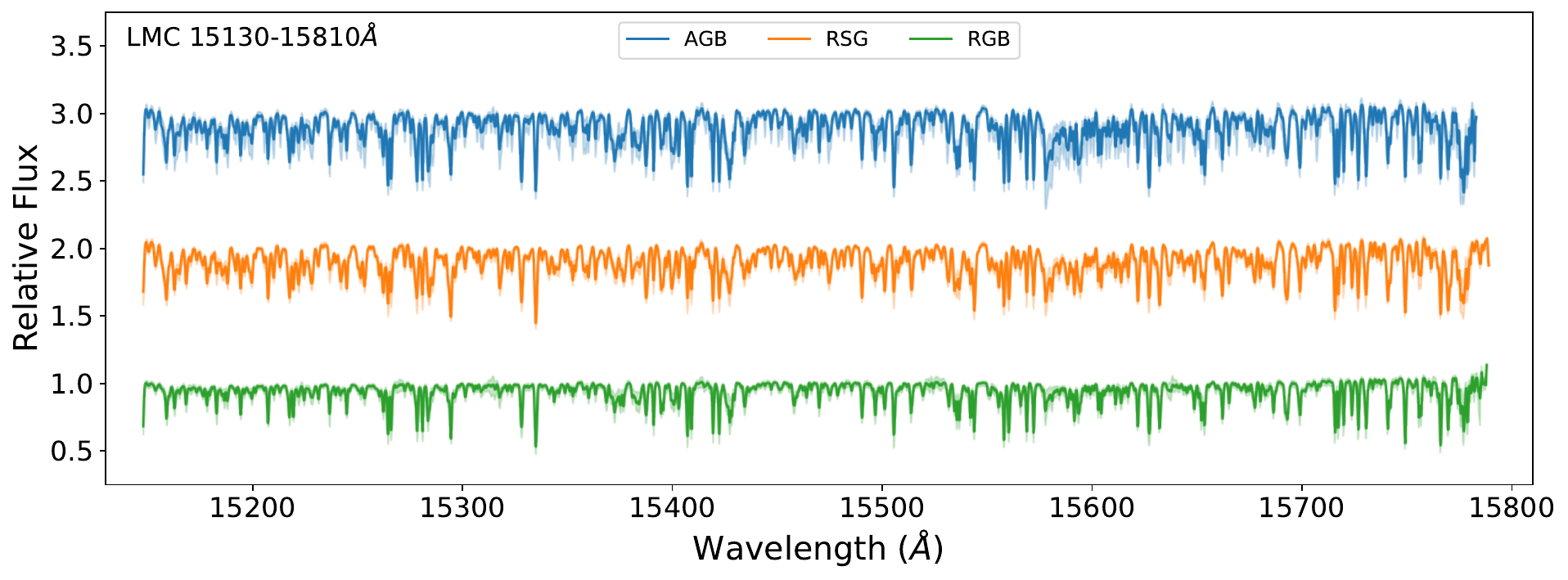}
\includegraphics[scale=0.56]{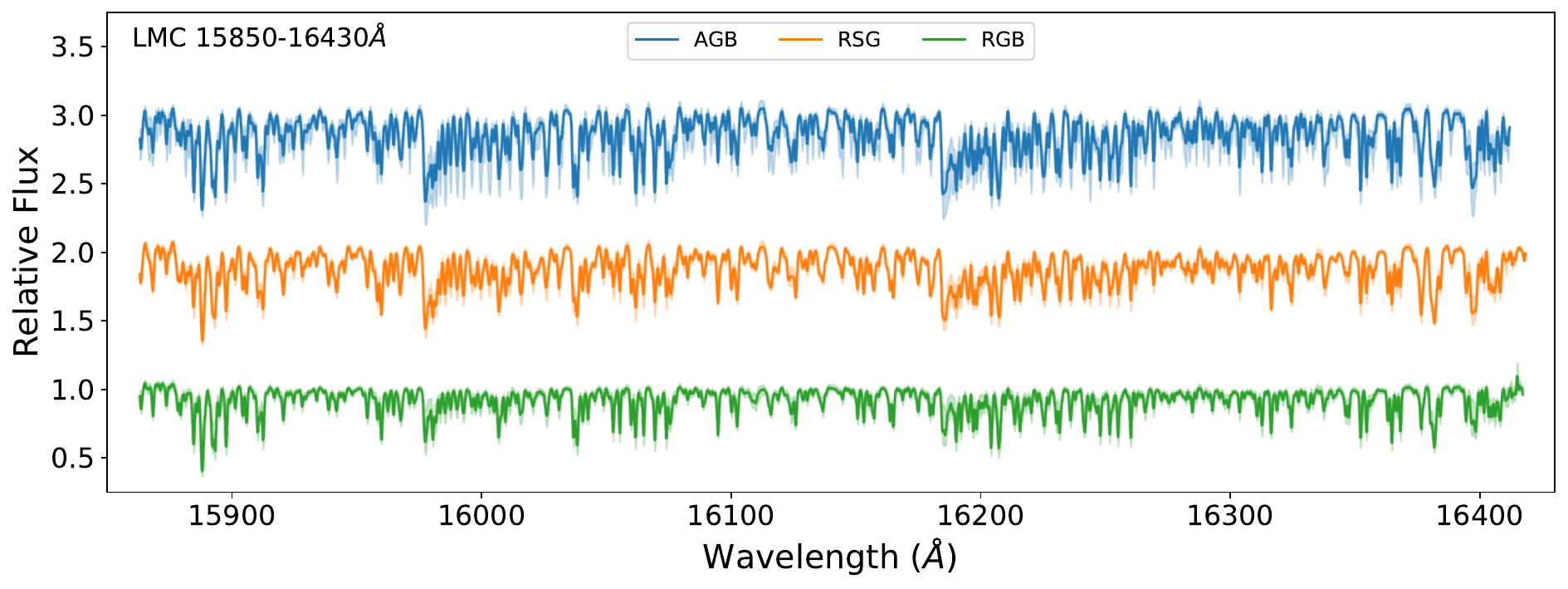}
\includegraphics[scale=0.56]{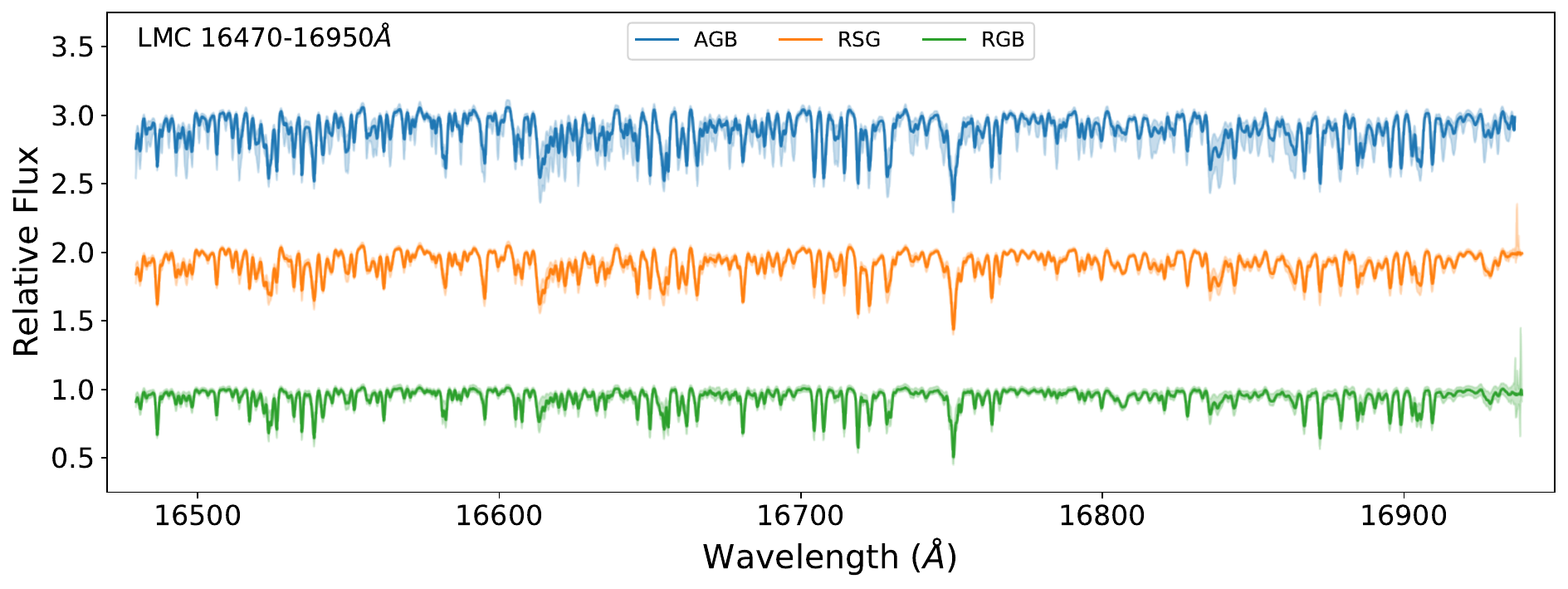}
\caption{The median spectra of AGB, RSG, and RGB populations. The shades indicate the 10\% and 90\% quantiles, respectively.
\label{fig:spectra_examples_lmc}}
\end{figure*}

Figure~\ref{fig:spectra_examples_lmc} shows the derived median (50\% of the relative intensity) reference spectra of the three different stellar populations, which is largely different from the poor spectra shown in Figure~\ref{fig:spectra_examples_poor_cagb}. Meanwhile, we also calculated 1\%, 10\%, 30\%, 70\%, 90\%, and 99\% quantile (of the relative intensity) reference spectra for each of the stellar population. As can be seen from the diagrams, the spectra of cool star populations are sharing very similar spectral characteristics, with the majority of the strong spectral features come from iron (Fe), hydroxyl radical (OH), cyanogen (CN), CO, etc. From the diagrams, the most prominent feature would be the weakening of relative intensity (as the pseudo-continuum normalized spectra are used in the analysis) of CO bandheads, e.g., $\rm ^{13}CO~3-0$, $\rm ^{12}CO~4-1$, $\rm ^{12}CO~5-2$, $\rm CO~6-3$, $\rm ^{12}CO~7-4$, $\rm ^{12}CO~8-5$, and $\rm ^{12}CO~9-6$, from AGBs to RSGs then to RGBs. As a consequence, many CO molecular bands also show such trend. In the meantime, the tendencies of other molecular bands are different. For example, generally, the intensity of CN bands are comparable between AGBs and RSGs but weaker in RGBs, while the intensity of OH bands are comparable between RSGs and RGBs but stronger in AGBs. Notice that, the occurrence of carbon species, like CN, in an O-rich environment is possible due to the dissociation of CO by the strong UV emission from the chromosphere \citep{Beck1992, Hofner2007}. Other metal lines, like Al, Mg, Si, etc., show roughly the same intensity among the three populations. The full diagnostic reference spectra are shown in the Section \ref{sec:ref_spec_appendix} in the Appendix, for which the atlas of spectral lines of APOGEE\footnote{https://github.com/sdss/apogee/blob/master/data/lines/\\atlas\_line\_ids\_apogee.txt} is adopted. Notice that, due to the complicated line structures in the H-band high resolution spectrum, not all lines are identified by the APOGEE spectral line atlas.

\section{Classification of Different Stellar Populations} \label{sec:classification}

\subsection{Spectral lines} \label{subsec:splines}

\begin{figure}
\center
\includegraphics[scale=0.45]{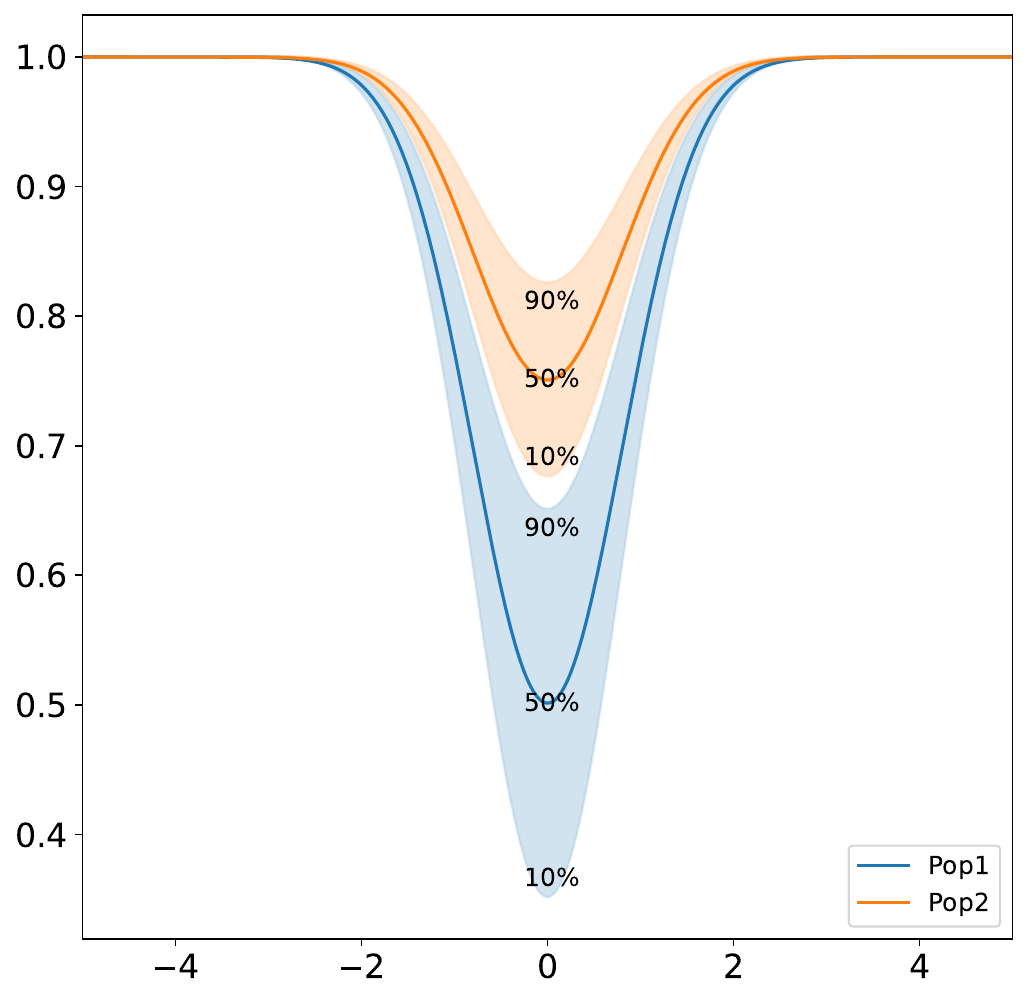}
\caption{A schematic figure shows a roughly 80\%$\sim$90\% isolation rate for the same spectral line from two different stellar populations.
\label{fig:line_sep_example}}
\end{figure}

\begin{figure*}
\center
\includegraphics[scale=0.56]{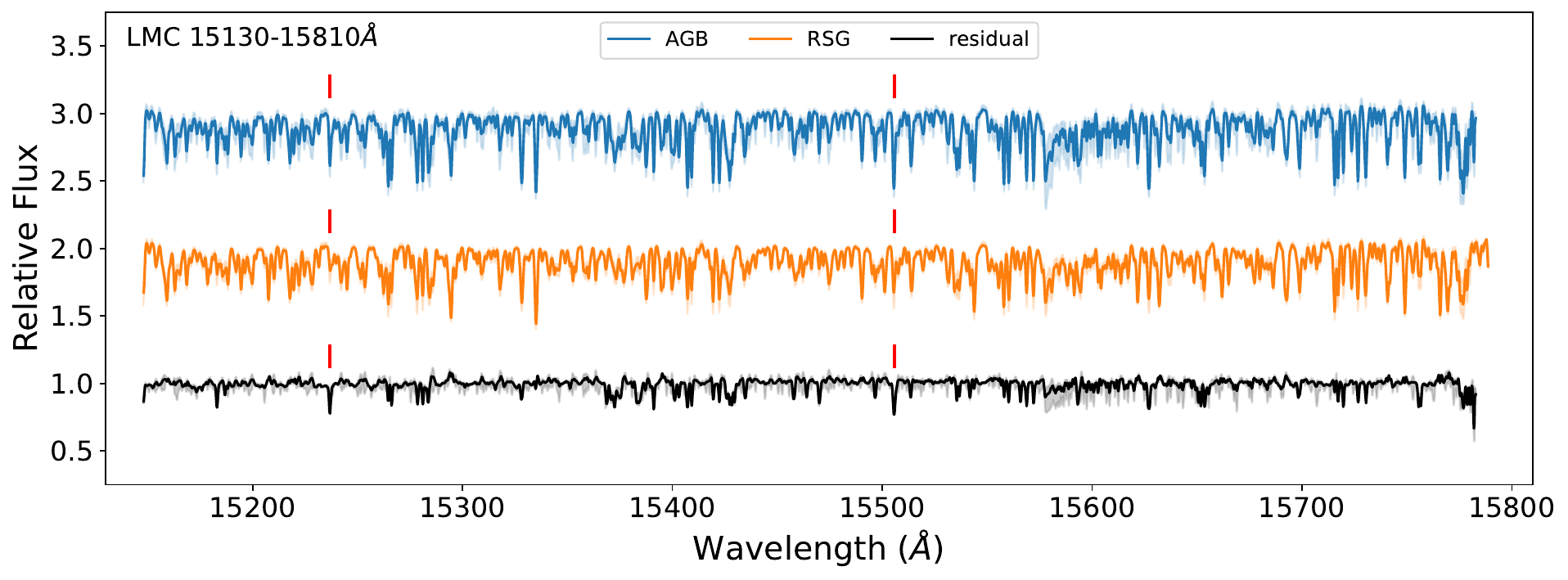}
\includegraphics[scale=0.56]{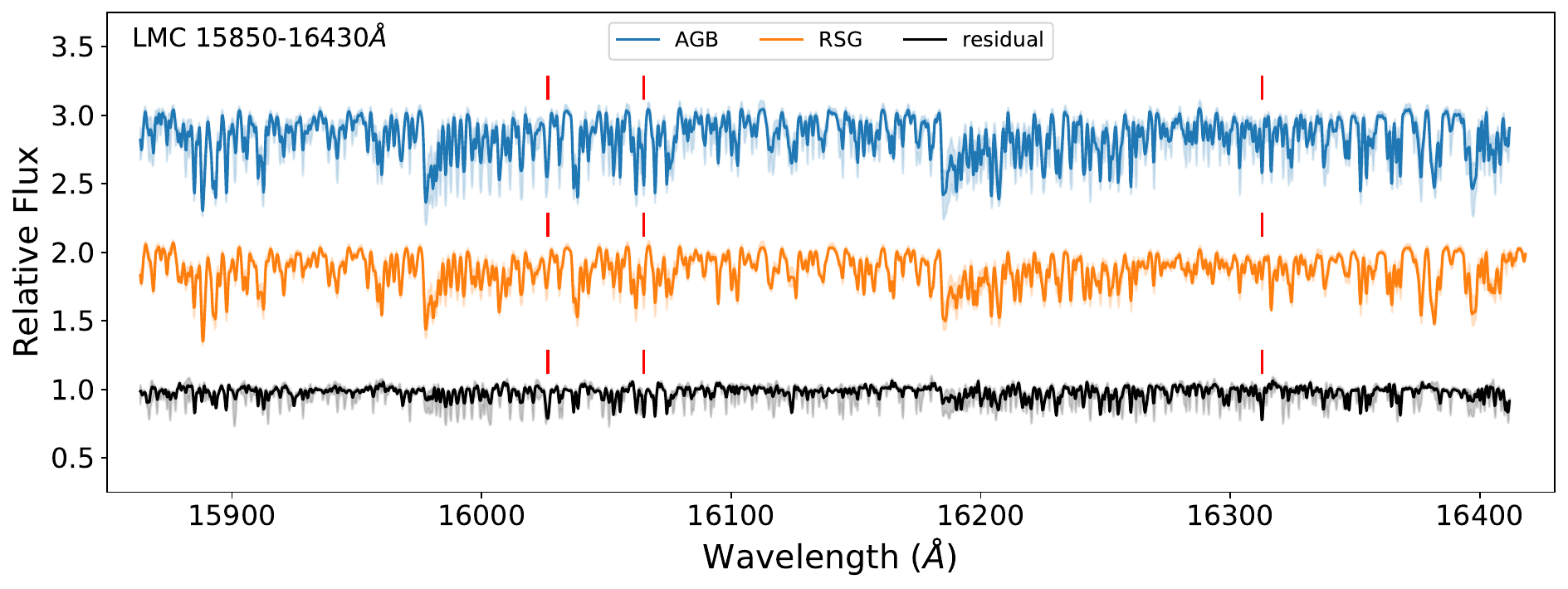}
\includegraphics[scale=0.56]{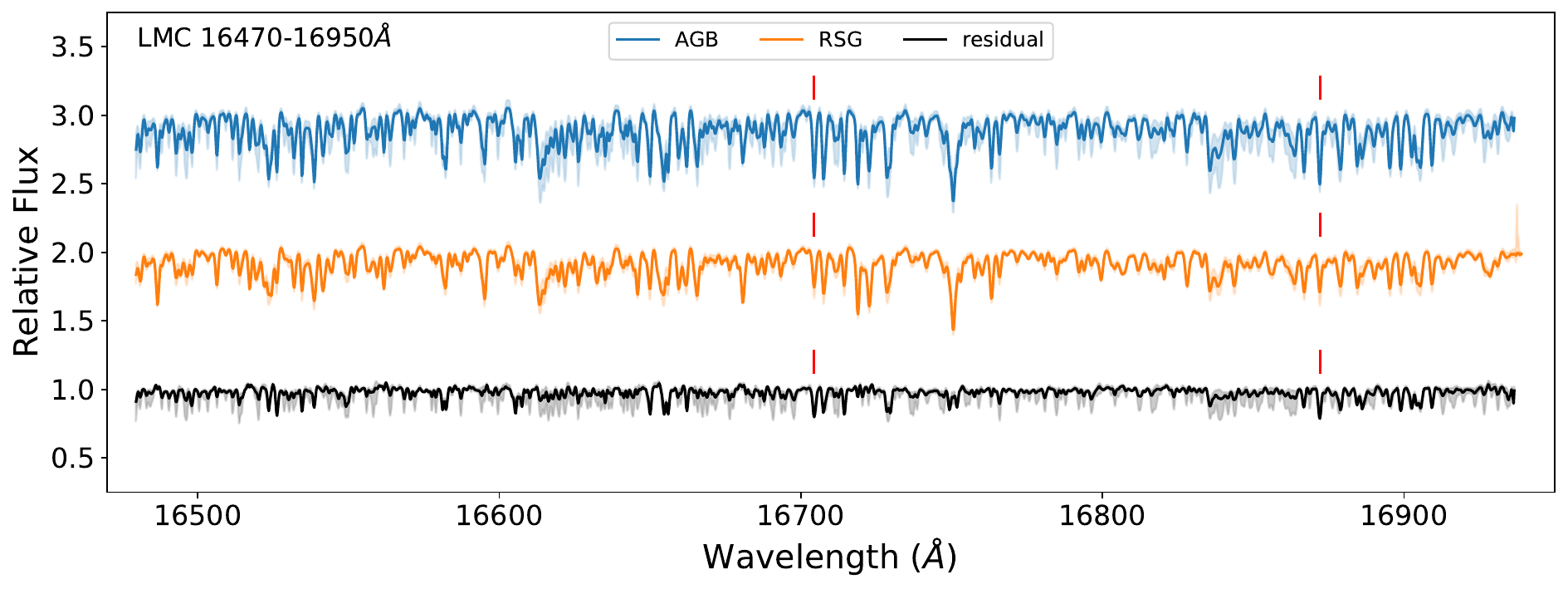}
\caption{The median reference spectra of AGBs and RSGs, and the residual spectrum. The red sticks mark the wavelengths with residual larger than 0.2. 
\label{fig:med_sp_line_agb_rsg}}
\end{figure*}

\begin{figure*}
\center
\includegraphics[scale=0.56]{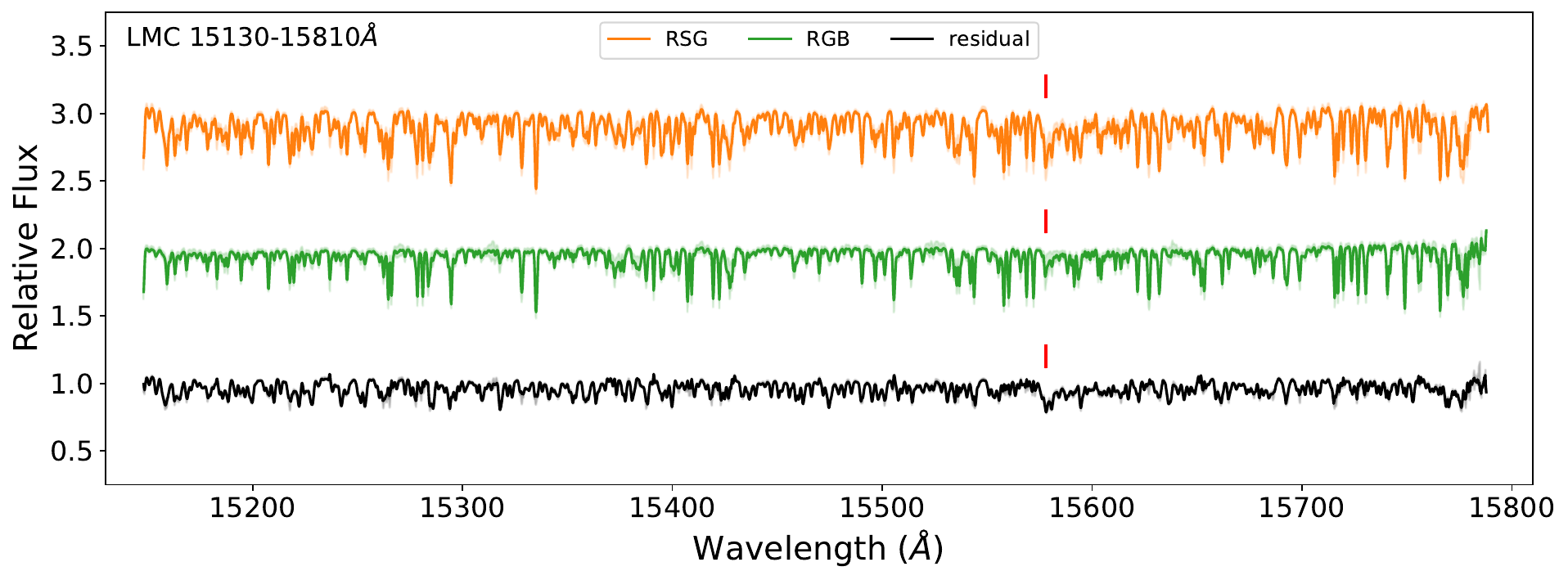}
\includegraphics[scale=0.56]{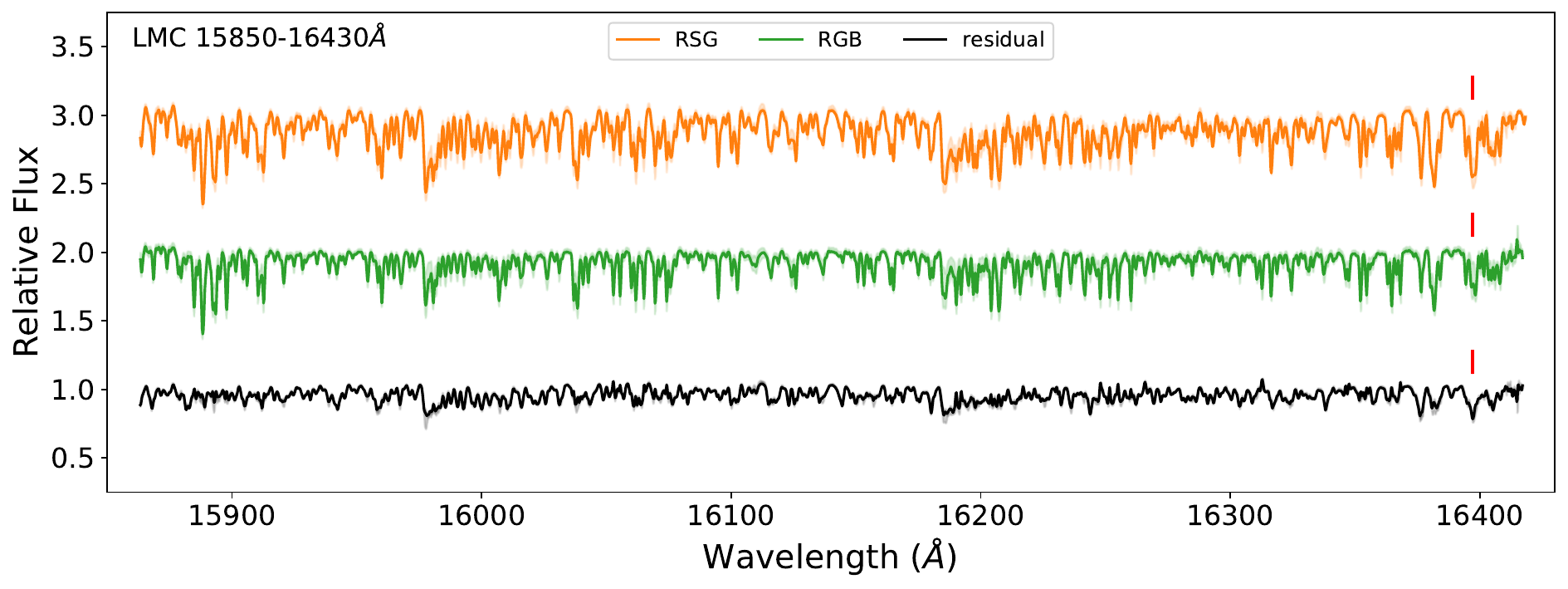}
\caption{Same as Figure~\ref{fig:med_sp_line_agb_rsg}, but for RSGs and RGBs.
\label{fig:med_sp_line_rsg_rgb}}
\end{figure*}

\begin{figure*}
\center
\includegraphics[scale=0.56]{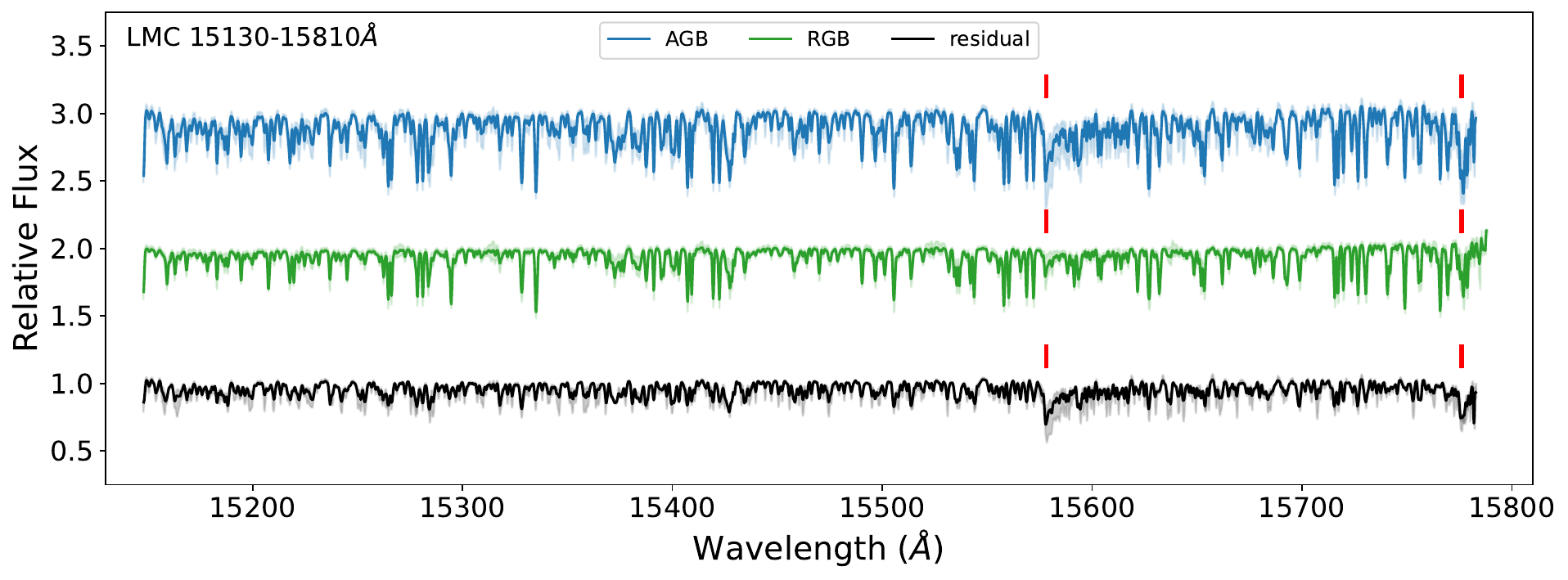}
\includegraphics[scale=0.56]{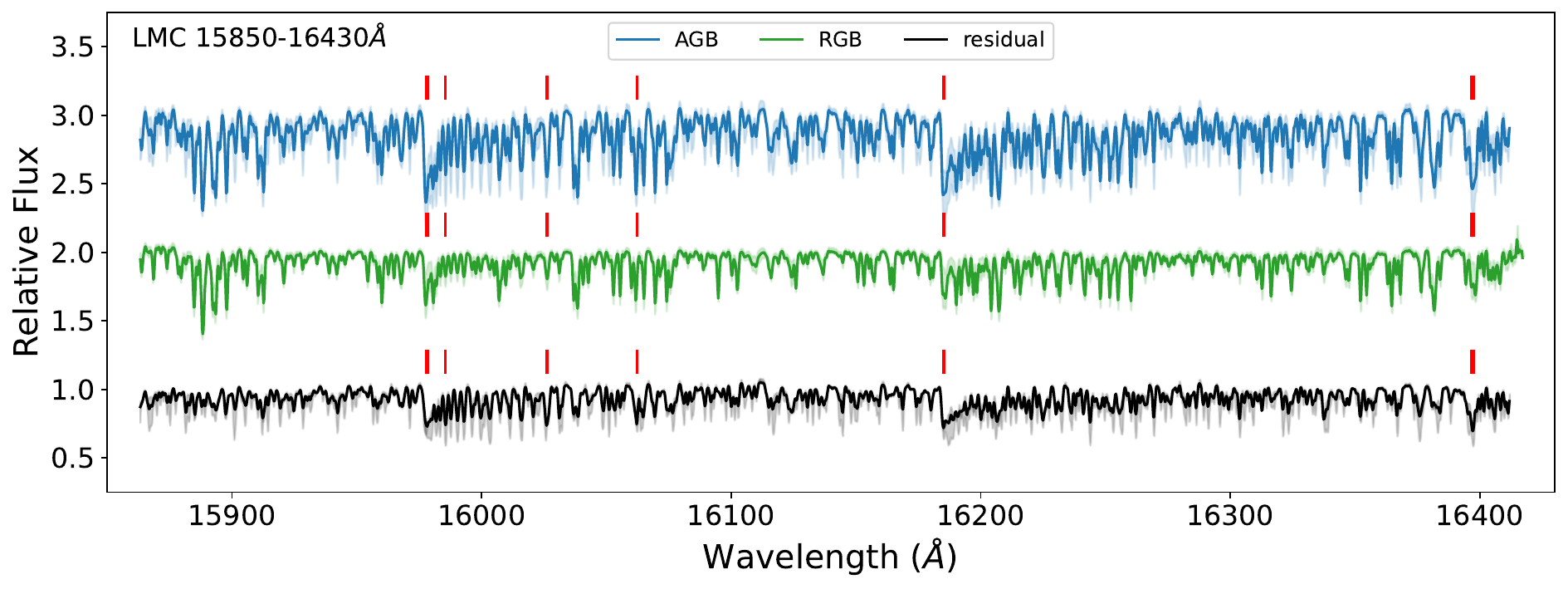}
\caption{Same as Figure~\ref{fig:med_sp_line_agb_rsg}, but for AGBs and RGBs. The red sticks mark the wavelengths with residual larger than 0.25.
\label{fig:med_sp_line_agb_rgb}}
\end{figure*}

Our initial goal was trying to find decisive spectral lines to separate the AGB, RSG, and RGB populations, in order to directly constrain the lower mass limit of RSG population. We believed that at least 80\%$\sim$90\% accuracy would be an acceptable criterion, e.g., the 90\% quantile line of population 1 is less than the 10\% quantile line of population 2, for which a schematic figure was shown in Figure~\ref{fig:line_sep_example}. However, this effort was failed that no such line was found. This is most likely due to the variation of certain line profiles among different stellar populations. That is to say, even the difference between \textit{median} reference spectra show some prominent spectral features, it is not guaranteed that the entire stellar population would fulfill the criteria, especially at the overlapping region. There are always (more than a few) outliers. Therefore, some indirect ways were adopted to achieve this goal as shown in the following sections. For interested readers, we also illustrated some strongest spectral features between different populations (almost all of them are molecular bands like OH or CO, etc.), based only on the residuals of \textit{median} reference spectra as shown in Figure~\ref{fig:med_sp_line_agb_rsg}, Figure~\ref{fig:med_sp_line_rsg_rgb}, and Figure~\ref{fig:med_sp_line_agb_rgb}. Notice that, the residuals were 0.2 for AGBs/RSGs and RSGs/RGBs, but increase to 0.25 for AGBs/RGBs, due to the large numbers of available lines. The zoomed-in regions for each spectral features are shown in the Section \ref{sec:speclines_appendix} in the Appendix.

\subsection{Chi-square and Cosine Similarity} \label{subsec:chi2_and_cs}

\begin{figure*}
\center
\includegraphics[scale=0.21]{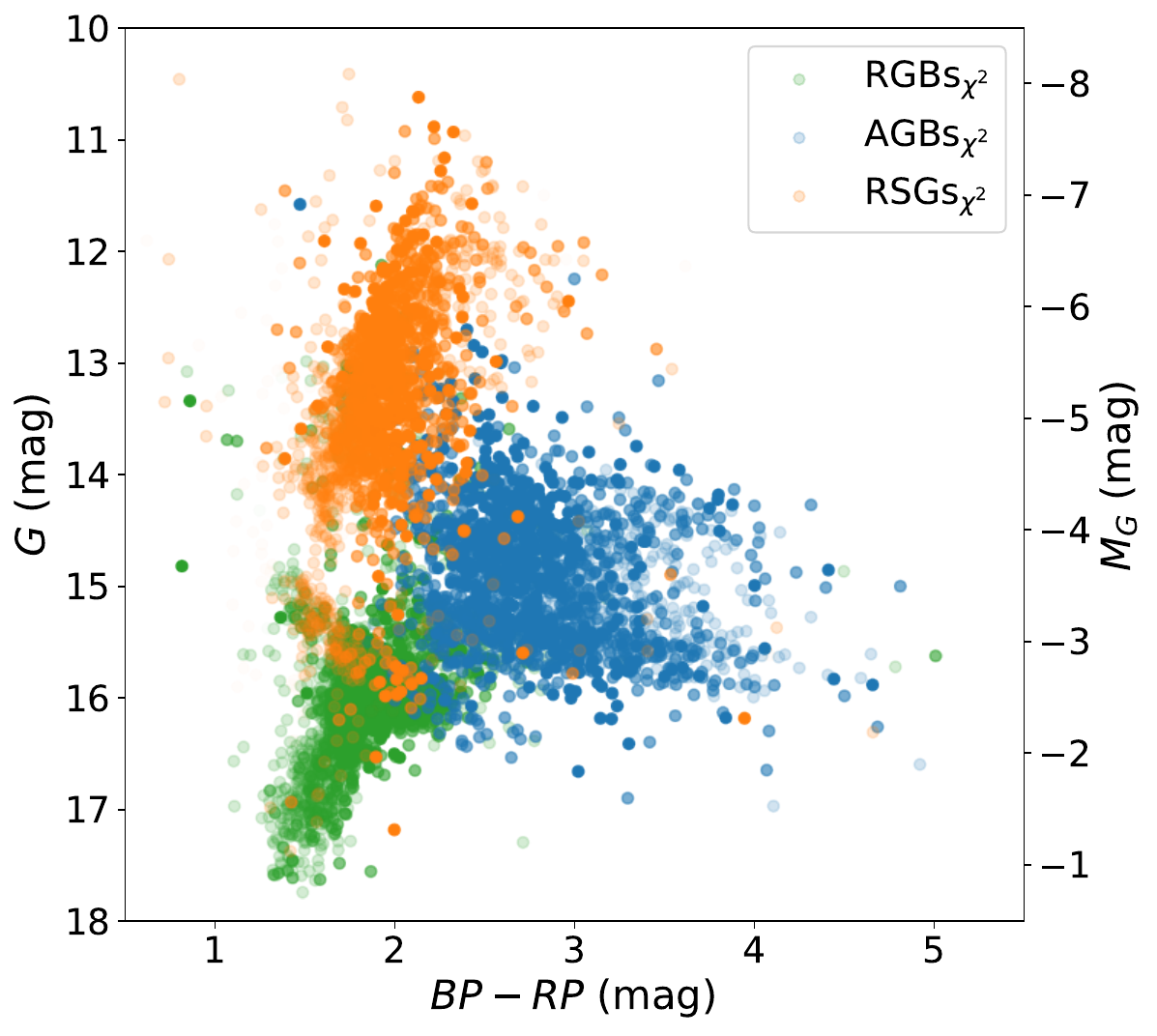}
\includegraphics[scale=0.21]{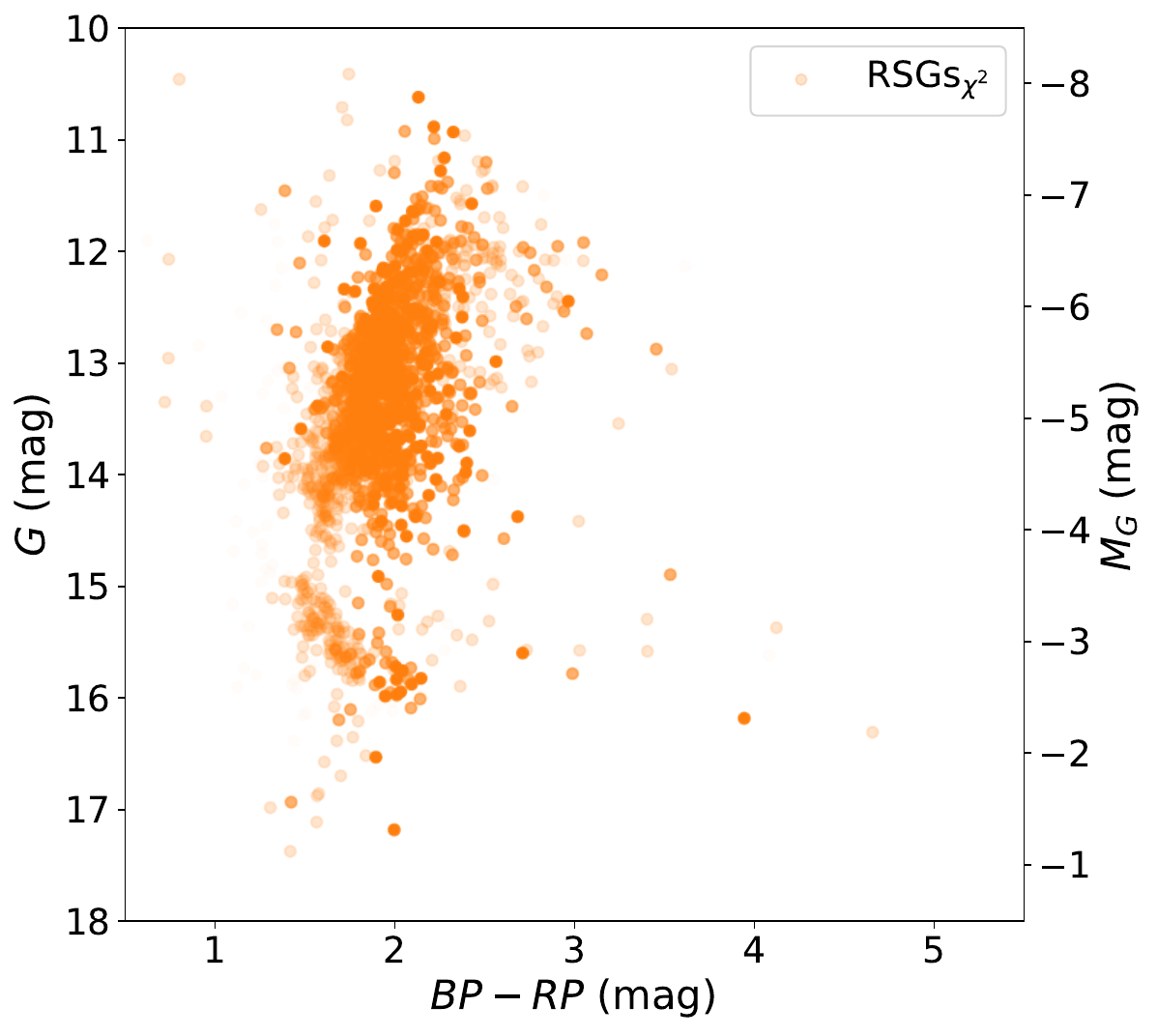}
\includegraphics[scale=0.21]{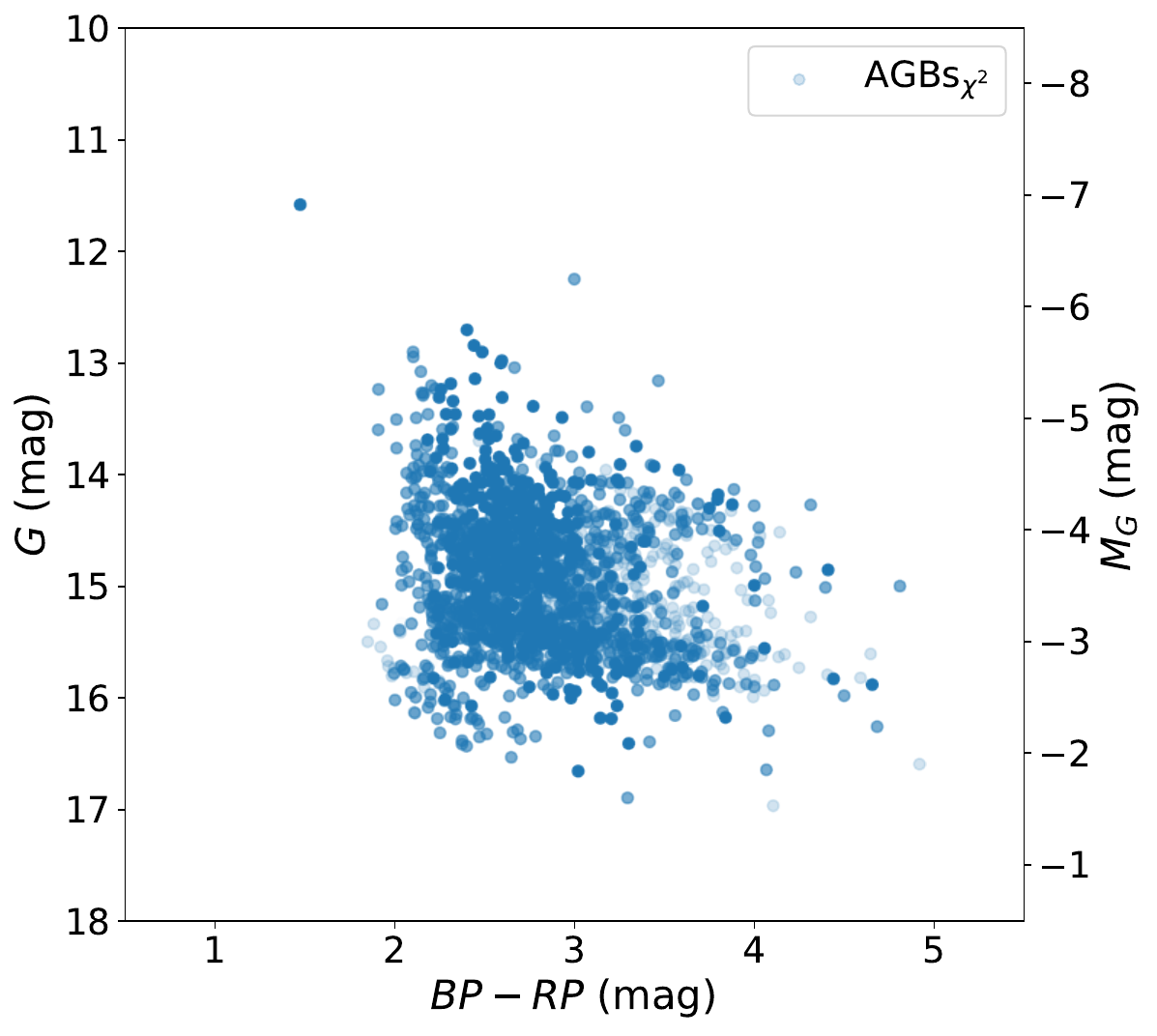}
\includegraphics[scale=0.21]{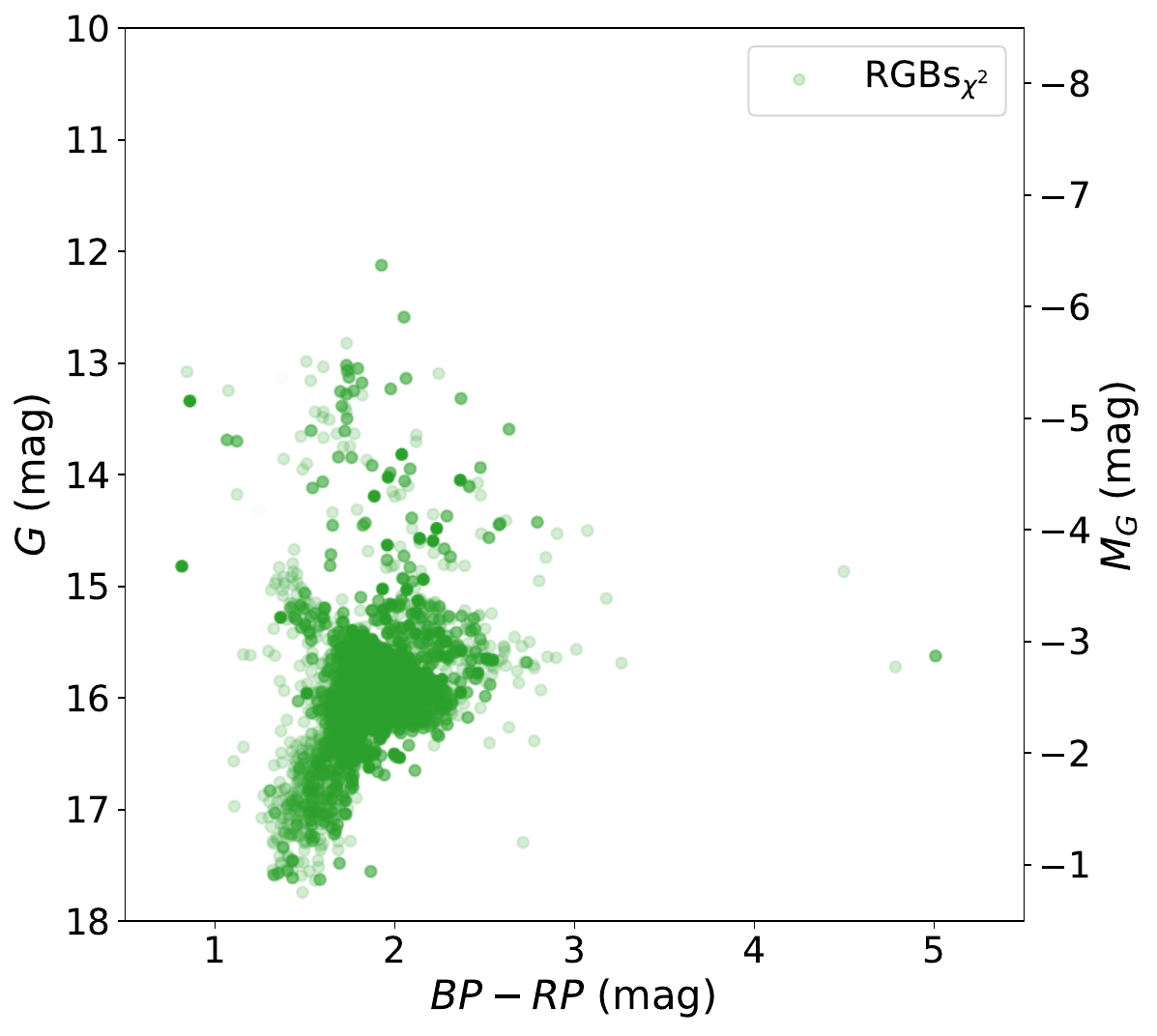}
\includegraphics[scale=0.21]{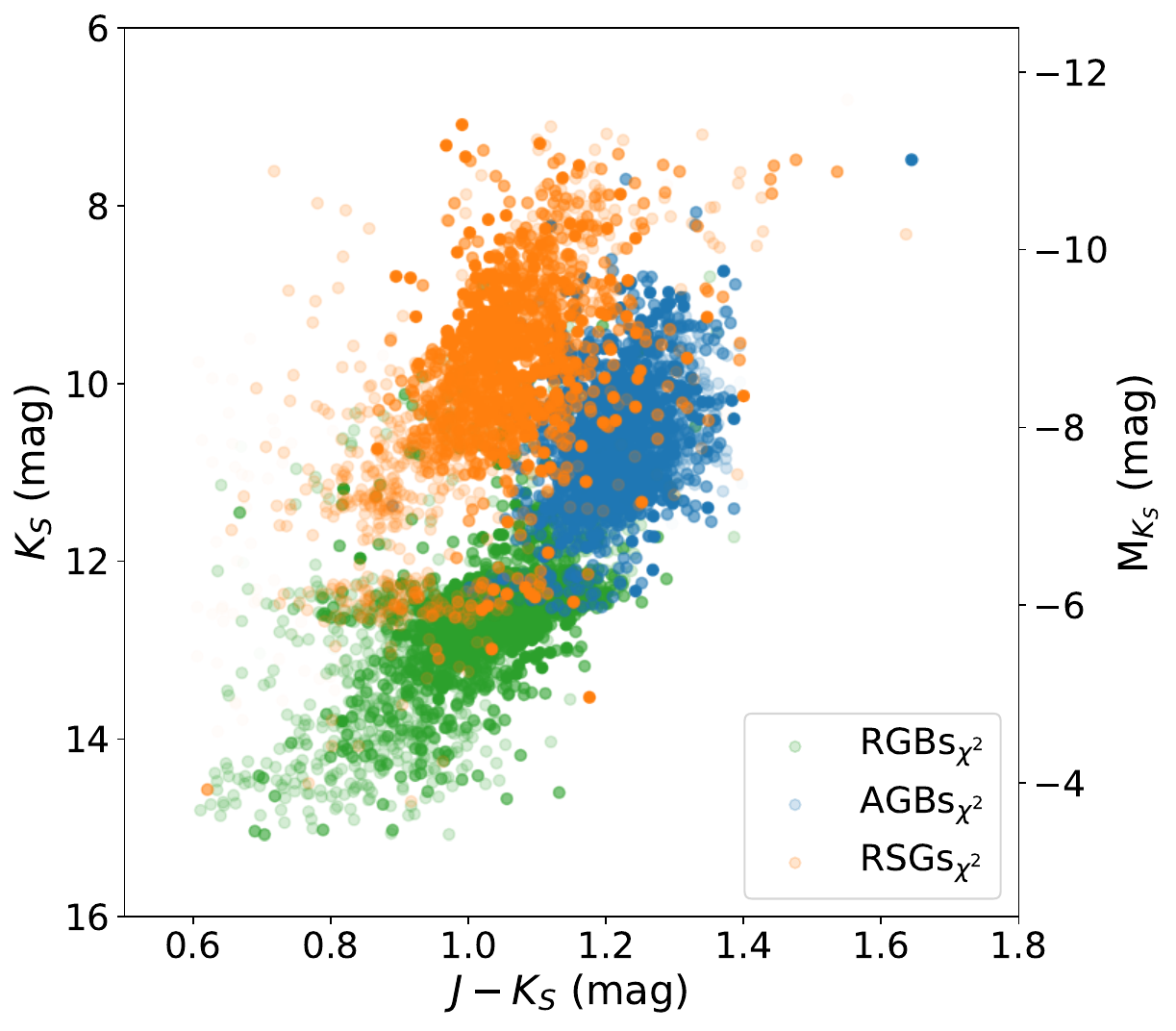}
\includegraphics[scale=0.21]{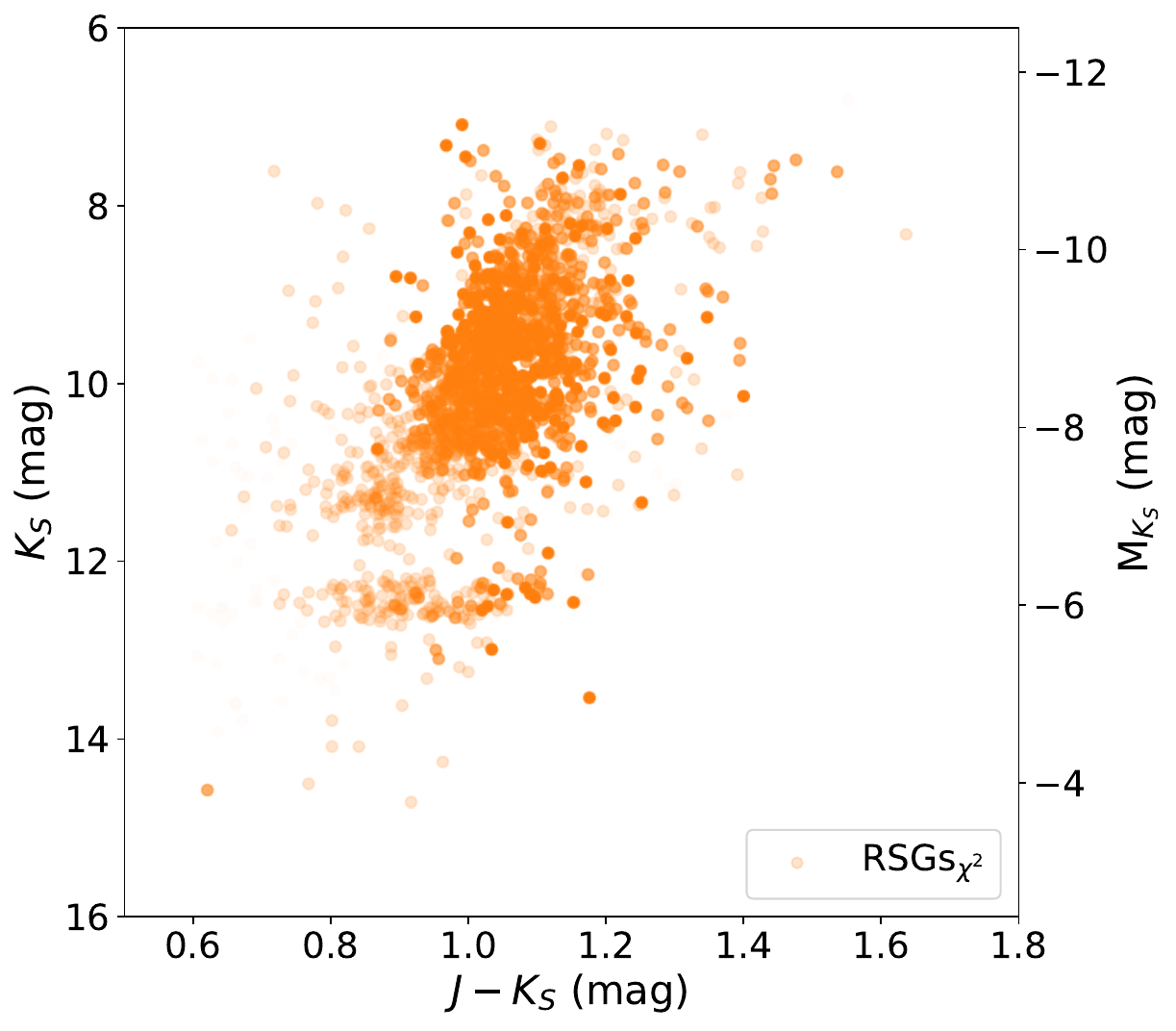}
\includegraphics[scale=0.21]{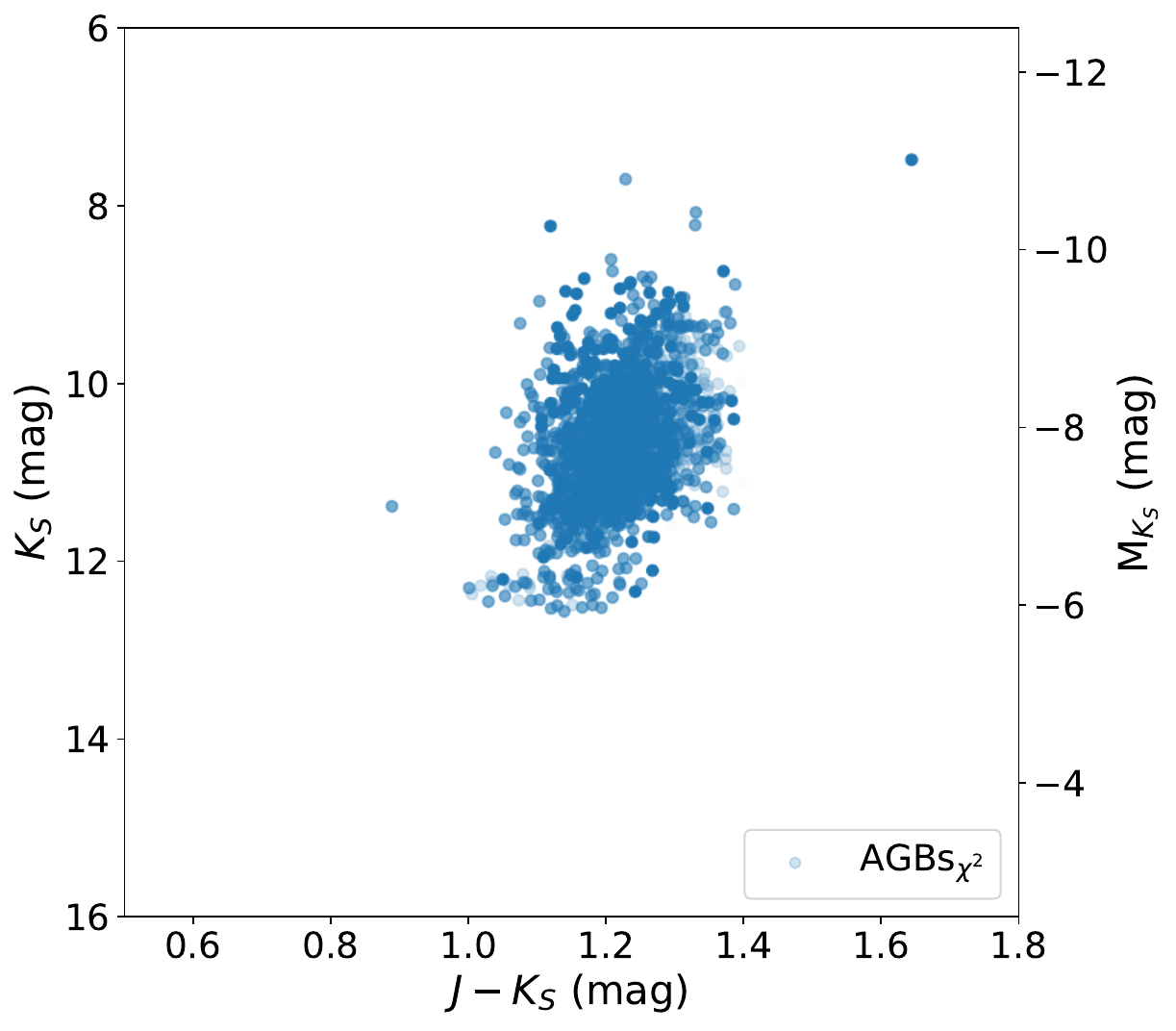}
\includegraphics[scale=0.21]{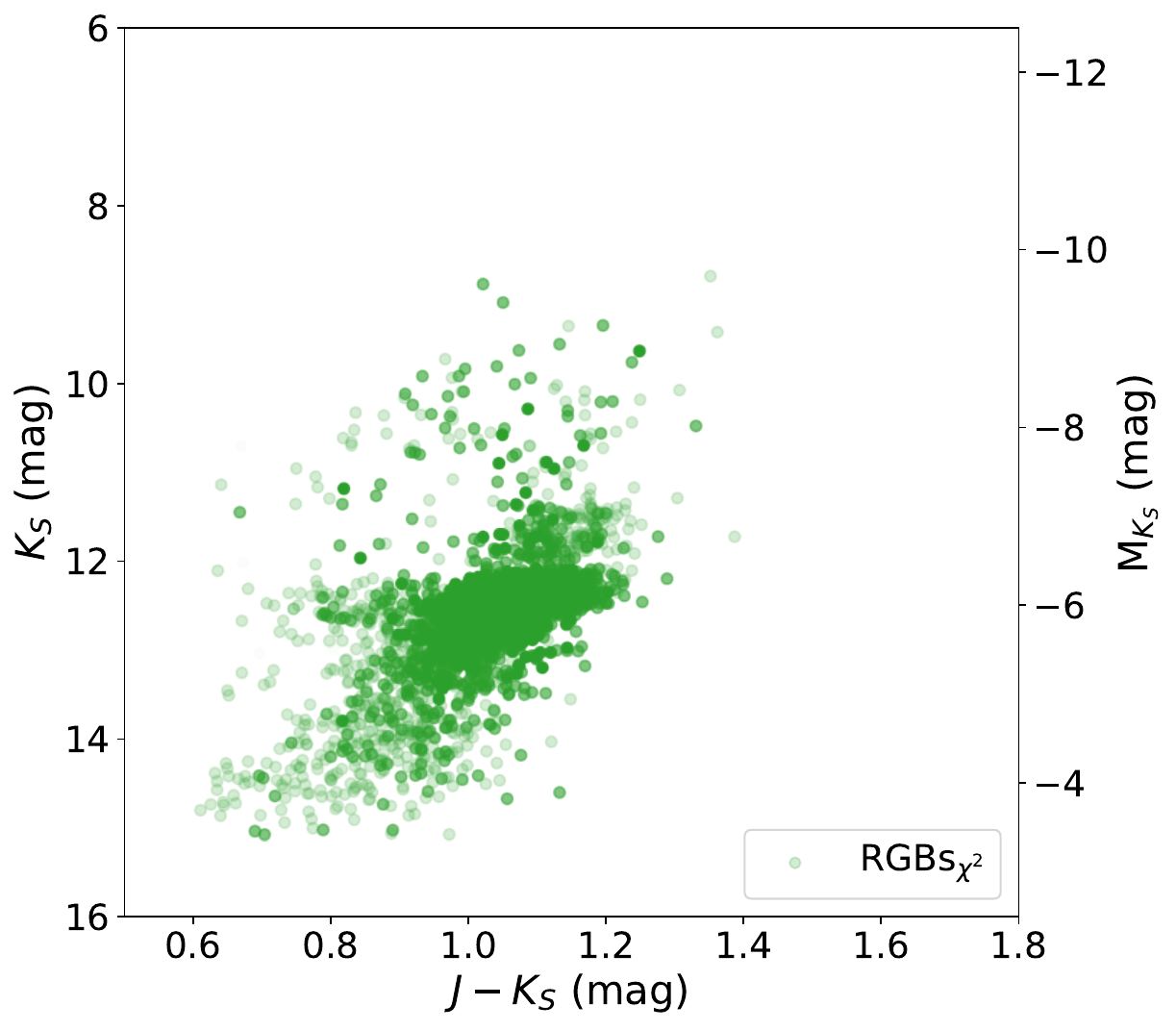}
\caption{The overall, RSG, AGB, and RGB populations (from the left to the right columns) on the \textit{Gaia} (upper row) and 2MASS (bottom row) CMDs based on the calculation of $\chi^2$ between the observed and reference APOGEE-2 spectra. The transparency represents the similarity between each target and reference spectra, e.g., the most solid color represents the similarity of the median (50\%) reference spectra, while the most transparent color represents the similarity of the 1\% or 99\% reference spectra.
\label{fig:pred_lmc_chi2}}
\end{figure*}

The most simple and straightforward way to classify the spectra is to calculate the minimum chi-square ($\chi^2$) between the observed and reference spectra. Here we used the classic $\chi^2$ formula,
\begin{equation}
\chi^2=\frac{1}{N}\frac{[f(Obs)-f(Ref)]^2}{f(Ref)},
\end{equation}
where $N$ represented the number of data points for each spectrum, $f(Obs)$ and $f(Ref)$ indicated the observed and reference spectra, respectively. The spectrum of each target was post-processed and then compared with 1\%, 10\%, 30\%, 50\%, 70\%, 90\%, and 99\% reference spectra of AGB, RSG, and RGB populations. The best classification for each target was chosen by the minimal $\chi^2$ of corresponding reference spectrum. Figure~\ref{fig:pred_lmc_chi2} shows the resultant RSG, AGB, and RGB populations on the 2MASS CMDs based on the calculation of $\chi^2$, while Table~\ref{tbl:sourcetable} shows the full information and classifications of all the 6,602 targets.

In addition, we also calculated the cosine similarity ($k$; $0 \leq k \leq 1$) between each target and reference spectra. The cosine similarity (CS) is the Euclidean L2-normalized dot product of vectors defined as:
\begin{equation}
k(A,B)=\frac{\langle A,B \rangle}{\|A\|*\|B\|}=\frac{\sum\limits_{i=1}\limits^n A_i B_i}{\sqrt{\sum\limits_{i=1}\limits^n A_i^2}\sqrt{\sum\limits_{i=1}\limits^n B_i^2}},
\end{equation}
where A and B are the observed and reference spectra, respectively. Basically, cosine similarity measurement is a measure of the cosine of the angle between the two non-zero vectors A and B, that the value is closer to one, the angle is smaller and the two vectors are more similar. The result is not shown here, since it is very similar to the result of $\chi^2$.

\begin{deluxetable}{ccccccc}
\tablecaption{The Full Information and Classifications of the 6,602 Targets in the Sample.\label{tbl:sourcetable}}
\tablewidth{0pt}
\tabletypesize{\scriptsize}
\tablehead{
\colhead{R.A.} & \colhead{Dec.} & \colhead{J} & \colhead{e\_J} & \colhead{...} & \colhead{Type$_{\chi^2}$} & \colhead{Type$_{ML}$} \\
\colhead{(deg)} & \colhead{(deg)} & \colhead{(mag)} & \colhead{(mag)} & \colhead{...} & \colhead{} & \colhead{} \\
} 
\startdata
46.902584 & -77.179604 & 15.442 & 0.052 & ... & RGB & RGB \\
47.334310 & -77.225609 & 14.238 & 0.032 & ... & RGB & RGB \\
47.480656 & -77.463486 & 13.998 & 0.033 & ... & RGB & RGB \\
...       & ...        & ...    & ...   & ... & ... & ... \\
\enddata
\tablecomments{This table is available in its entirety in machine-readable form.}
\end{deluxetable}

\subsection{Machine Learning} \label{subsec:ml}

A more sophisticated and popular way to classify the stellar populations is the machine learning (ML). We adopted a LeNet-like CNN model to predict the stellar classes. The first three convolution-ELU blocks aimed to identify the spectral features, followed by the pooling, flattening, and linear layers to shrink data length to the number of classes. At the network's end, a Softmax function turned the output into a one-hot prediction.

\begin{figure}
\center
\includegraphics[scale=0.75]{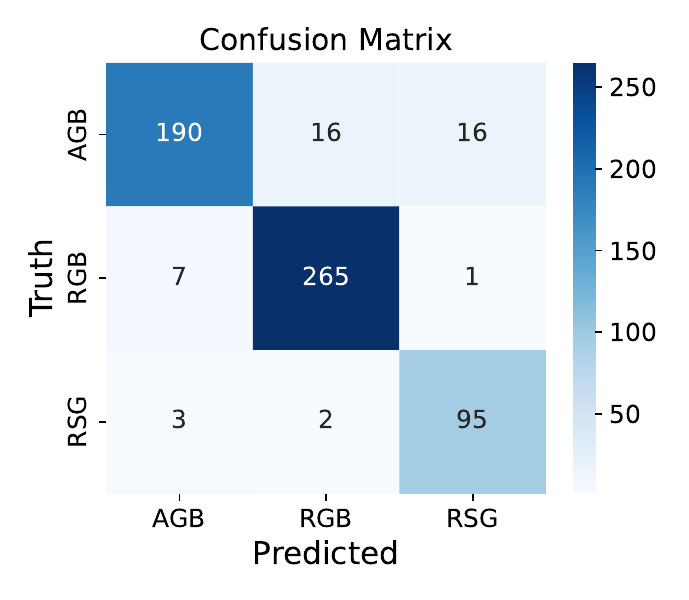}
\caption{The confusion matrix for the ML reference data set.
\label{fig:confusion_matrix}}
\end{figure}

The APOGEE spectra were post-processed as mentioned before, and the labels of AGBs, RGBs, and RSGs are coded as one-hot vectors, i.e., [1,0,0], [0,1,0], [0,0,1], respectively. Our reference data set contains 2,083 AGBs, 2,762 RGBs, and 1,096 RSGs, classified with multi-band photometry as mentioned in Section 2 and 3. We randomly selected 90\% of them for training and the rest for test. The model accuracy converged to 92.44\% after 50 epoch training. Figure~\ref{fig:confusion_matrix} shows the resultant confusion matrix. The implementation code based on PyTorch is shown in Section \ref{sec:ml_code_appendix} in the Appendix. Figure~\ref{fig:pred_lmc_ml} shows the predicted stellar populations on the 2MASS CMDs based on the ML. It can be seen that there is a general agreement between the results from ML and $\chi^2$.

\begin{figure*}
\center
\includegraphics[scale=0.21]{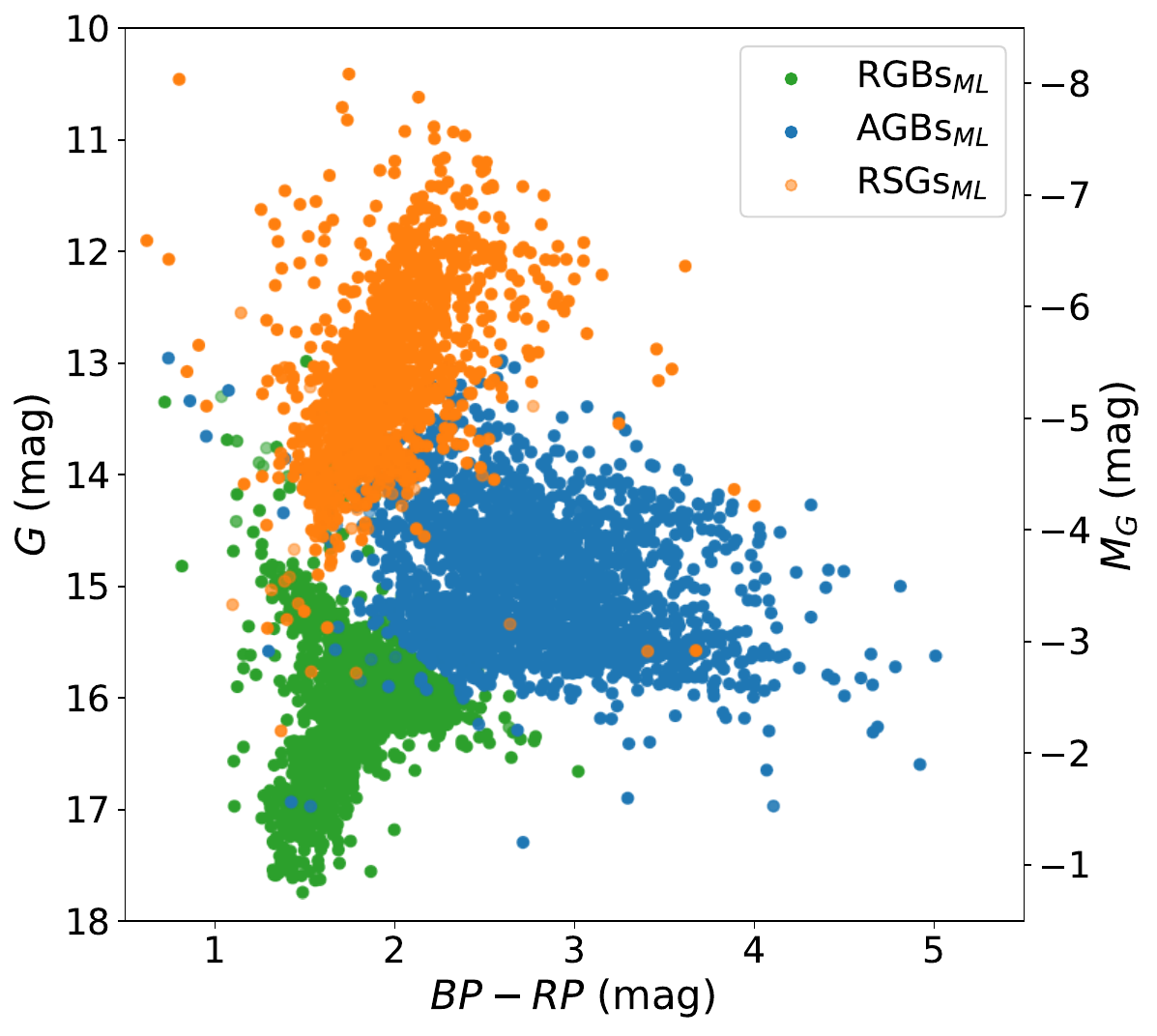}
\includegraphics[scale=0.21]{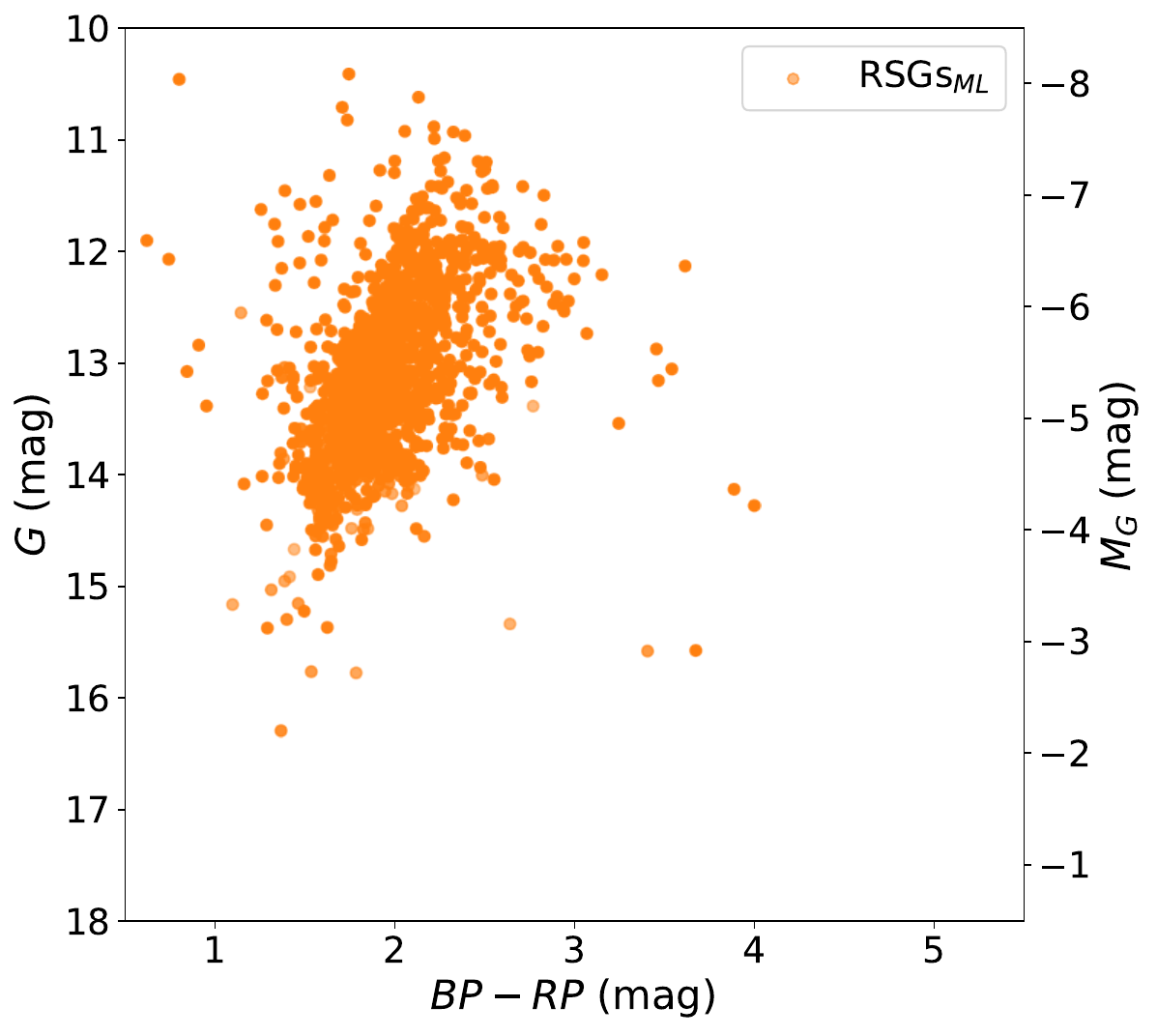}
\includegraphics[scale=0.21]{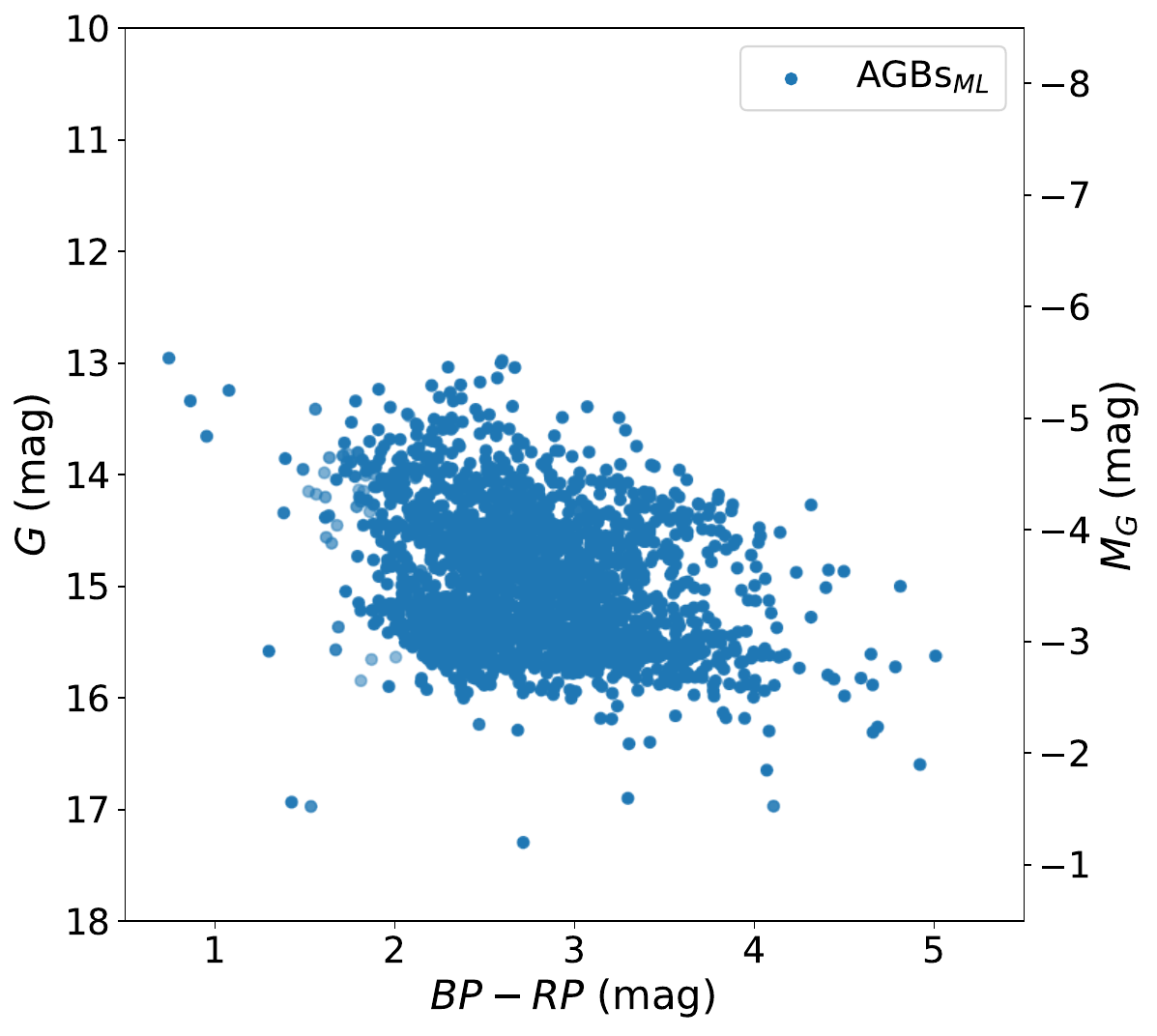}
\includegraphics[scale=0.21]{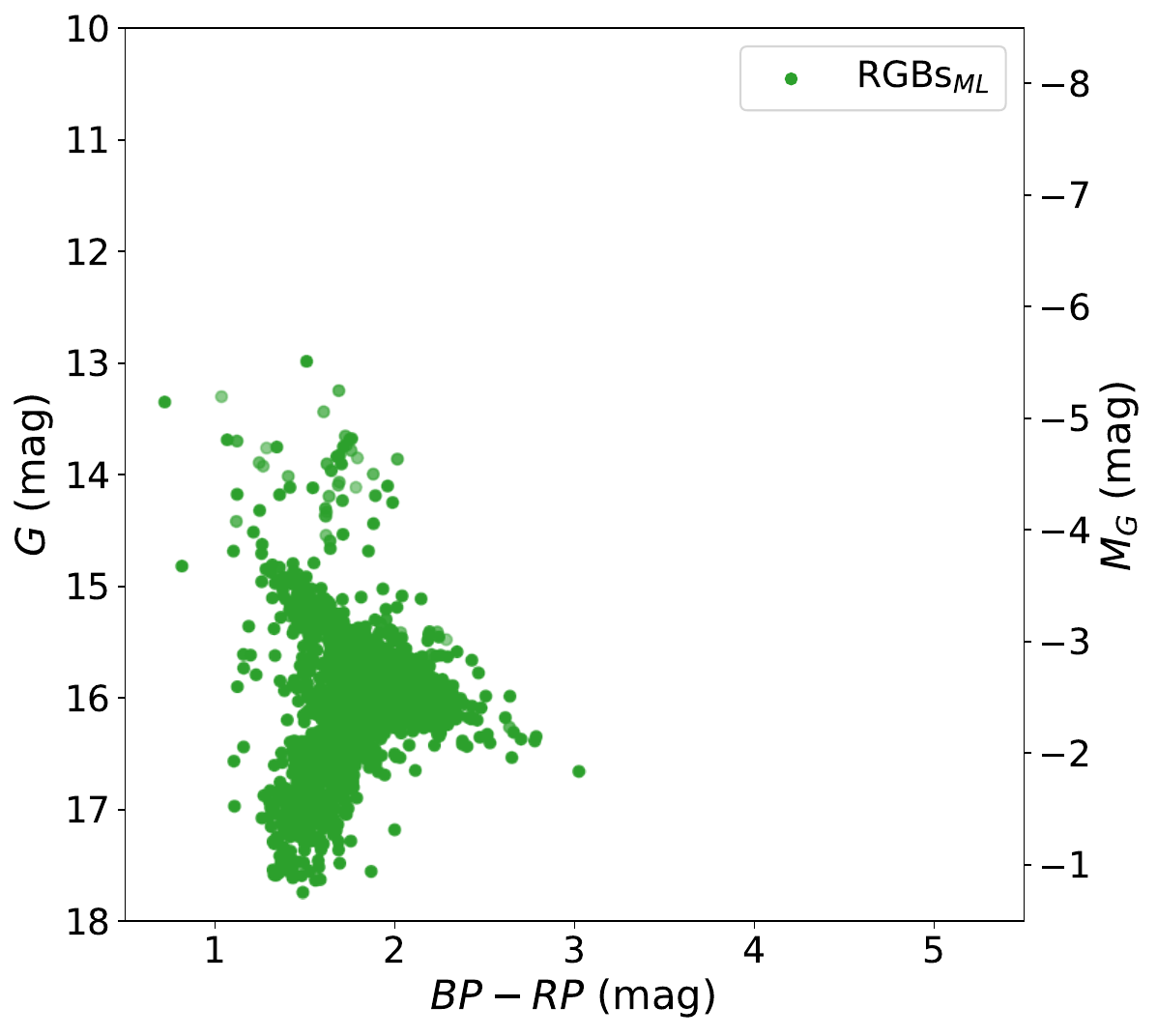}
\includegraphics[scale=0.21]{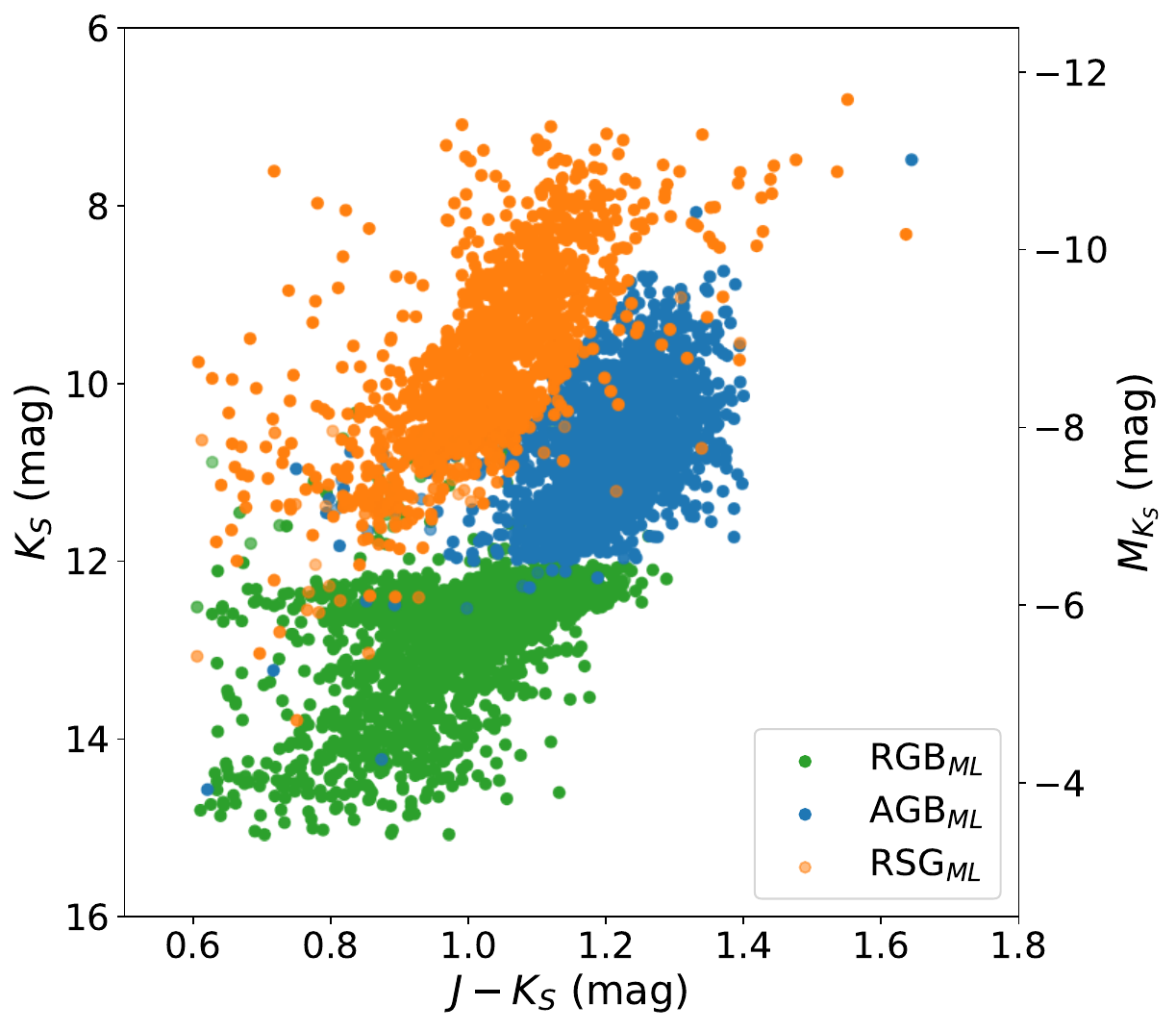}
\includegraphics[scale=0.21]{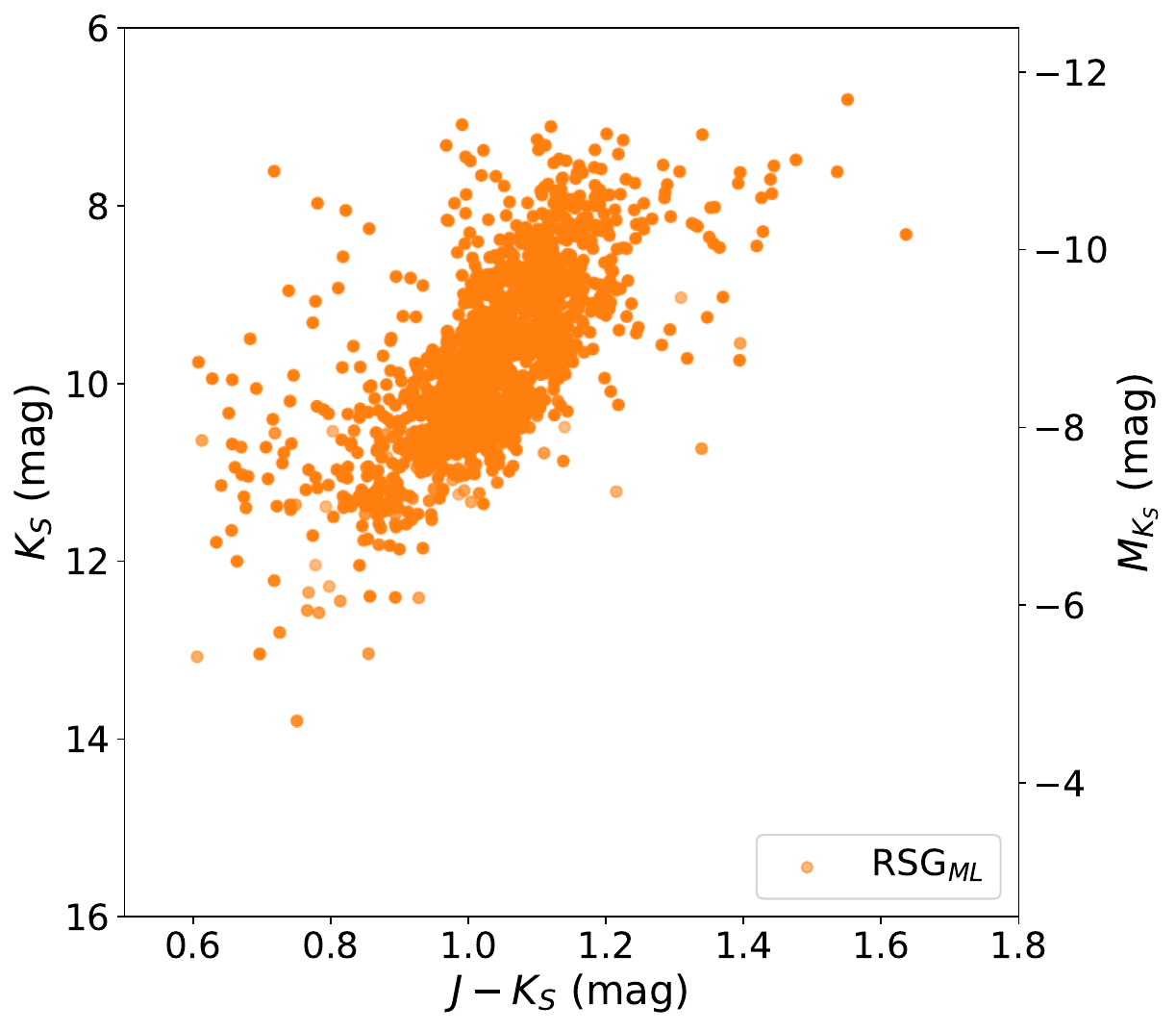}
\includegraphics[scale=0.21]{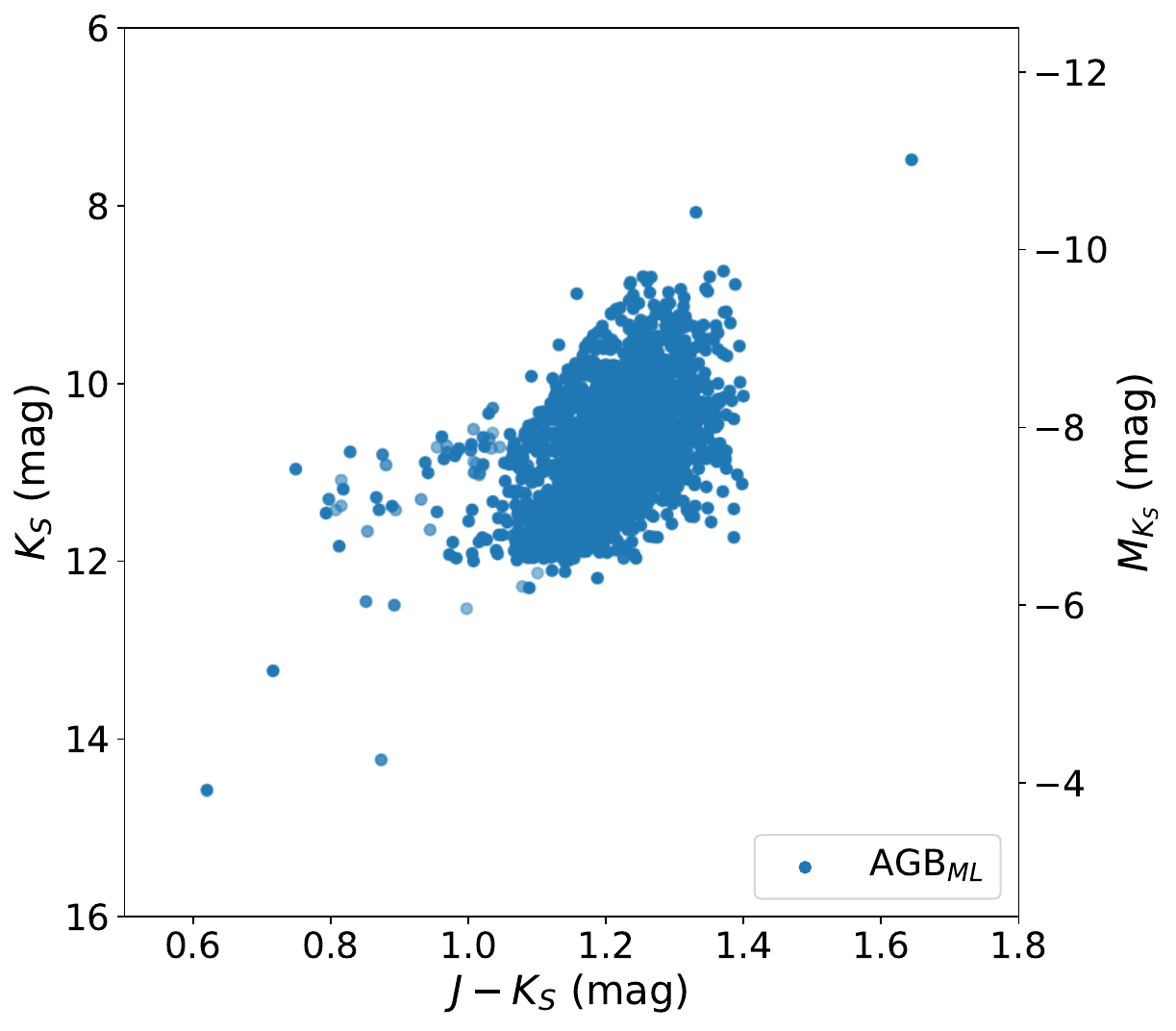}
\includegraphics[scale=0.21]{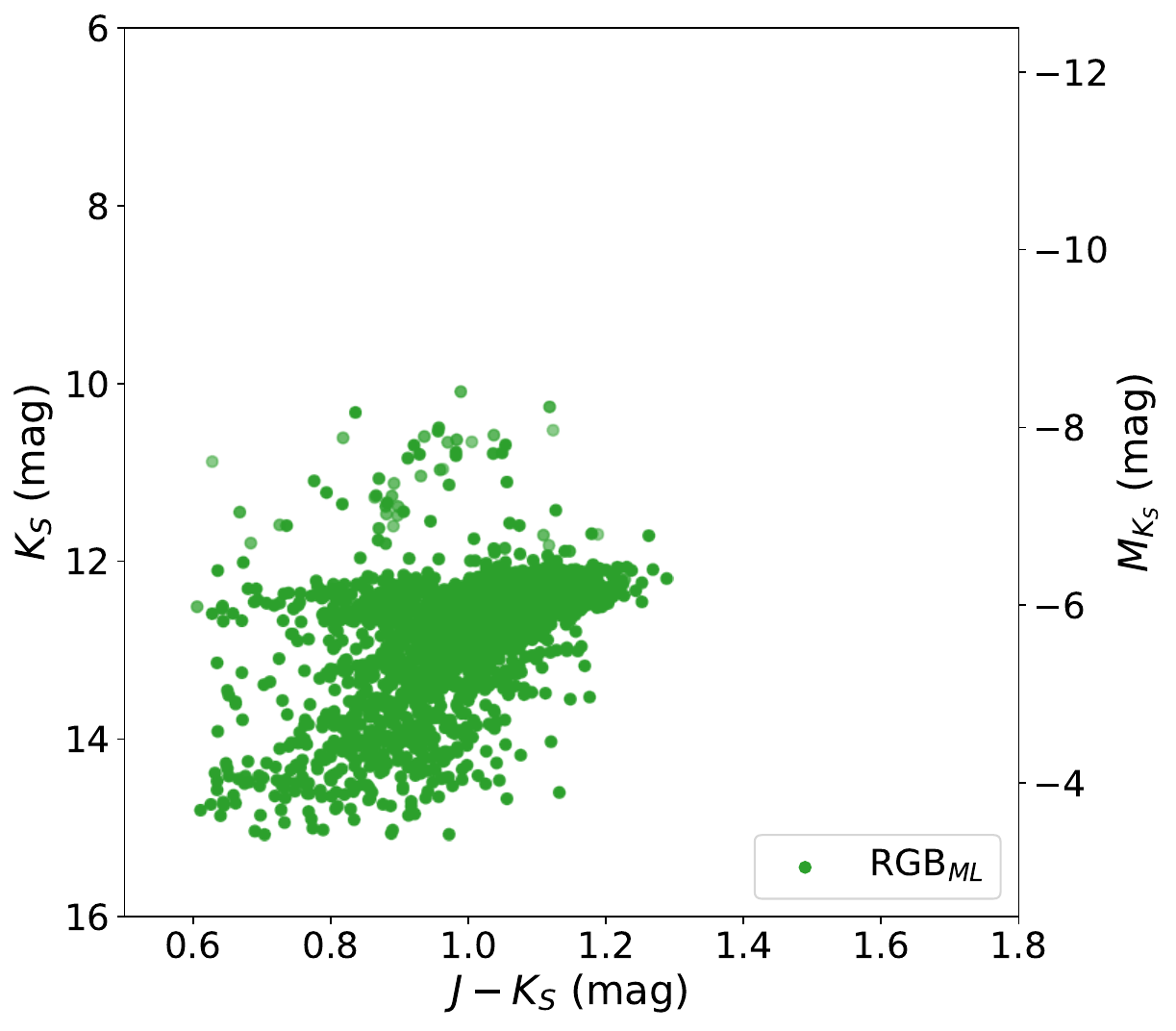}
\caption{Same as Figure~\ref{fig:pred_lmc_chi2} but for the ML. The transparency represents the probability of the best predicted population.
\label{fig:pred_lmc_ml}}
\end{figure*}

We also tried other unsupervised learning methods, like K-Means, Bisecting K-Means, Affinity Propagation, MeanShift, SpectralClustering, etc., to classified the spectra but didn't get good enough results.

\subsection{Equivalent Width} \label{subsec:ews}

\begin{figure}
\center
\includegraphics[scale=0.35]{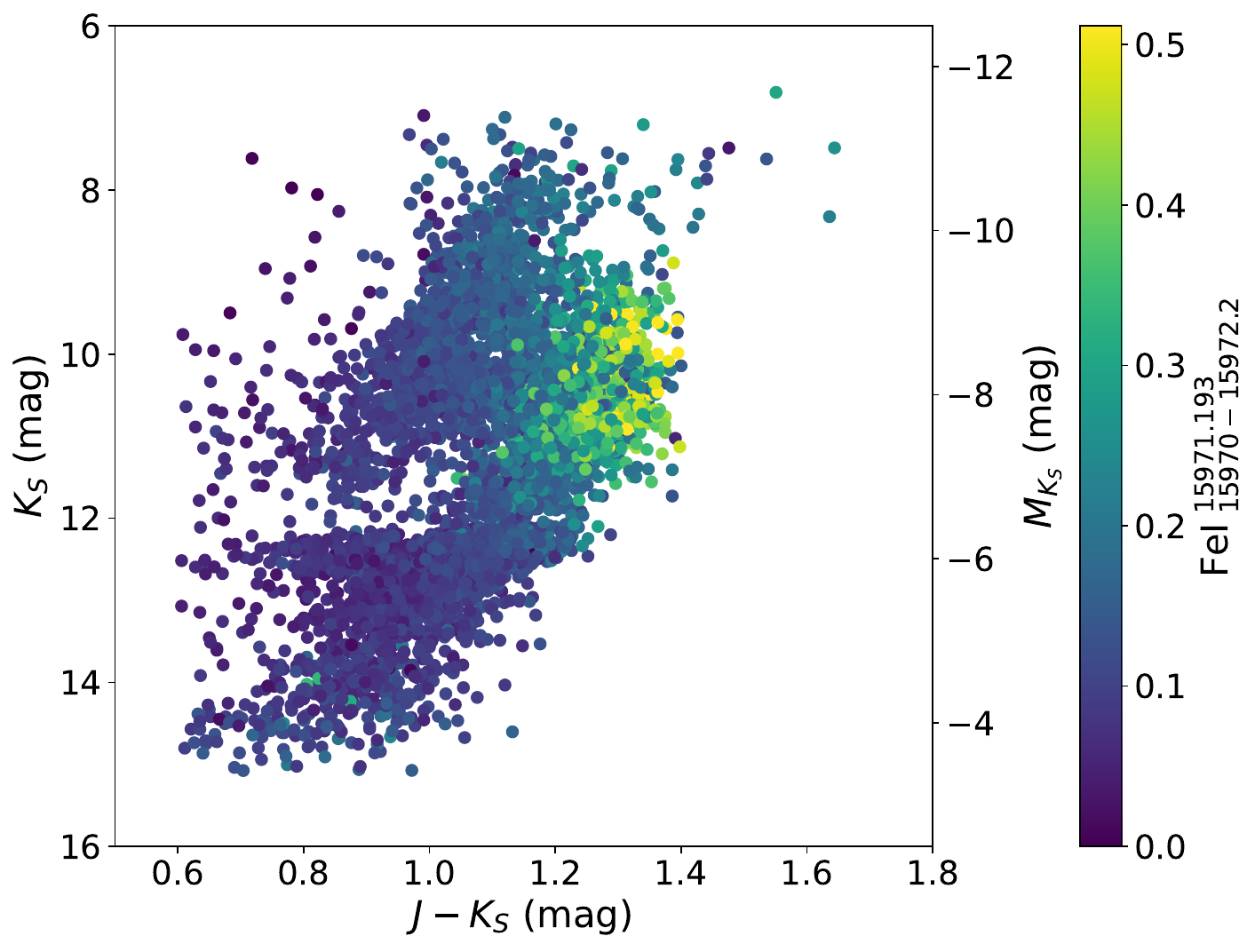}
\caption{2MASS CMD color coded with the calculated EW for Fe \uppercase\expandafter{\romannumeral1} line at 15971.193\AA.
\label{fig:ew}}
\end{figure}

\begin{figure*}
\center
\includegraphics[scale=0.3]{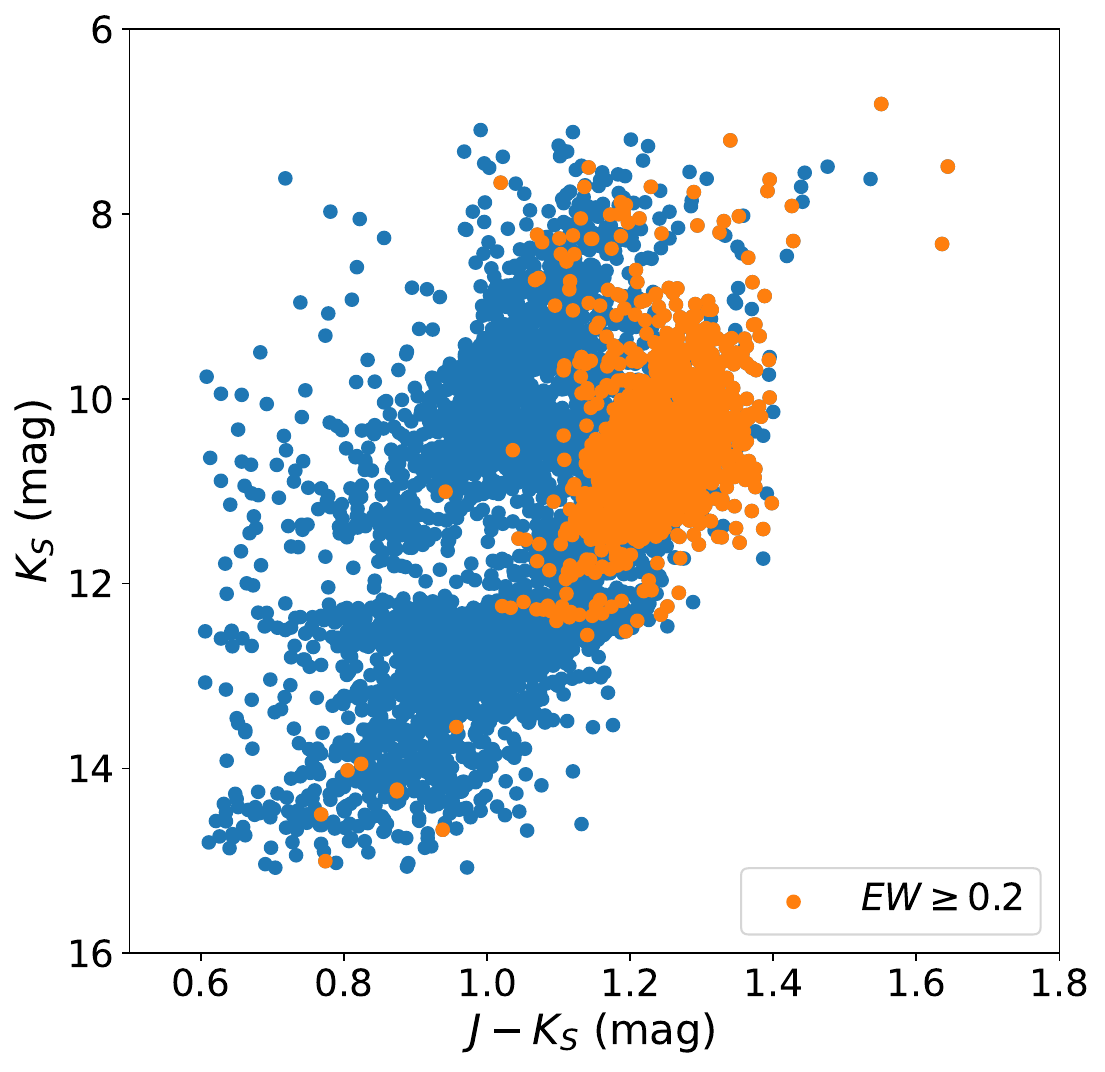}
\includegraphics[scale=0.3]{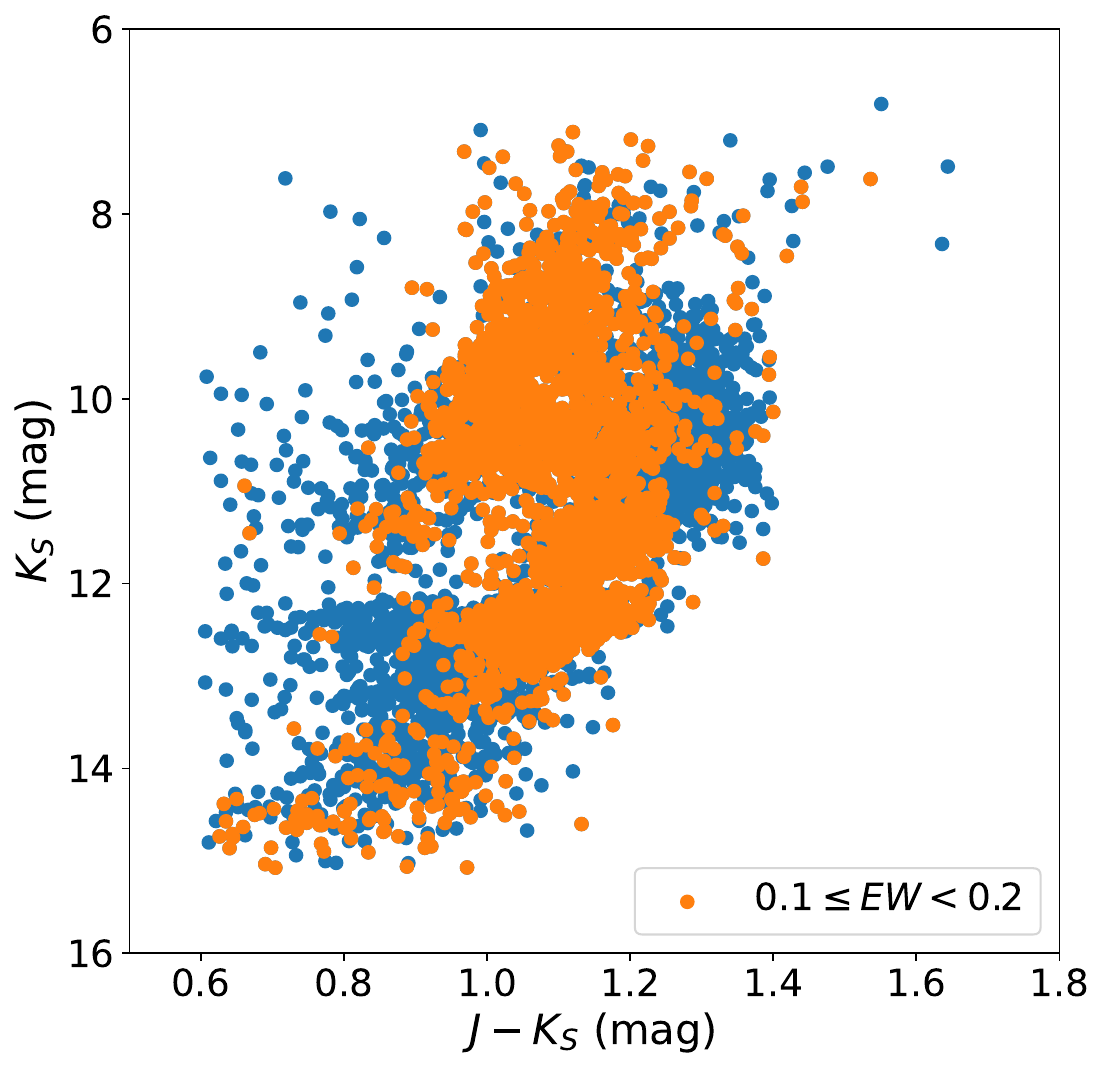}
\includegraphics[scale=0.3]{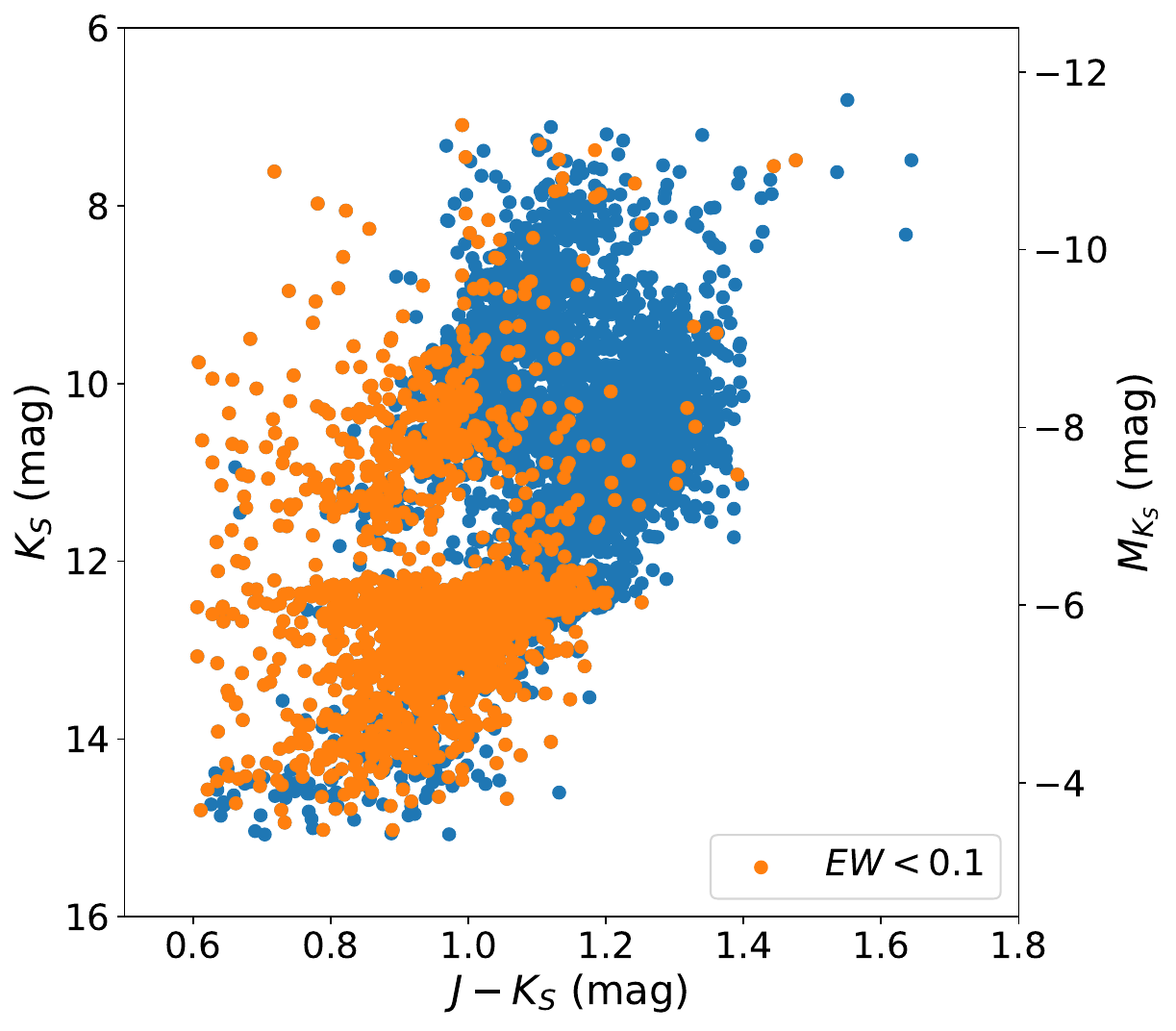}
\caption{Same as Figure~\ref{fig:ew} but for the distribution of piecewise EW as shown in each panel.
\label{fig:ew_cmd}}
\end{figure*}

Moreover, we also calculated the equivalent widths (EWs) of 38 different spectral lines or blended line complex as listed in Table~\ref{tbl:ewtable}, trying to separate different stellar populations. Notice that, the calculation was done only for lines or line complexes with obvious intensity gradient along AGBs, RSGs, and RGBs by visually inspecting the reference spectra. Meanwhile, the precise EW measurements are also affected by the normalization of the spectra, where this effect is minimized as long as the the spectra are post-processed in the same way as mentioned above. Figure~\ref{fig:ew} shows a typical example of 2MASS CMD color coded with the calculated EW for one Fe \uppercase\expandafter{\romannumeral1} line at 15971.193\AA, while Figure~\ref{fig:ew_cmd} shows the same CMDs with the distribution of piecewise EWs. As can be seen from the diagrams, the piecewise EWs do not represent the actual boundaries between different stellar populations, but start to blend at some points. This is also true for all the rest EWs as shown in Section \ref{sec:ews_appendix} in the Appendix.

\begin{deluxetable}{cccc}
\tablecaption{Parameters for the Calculated Equivalent Width of Spectral Lines or Line Complexes\label{tbl:ewtable}}
\tablewidth{0pt}
\tabletypesize{\scriptsize}
\tablehead{
\colhead{Characteristic Wavelengths} & \colhead{Lower Limit} & \colhead{Upper Limit} & \colhead{Name}
} 
\startdata
15173.415 &  15172.5 &  15175 &  NiI \\
15183.061 &  15182 &  15183.8 &  Complex \\
15236.855 &  15235.5 &  15239 &  Complex \\
15283.659 &  15282.5 &  15287.5 &  Complex \\
15372.600 &  15369.8 &  15378.5 &  Complex \\
15434.314 &  15432 &  15438.8 &  Complex \\
15496.704 &  15493 &  15503 &  Complex \\
15505.484 &  15504 &  15510 &  Complex \\
15535.504 &  15533 &  15538.8 &  Complex \\
15543.662 &  15538.8 &  15546 &  Complex \\
15577.844 &  15576.8 &  15587 &  $^{13}$CO 3-0 \\
15594.640 &  15592.5 &  15596.8 &  Complex \\
15715.541 &  15713.5 &  15718.5 &  Complex \\
15912.608 &  15907 &  15915.5 &  Complex \\
15971.193 &  15970 &  15972.2 &  FeI \\
15977.593 &  15975.8 &  15994 &  $^{12}$CO 5-2 \\
16015.827 &  16014.5 &  16017.5 &  Complex \\
16026.230 &  16017.5 &  16028 &  Complex \\
16031.101 &  16030 &  16033 &  Complex \\
16089.456 &  16088 &  16090.5 &  FeI \\
16125.952 &  16120.8 &  16127.5 &  Complex \\
16184.876 &  16183 &  16188 &  CO 6-3 \\
16269.394 &  16268 &  16270.5 &  Unknown \\
16303.595 &  16301.5 &  16306.5 &  CO \\
16396.886 &  16393.5 &  16400.8 &  Complex \\
16492.761 &  16491.5 &  16499.2 &  Complex \\
16506.438 &  16505.5 &  16507.5 &  FeI \\
16511.456 &  16510.8 &  16512.8 &  CO \\
16513.737 &  16512.8 &  16516 &  Unknown \\
16549.138 &  16547.5 &  16551 &  CO \\
16568.583 &  16567.5 &  16572.8 &  Complex \\
16610.067 &  16609.2 &  16611 &  CO \\
16613.050 &  16611.5 &  16618.8 &  $^{12}$CO 8-5 \\
16626.138 &  16625.2 &  16628 &  Unknown \\
16654.415 &  16651.2 &  16657.2 &  Complex \\
16693.345 &  16692.2 &  16696 &  Complex \\
16697.728 &  16696 &  16699 &  CO \\
16835.553 &  16834.2 &  16840.8 &  $^{12}$CO 9-6 \\
\enddata
\end{deluxetable}

\subsection{Specific Line Ratios}

\begin{figure*}
\center
\includegraphics[scale=0.4]{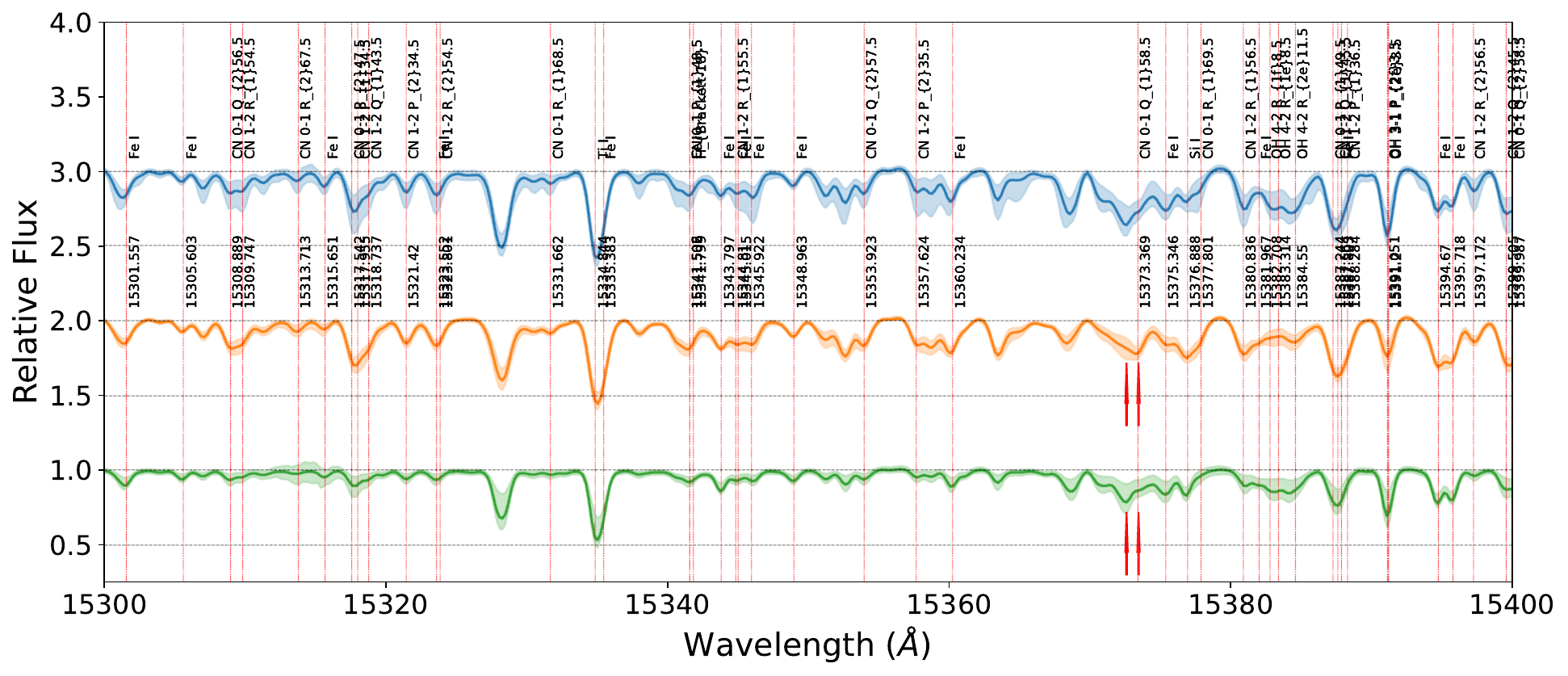}
\includegraphics[scale=0.4]{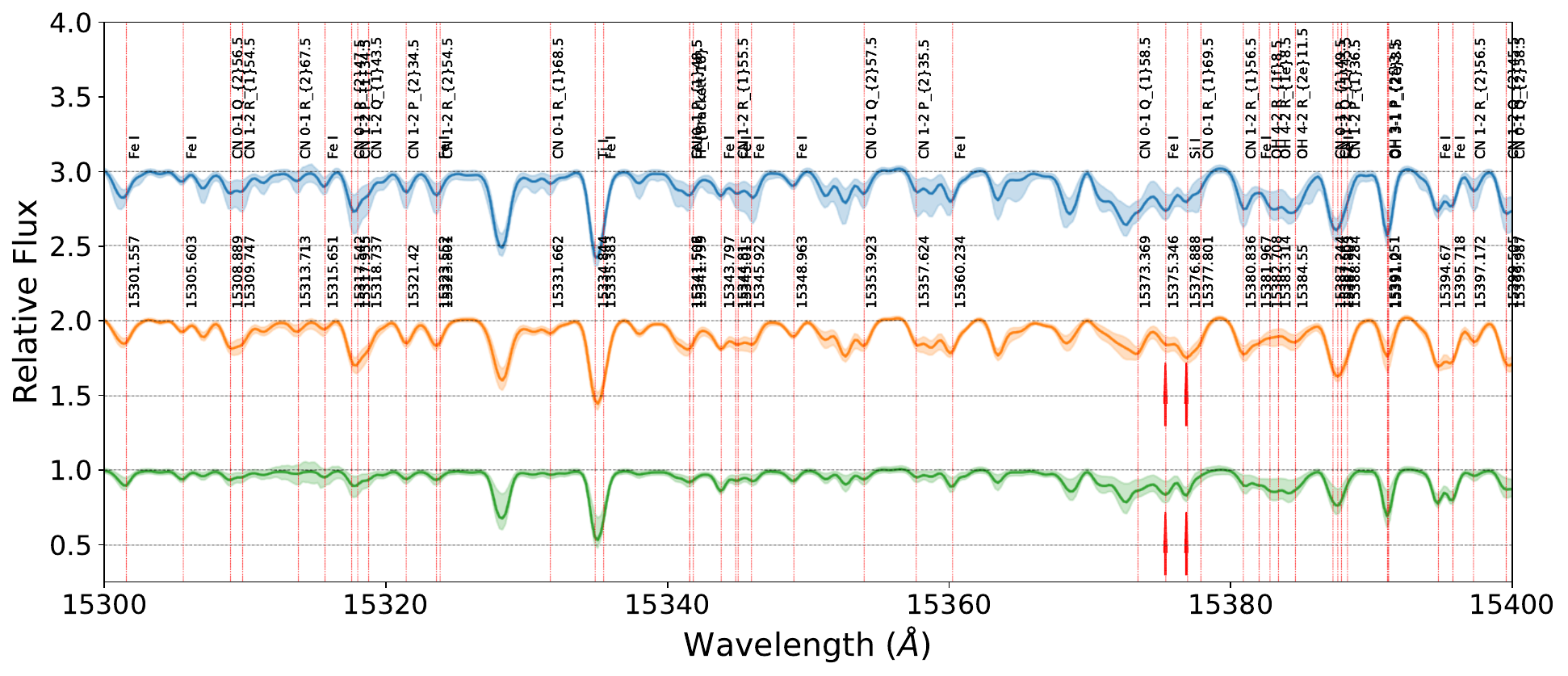}
\caption{Examples of the identified spectral features (red arrows) with different line ratios (Ratio 1, upper; Ratio 2, lower) between RSG and other populations.
\label{fig:lr}}
\end{figure*}

\begin{figure}
\center
\includegraphics[scale=0.41]{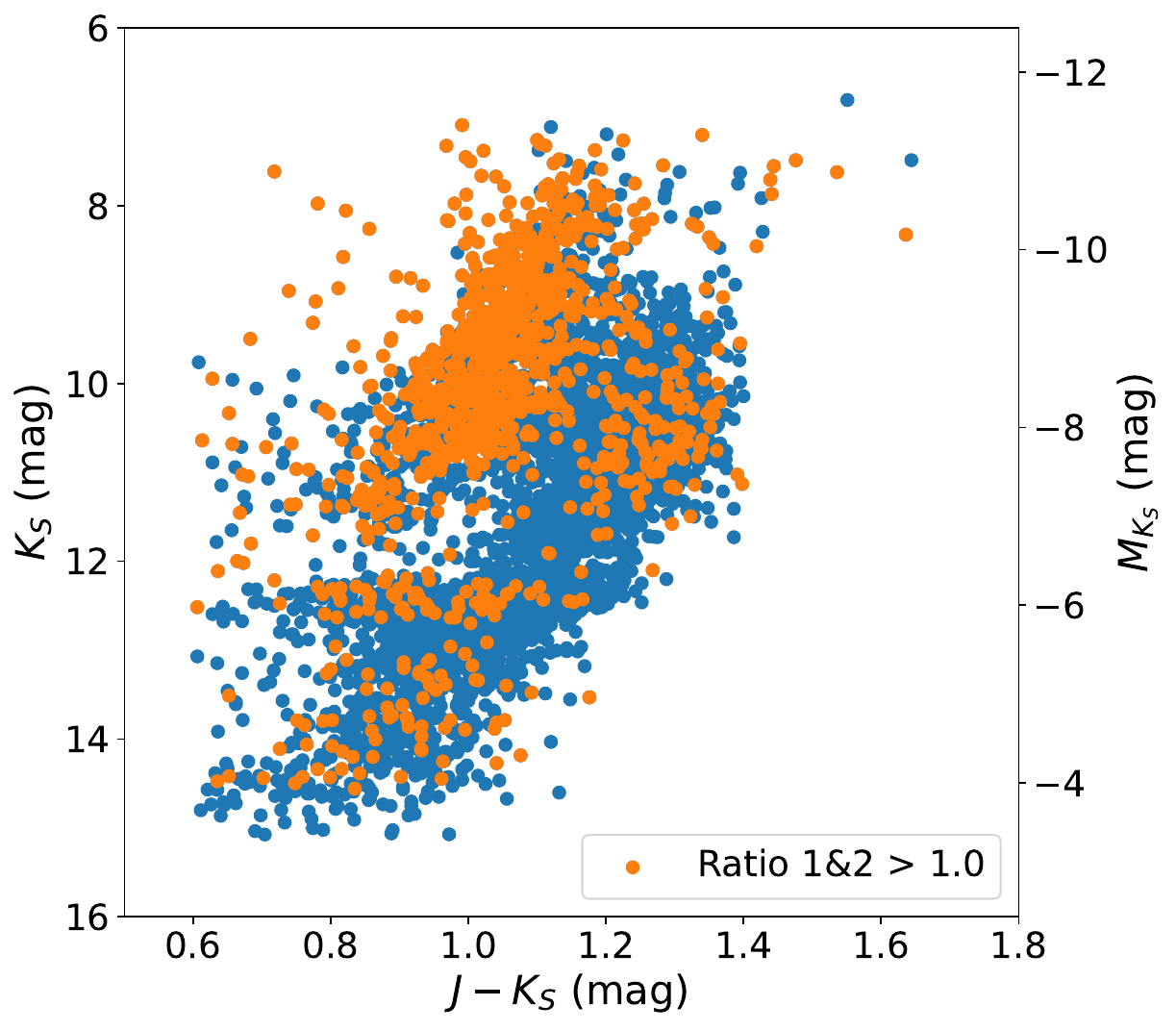}
\caption{The resultant RSG population on 2MASS CMD based on both the line ratios from Figure~\ref{fig:lr}.
\label{fig:lr_cmd}}
\end{figure}

Finally, we also tentatively calculated some specific line ratios (LRs) in order to separate RSG from other stellar populations. Such LRs are based on specific line structures identified by visual inspection of the reference spectra, which presents different patterns in the RSG than the others. Figure~\ref{fig:lr} shows examples of likely promising spectral feature with different LRs between RSG and other populations, and Figure~\ref{fig:lr_cmd} shows the resultant 2MASS CMD. Unfortunately, like the EW, the LR does not fully separate the RSG population from others, even with combinations of different LRs.

\section{Discussion}  \label{sec:discussion}

As can be seen from both Figure~\ref{fig:pred_lmc_chi2} and Figure~\ref{fig:pred_lmc_ml} that, even using different methods (e.g., $\chi^2$ or ML), based on the APOGEE-2 H-band spectra, the derived lower mass limit of the RSG population is still able to reach to the $\rm K_S$-TRGB ($\sim$12.0 mag). This is a strong support for previous studies based only on the photometric data (e.g., \citealt{Boyer2011, Yang2018, Yang2019, Neugent2020, Yang2020, Massey2021, Yang2021a, Yang2021b}), proving that the RSG population indeed has a faint lower mass limit. In other word, this indicates a luminosity as low as about $10^{3.5}~L_{\sun}$, which corresponds to a stellar radius only about $100~R_{\sun}$ (see also Figure~\ref{fig:hrd}). Such low luminosity is obviously lower than the traditional view of $10^{4.0}~L_{\sun}$ for the RSGs by half an order of magnitude. 

Given the mass-luminosity relation of $L/L_\sun =f(M/M_\sun)^3$ with $f\approx15.5\pm3$ (averaged from $f=12.5$ for MESA and $f=18.5$ for Geneva; \citealt{Mauron2011, Kee2021}), it infers a \textit{current} stellar mass of $5.9\pm0.4~M_\sun$. It is needed to be emphasized that, when we are talking about the $M_{min}$ of the RSG population, it typically refers to the initial mass. Meanwhile, there are also many other definitions, like most expected mass, final mass, and so on. What we can measure is actually the \textit{current} mass, so that the mass loss of the star up to now has to be taken into account, in order to deduce the initial mass. Fortunately, the mass-loss rate (MLR) of RSGs decreases along with decreasing of luminosity, meaning that the faintest RSGs usually are optically thin and have very low MLRs of about $10^{-8}~M_\sun~yr^{-1}$ or less \citep{Riebel2012, Boyer2012, Matsuura2013, Srinivasan2016, Wang2021, Yang2023}. Assuming an average age of 20~$Myr$, such MLR will only cause about 0.2~$M_\sun$ mass loss. In that sense, the $M_{min}$ of the RSG population would be about $6.1\pm0.4~M_\sun$, which is again much lower than the traditional limit of $8~M_\sun$ for the RSGs. Moreover, the recent stellar evolutionary models already consider $7~M_\sun$ stars evolve to the RSG phase at the LMC metallicity \citep{Eggenberger2021}. Thus, we believe that, based on our sample, this is the first large-scale spectroscopic evidence, indicating that the lower mass limit of RSG population is around $6~M_\sun$. 

Still, there are some limitations should be kept in mind. Firstly, the estimated $M_{min}$ of RSG population is model-dependent due to the mapping of observed luminosity into stellar mass based on the mass-luminosity relation. Using different relations may lead to slightly different results, but shall not deviate too much from our result, since the uncertainties of the $M_{min}$ are already determined from two different popular stellar evolutionary models (it also has to admit that our knowledge of massive stars is still quite poor). Secondly, the luminosity of our sample is derived based on comparison of difference BCs, for which the results agree within $\pm0.1~dex$ as mentioned before. This may bring an uncertainty of $0.5~M_\sun$, but comparable with the error from the mass-luminosity relation ($0.4~M_\sun$). Using different methods to derive the luminosity, e.g., by integrating over the entire spectral energy distribution, will yield very similar result \citep{Yang2023}. Thirdly, we constrained the sample by using astrometric information from \textit{Gaia} DR3 in order to obtain the true membership of the LMC. Still, there was a very small possibility that some foreground contamination was unremoved ($<0.5\%$) based on the radius velocity measurements from both APOGEE and \textit{Gaia} but could be ignored. Fourthly, as mentioned above, due to the observational strategy of APOGEE-2, there is an obvious gap between the $\rm K_S$-TRGB and the faint end of the RSG branch. For example, there are only 49 targets with $11.5 \leq K_S \leq 12.0$ mag and $J-K_S\leq1.0$ in our sample. This is mostly due to that stars fainter than $H=11.8$ mag in the RSG branch have very low priorities to be observed \citep{Neugent2012, Santana2021}. Fortunately, the number of remaining targets are still good enough for us to make the conclusion. Further H-band spectroscopic observations that fill in this gap may help to strengthen our result. Fifthly, the wavelength coverage of APOGEE-2 H-band is only about 0.2 $\mu$m, even though the resolution is high (R$\sim$22,500). It means that other wavelengths may contain important information to distinguish different stellar populations. Sixthly, the major difference between $\chi^2$ and ML occurs at the edges of different stellar populations, e.g., there are 201 and 16 targets are classified as RSGs below the $\rm K_S$-TRGB by $\chi^2$ and ML, respectively. This is mainly due to the blurred boundaries as mentioned in \citet{Yang2020}, for which there is a continuum with similarity and overlapping between these stellar populations in photometry, spectroscopy, and variability. Even if the reference spectra used in $\chi^2$ or the convolutional layers used in ML are built based on the statistics of different populations, it still can not compensate such edge effect. Meanwhile, similar but more serious issue is also found in the EW as shown in Figure~\ref{fig:ew_cmd}. Considerable part of each population is contaminated by others, because of the transition region of corresponding EW. The situation for the line ratios seems better than the EW, but is still suffered from incompleteness (at both upper and lower mass limits) and contamination.

Since stars with initial mass down to about $6~M_\sun$ could evolve to the RSG phase, their destinies would be a great question. The most common fate of RSGs is to explode as hydrogen-rich Type \uppercase\expandafter{\romannumeral2}-P core-collapse supernovae (CCSNe; \citealt{Smartt2009, Smartt2015}). Meanwhile, RSGs may also evolve bluewards and spend short periods of time as yellow supergiant stars (YSGs), blue supergiant stars (BSGs), or even Wolf-Rayet stars (WRs) before the final SN explosion \citep{Ekstrom2012, Meynet2015, Davies2018}. Alternatively, some RSGs may directly collapse into a black hole without the fabulous SN explosion (``failed supernova''; \citealt{Kochanek2008, Adams2017}). Some of them may be also related to the intermediate-luminosity optical transients (ILOTs; \citealt{Berger2009}) However, the fate of those very low mass RSGs is still elusive from both theoretical and observational points of view. It is very interesting to see how do these low mass RSGs end their life, since based on the initial mass function, there will be a much larger population of faint and diverse RSGs than the bright ones.

\begin{figure}
\center
\includegraphics[scale=0.45]{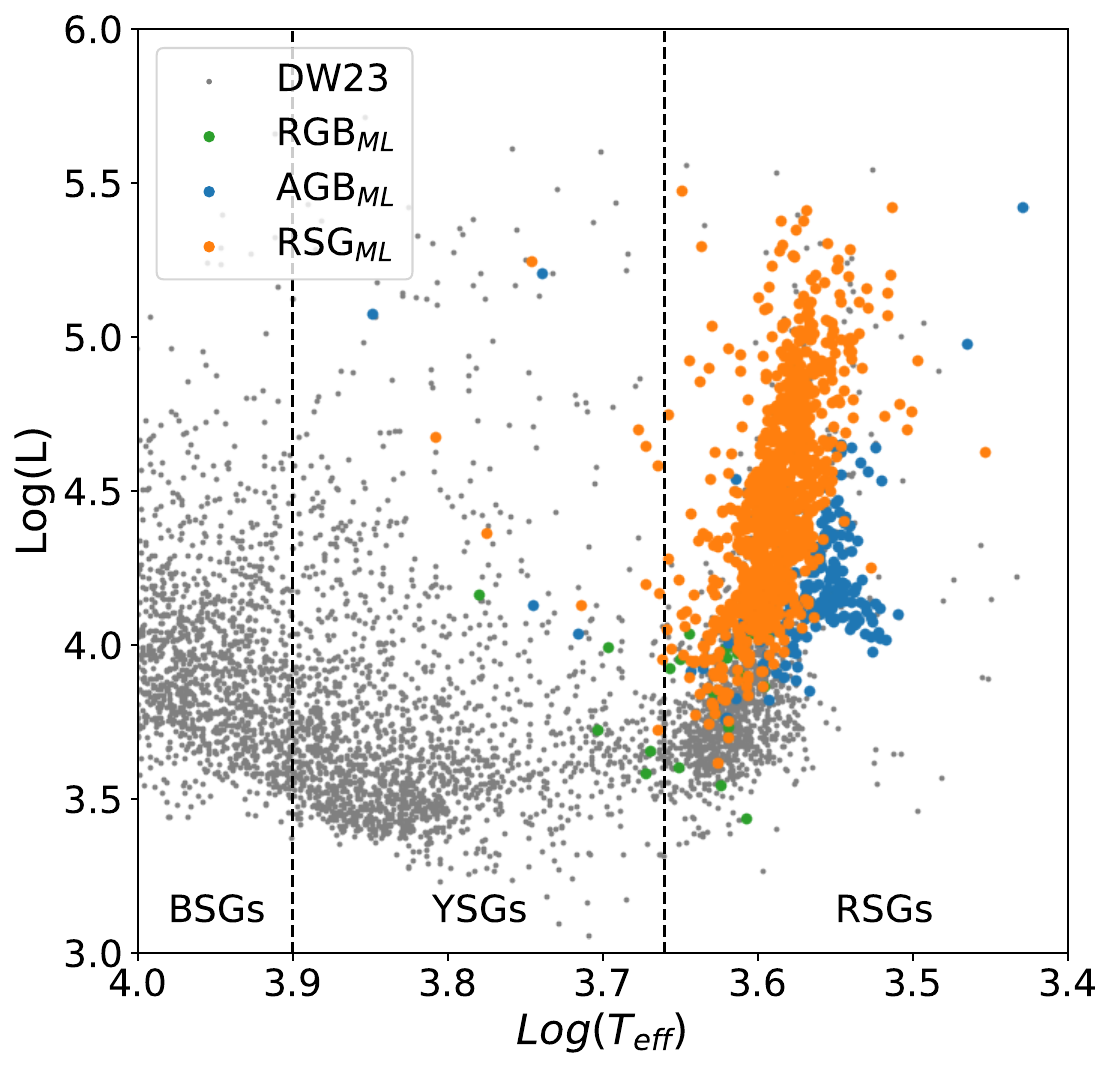}
\caption{H-R diagram of \citet{DornWallenstein2023} (gray dots). There are 1,131 common targets between our study and \citet{DornWallenstein2023}, for which our results of ML method is applied. The simple classification of BSGs, YSGs, and RSGs according to the $T_{\rm eff}$ is adopted from \citet{DornWallenstein2023} shown as the vertical dashed lines.
\label{fig:dw23_apg}}
\end{figure}

Moreover, we also compared our result with \citet{DornWallenstein2023}, who utilized the \textit{Gaia} DR3 photometry and XP spectra to derive luminosity and $T_{\rm eff}$ of $\sim$5000 cool supergiants by using a simple and easily interpretable ML model. The crossmatching between our sample and \citet{DornWallenstein2023} with a search radius of 1'' resulted in 1,131 targets as shown in Figure~\ref{fig:dw23_apg}. Unfortunately, since there was only simple classification of blue supergiants (BSGs), yellow supergiants (YSGs), and RSGs according to the $T_{\rm eff}$ in \citet{DornWallenstein2023}, we were able to tentatively show some general differences. As can be seen from Figure~\ref{fig:dw23_apg}, based on our ML method, there are 924 RSGs, 180 AGBs, and 27 RGBs. The RGBs mostly appear at the low luminosity end, while the AGBs occupy the red end, as expected. Especially, the intermediate-luminosity red ones (previously identified as RSGs around $3.5 < \log T_{\rm eff} < 3.6$ and $4.0 < \log L < 4.5$ from \citealt{Neugent2012}) are most likely the AGBs based on their spectral features from APOGEE.

Besides our main goal, there are also other interesting aspects arise from the study. For example, the decreasing relative intensity of CO bands from AGBs to RSGs then to RGBs is possibly related to the molecular atmosphere in a spherical shells (MOLsphere) in the evolved cool luminous stars \citep{Tsuji2000, Tsuji2008, Ohnaka2011, Hadjara2019}. Such trend seems correlate with both luminosity and $T_{\rm eff}$, that the intensity increases from the hot faint region towards the cool bright part of the CMD. Meanwhile, the $T_{\rm eff}$ may be the most dominate factor, as in the majority cases, targets with highest relative intensities appear at the red end of O-AGBs. This is also true for almost all calculated EWs, including some metal lines of Fe \uppercase\expandafter{\romannumeral1} (at 16089.456\AA and 16506.438\AA) and Ni \uppercase\expandafter{\romannumeral1} (at 15173.415\AA). However, other molecular bands, like OH or CN, may show different tendency, such as the RSGs and RGBs have similar strength of OH lines (see also the Appendix).

\section{Summary}  \label{sec:summary}

In order to investigate the lower mass limit of the RSG population, we assemble a spectroscopic evolved cool star sample in the LMC based on the H-band high resolution spectra from SDSS-IV/APOGEE-2. The initial sample is taken from the APOGEE-2 summative catalog, and then have been cleared by using both the astrometric parameters from \textit{Gaia} DR3 and quality flags from APOGEE-2. We fellow previous studies to separate our sample into three stellar populations of RSGs, AGBs, and RGBs on the 2MASS CMD. A ``genuine'' RSG population is selected as brighter than $K_{\rm S}=10.5$ mag ($L\gtrsim10^{4.0}~L_{\sun}$) by comparing different BCs from the literature. 

The pseudo-continuum normalized and best fit spectra from APOGEE-2 are post-processed in order to removed bad data points. Meanwhile, we find a bunch of stars with abnormal spectra, which are mainly belonged to the C-AGBs. Those targets are removed, which results in 6,602 targets in the final sample including RSGs, O-AGBs, and RGBs. The reference spectra are then built according to the quantile range ($1\%\sim99\%$) of each stellar population. From the reference spectra, it indicates that the cool star populations are sharing very similar spectral characteristics, with the majority of the strong spectral features come from Fe, OH, CN, CO, etc, while the most prominent spectral feature is the weakening of relative intensity of CO bandheads,

We use five different methods, e.g., $\chi^2$, CS, ML, EW, and LR, in order to separate different stellar populations. Among them, The ML and $\chi^2$ (as well as CS) provide the best and relatively consistent prediction of certain population, which is consistent with previous results based only on the photometric data. Meanwhile, EW and LR are less precise due to the edge effect and contamination between different populations. 

According to our analysis, the derived lower mass limit of the RSG population is able to reach to the $\rm K_S$-TRGB ($\sim12.0$ mag), indicating a luminosity as low as about $10^{3.5}~L_{\sun}$. Such low luminosity corresponds to a stellar radius only about $100~R_{\sun}$. Given the mass-luminosity relation of $L/L_\sun =f(M/M_\sun)^3$ with $f\approx15.5\pm3$ and taking into account of the mass loss of faint RSGs up to now, the minimal initial mass of the RSG population would be about $6.1\pm0.4~M_\sun$, which is much lower than the traditional limit of $8~M_\sun$ for the massive stars. Based on our sample, this is the first spectroscopic evidence, indicating that the lower mass limit of RSG population is around $6~M_\sun$. Moreover, the limitations of our studies are also discussed and the fate of such low mass RSGs is still elusive. Future observations focusing on these low mass RSGs are strongly needed, in order to further constrain the stellar evolutionary and supernova models.

\section*{Acknowledgments}

This study has received funding from the National Natural Science Foundation of China (Grant No.12373048, 12133002, 12203025 and 12003046). We acknowledge the science research grants from the China Manned Space Project with No.CMS-CSST-2021-A08. This work is also supported by National Key R\&D Program of China No.2019YFA0405501. S.W. also acknowledge support from the Youth Innovation Promotion Association of the CAS (grant No. 2023065). 

Funding for the Sloan Digital Sky Survey IV has been provided by the Alfred P. Sloan Foundation, the U.S. Department of Energy Office of Science, and the Participating Institutions.

SDSS-IV acknowledges support and resources from the Center for High Performance Computing at the University of Utah. The SDSS website is www.sdss4.org.

SDSS-IV is managed by the Astrophysical Research Consortium for the Participating Institutions of the SDSS Collaboration including the Brazilian Participation Group, the Carnegie Institution for Science, Carnegie Mellon University, Center for Astrophysics | Harvard \& Smithsonian, the Chilean Participation Group, the French Participation Group, Instituto de Astrof\'isica de Canarias, The Johns Hopkins University, Kavli Institute for the Physics and Mathematics of the Universe (IPMU) / University of Tokyo, the Korean Participation Group, Lawrence Berkeley National Laboratory, Leibniz Institut f\"ur Astrophysik Potsdam (AIP),  Max-Planck-Institut f\"ur Astronomie (MPIA Heidelberg), Max-Planck-Institut f\"ur Astrophysik (MPA Garching), Max-Planck-Institut f\"ur Extraterrestrische Physik (MPE), National Astronomical Observatories of China, New Mexico State University, New York University, University of Notre Dame, Observat\'ario Nacional / MCTI, The Ohio State University, Pennsylvania State University, Shanghai Astronomical Observatory, United Kingdom Participation Group, Universidad Nacional Aut\'onoma de M\'exico, University of Arizona, University of Colorado Boulder, University of Oxford, University of Portsmouth, University of Utah, University of Virginia, University of Washington, University of Wisconsin, Vanderbilt University, and Yale University.

This work has made use of data from the European Space Agency (ESA) mission {\it Gaia} (\url{https://www.cosmos.esa.int/gaia}), processed by the {\it Gaia} Data Processing and Analysis Consortium (DPAC, \url{https://www.cosmos.esa.int/web/gaia/dpac/consortium}). Funding for the DPAC has been provided by national institutions, in particular the institutions participating in the {\it Gaia} Multilateral Agreement.

This research has made use of the SIMBAD database and VizieR catalog access tool, operated at CDS, Strasbourg, France.

\software{astropy \citep{Astropy2013, Astropy2018}, 
Tool for OPerations on Catalogues And Tables (TOPCAT; \citealt{Taylor2005})
          }

\appendix

\section{Full diagnostic reference spectra of AGBs, RSGs, and RGBs} \label{sec:ref_spec_appendix}

\begin{figure*}
\center
\includegraphics[scale=0.4]{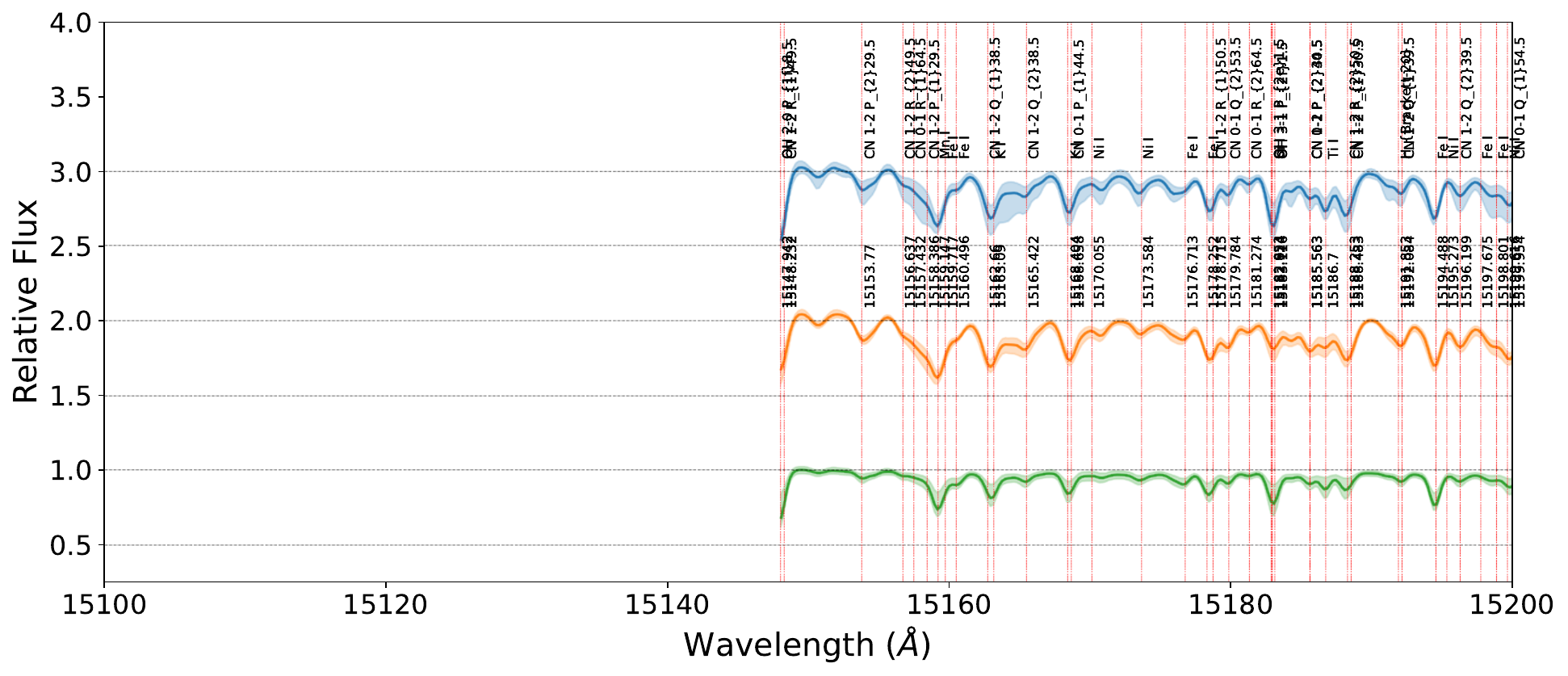}
\includegraphics[scale=0.4]{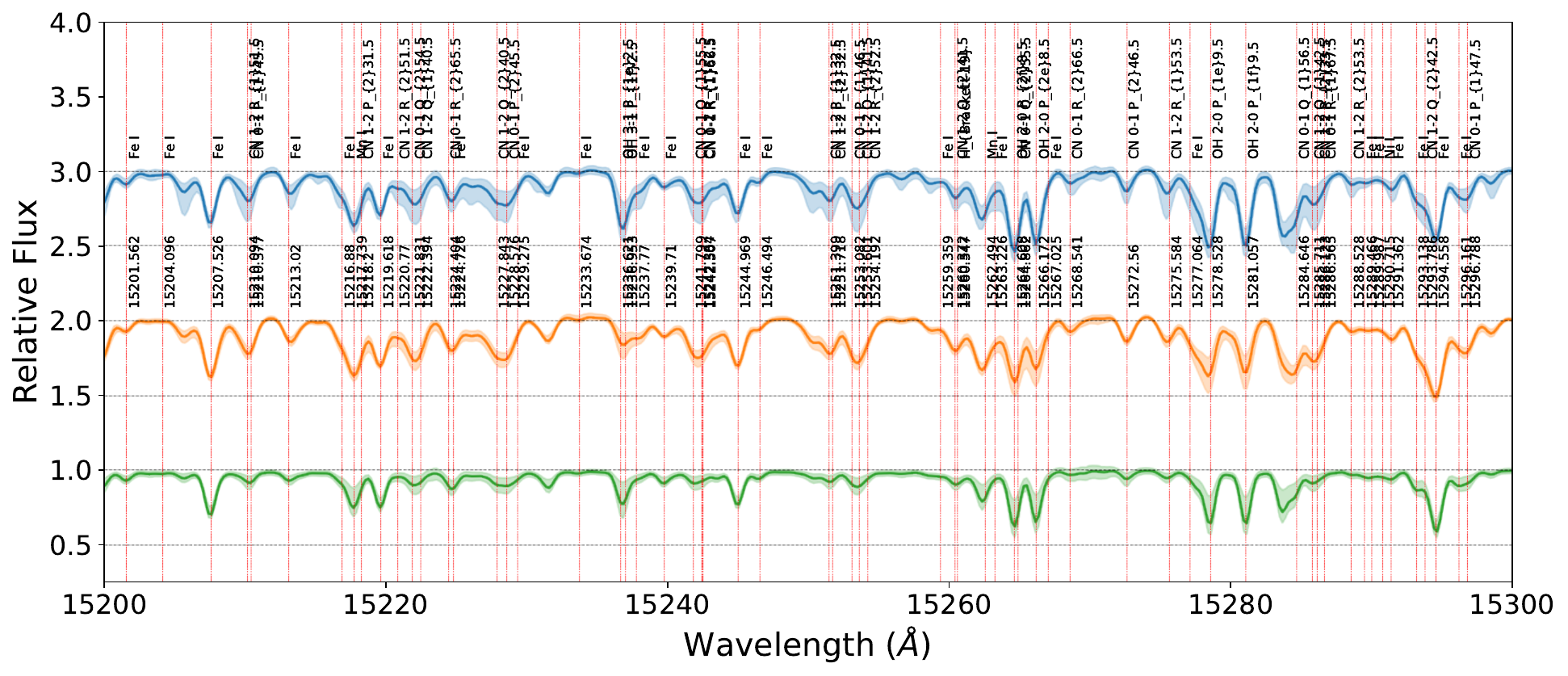}
\includegraphics[scale=0.4]{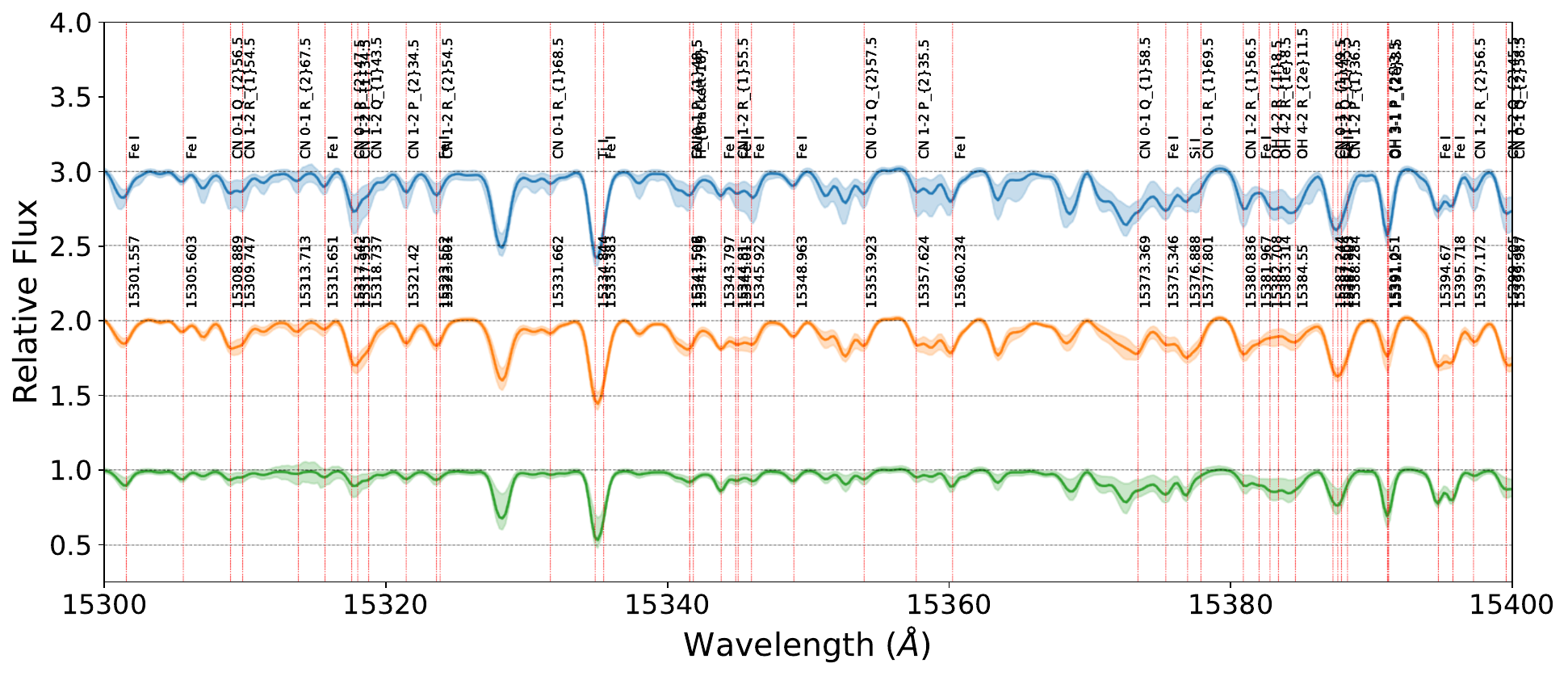}
\includegraphics[scale=0.4]{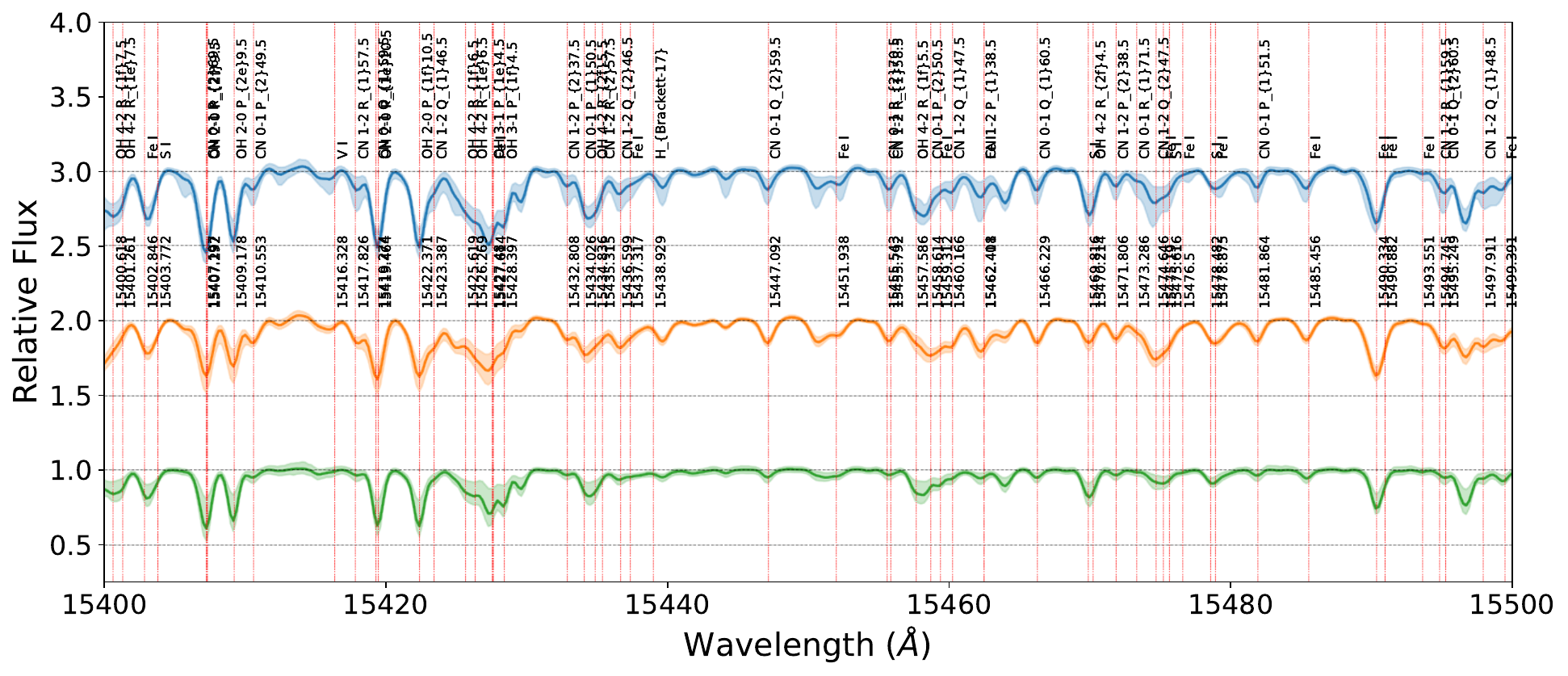}
\caption{The diagnostic reference spectra of AGB, RSG, and RGB between 15,100 and 15,500\AA. Color convention is the same as Figure~\ref{fig:spectra_examples_lmc}.
\label{fig:ref_spec_1}}
\end{figure*}

\begin{figure*}
\center
\includegraphics[scale=0.4]{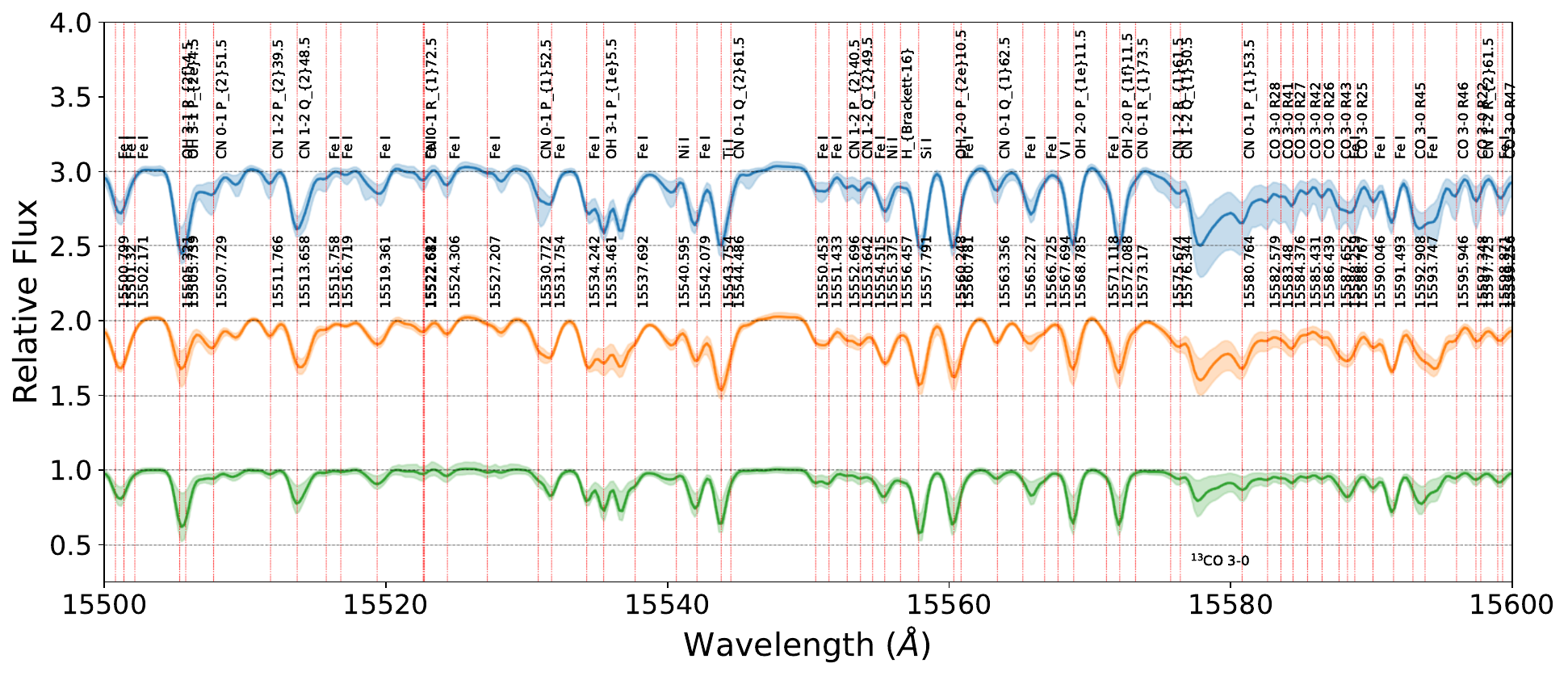}
\includegraphics[scale=0.4]{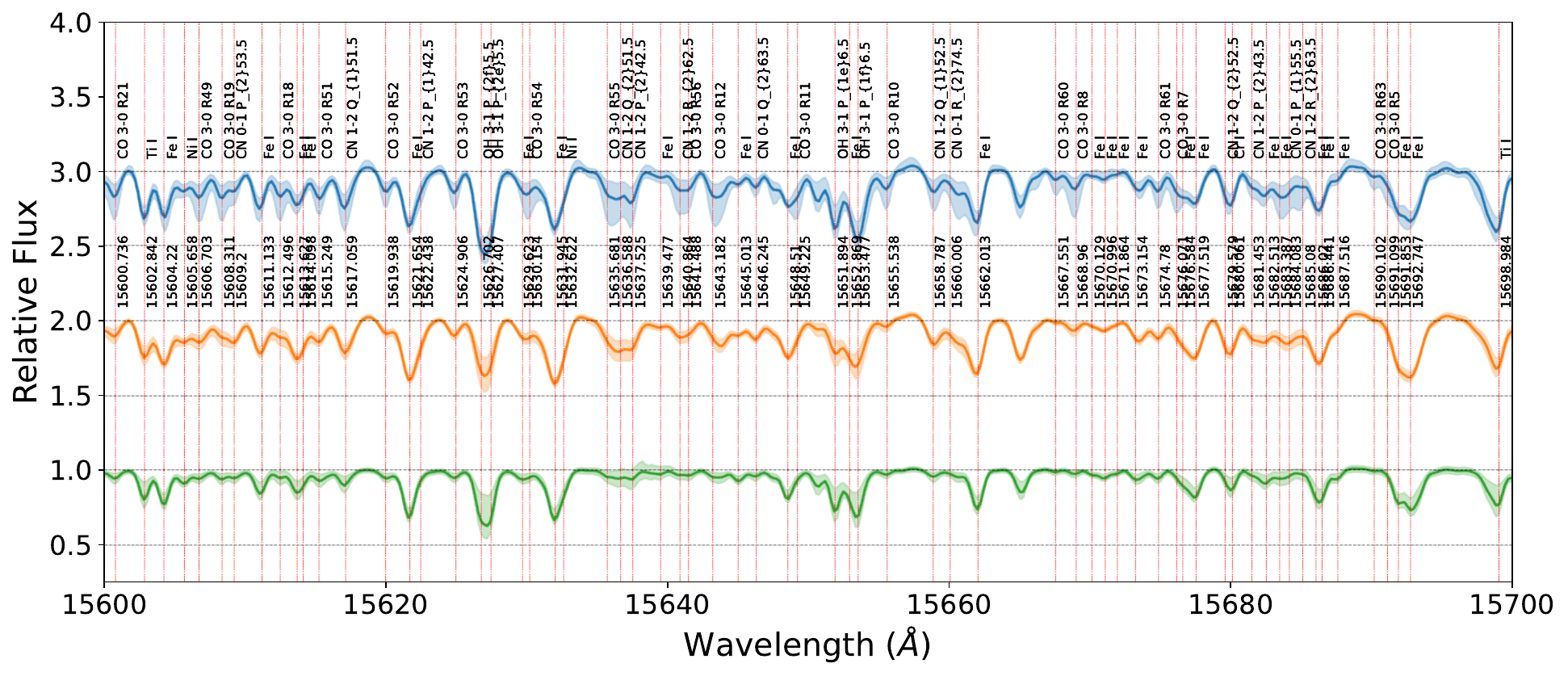}
\includegraphics[scale=0.4]{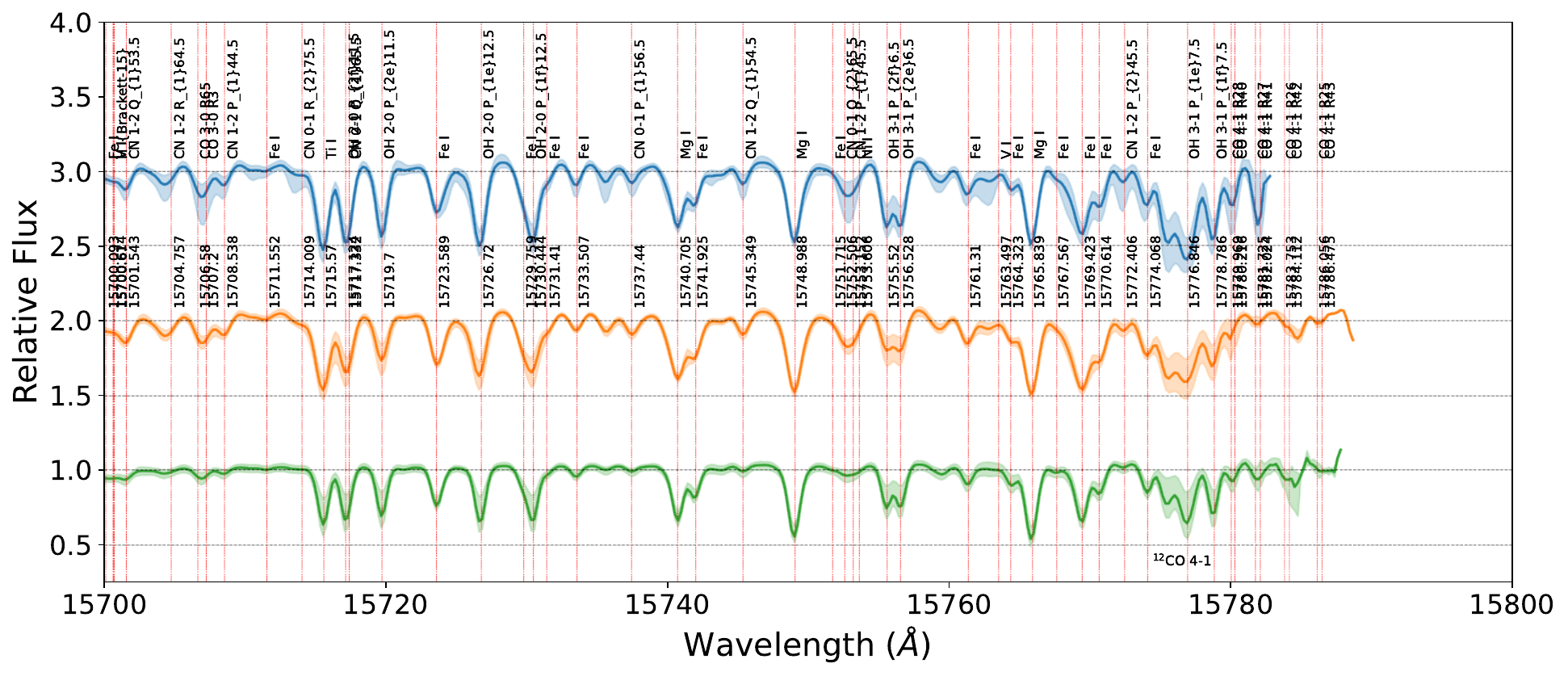}
\includegraphics[scale=0.4]{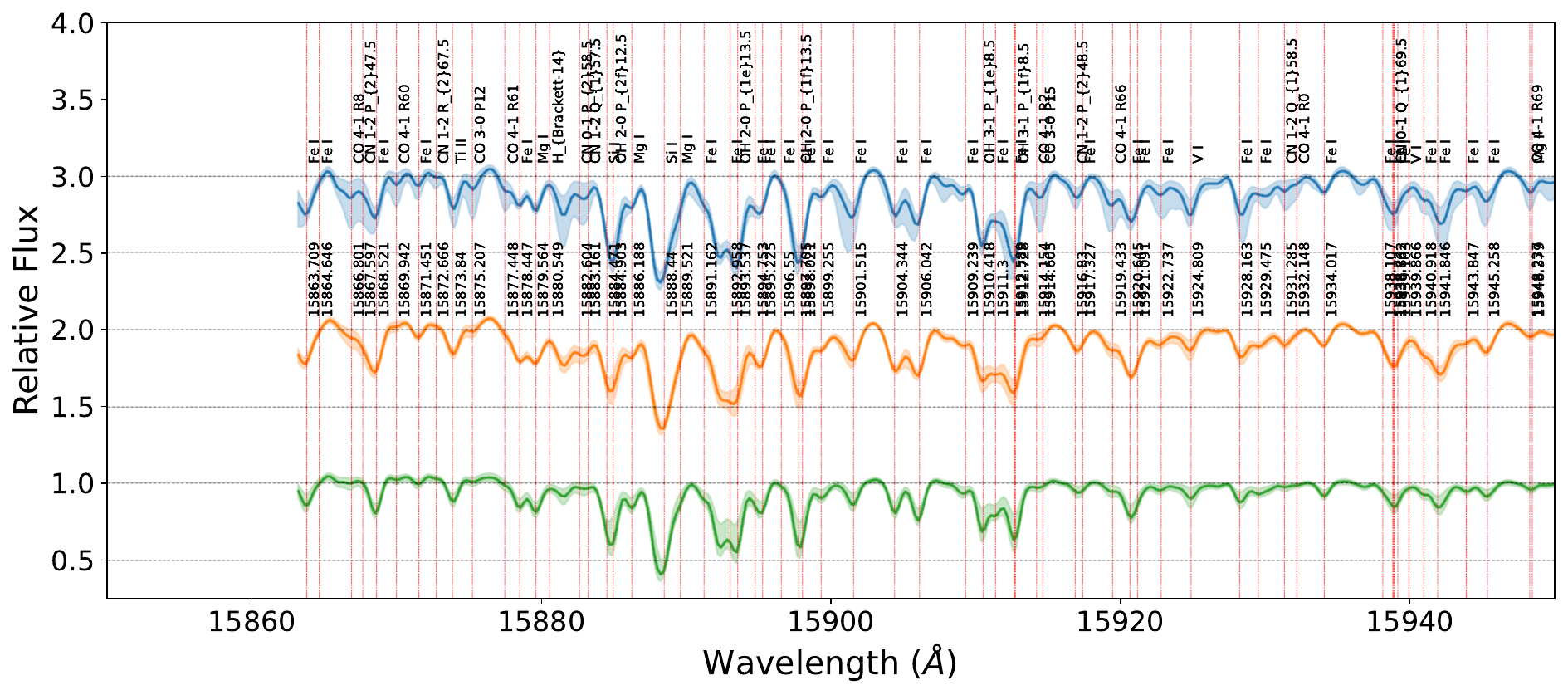}
\caption{Same as Figure~\ref{fig:ref_spec_1}, but for 15,500 to 15,950\AA.
\label{fig:ref_spec_2}}
\end{figure*}

\begin{figure*}
\center
\includegraphics[scale=0.4]{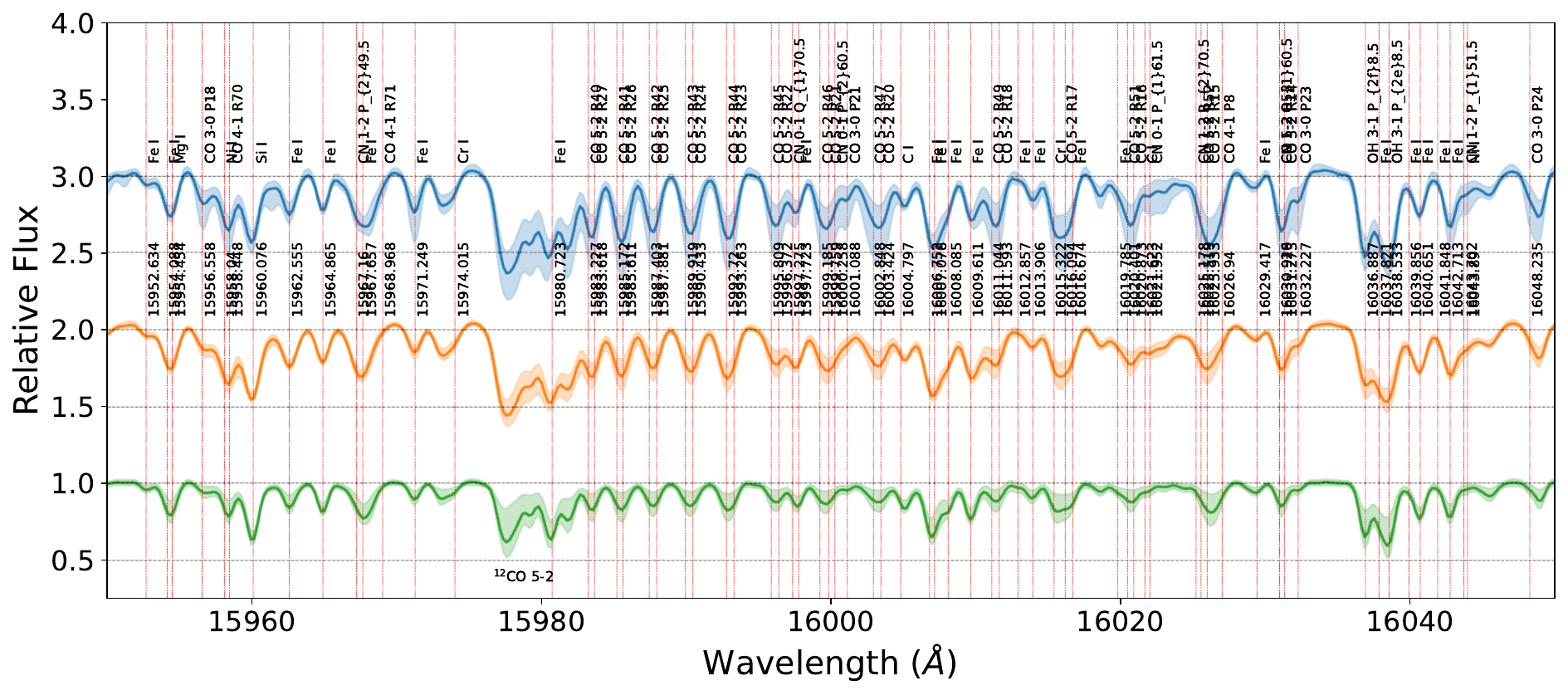}
\includegraphics[scale=0.4]{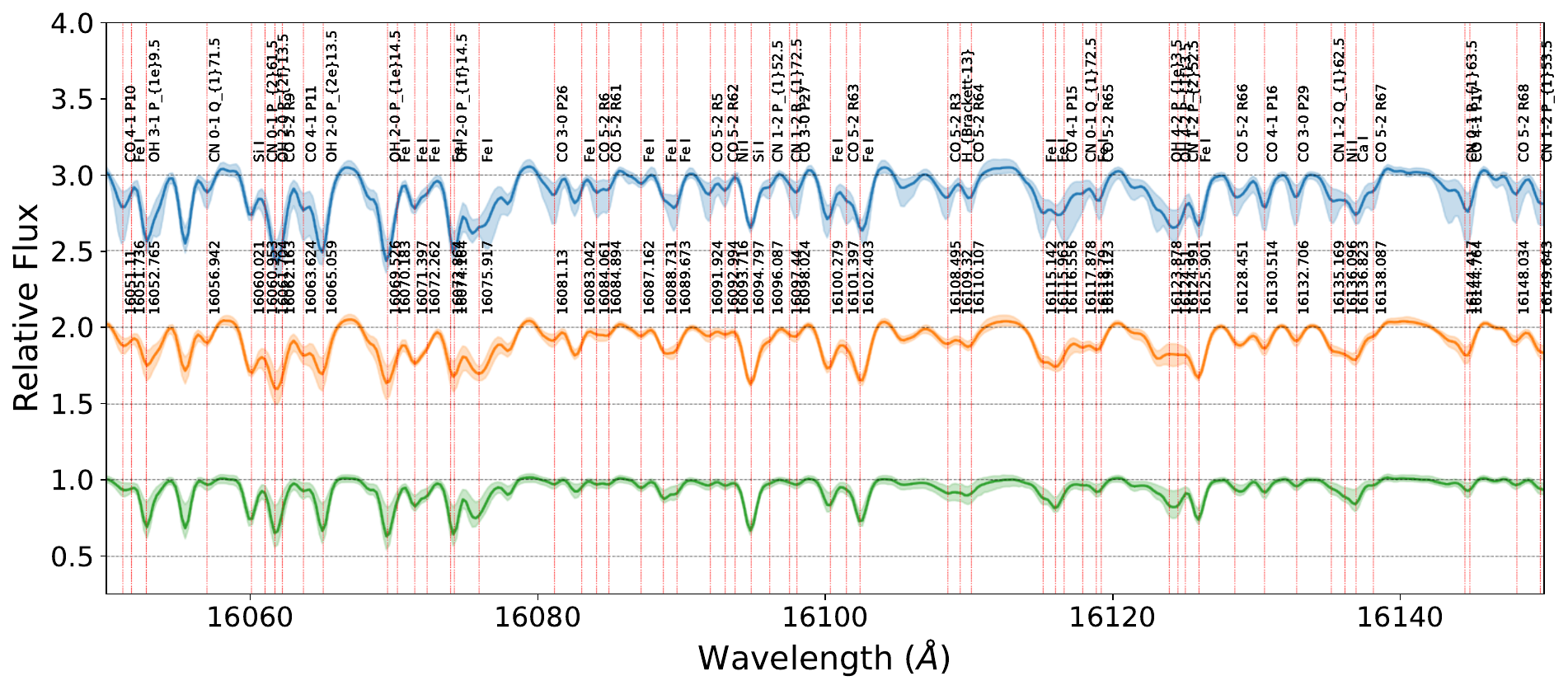}
\includegraphics[scale=0.4]{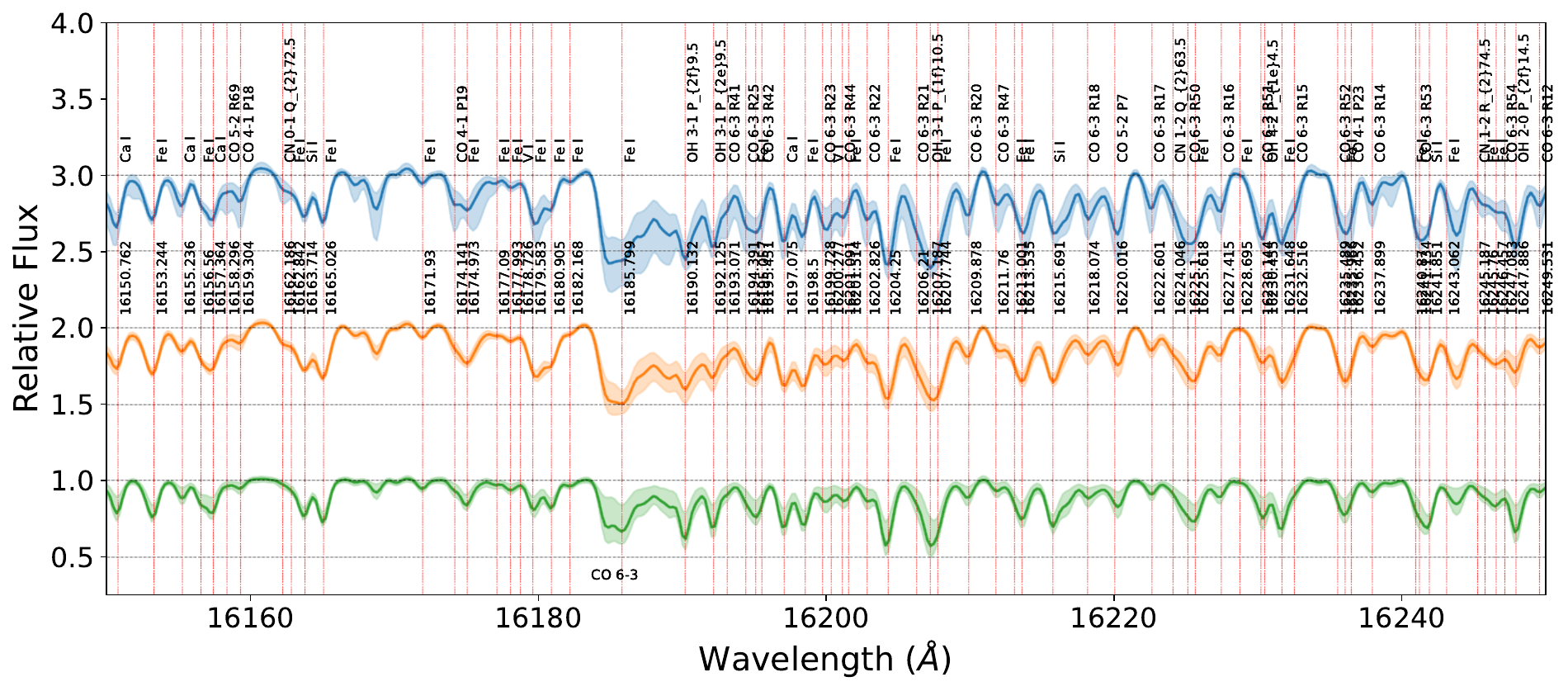}
\includegraphics[scale=0.4]{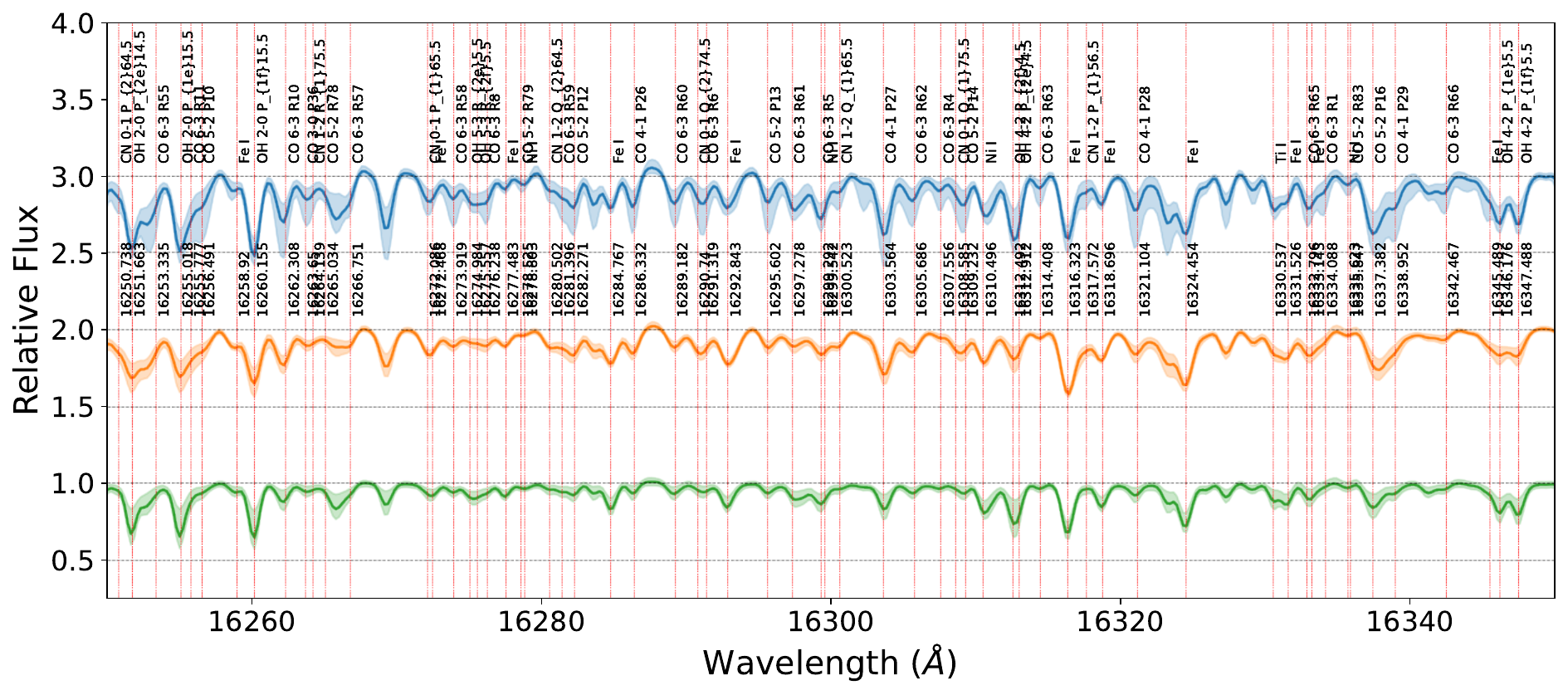}
\caption{Same as Figure~\ref{fig:ref_spec_1}, but for 15,950 to 16,350\AA.
\label{fig:ref_spec_3}}
\end{figure*}

\begin{figure*}
\center
\includegraphics[scale=0.4]{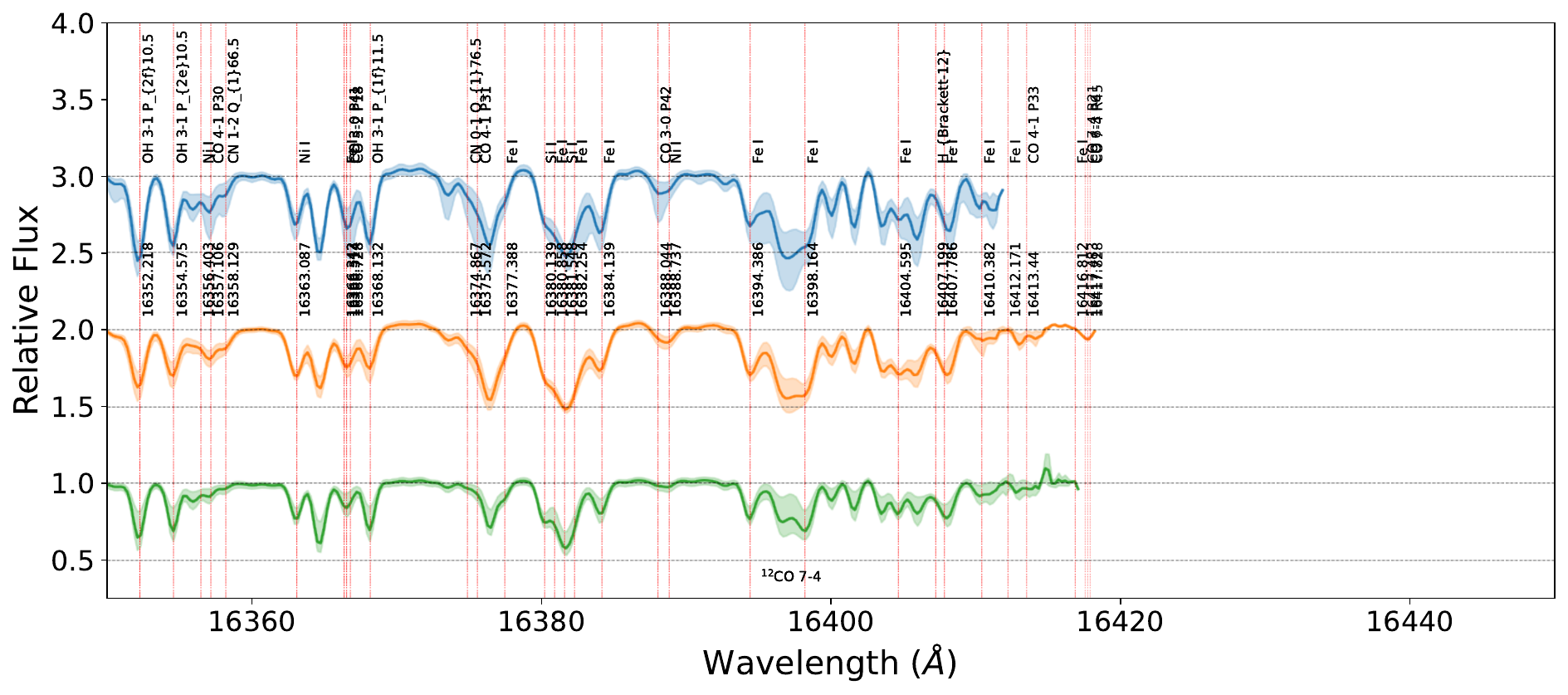}
\includegraphics[scale=0.4]{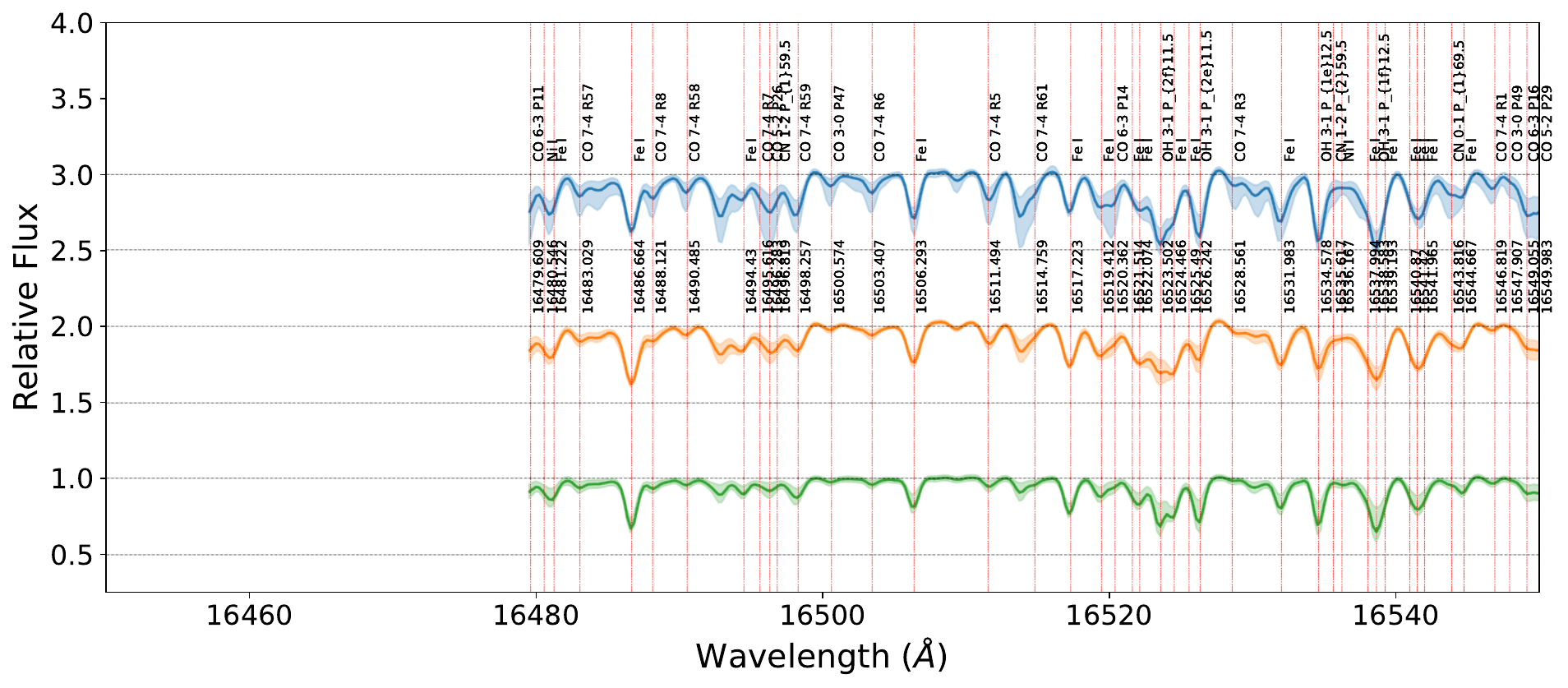}
\includegraphics[scale=0.4]{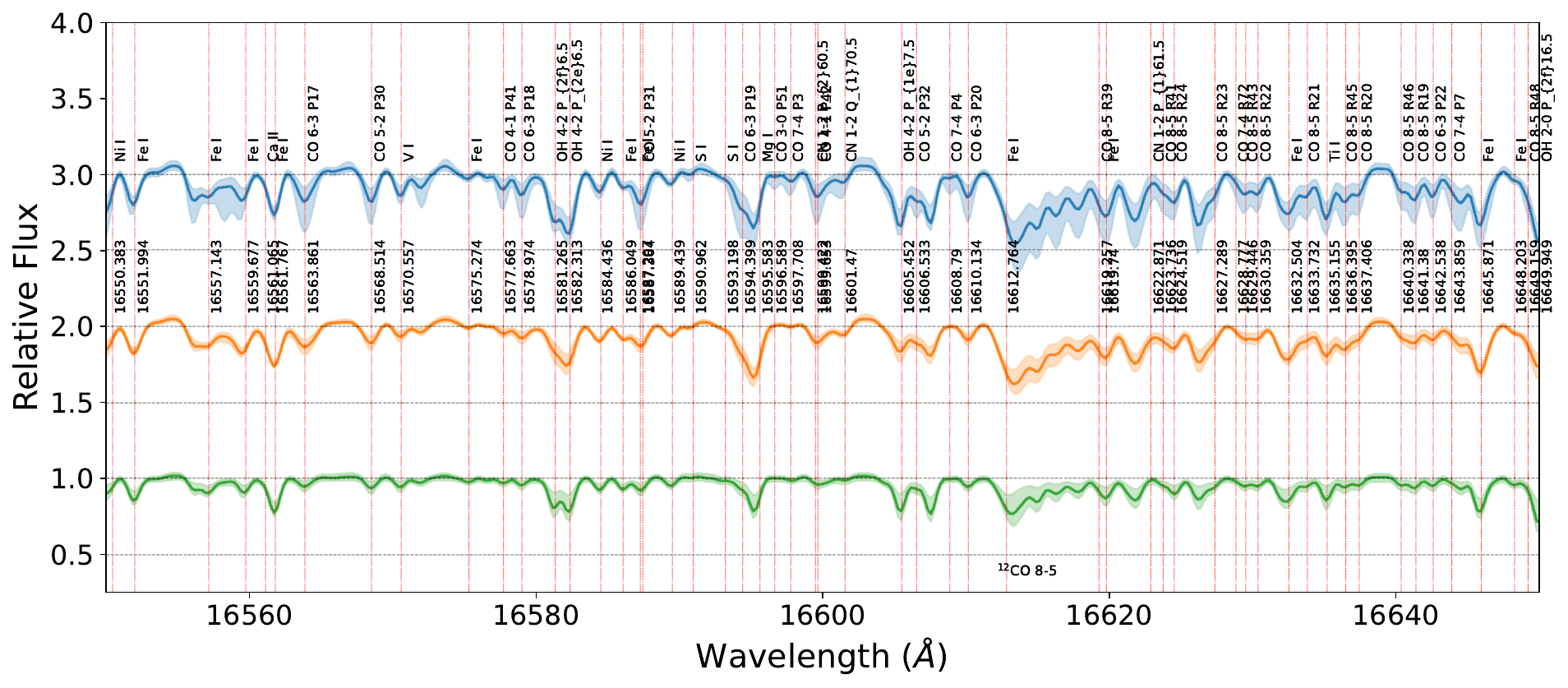}
\includegraphics[scale=0.4]{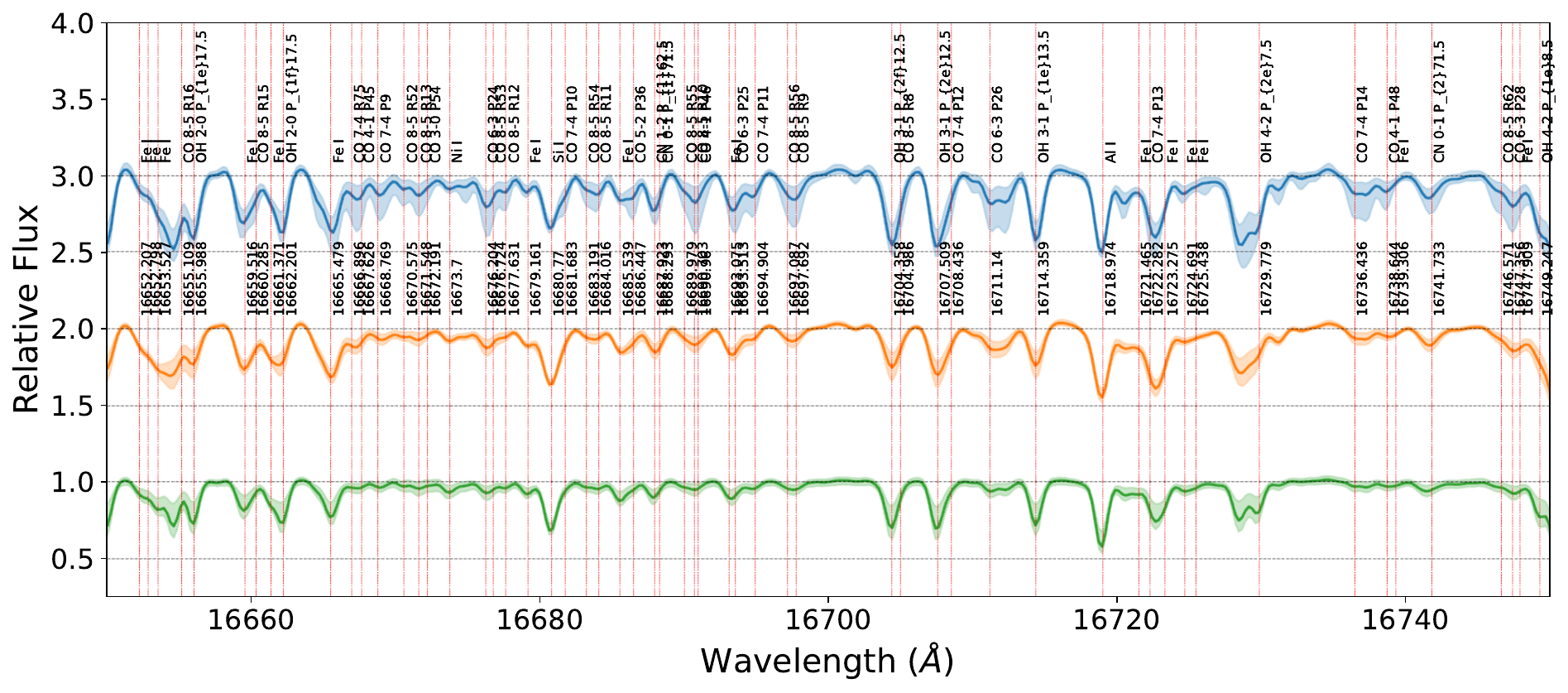}
\caption{Same as Figure~\ref{fig:ref_spec_1}, but for 16,350 to 16,750\AA.
\label{fig:ref_spec_4}}
\end{figure*}

\begin{figure*}
\center
\includegraphics[scale=0.4]{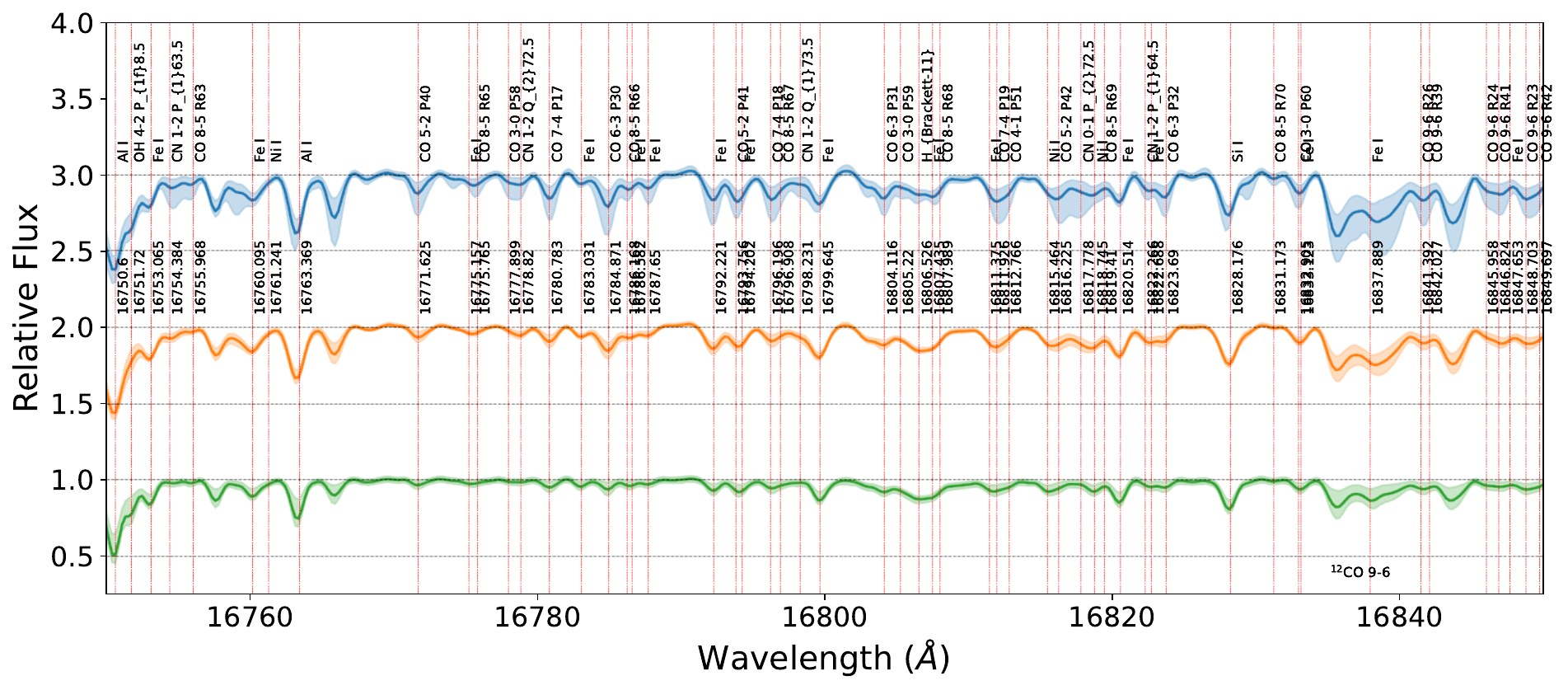}
\includegraphics[scale=0.4]{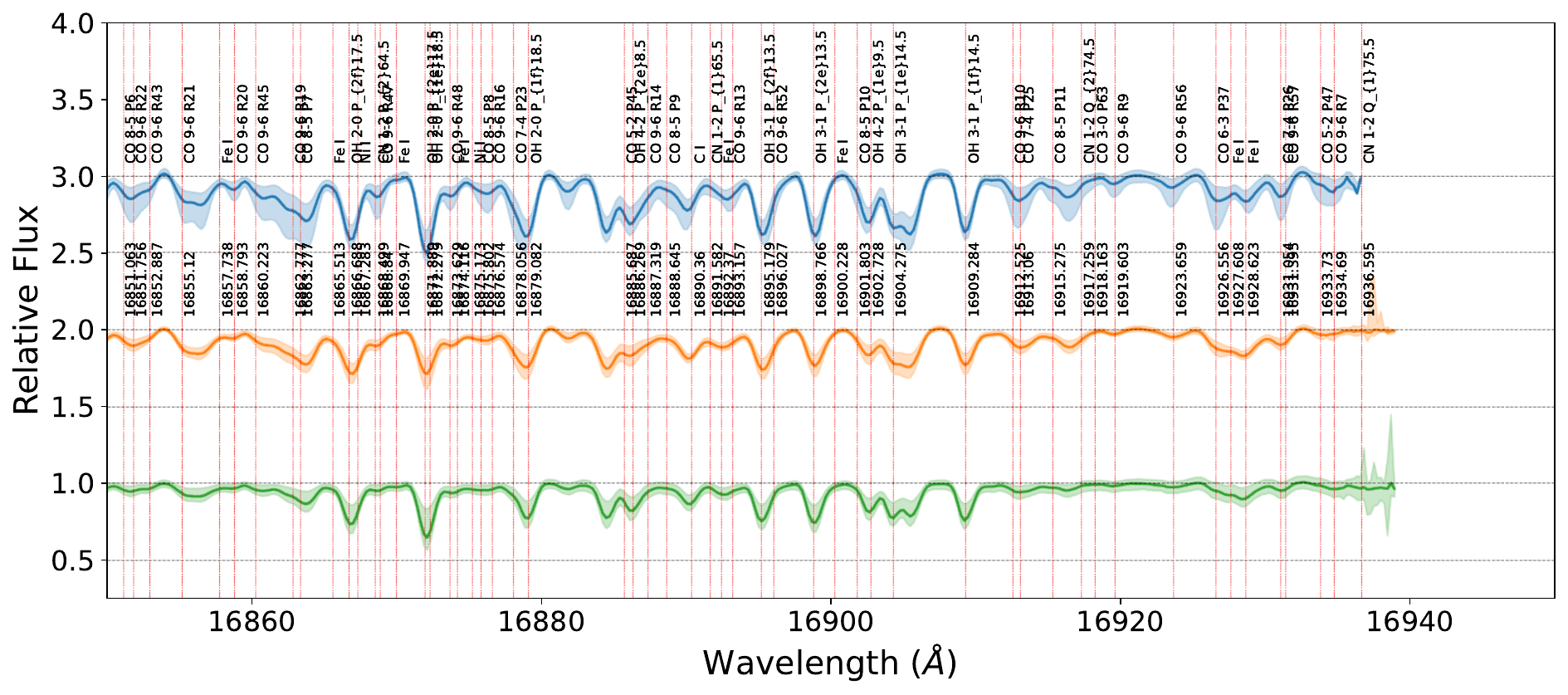}
\caption{Same as Figure~\ref{fig:ref_spec_1}, but for 16,750 to 16,950\AA.
\label{fig:ref_spec_5}}
\end{figure*}

Figure~\ref{fig:ref_spec_1}, Figure~\ref{fig:ref_spec_2}, Figure~\ref{fig:ref_spec_3}, Figure~\ref{fig:ref_spec_4}, and Figure~\ref{fig:ref_spec_5} show the full diagnostic reference spectra of AGBs, RSGs, and RGBs from APOGEE. The identification of each spectral line is based on the atlas of spectral lines of APOGEE. 

\section{Zoomed-in spectral lines between AGBs, RSGs, and RGBs} \label{sec:speclines_appendix}

\begin{figure*}
\center
\includegraphics[scale=0.3]{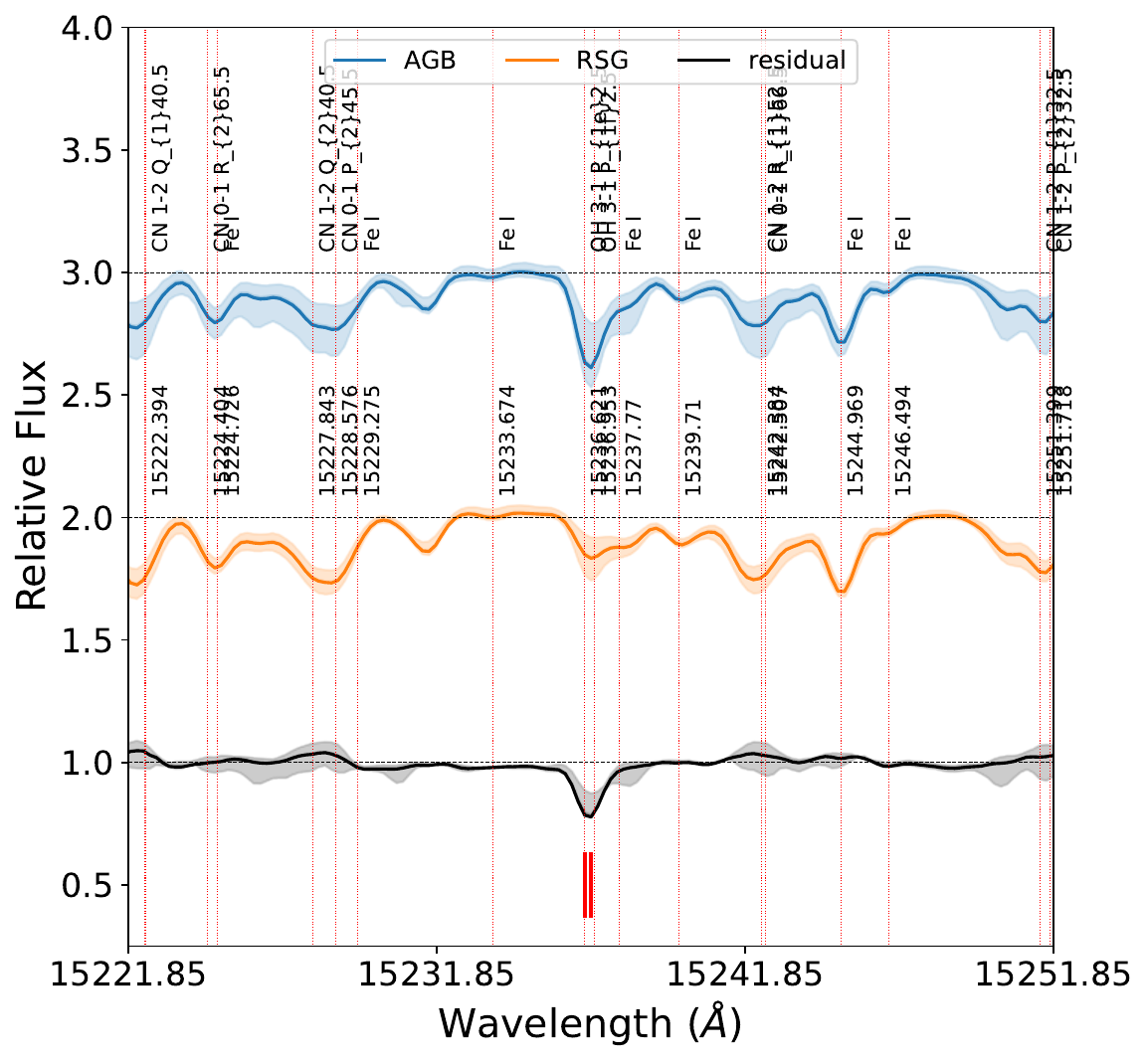}
\includegraphics[scale=0.3]{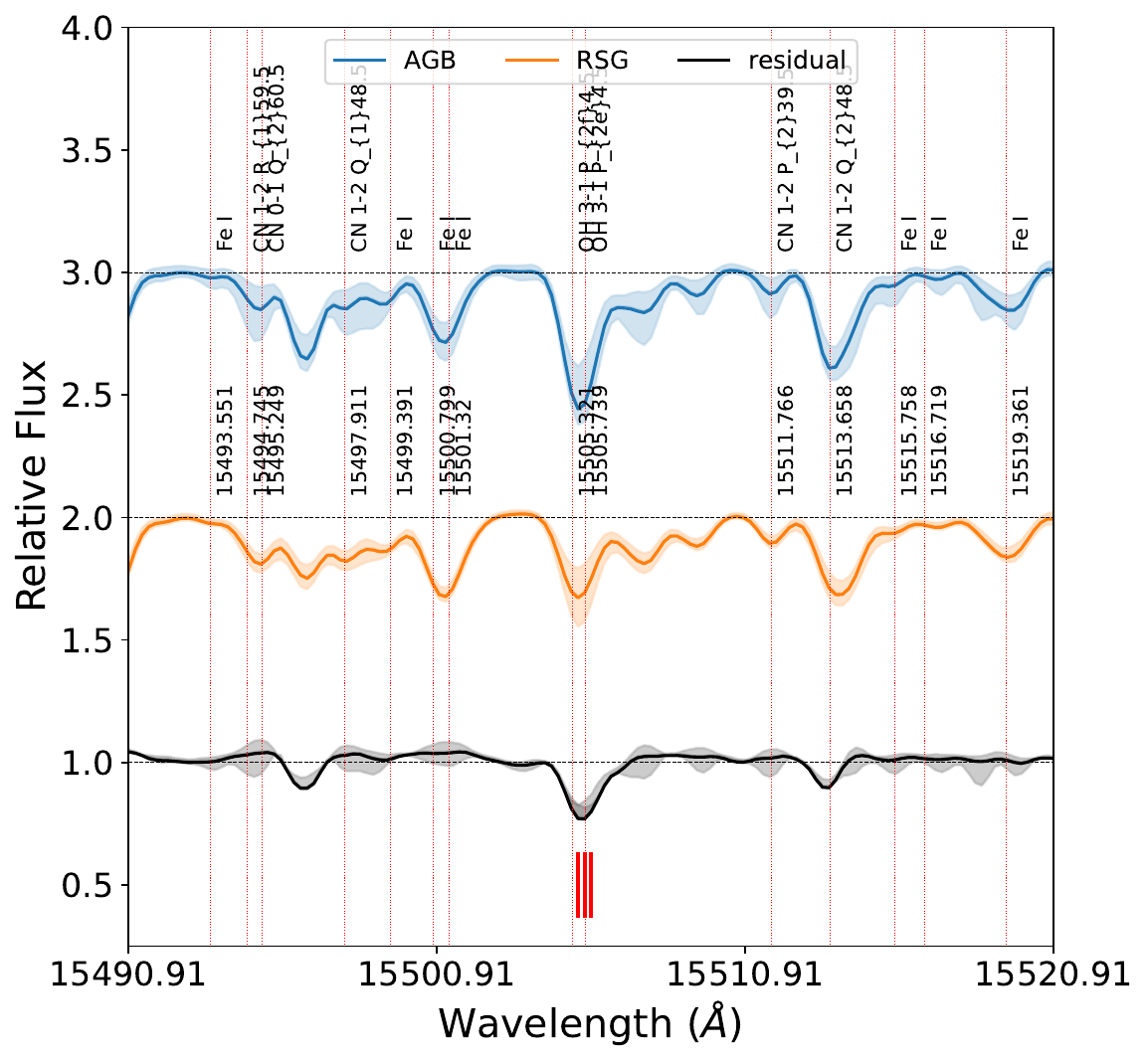}
\includegraphics[scale=0.3]{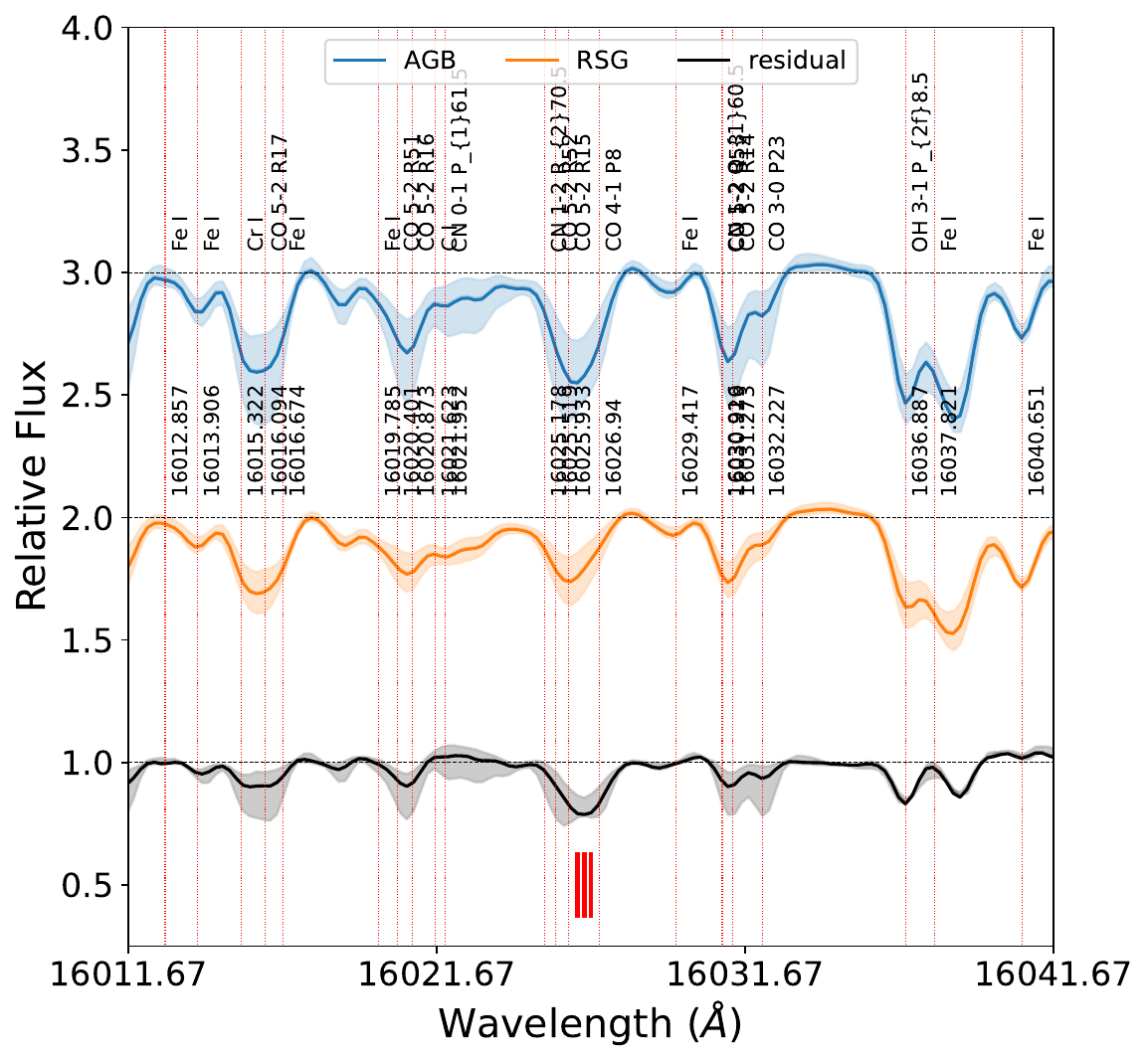}
\includegraphics[scale=0.3]{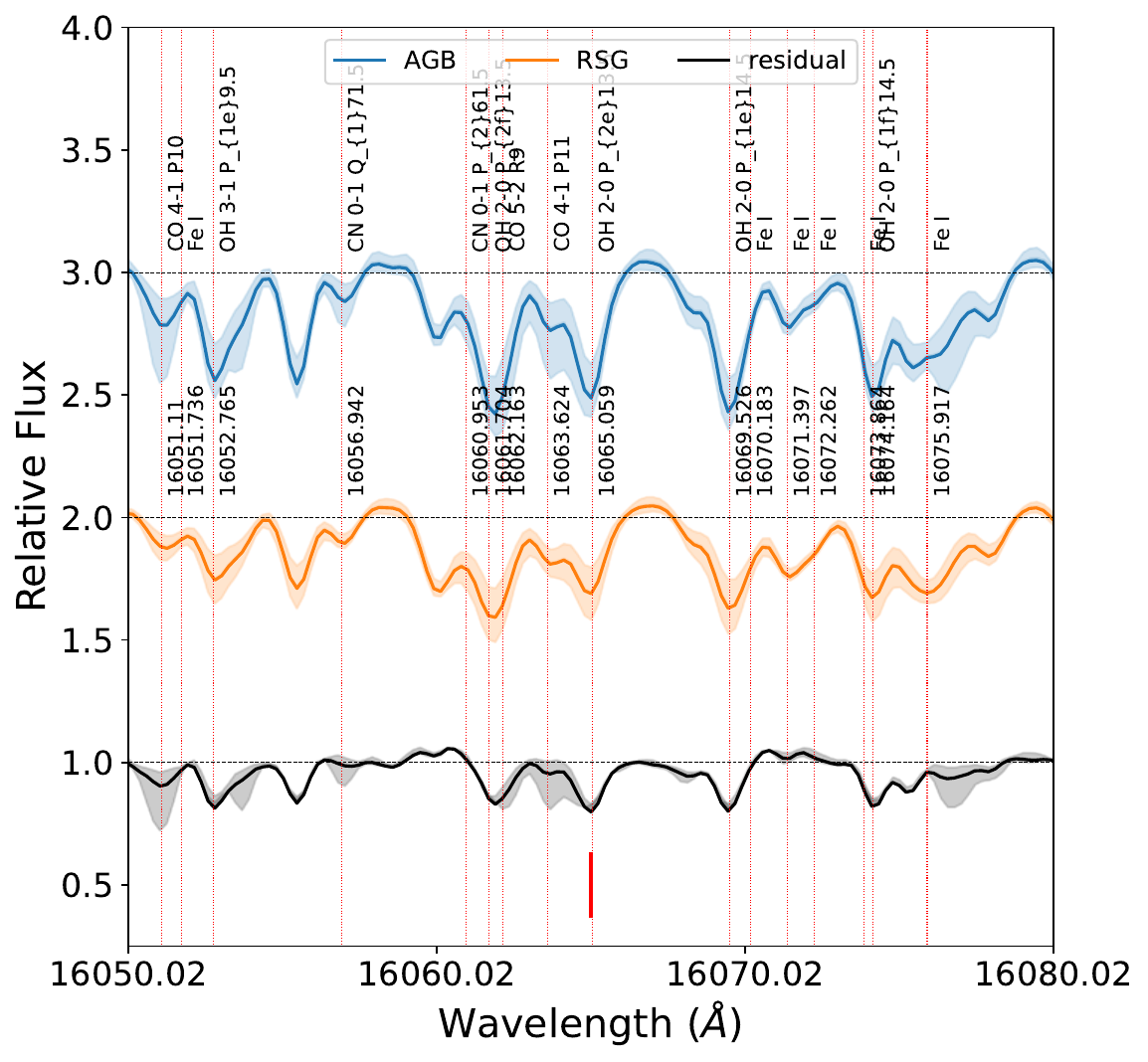}
\includegraphics[scale=0.3]{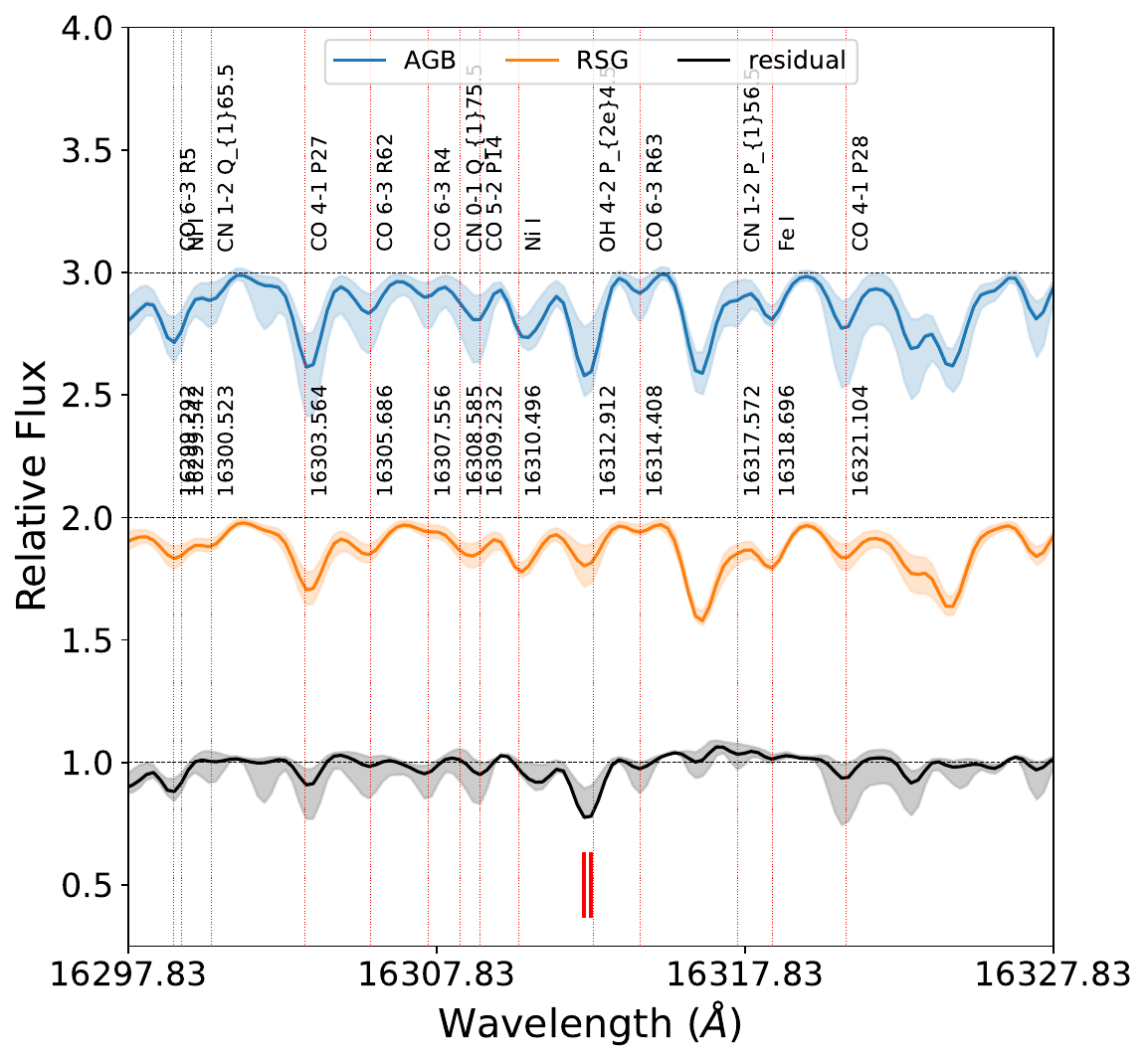}
\includegraphics[scale=0.3]{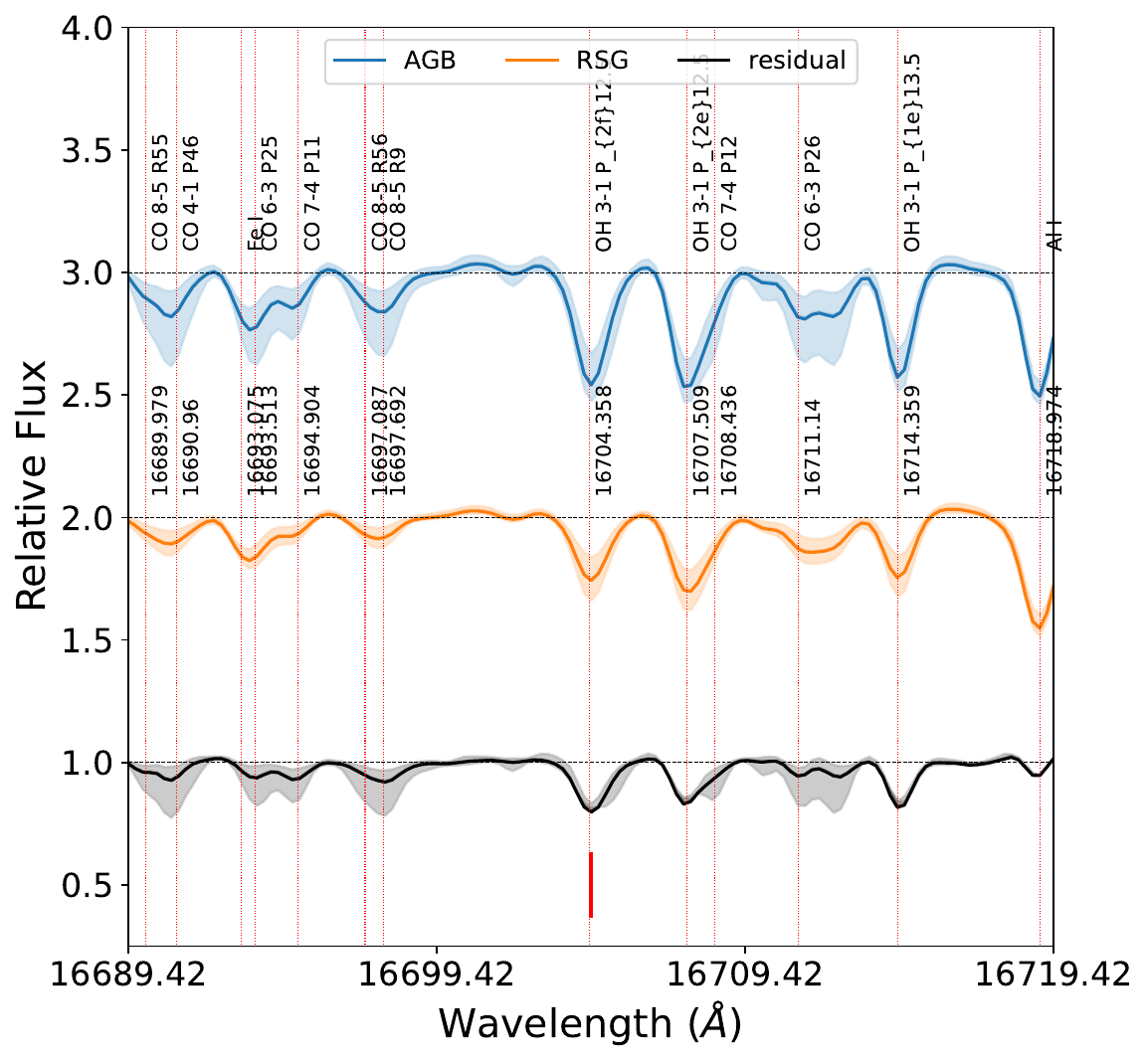}
\includegraphics[scale=0.3]{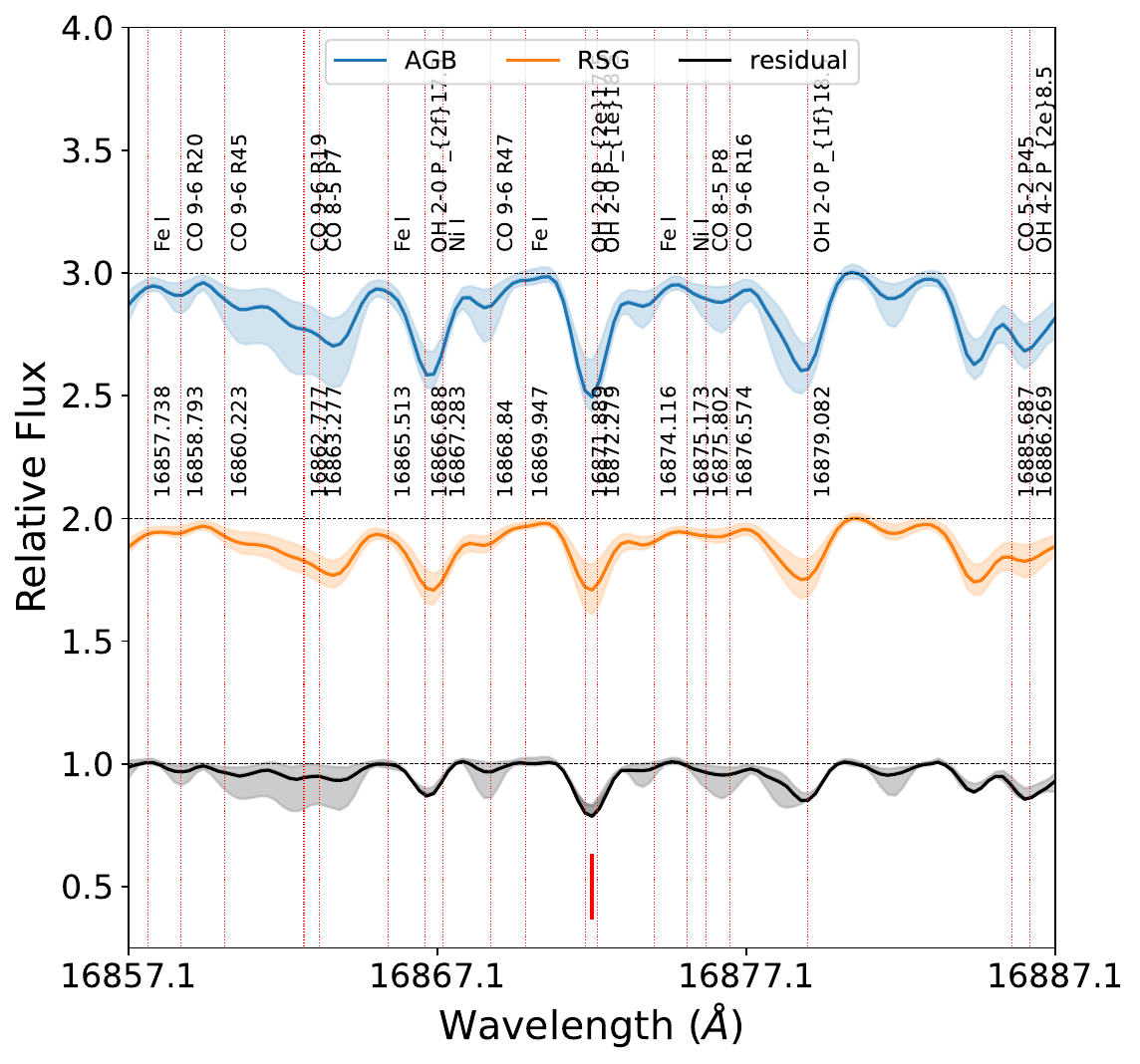}
\caption{Zoomed-in regions of strong spectral features between AGBs and RSGs. The red sticks mark the wavelengths with residual larger than 0.2.
\label{fig:spectral_lines1}}
\end{figure*}

\begin{figure*}
\center
\includegraphics[scale=0.3]{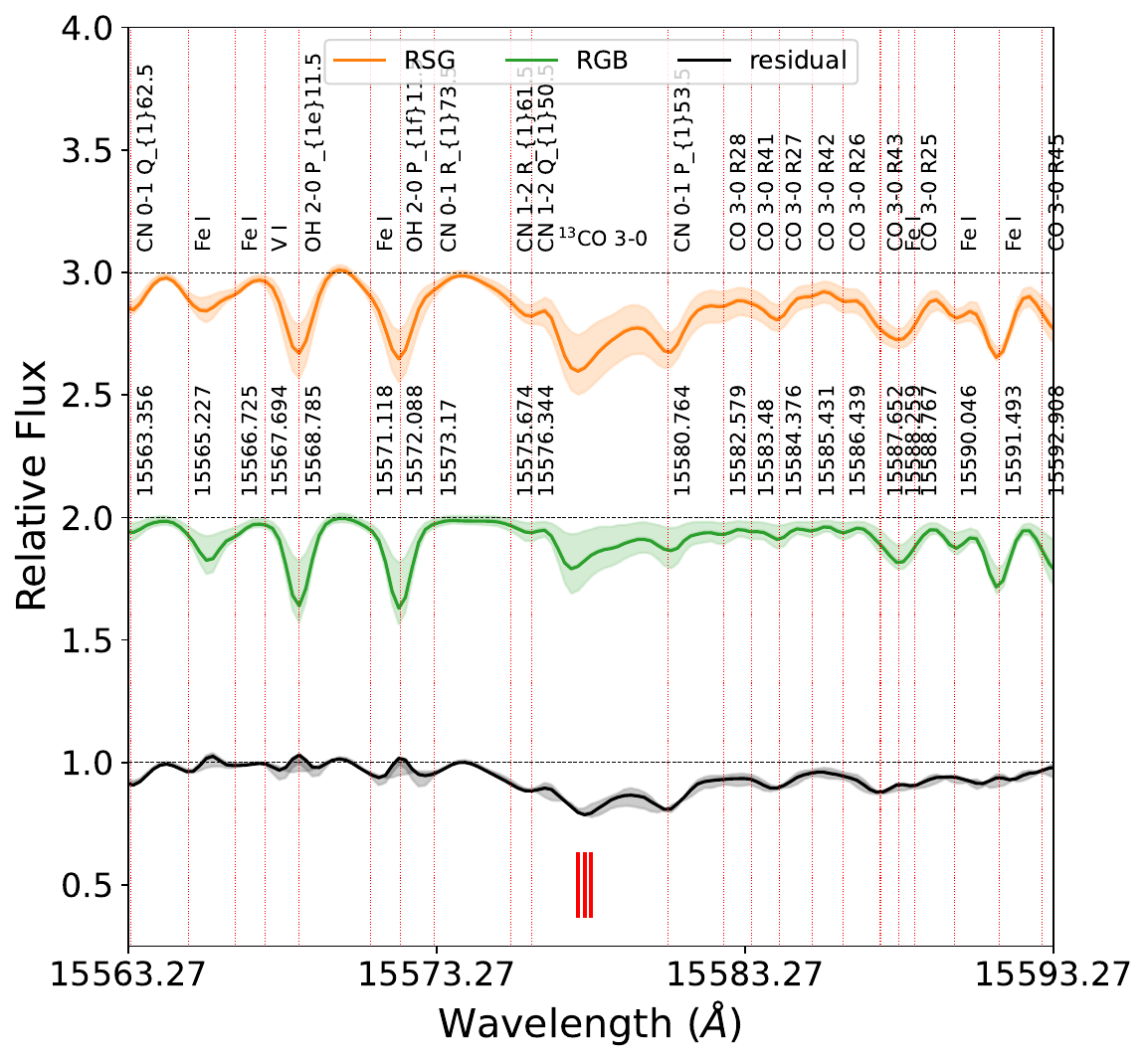}
\includegraphics[scale=0.3]{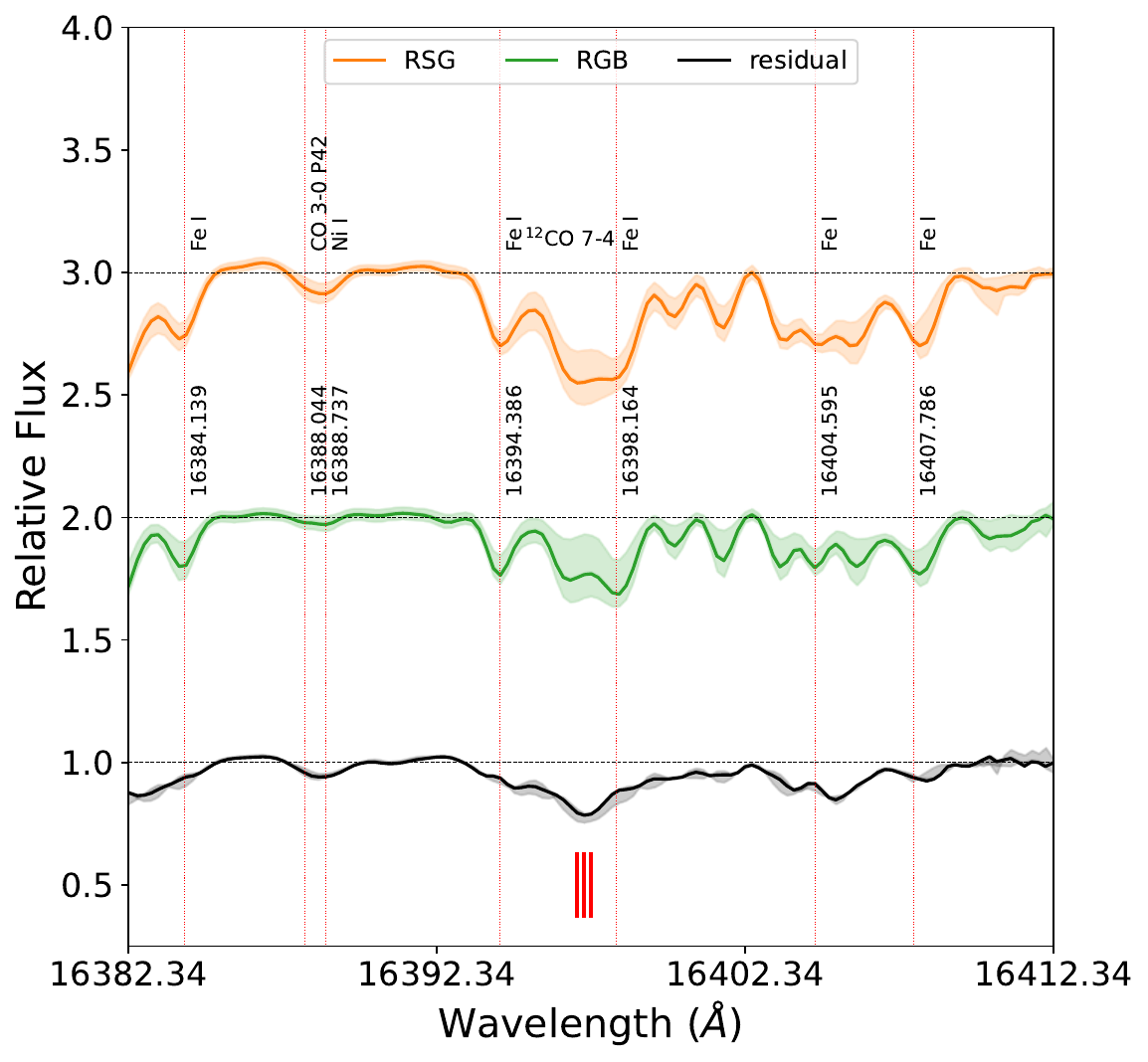}
\caption{Same as Figure~\ref{fig:spectral_lines2}, but for RSGs and RGBs.
\label{fig:spectral_lines2}}
\end{figure*}

\begin{figure*}
\center
\includegraphics[scale=0.3]{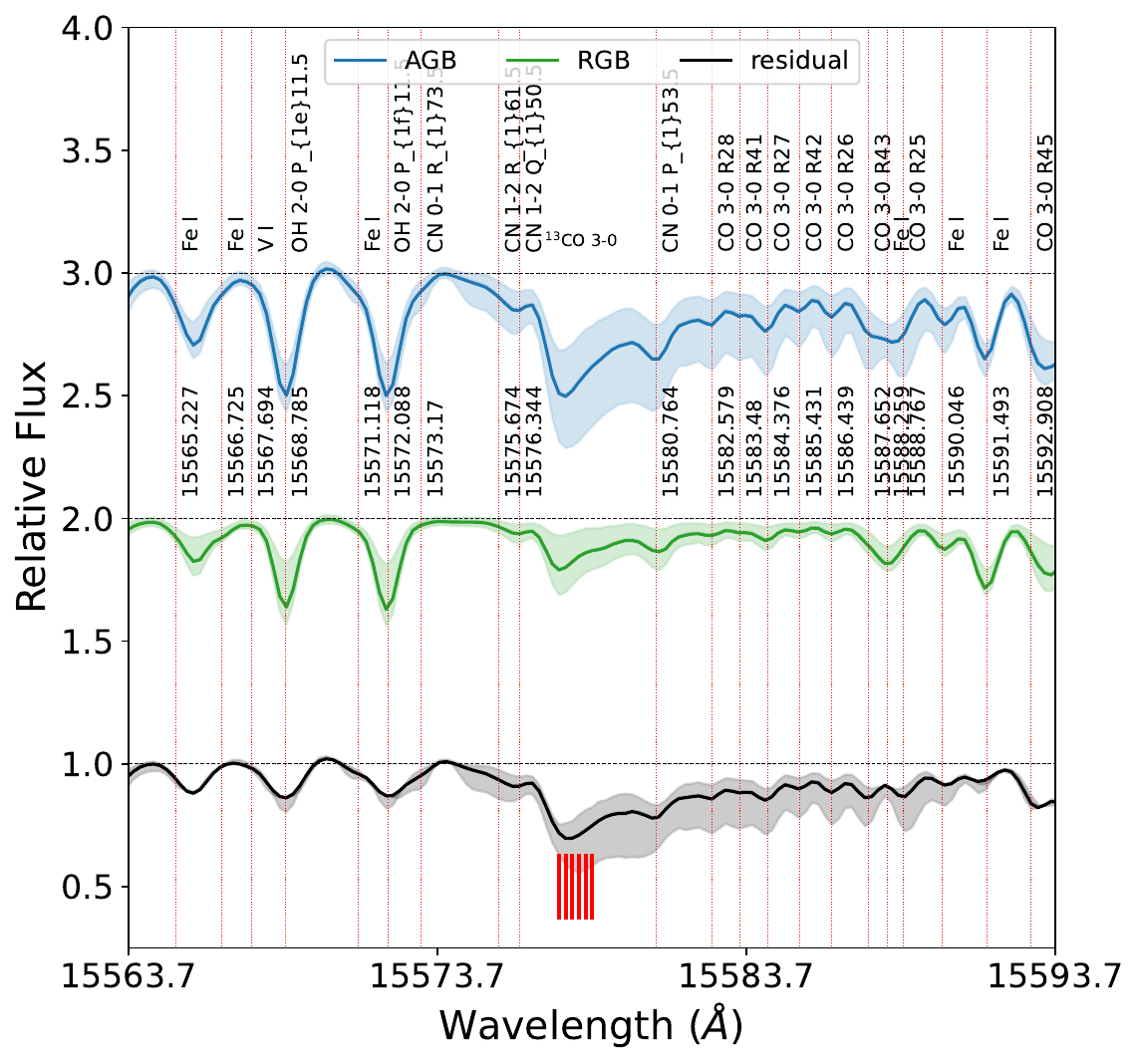}
\includegraphics[scale=0.3]{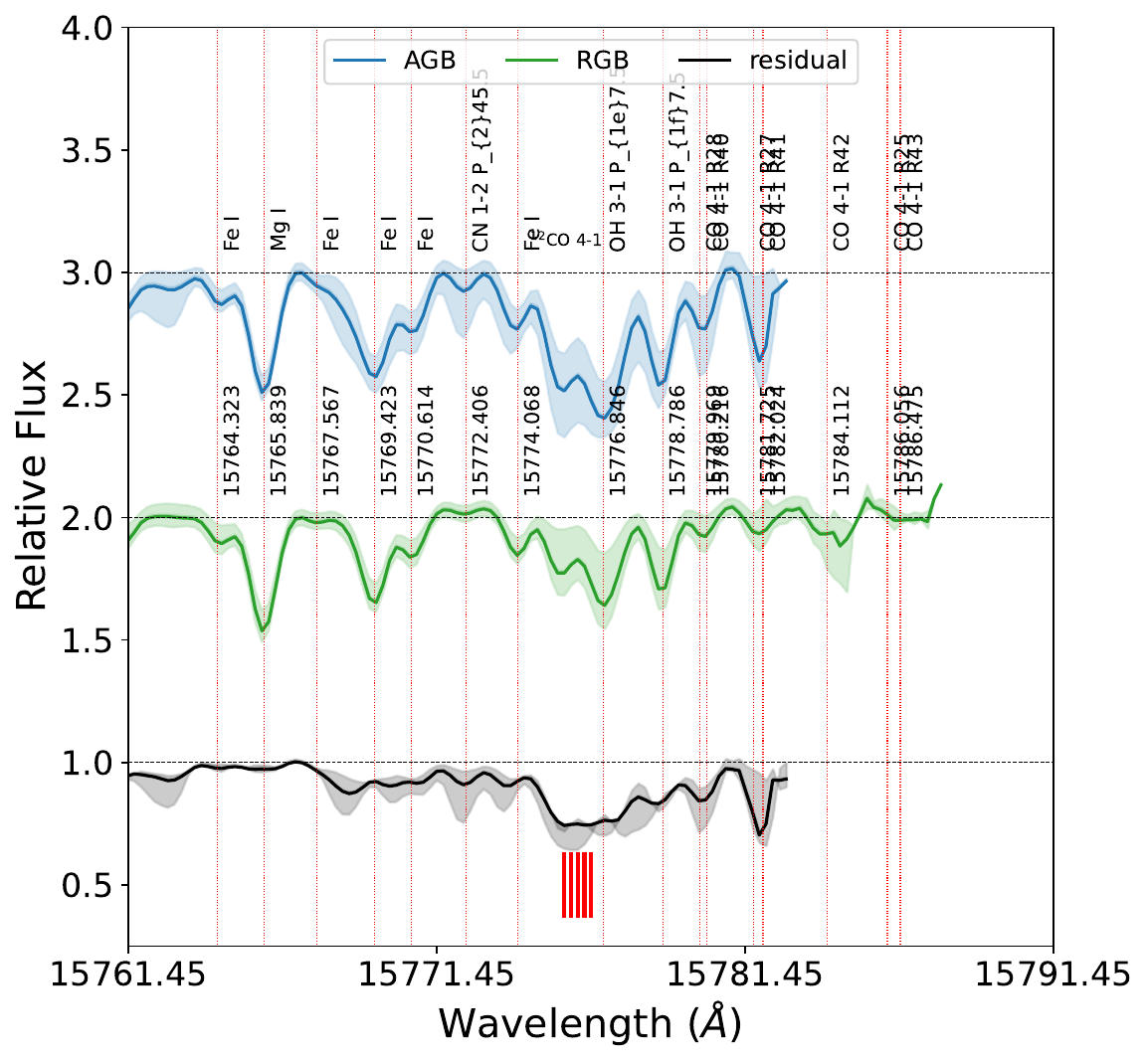}
\includegraphics[scale=0.3]{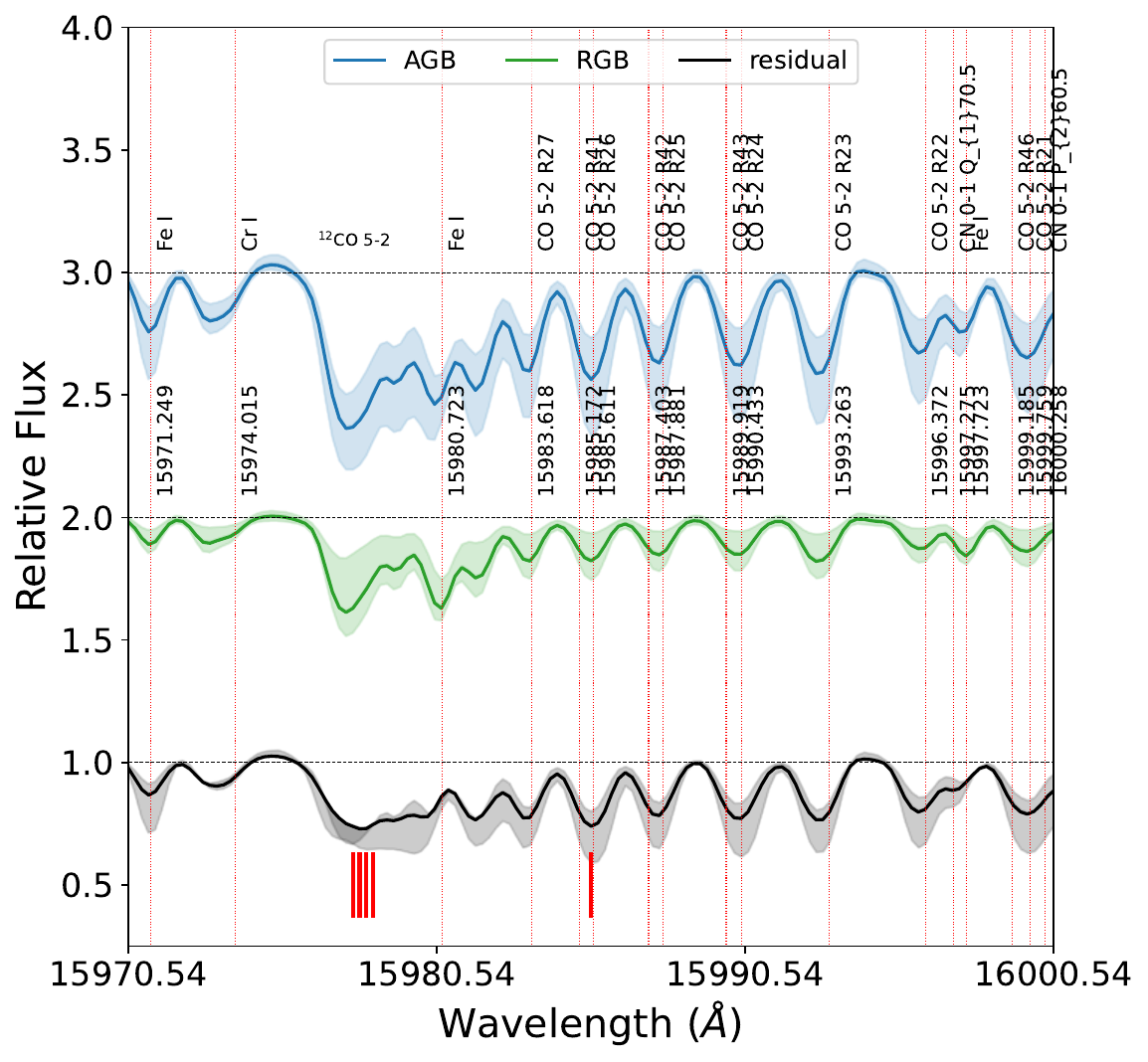}
\includegraphics[scale=0.3]{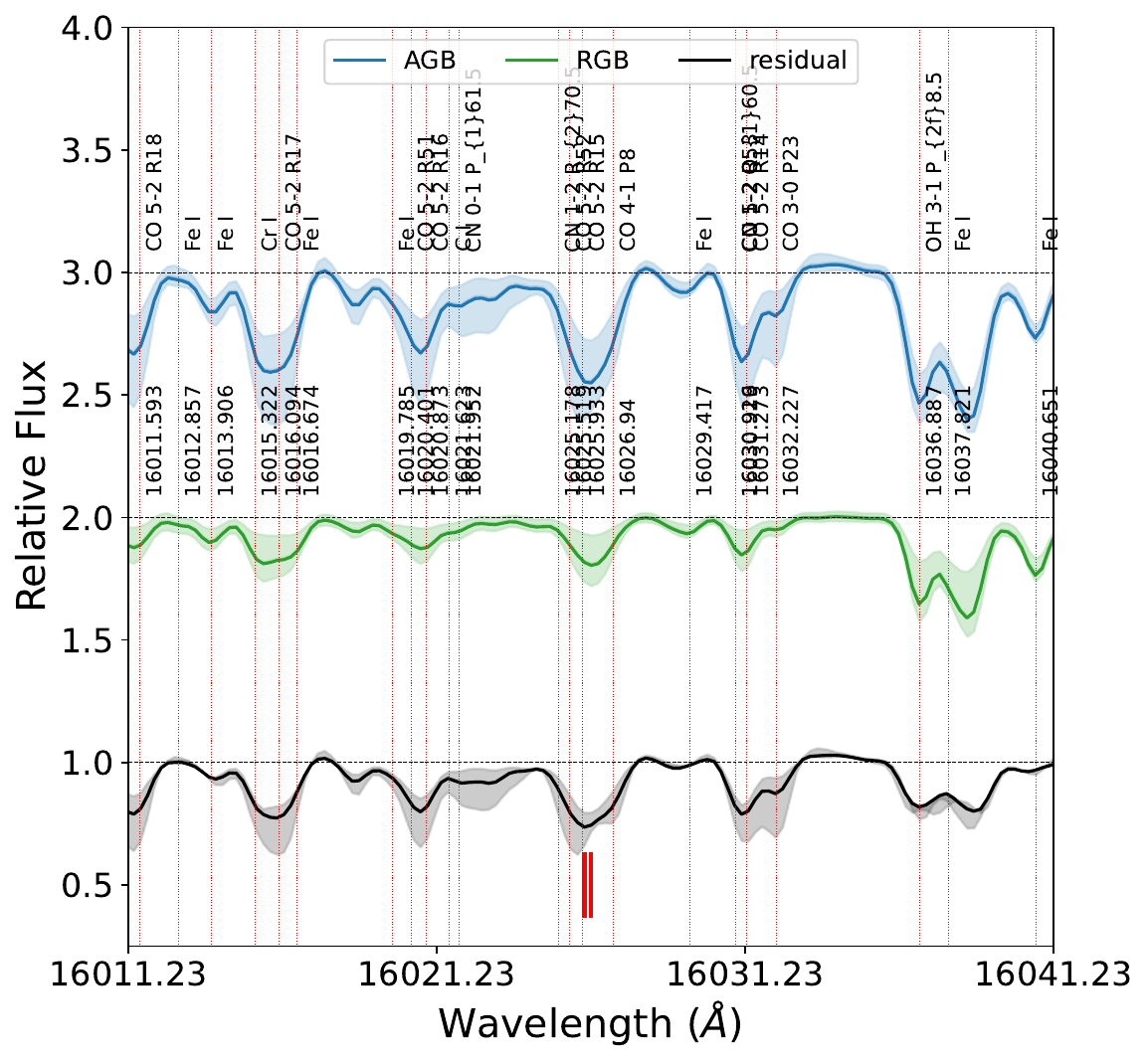}
\includegraphics[scale=0.3]{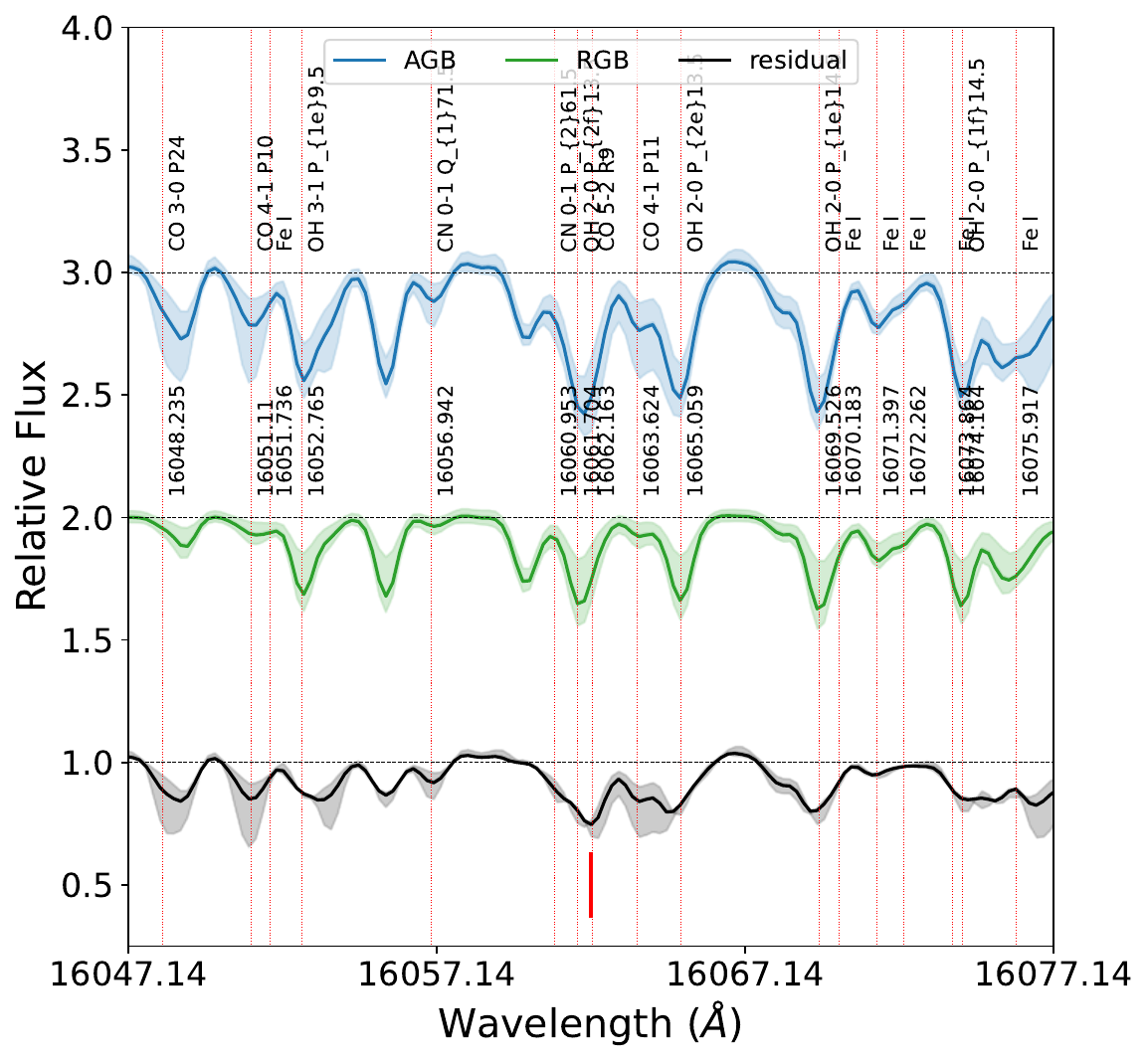}
\includegraphics[scale=0.3]{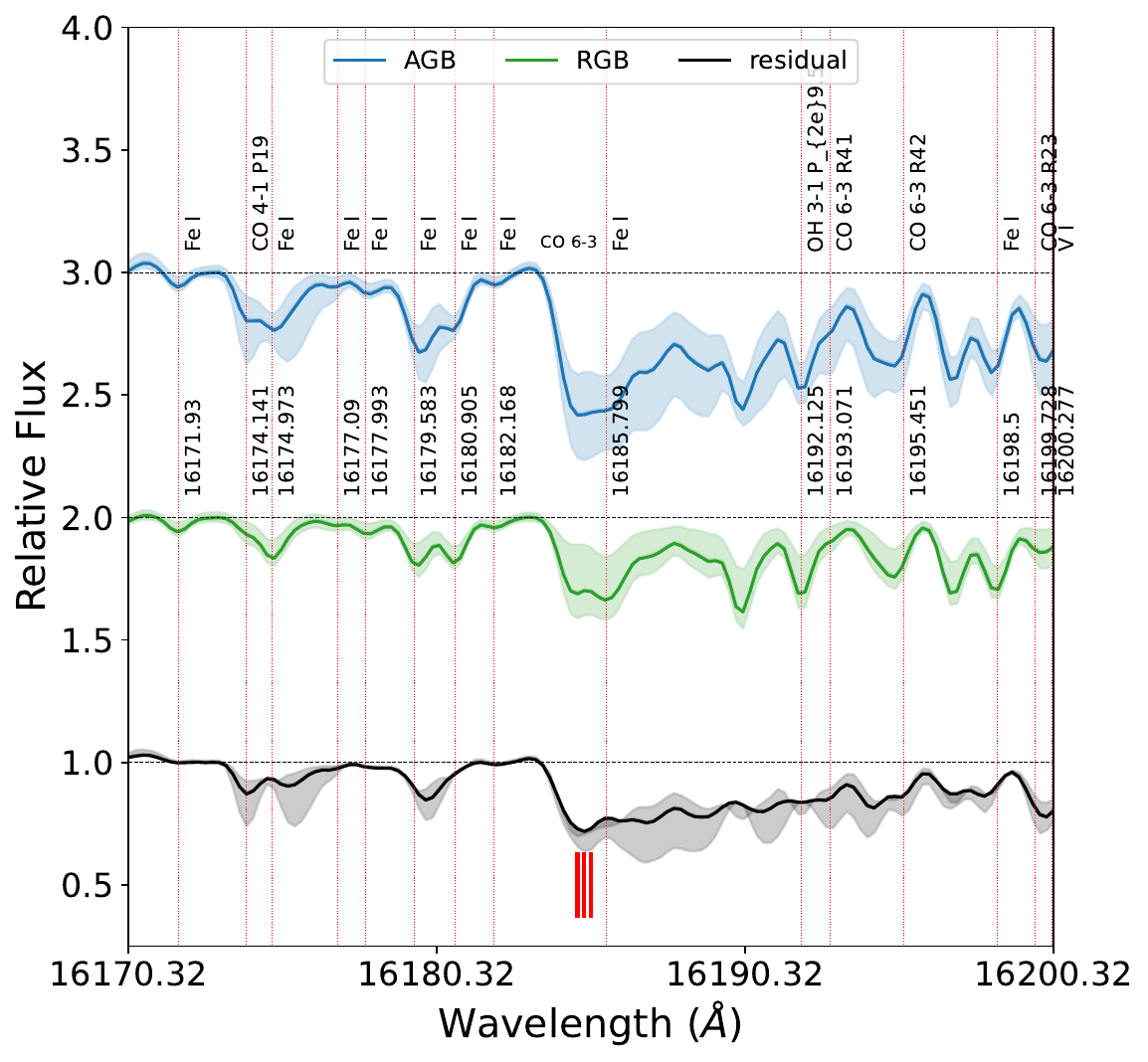}
\includegraphics[scale=0.3]{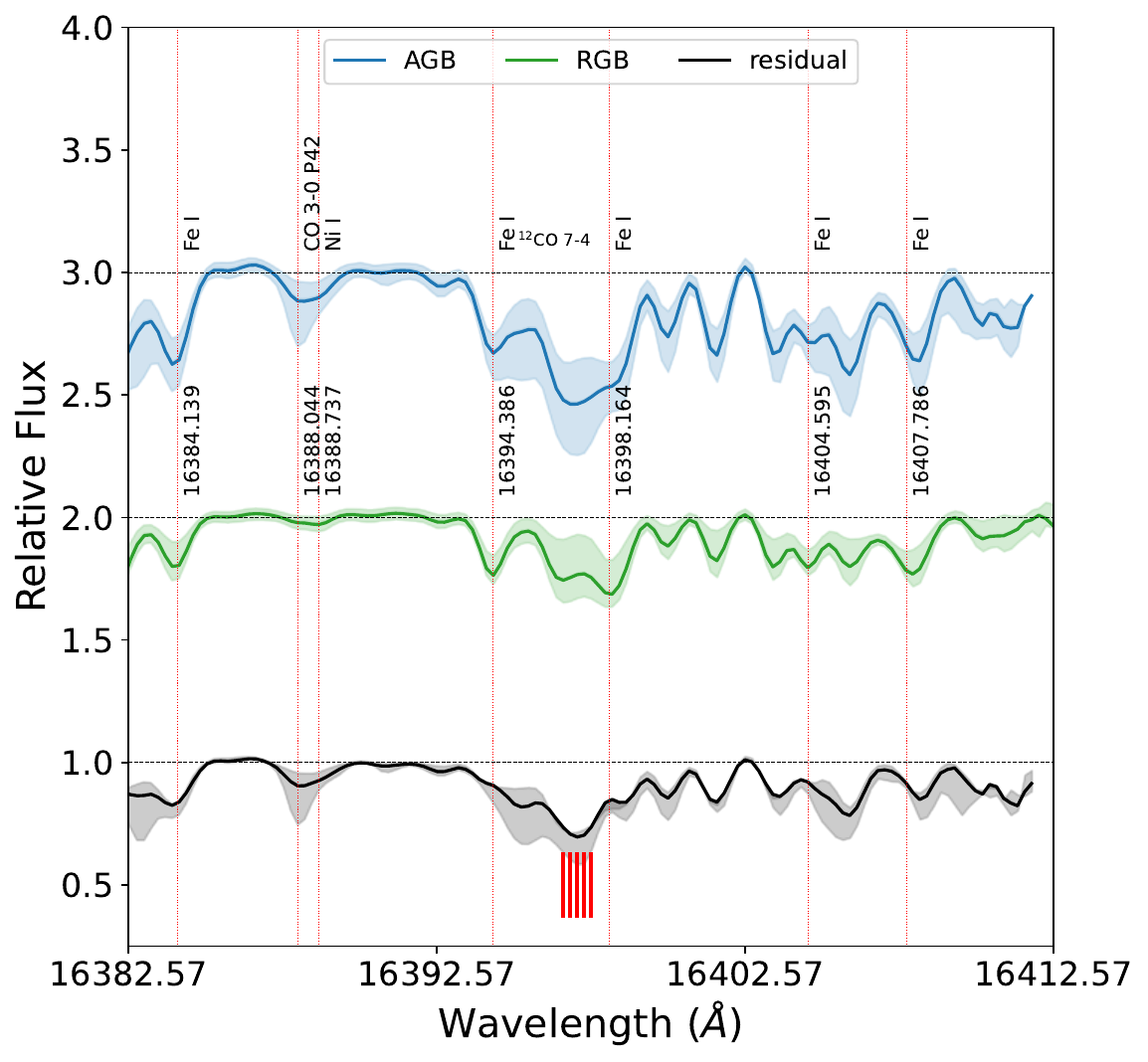}
\caption{Same as Figure~\ref{fig:spectral_lines2}, but for AGBs and RGBs. The red sticks mark the wavelengths with residual larger than 0.25.
\label{fig:spectral_lines3}}
\end{figure*}

Figure~\ref{fig:spectral_lines1}, Figure~\ref{fig:spectral_lines2}, and Figure~\ref{fig:spectral_lines3} show the zoomed-in regions of strong spectral features between AGBs and RSGs, RSGs and RGBs, AGBs and RGBs.

\section{Implementation of Machine Learning Code} \label{sec:ml_code_appendix}

The implementation code based on PyTorch is shown below.

\begin{lstlisting}[language=Python]
import torch.nn as nn
	model = nn.Sequential(
	nn.Conv1d(in_channels=1, out_channels=32, kernel_size=5, padding=1, stride=3),
	nn.ELU(),
	nn.Conv1d(in_channels=32, out_channels=32, kernel_size=3, padding=1, stride=2),
	nn.ELU(),
	nn.Conv1d(in_channels=32, out_channels=16, kernel_size=3, padding=1, stride=2),
	nn.BatchNorm1d(num_features=16),
	nn.ELU(),
	nn.AvgPool1d(kernel_size=3, stride=3),
	nn.Flatten(),
	nn.Linear(3808, 256),
	nn.BatchNorm1d(num_features=256),
	nn.Tanh(),
	nn.Linear(256, 3),
	nn.Softmax(dim=1)
)
\end{lstlisting}

\section{Equivalent Widths of Line or Lines Complexes} \label{sec:ews_appendix}

\begin{figure*}
\center
\includegraphics[scale=0.24]{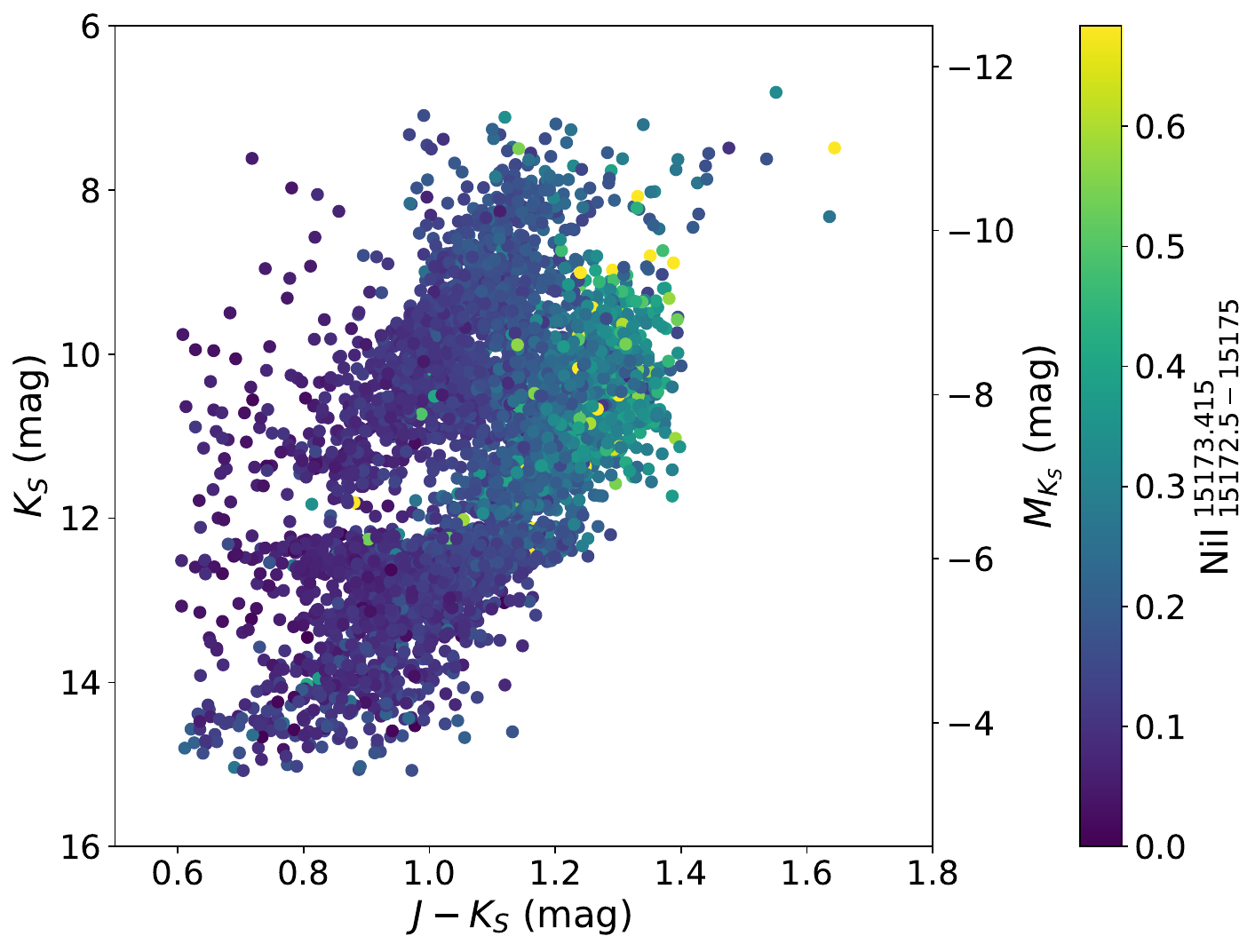}
\includegraphics[scale=0.24]{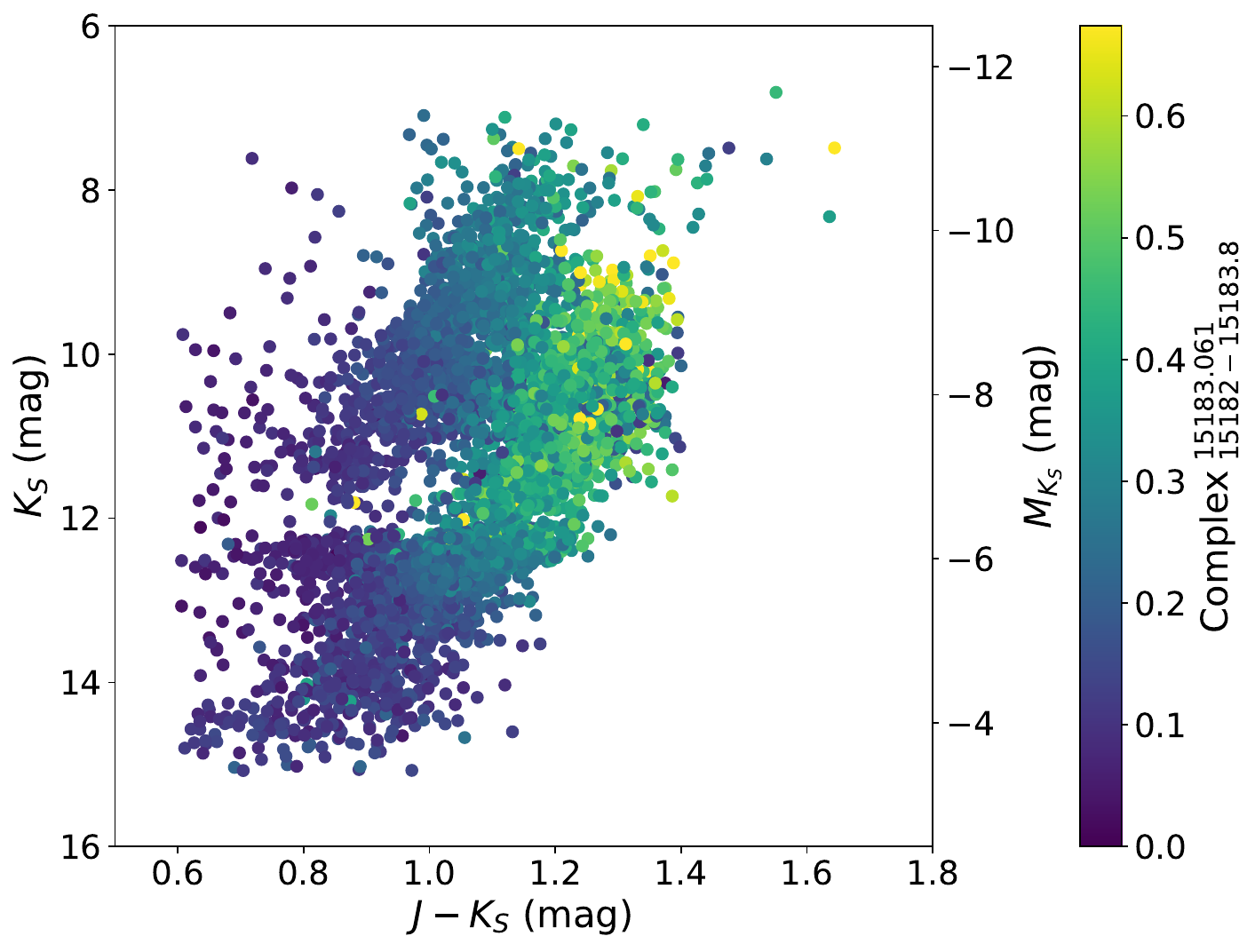}
\includegraphics[scale=0.24]{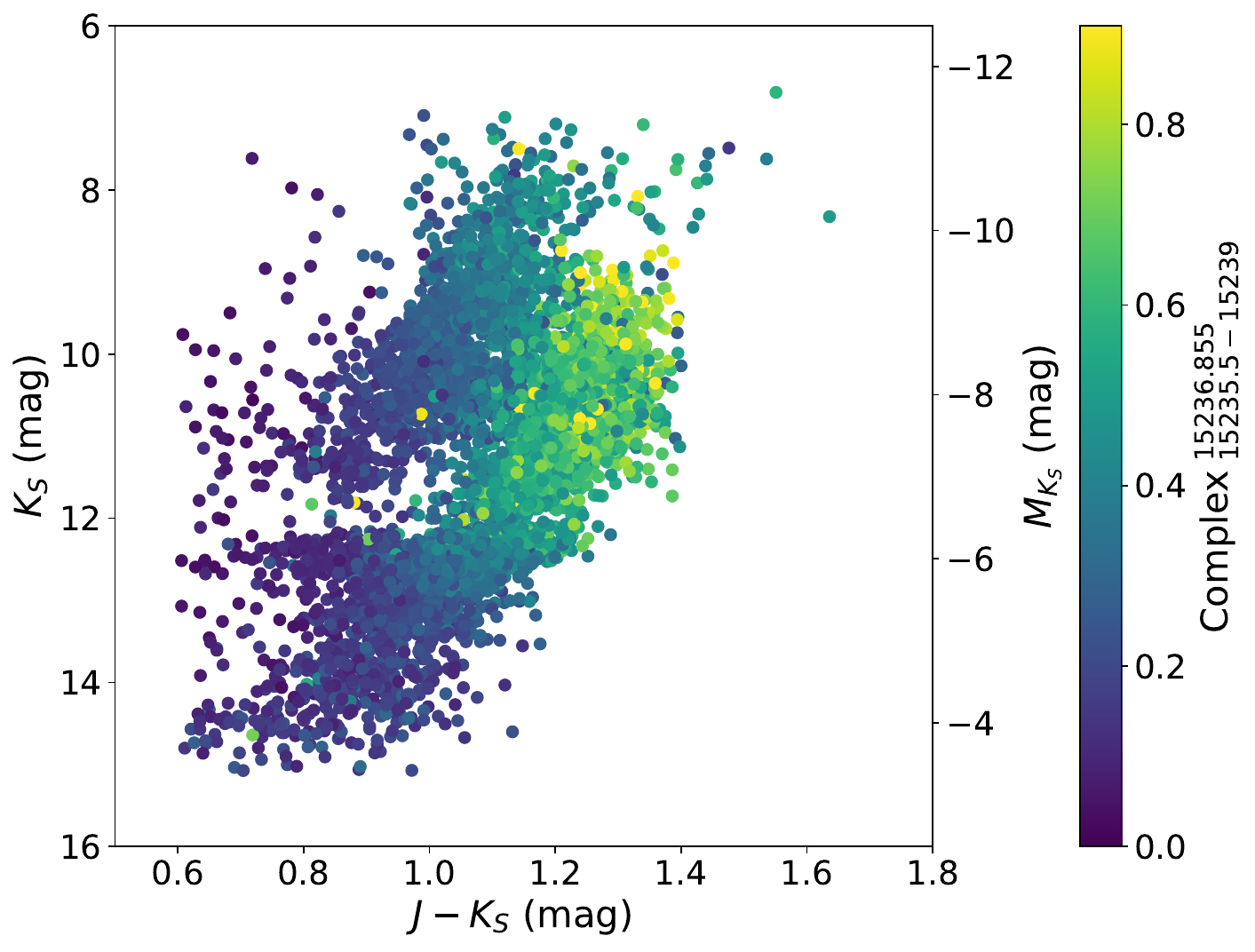}
\includegraphics[scale=0.24]{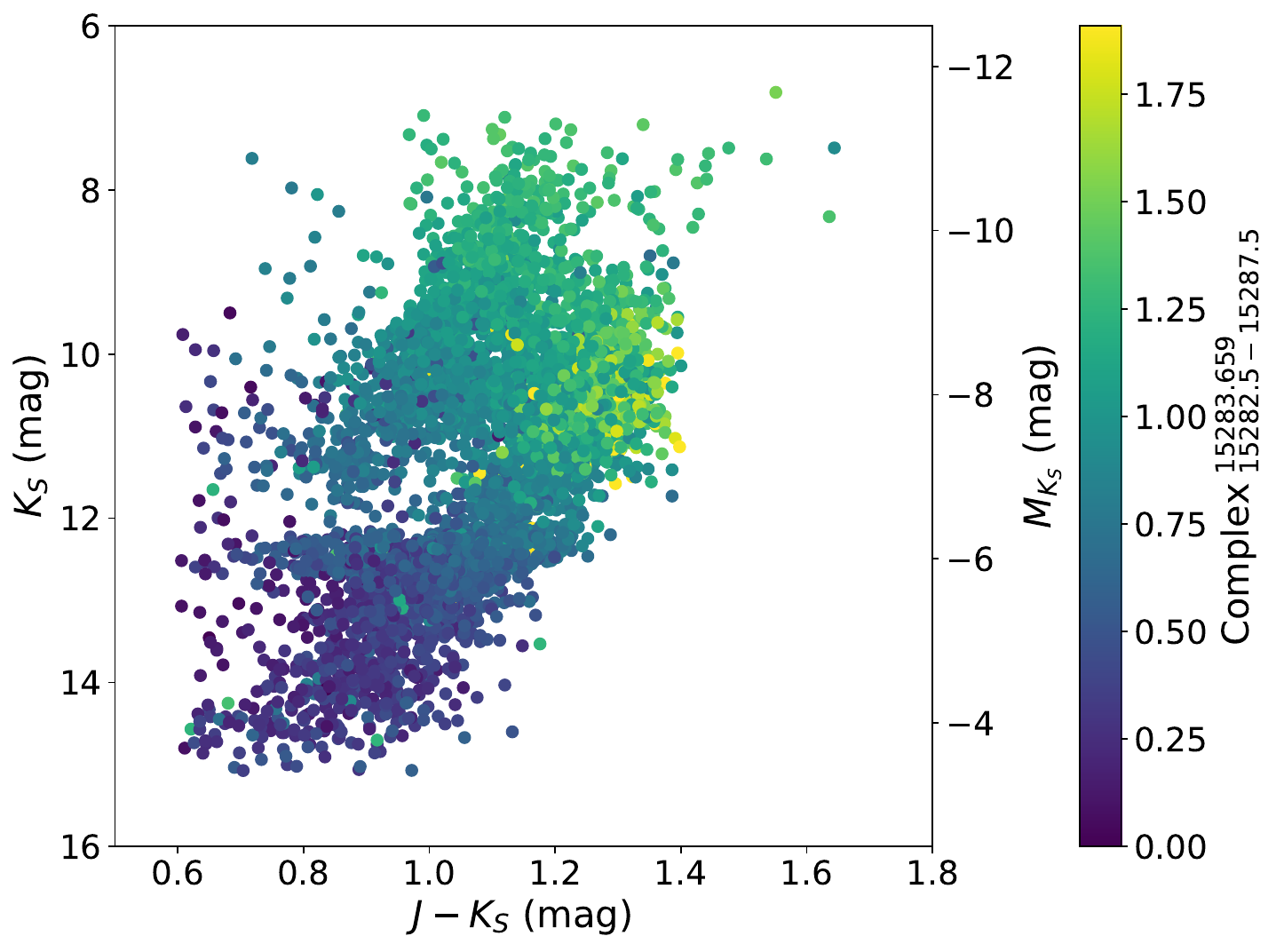}
\includegraphics[scale=0.24]{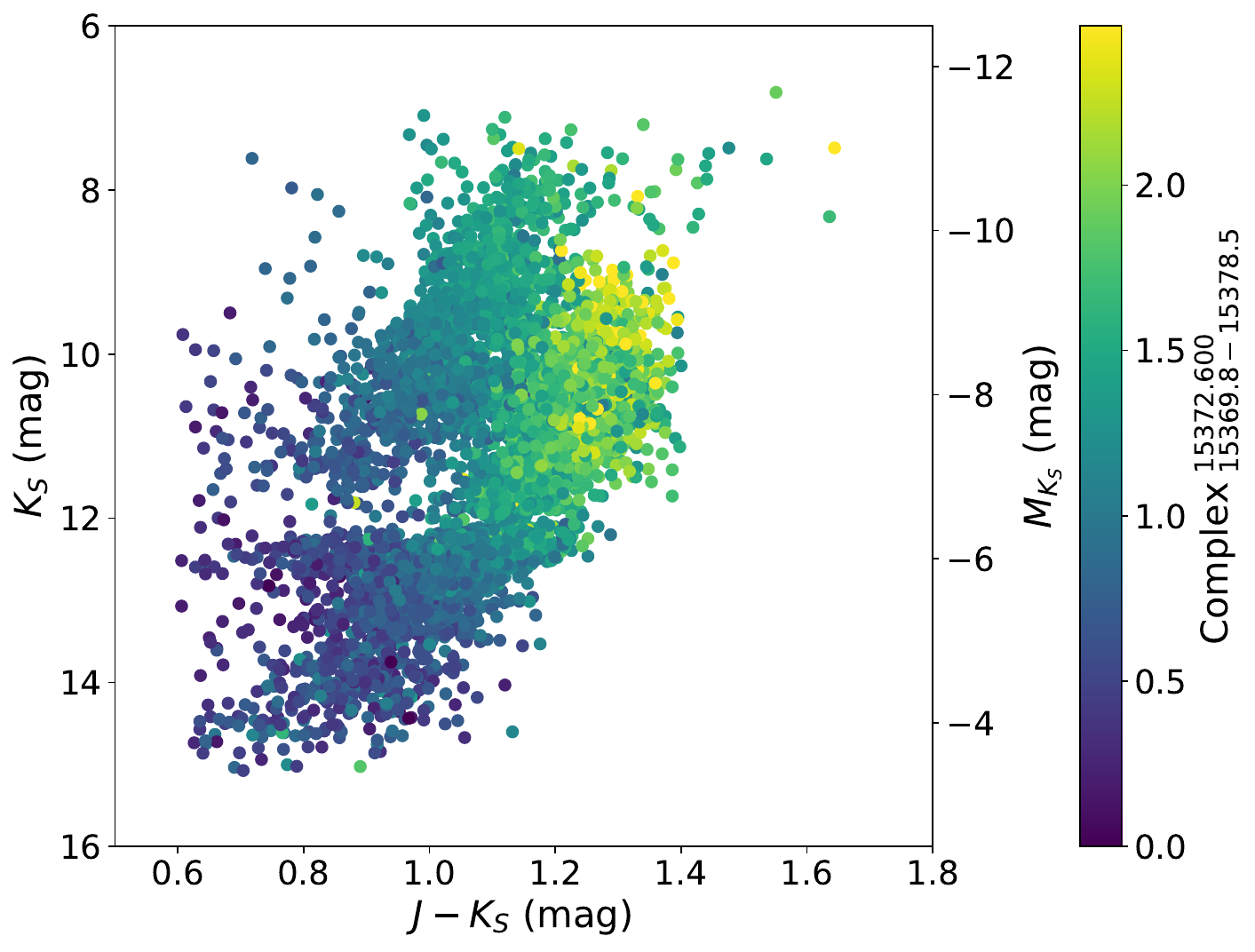}
\includegraphics[scale=0.24]{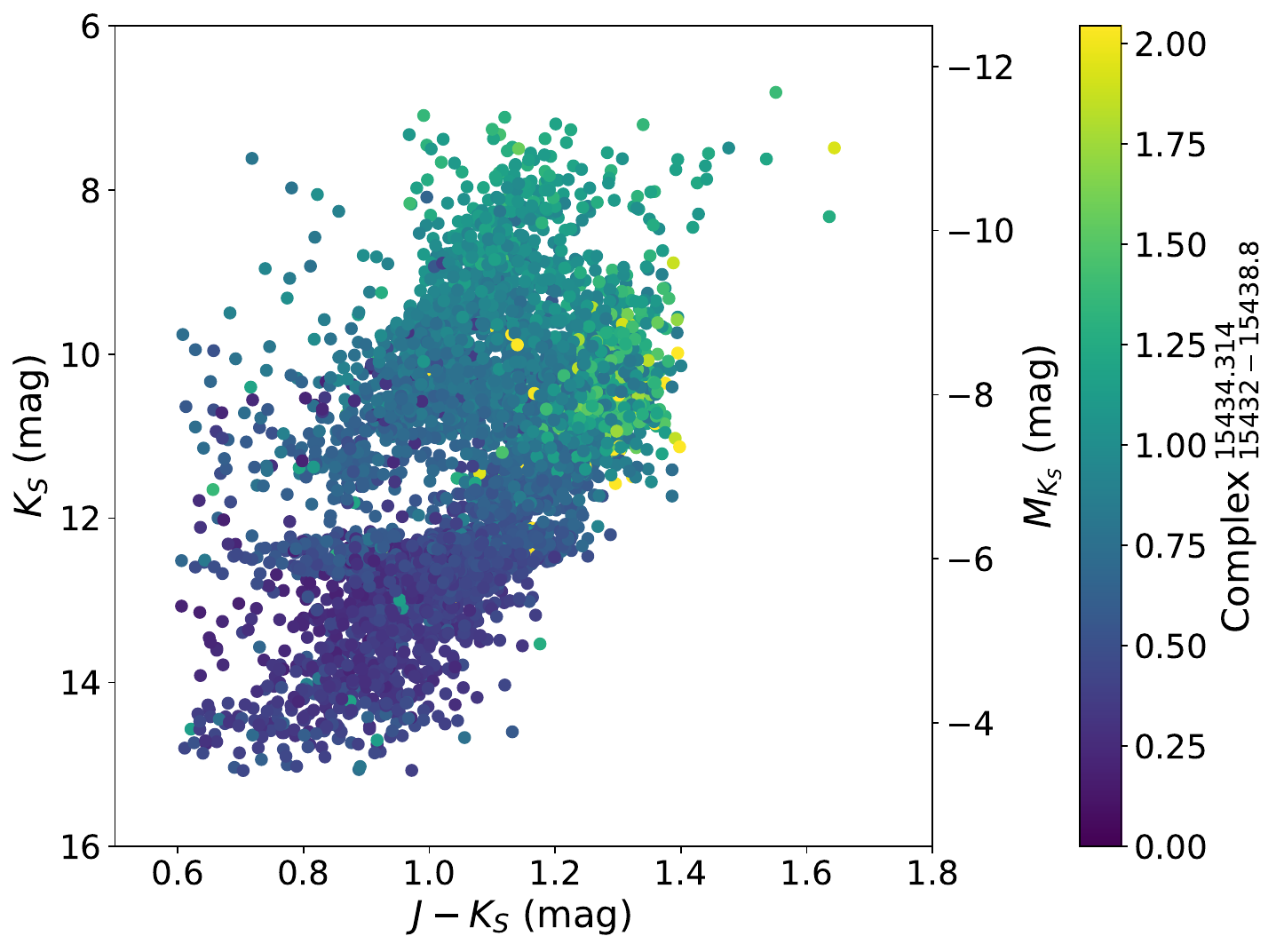}
\includegraphics[scale=0.24]{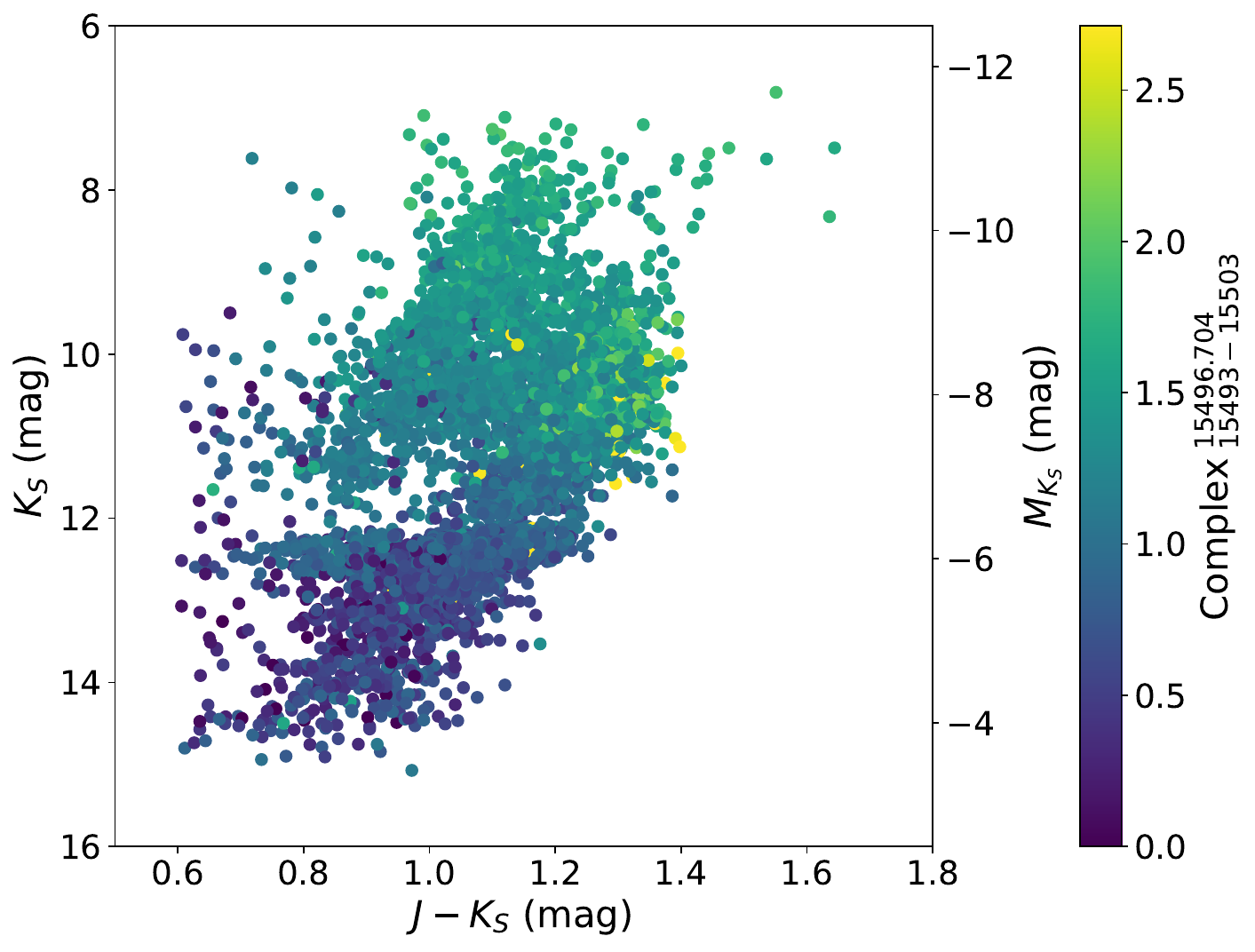}
\includegraphics[scale=0.24]{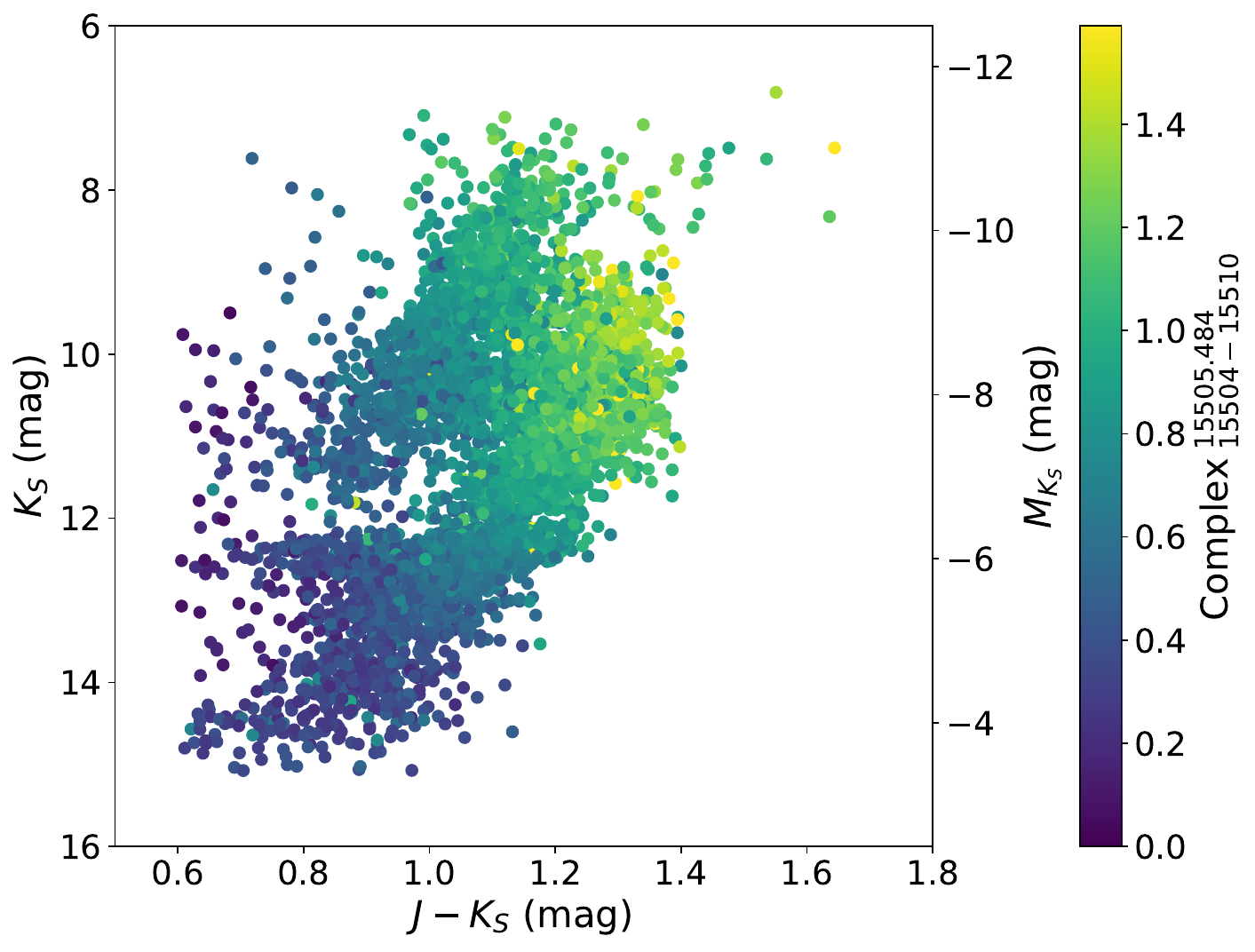}
\includegraphics[scale=0.24]{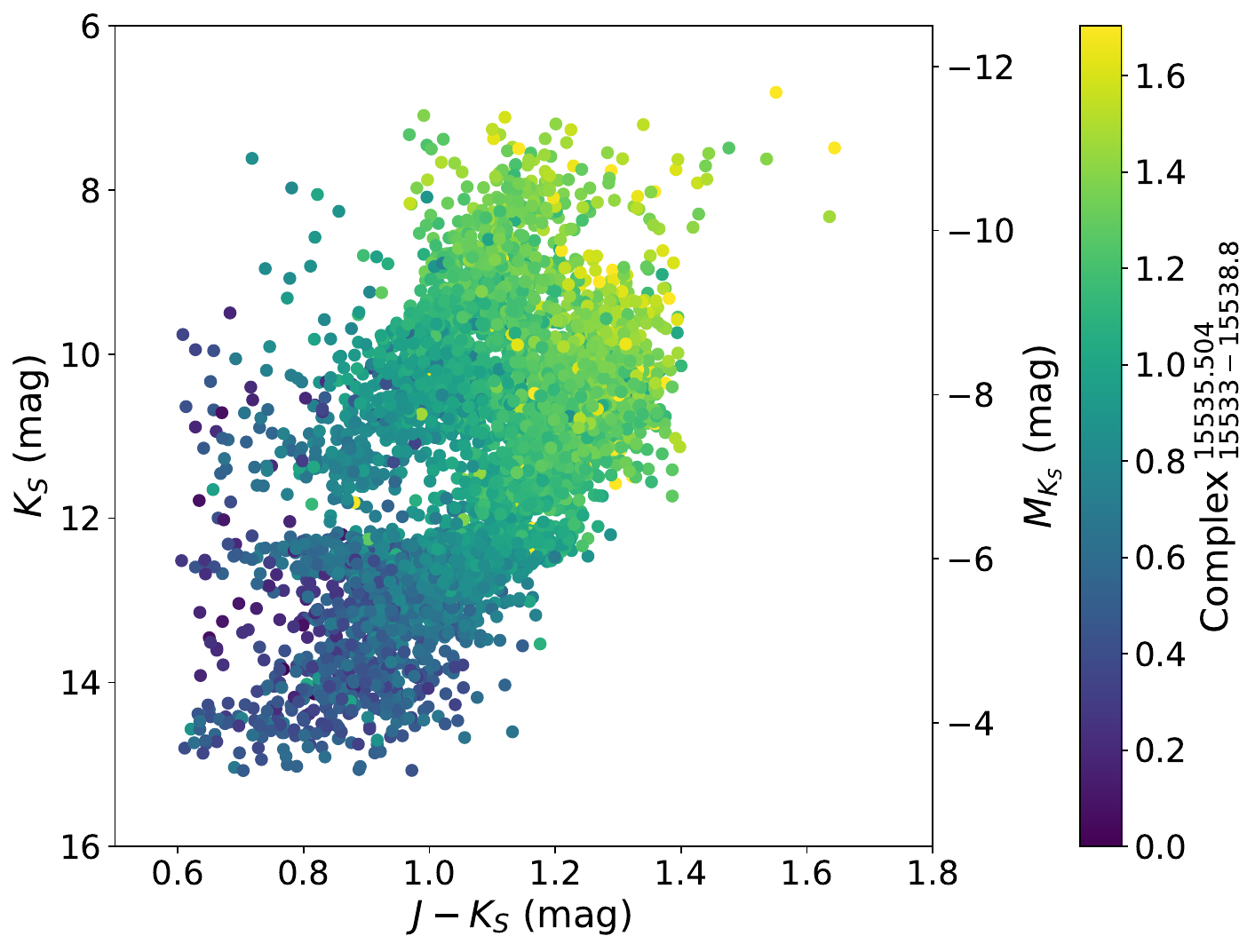}
\includegraphics[scale=0.24]{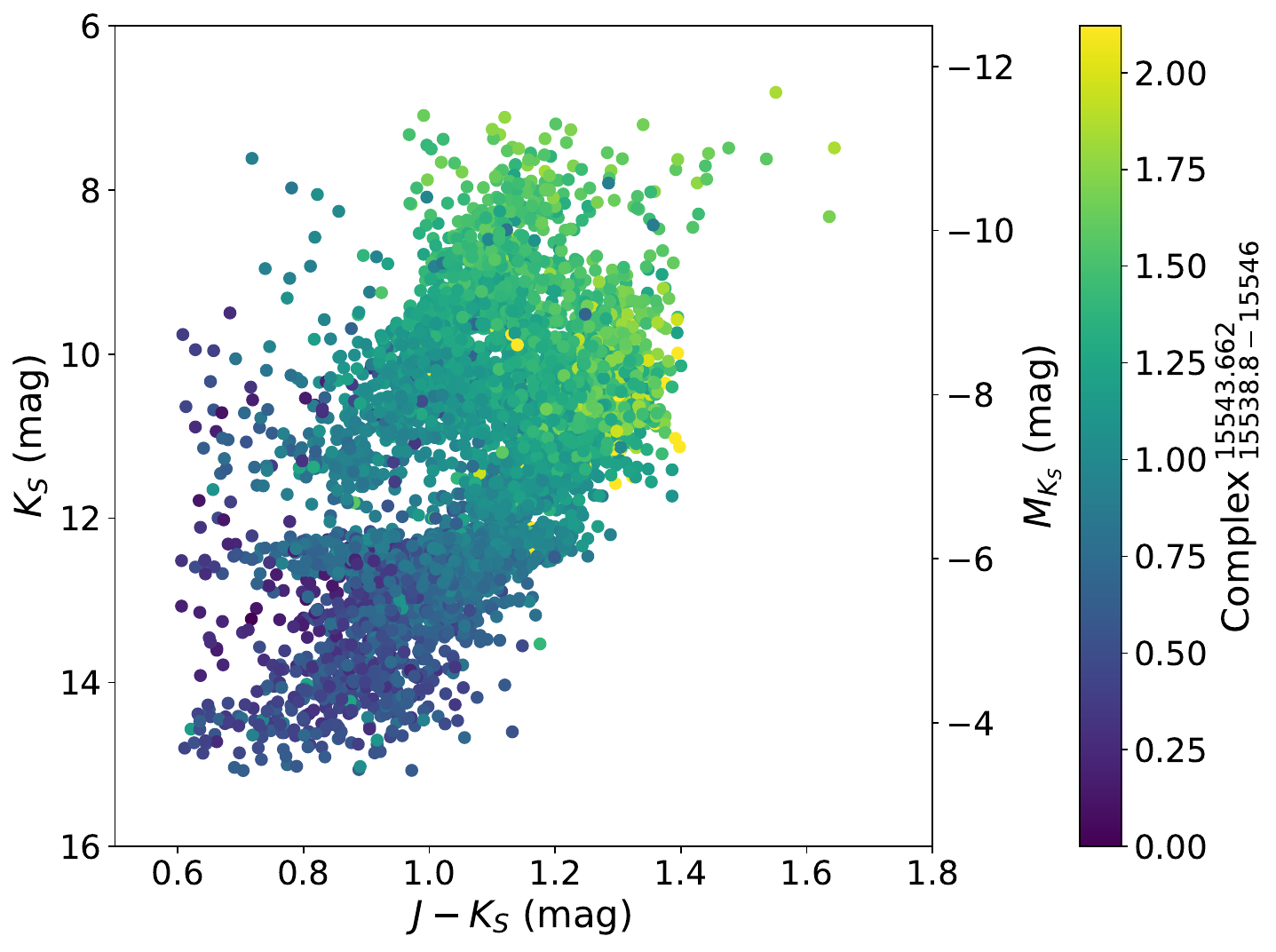}
\includegraphics[scale=0.24]{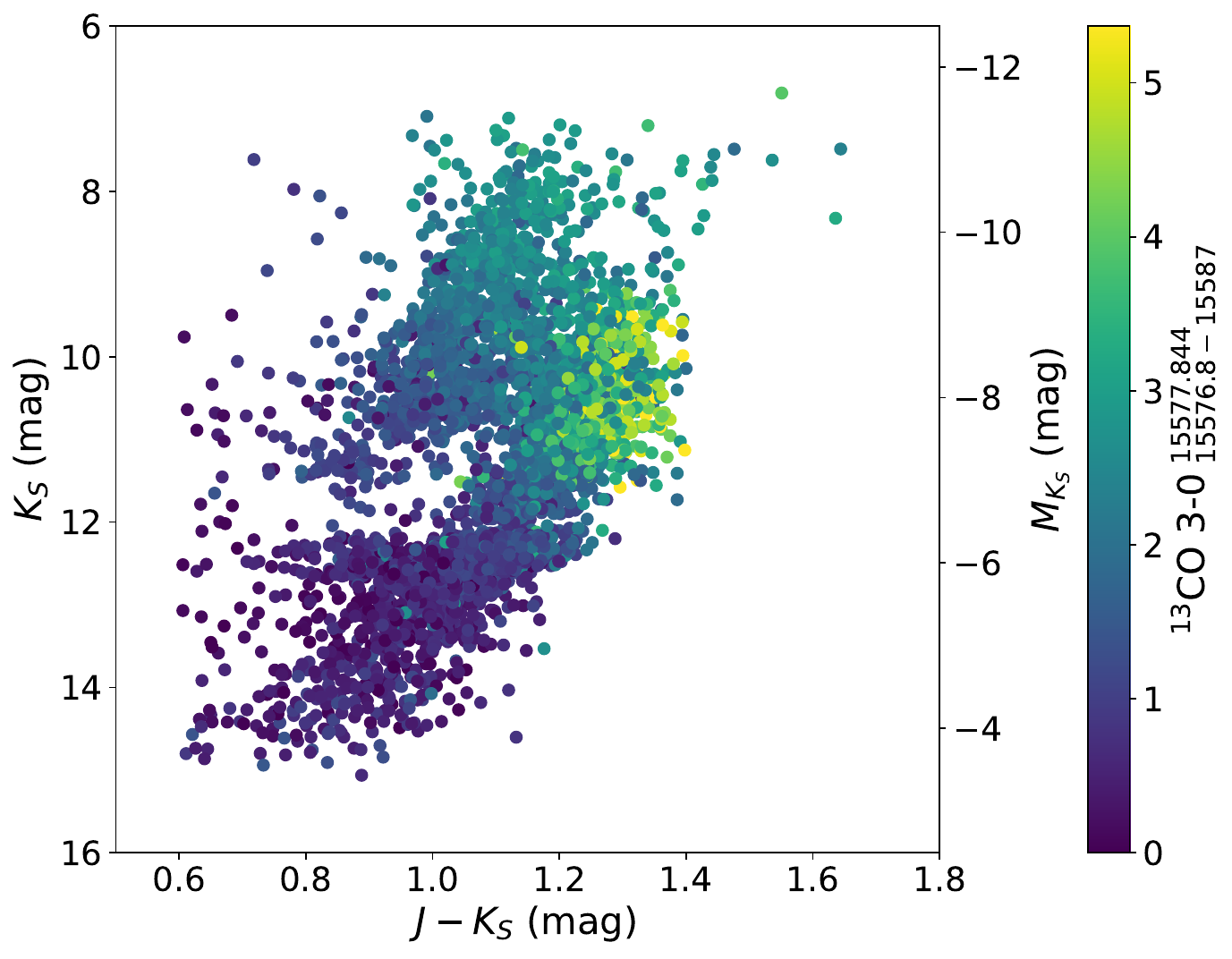}
\includegraphics[scale=0.24]{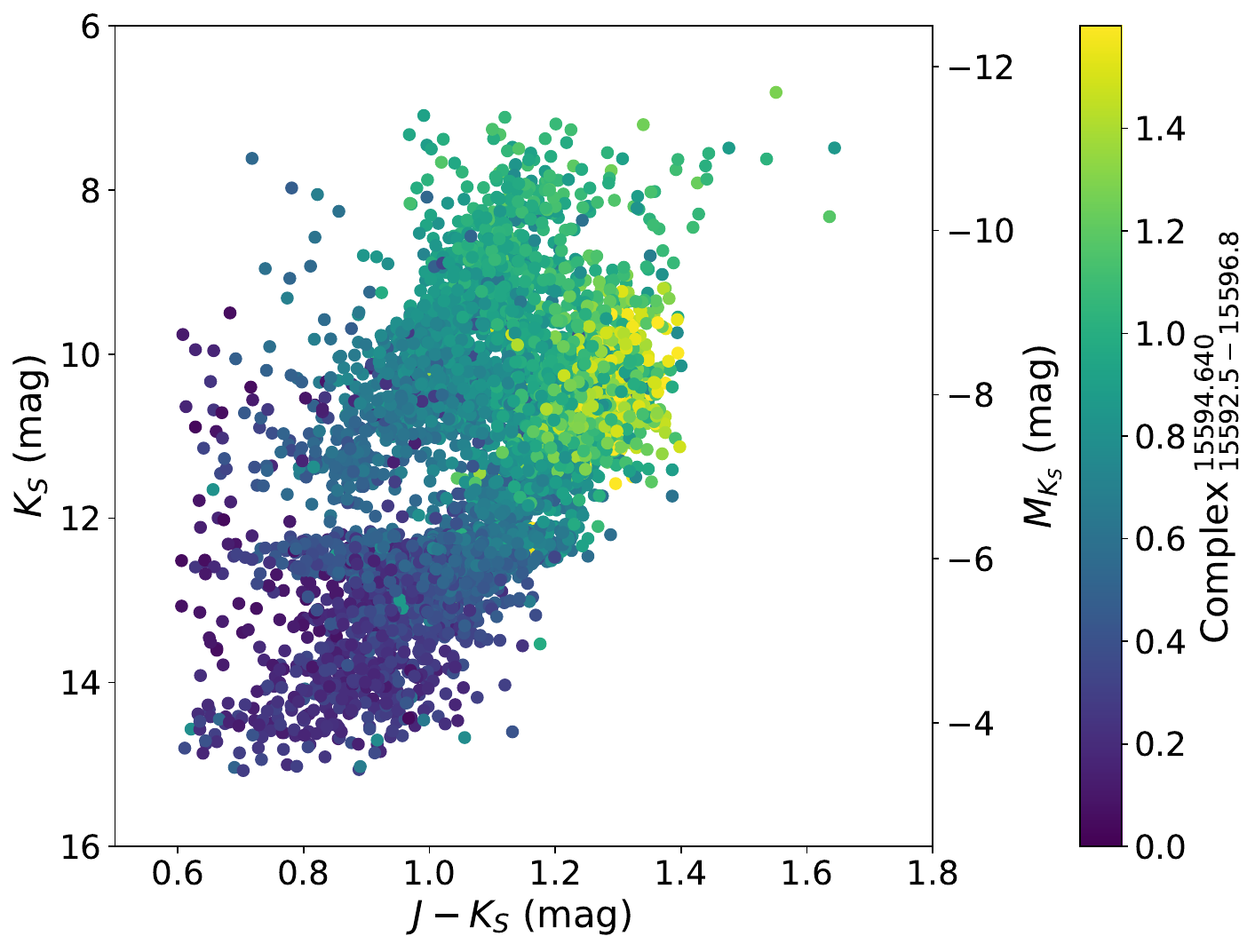}
\includegraphics[scale=0.24]{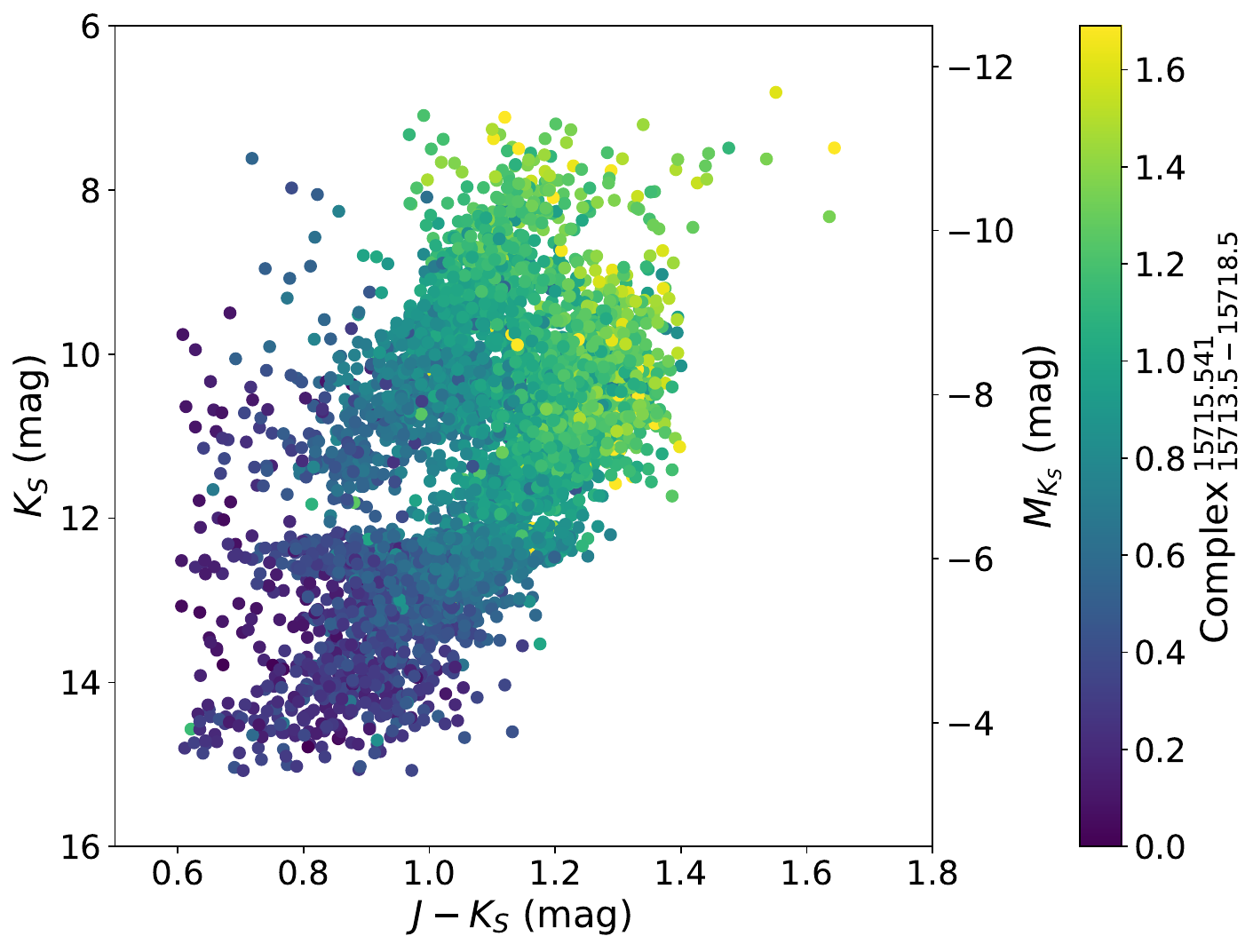}
\includegraphics[scale=0.24]{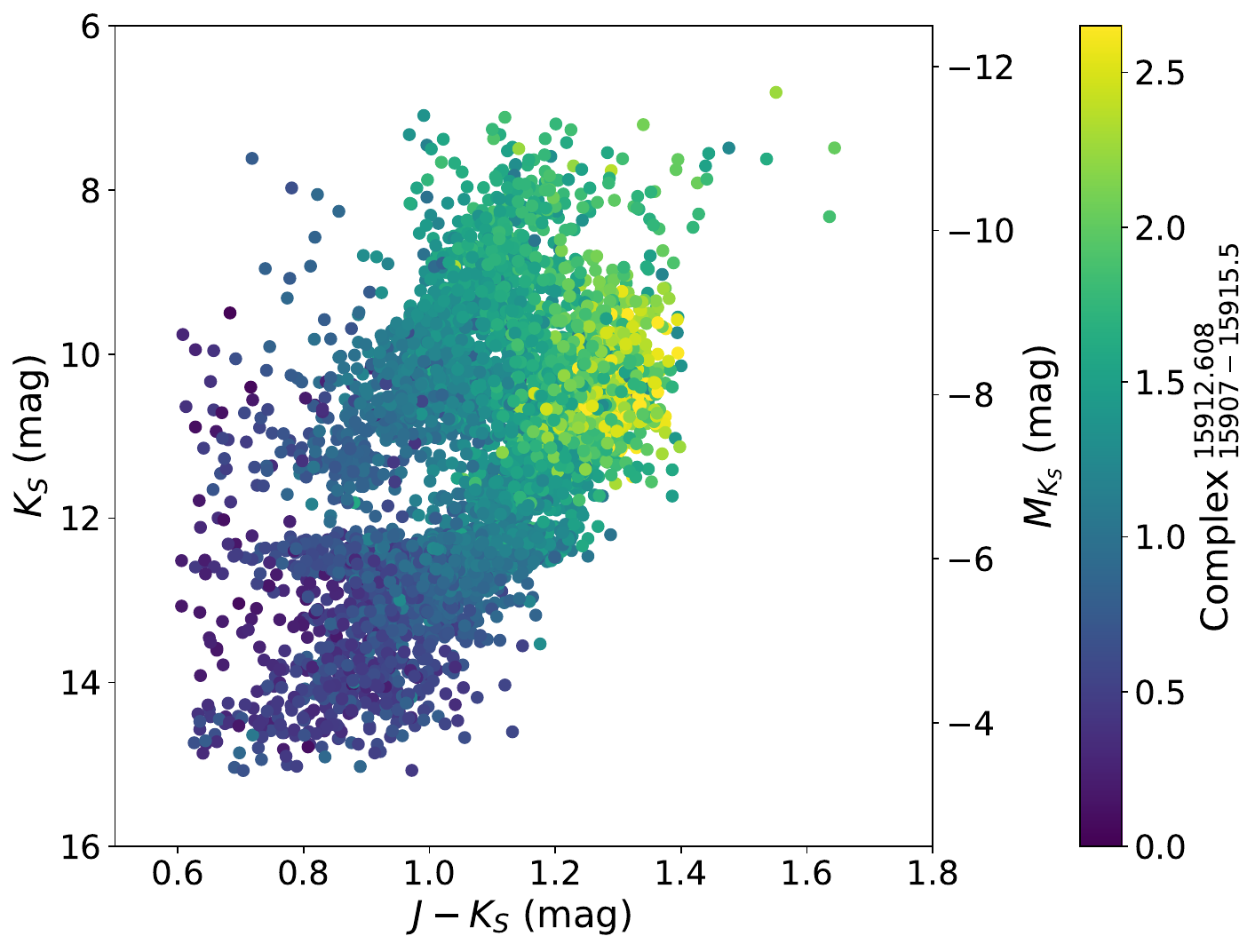}
\includegraphics[scale=0.24]{ew_FeI_15971.193.pdf}
\caption{Same as Figure~\ref{fig:ew}, but for the first 15 spectral lines or line complexes in Table~\ref{tbl:ewtable}. 
\label{fig:ew1}}
\end{figure*}

\begin{figure*}
\center
\includegraphics[scale=0.24]{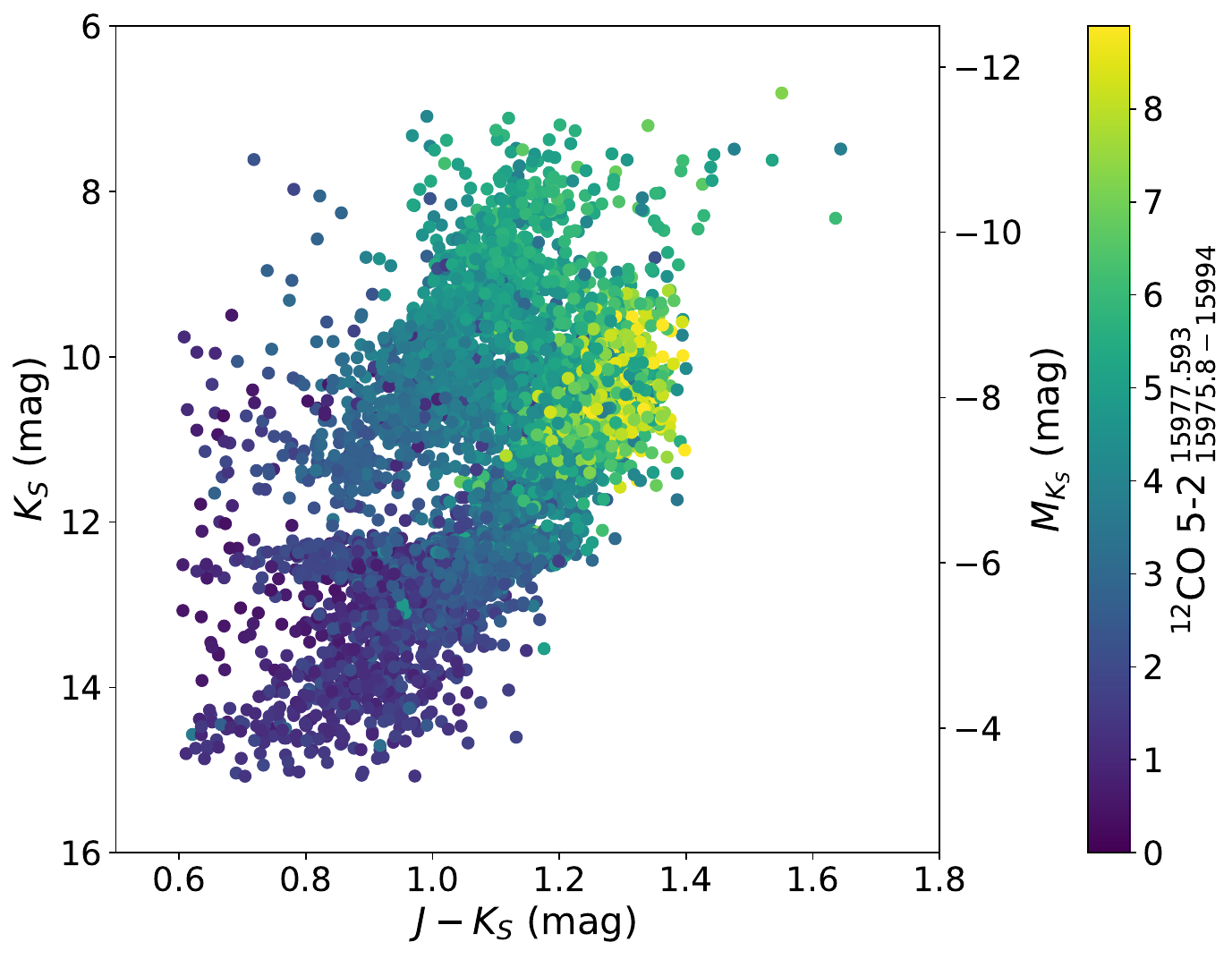}
\includegraphics[scale=0.24]{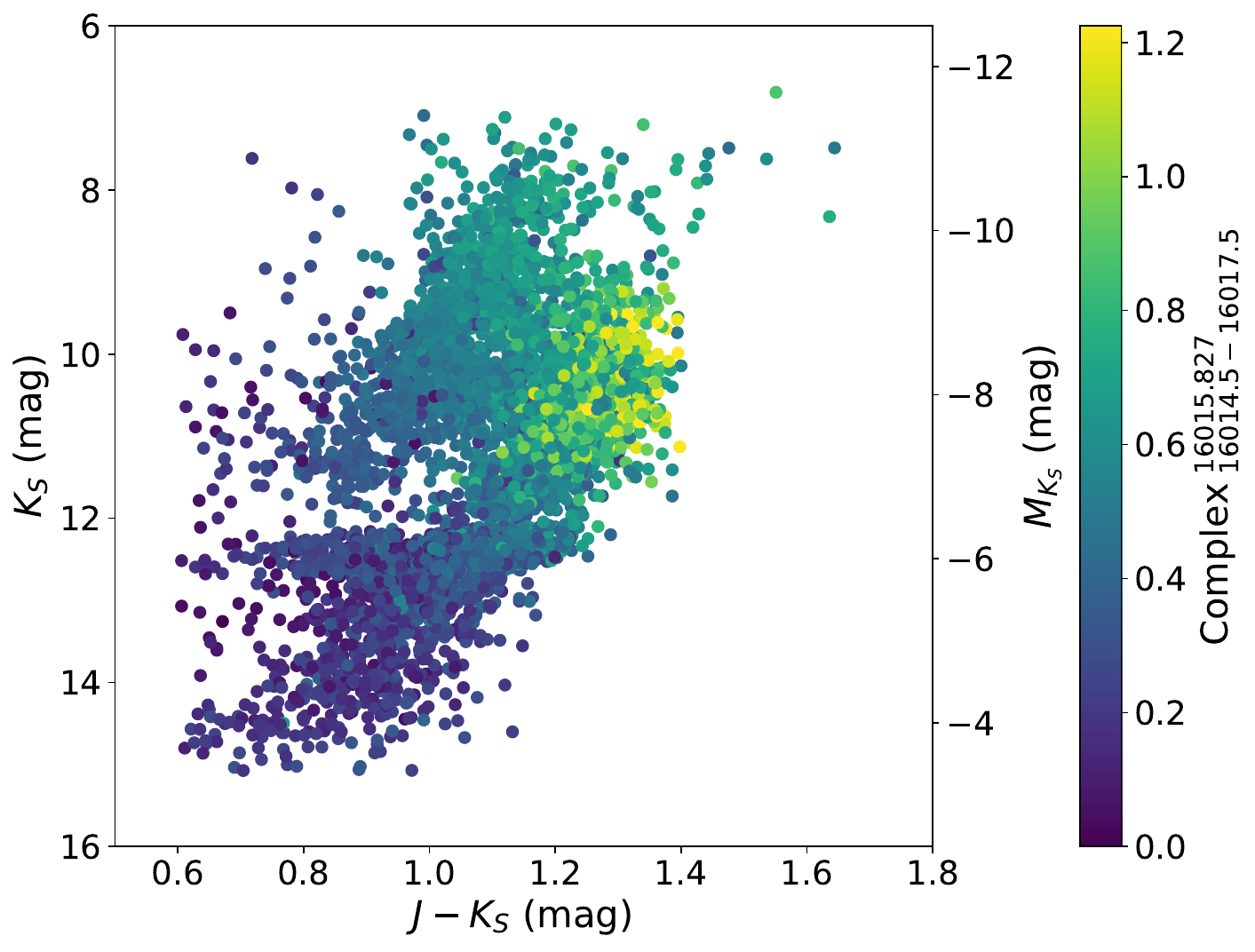}
\includegraphics[scale=0.24]{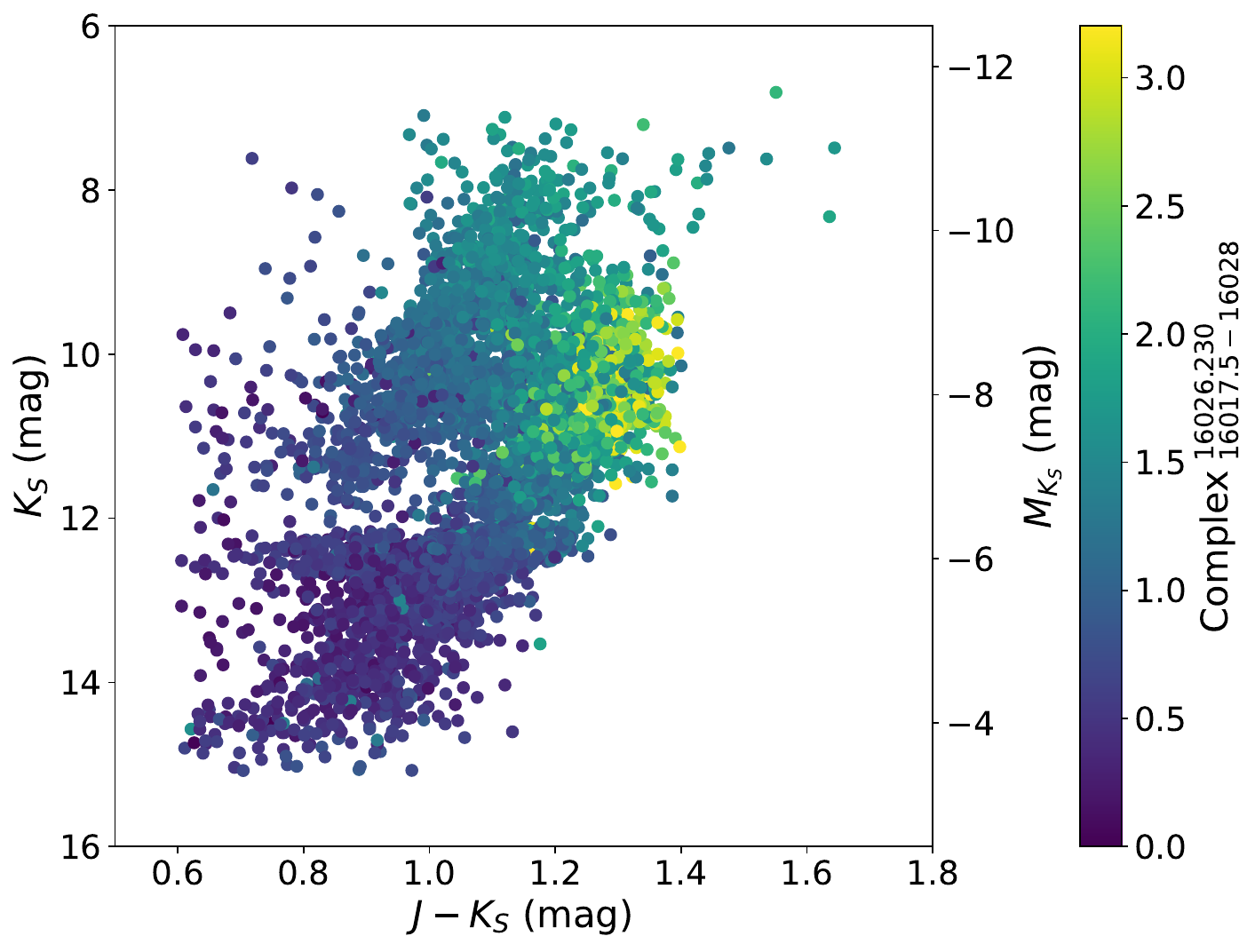}
\includegraphics[scale=0.24]{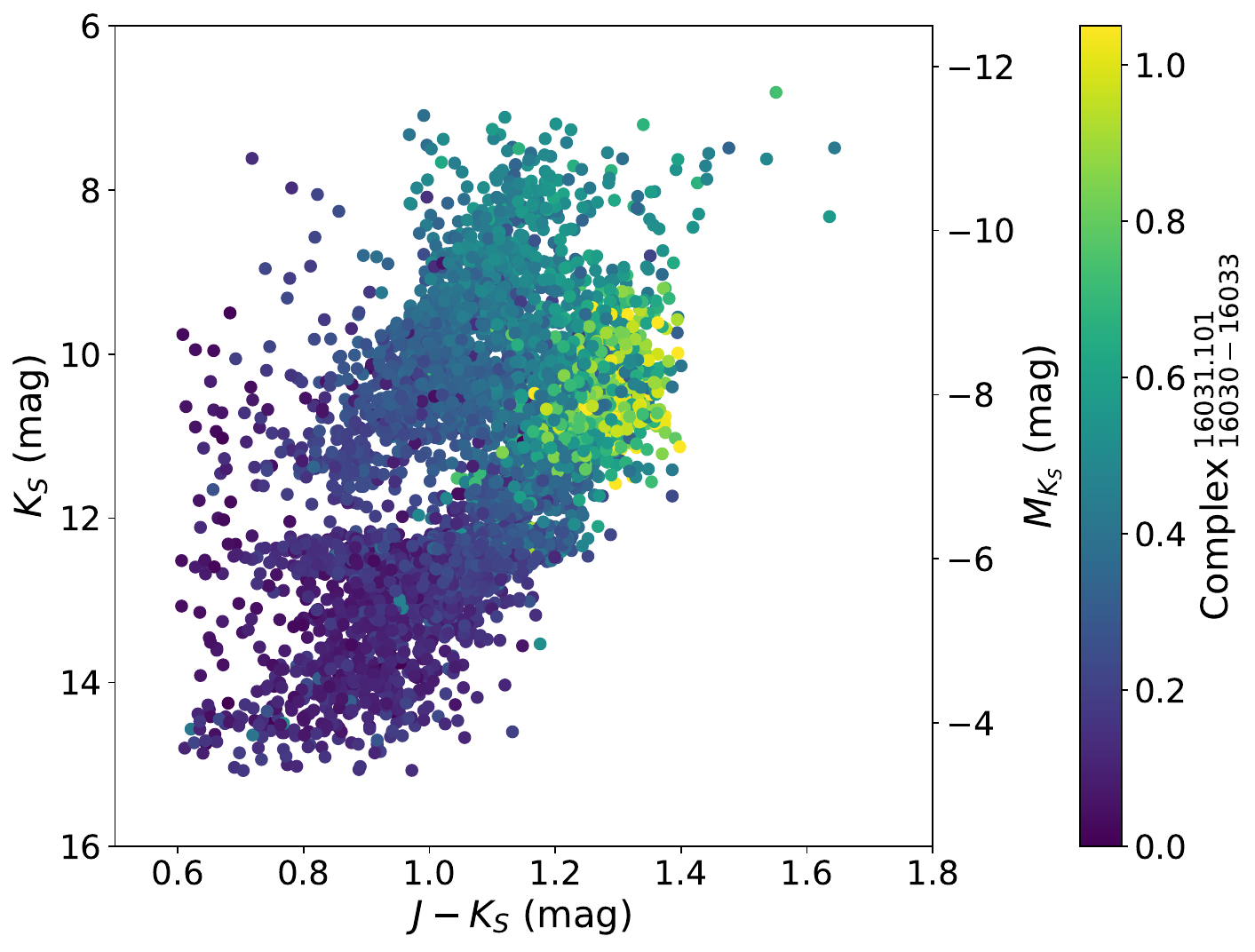}
\includegraphics[scale=0.24]{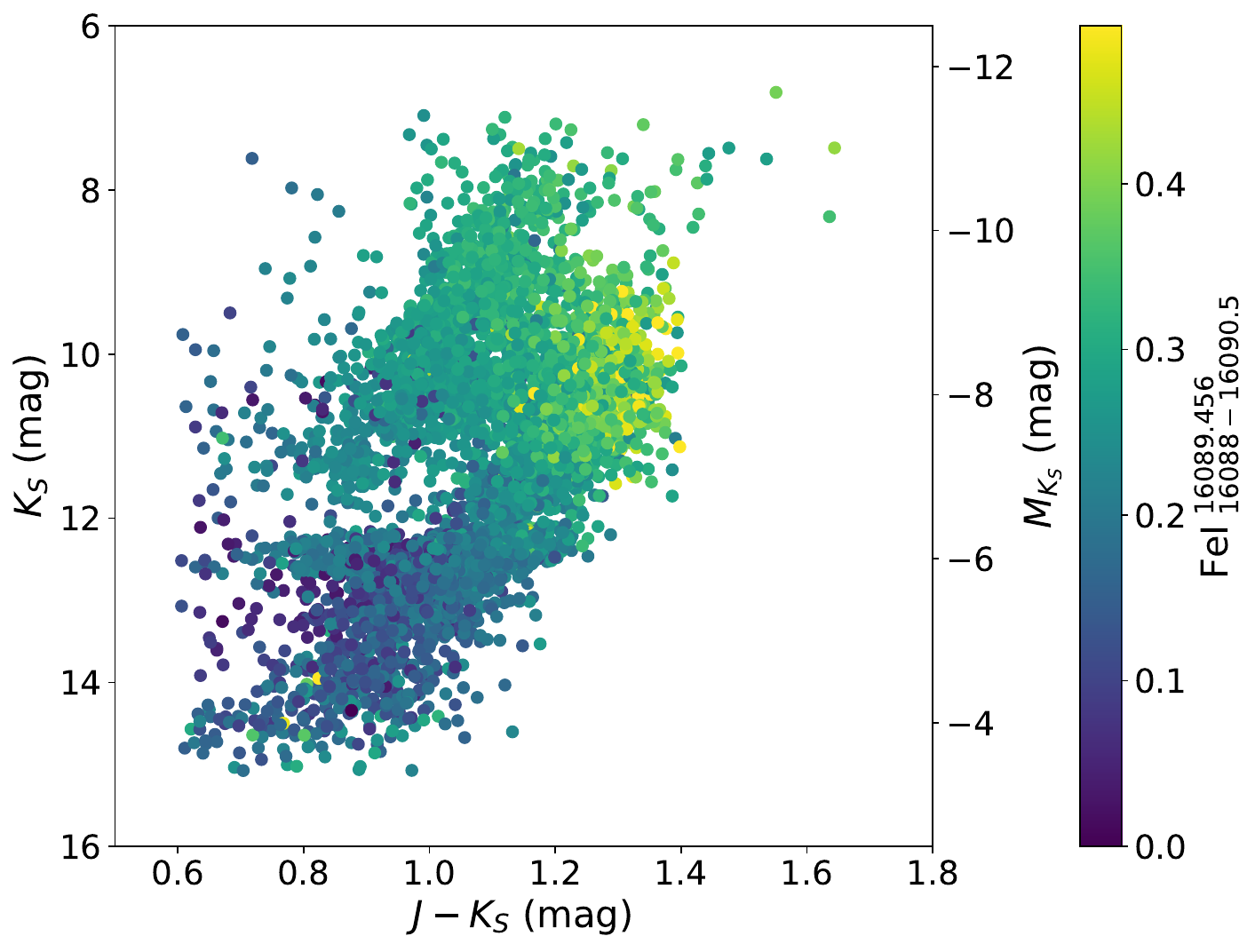}
\includegraphics[scale=0.24]{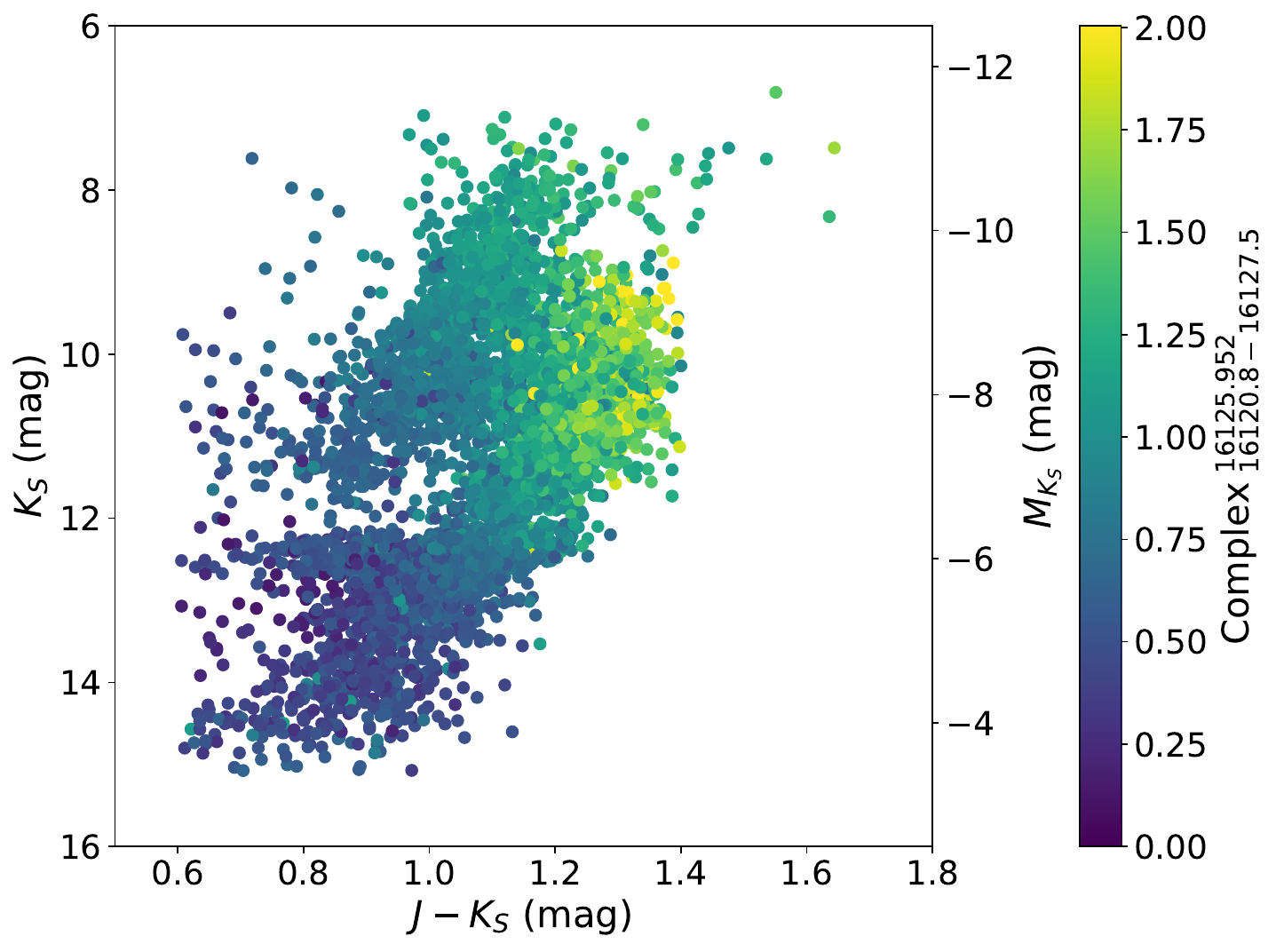}
\includegraphics[scale=0.24]{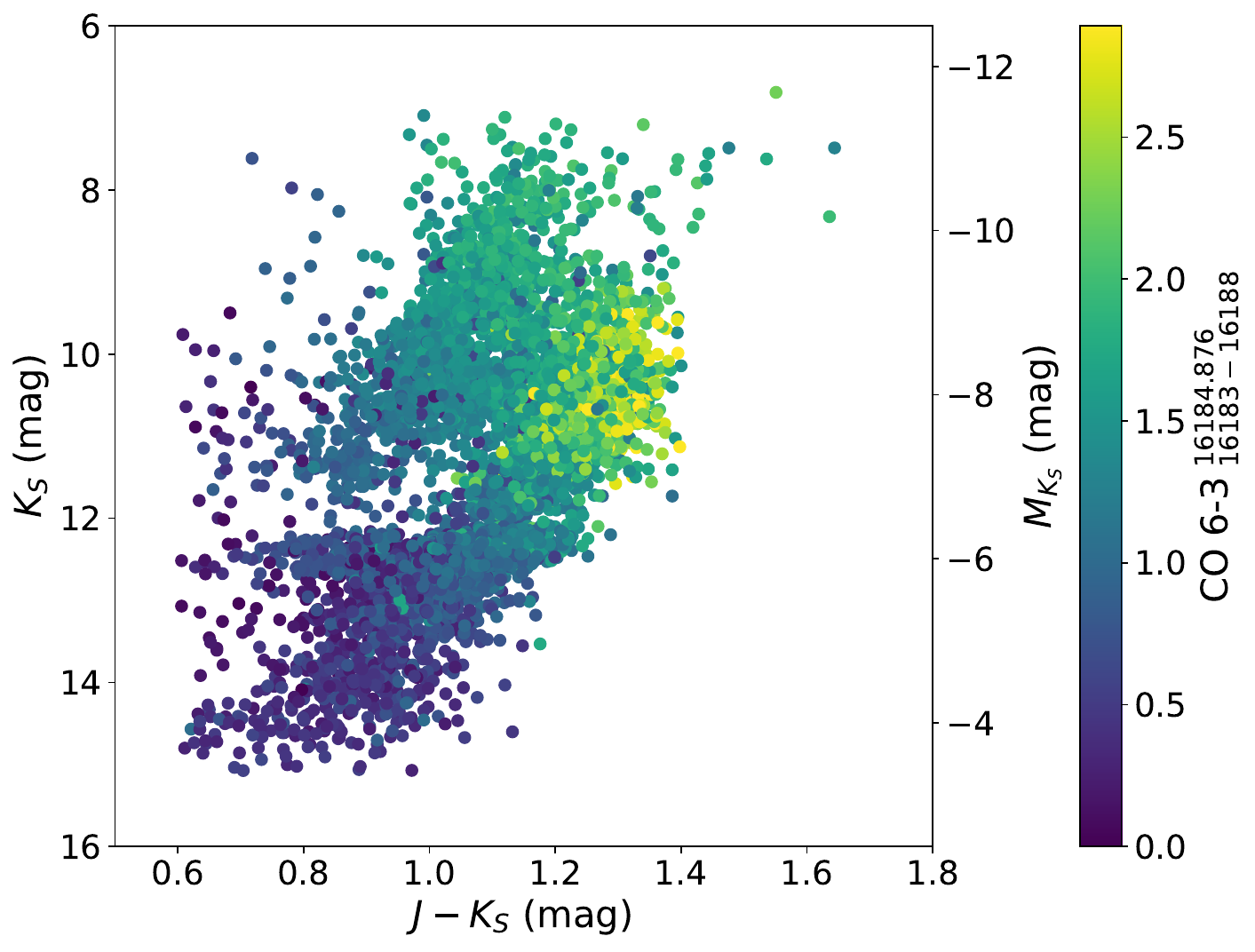}
\includegraphics[scale=0.24]{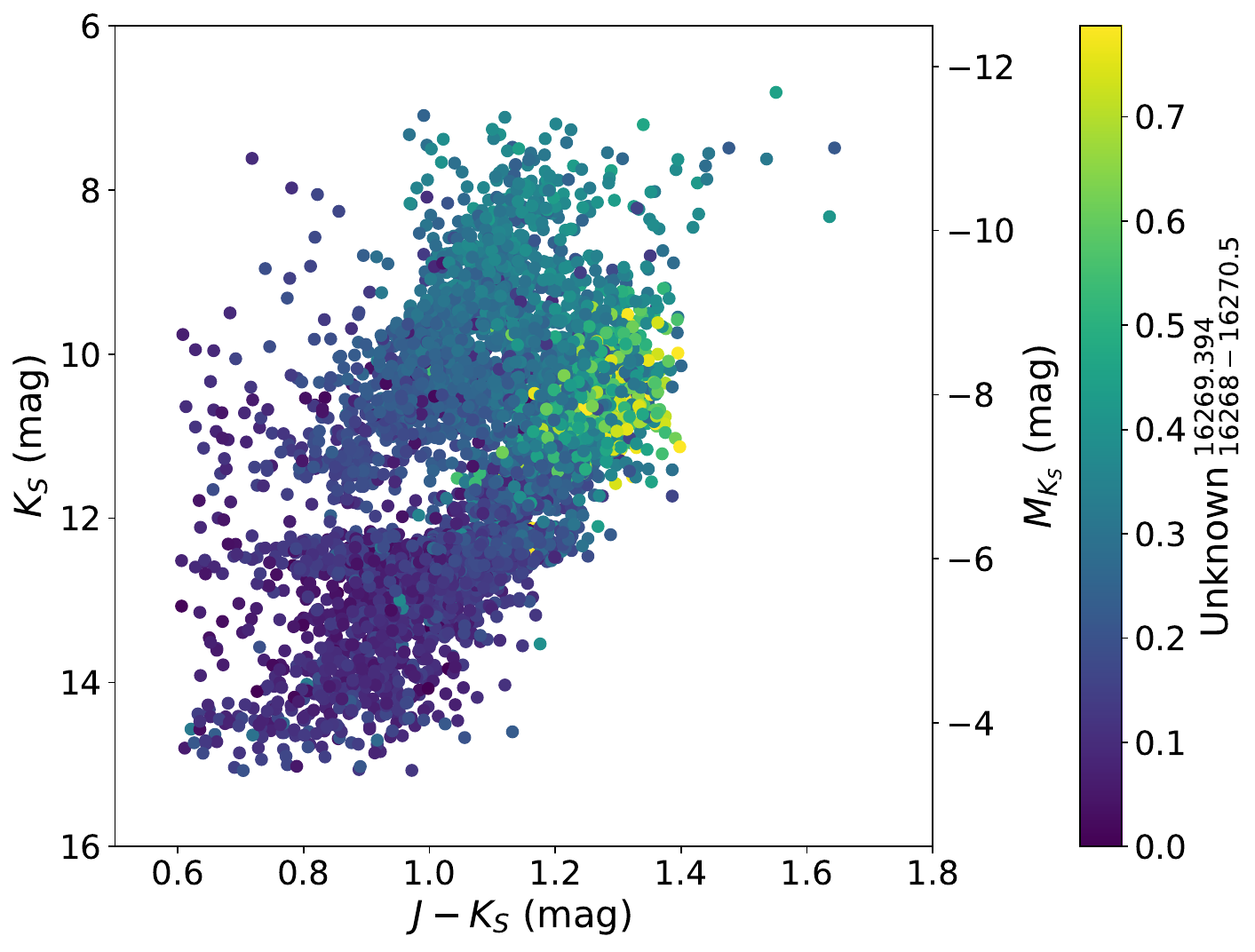}
\includegraphics[scale=0.24]{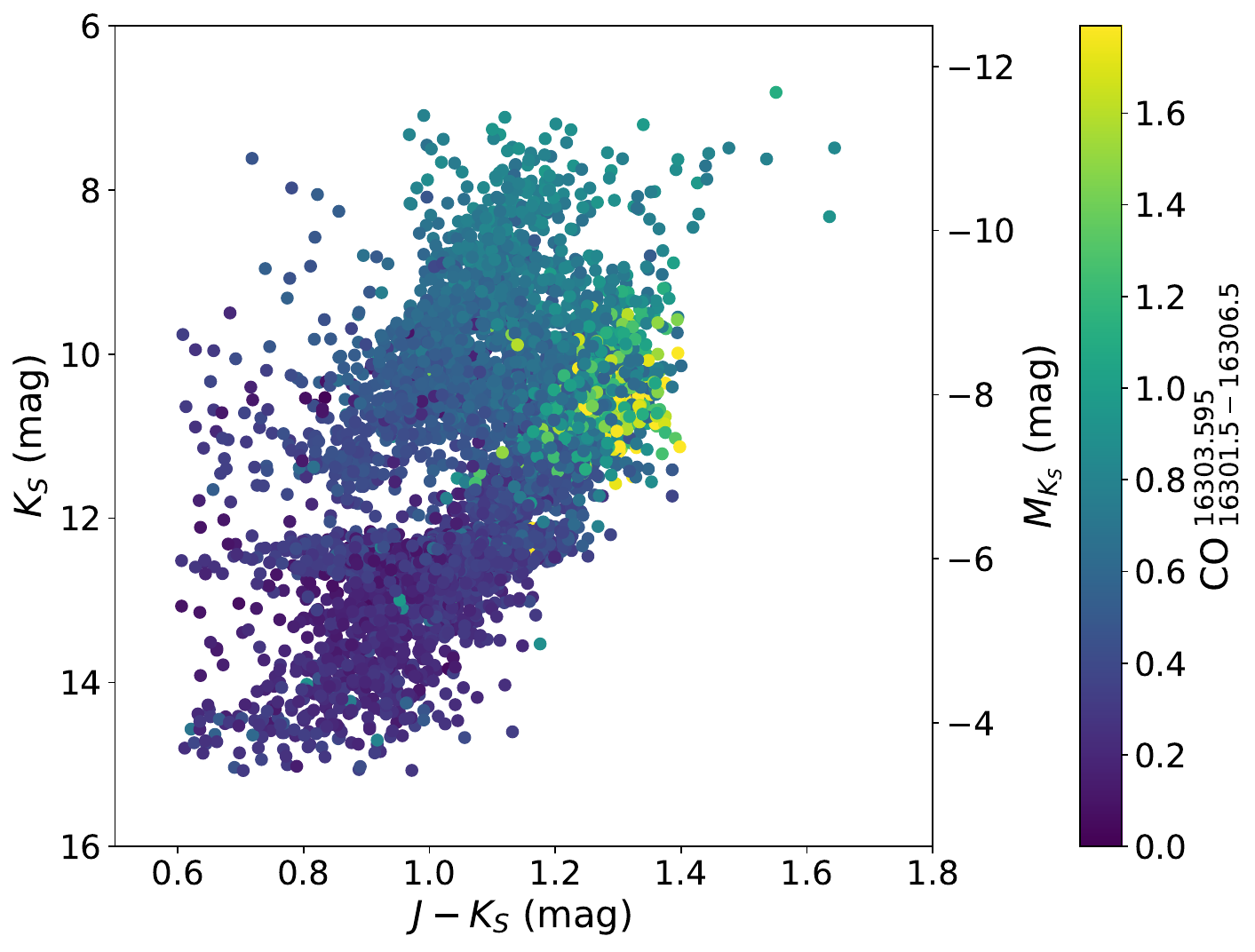}
\includegraphics[scale=0.24]{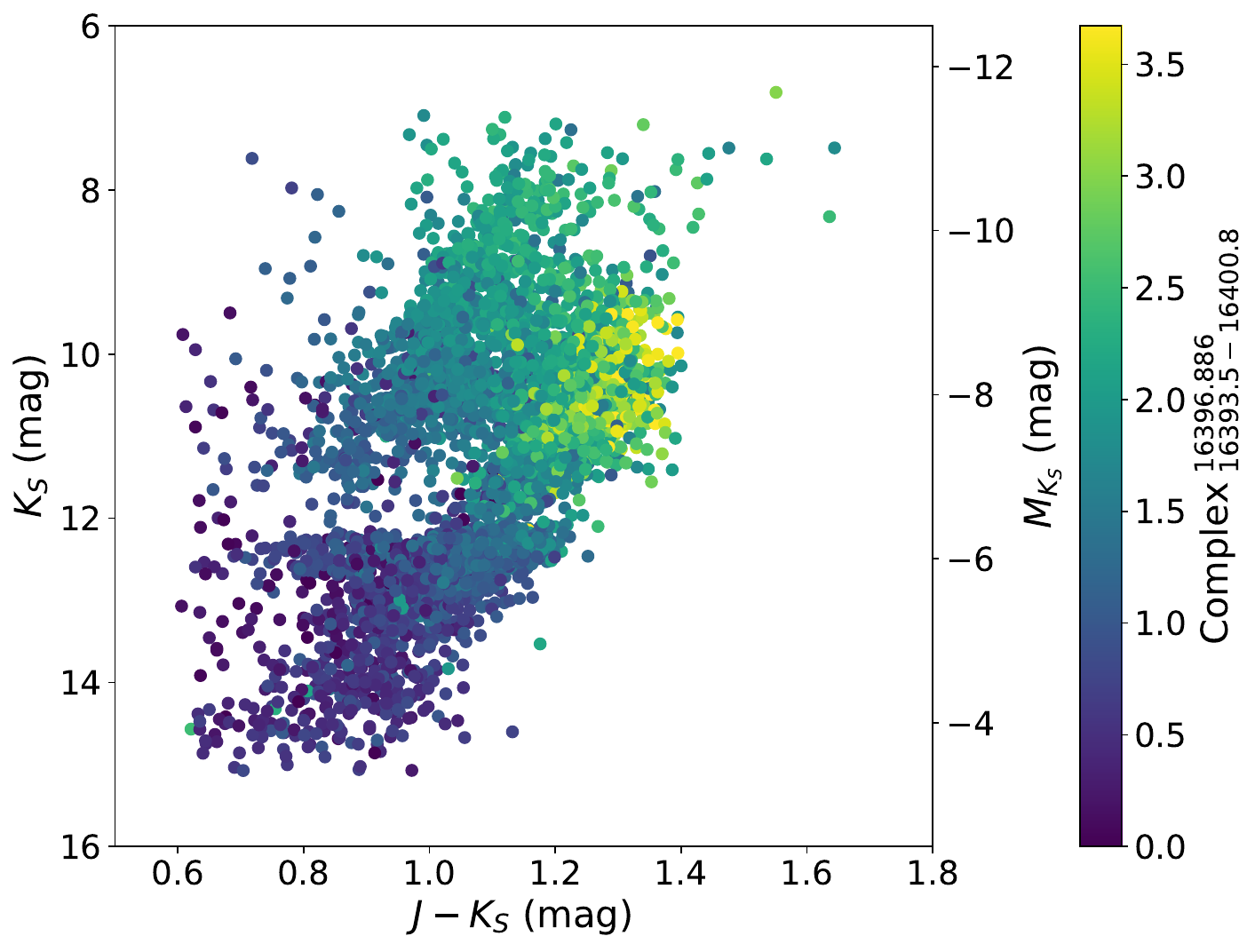}
\includegraphics[scale=0.24]{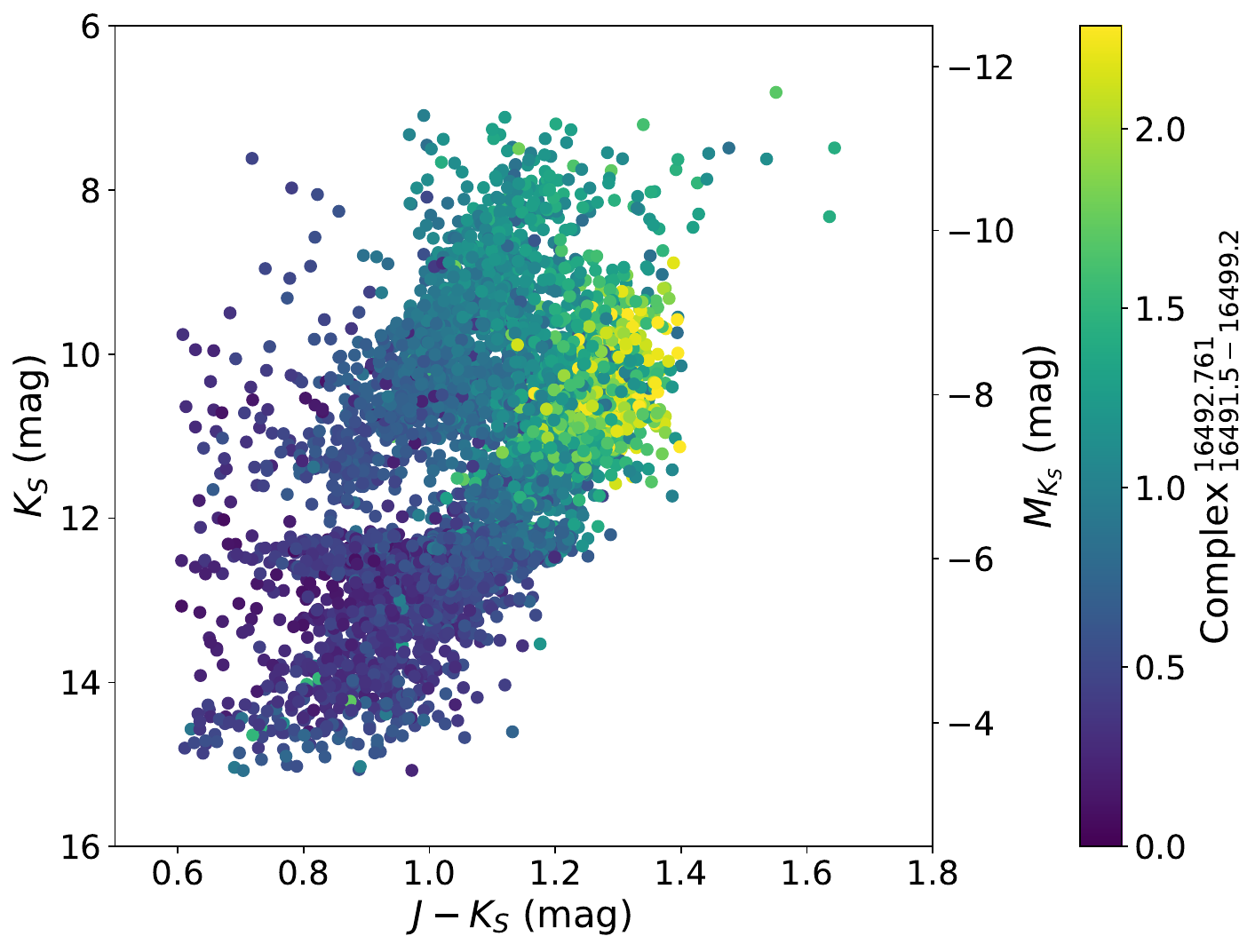}
\includegraphics[scale=0.24]{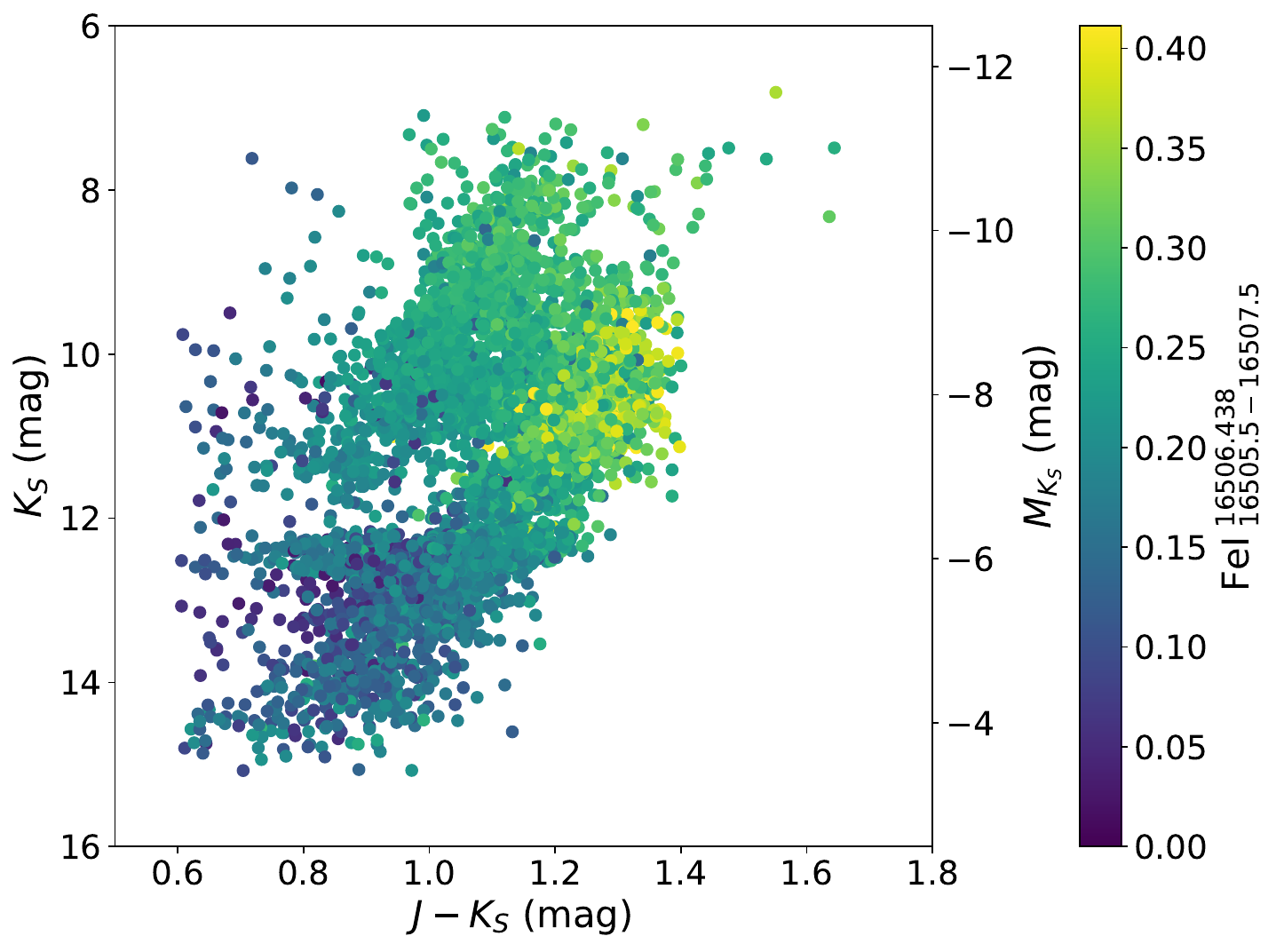}
\includegraphics[scale=0.24]{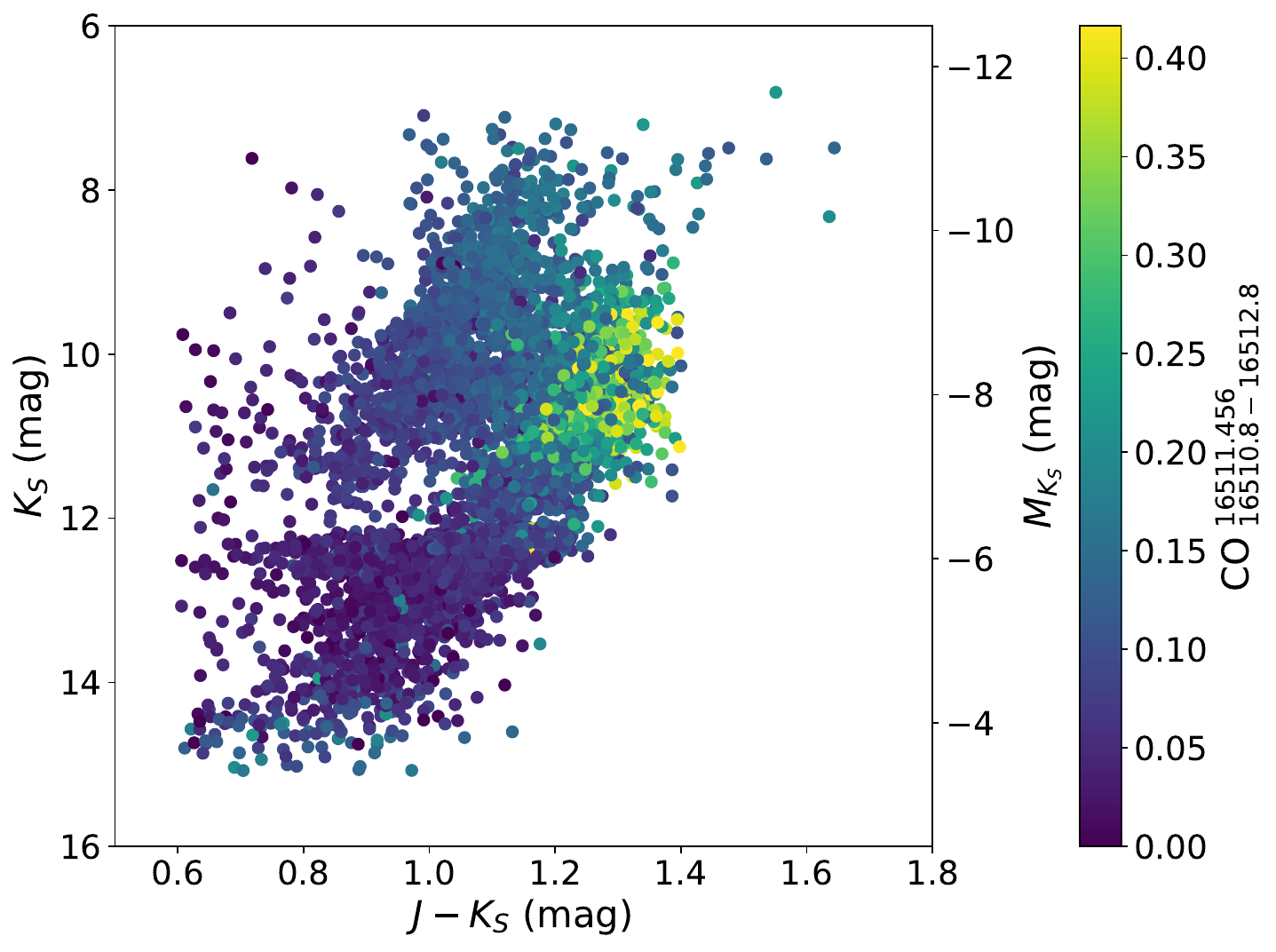}
\includegraphics[scale=0.24]{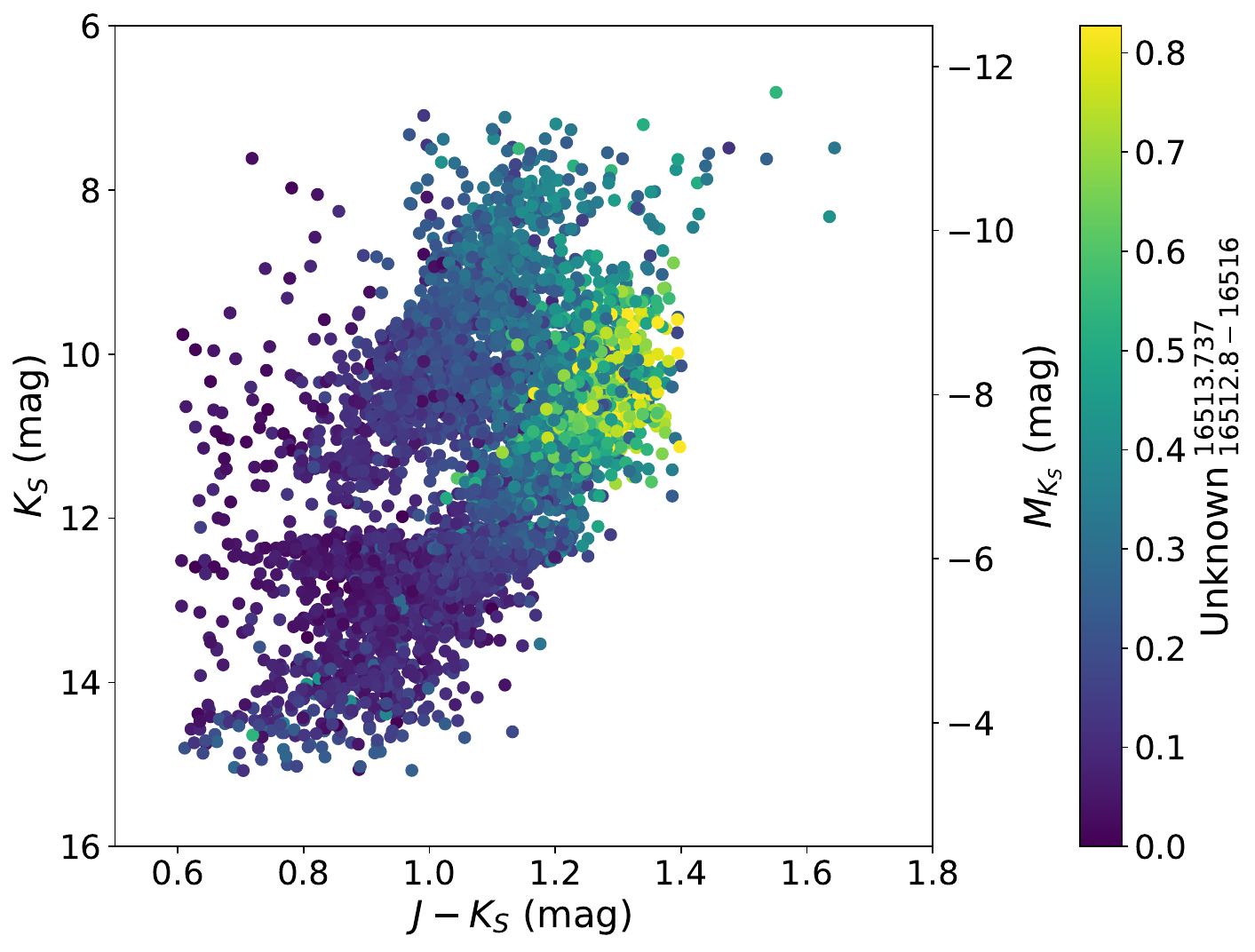}
\includegraphics[scale=0.24]{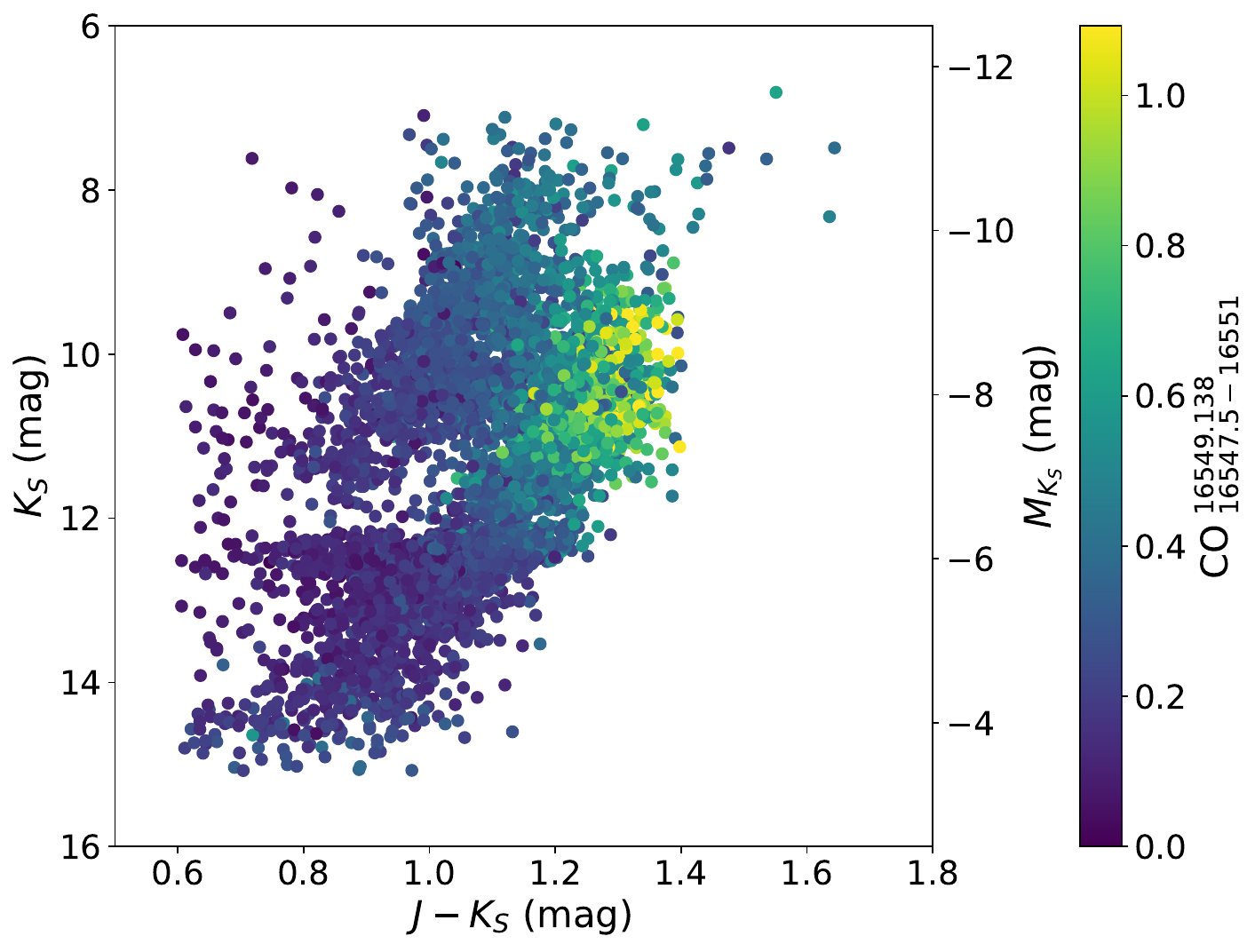}
\caption{Same as Figure~\ref{fig:ew}, but for the next 15 spectral lines or line complexes in Table~\ref{tbl:ewtable}. 
\label{fig:ew2}}
\end{figure*}

\begin{figure*}
\center
\includegraphics[scale=0.24]{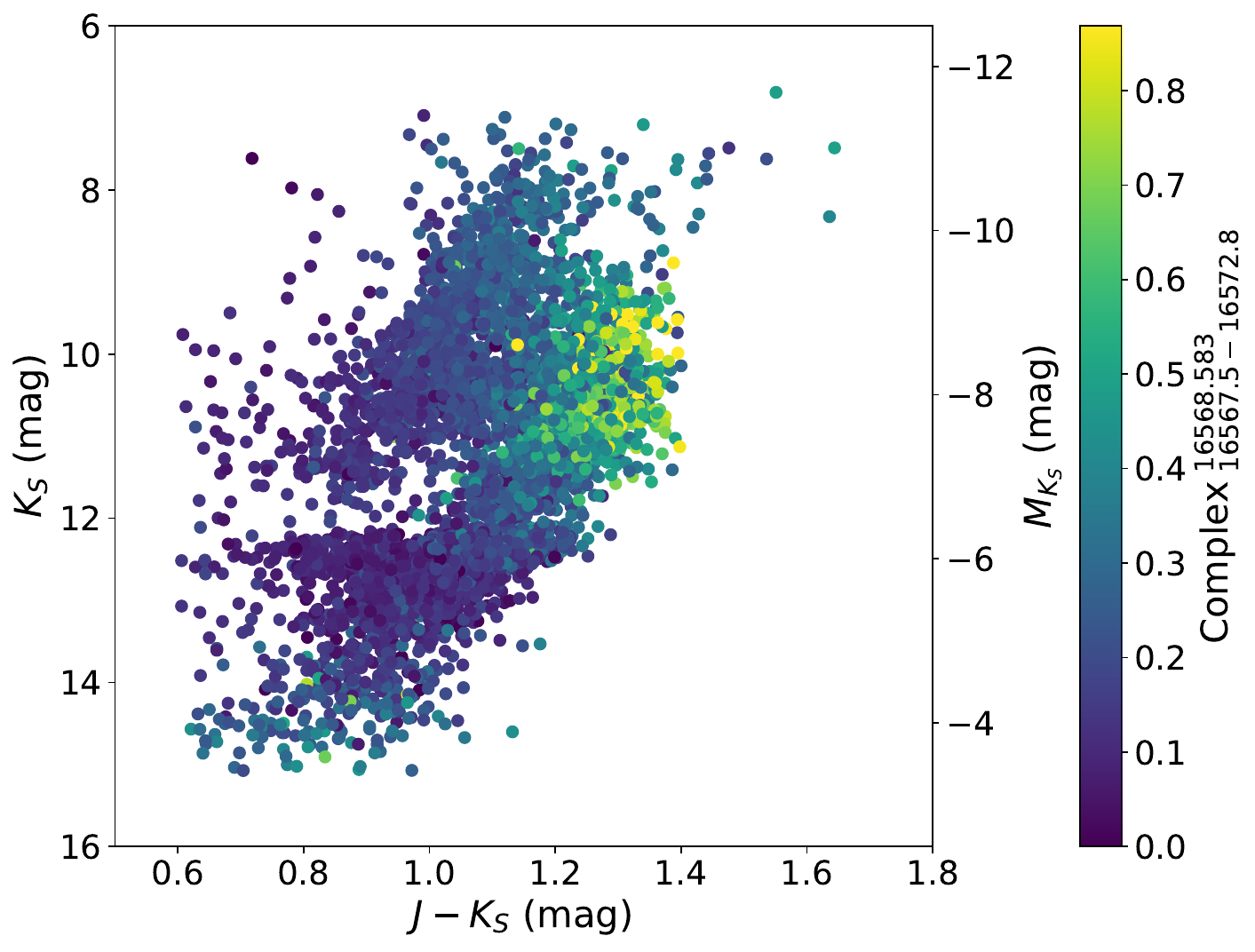}
\includegraphics[scale=0.24]{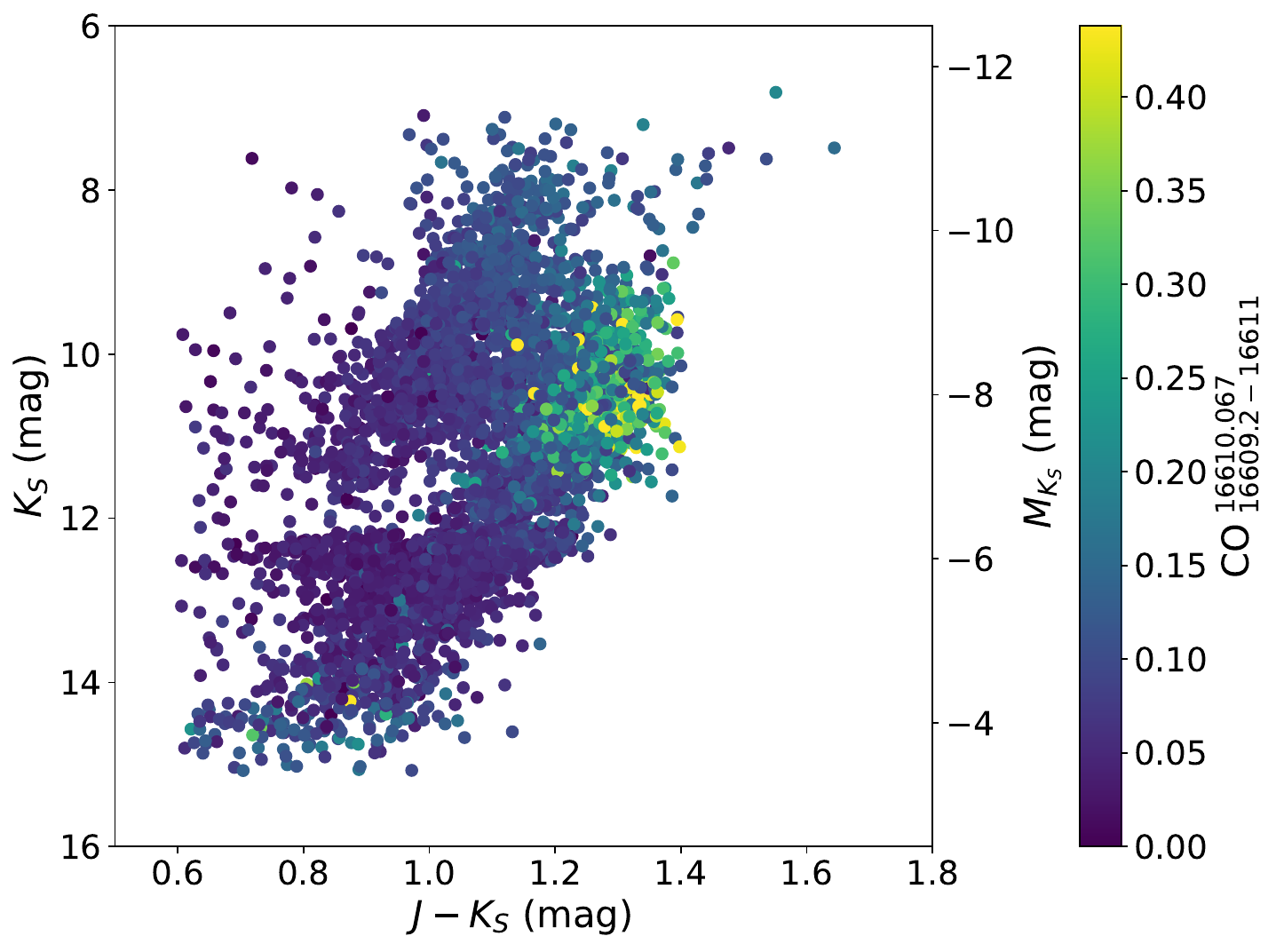}
\includegraphics[scale=0.24]{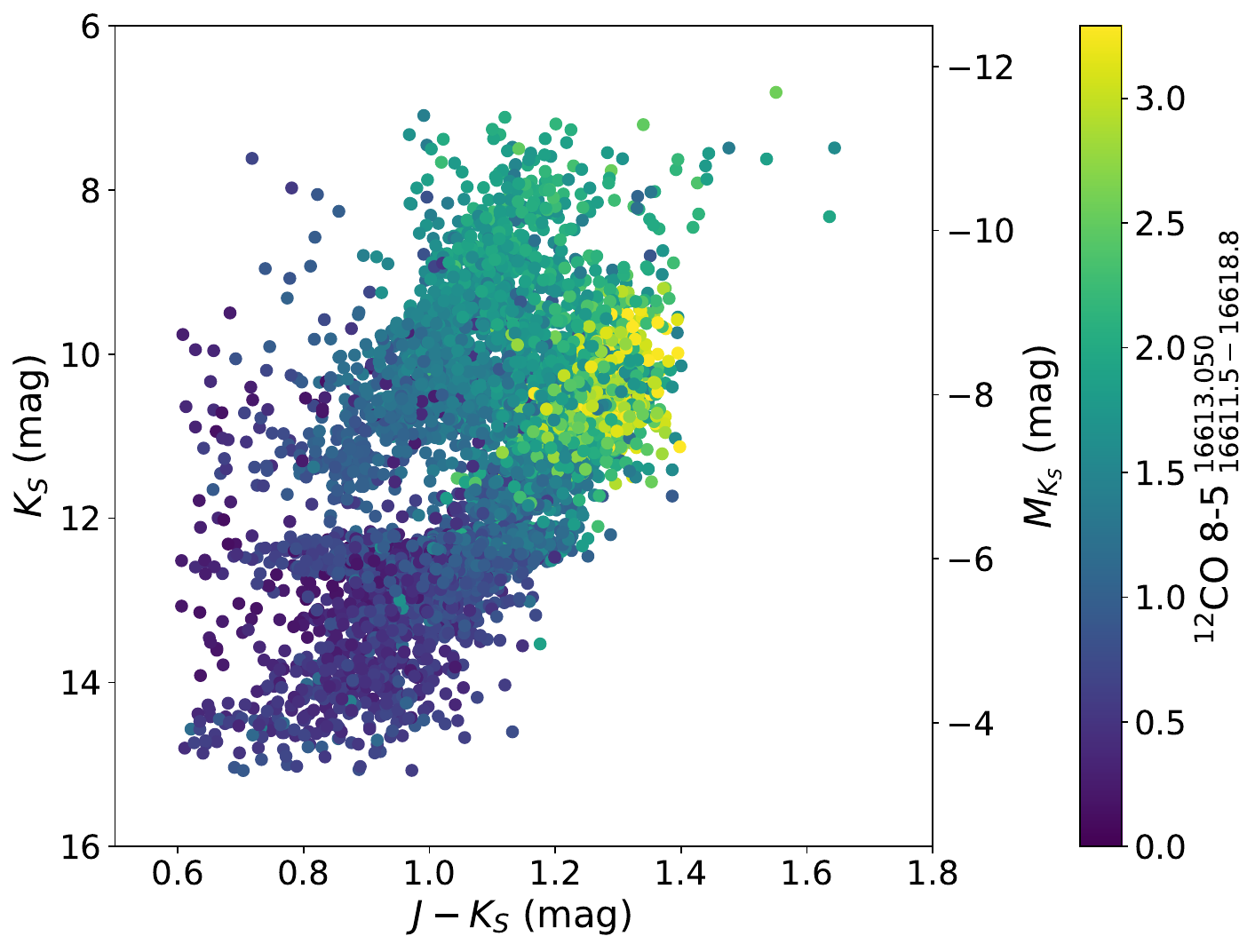}
\includegraphics[scale=0.24]{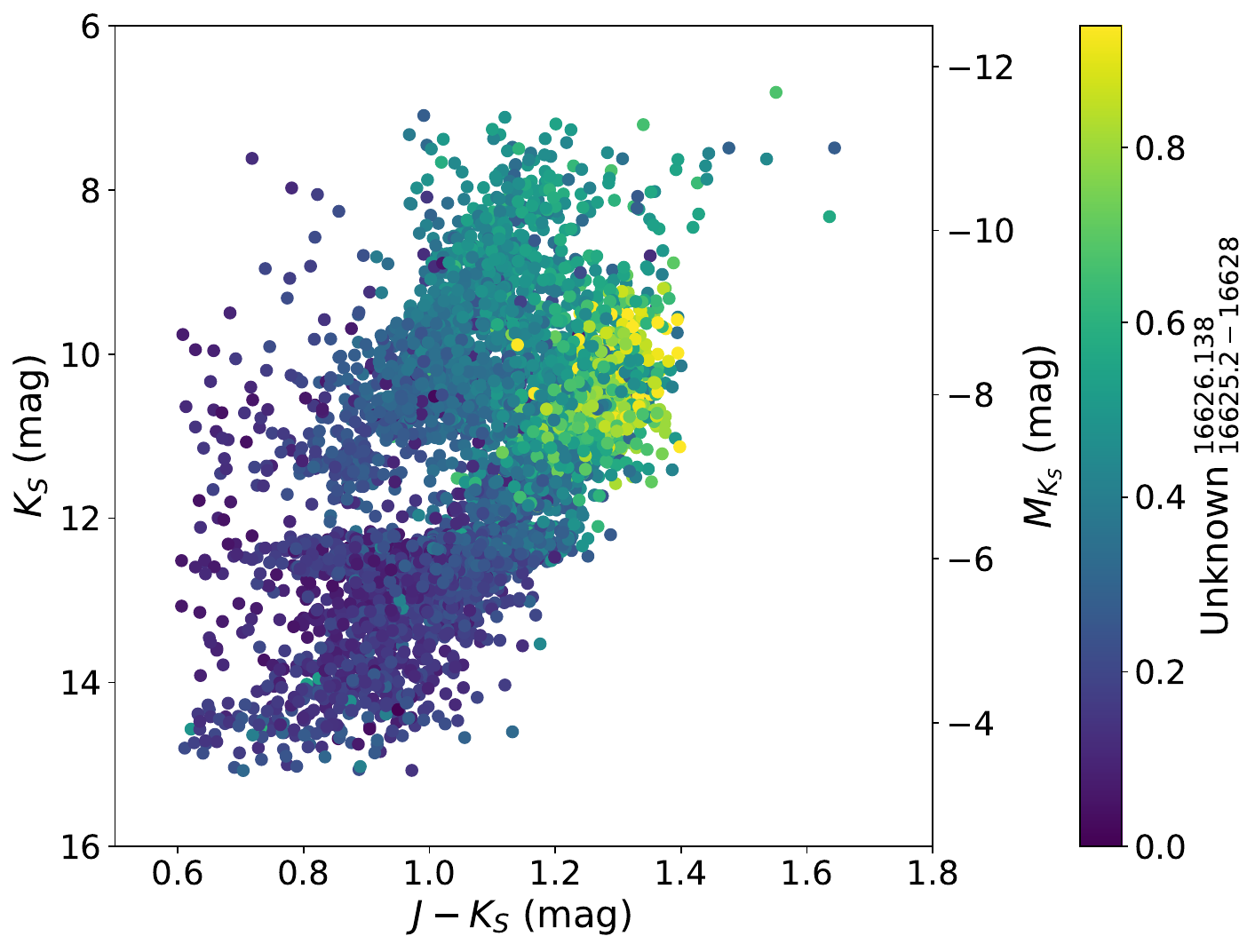}
\includegraphics[scale=0.24]{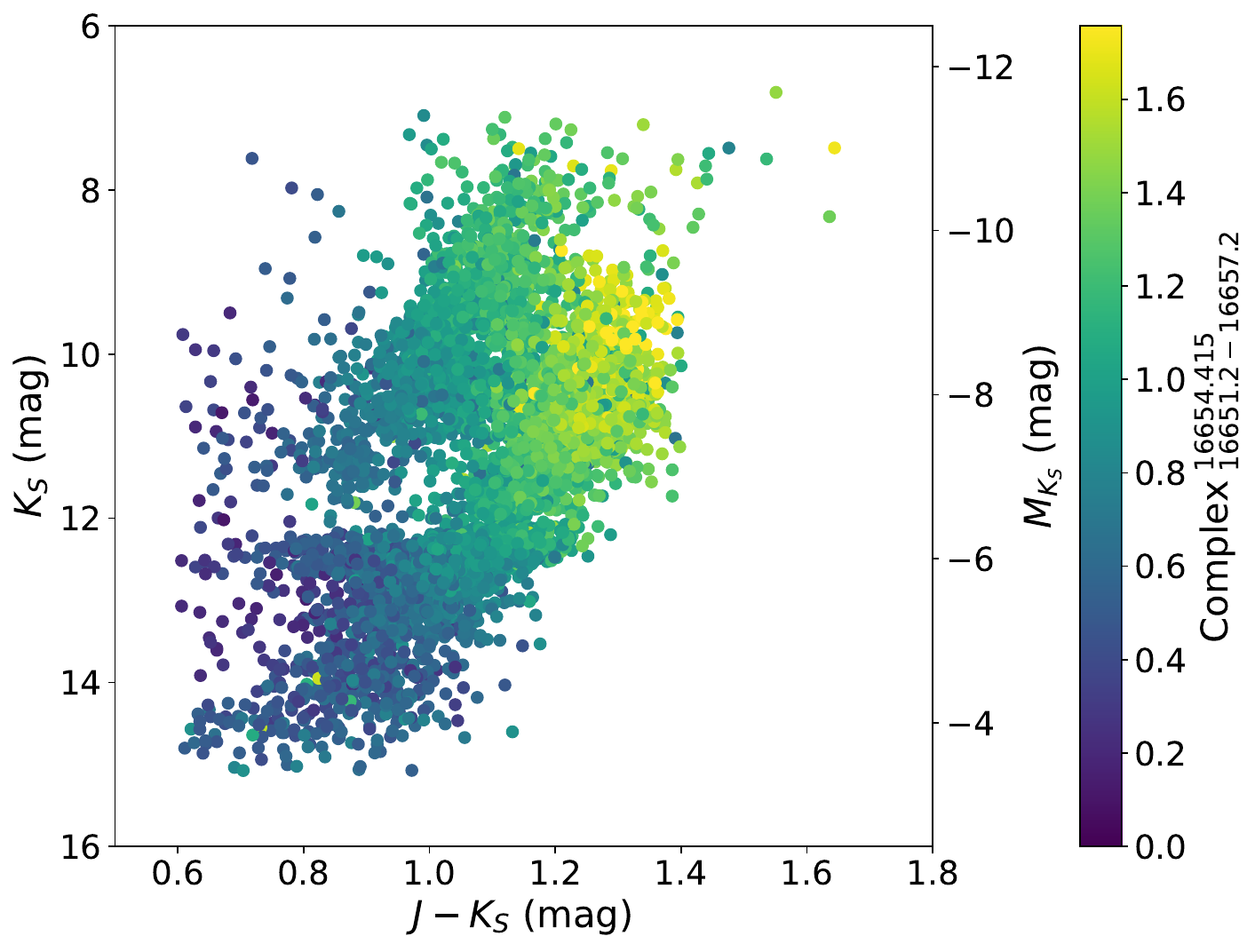}
\includegraphics[scale=0.24]{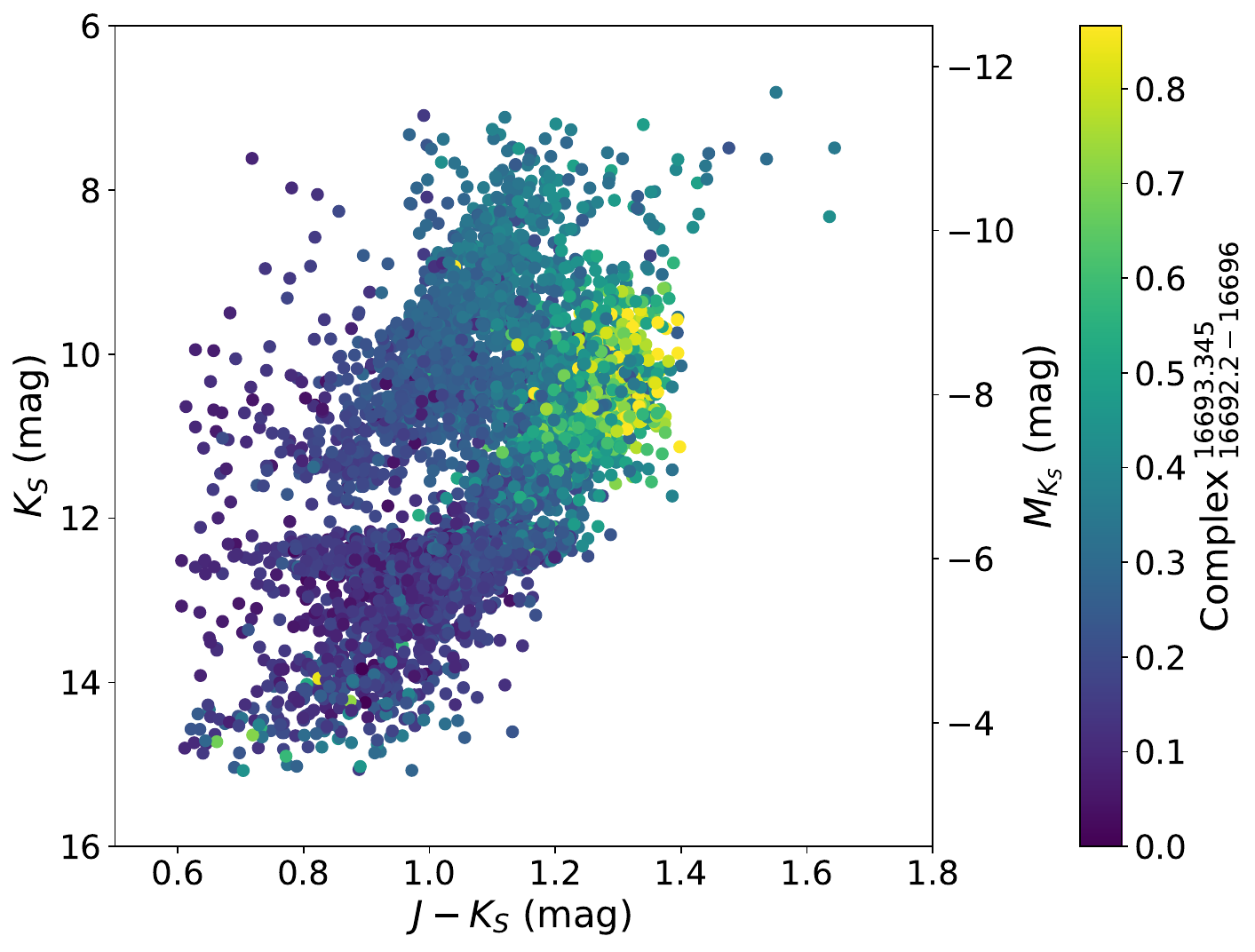}
\includegraphics[scale=0.24]{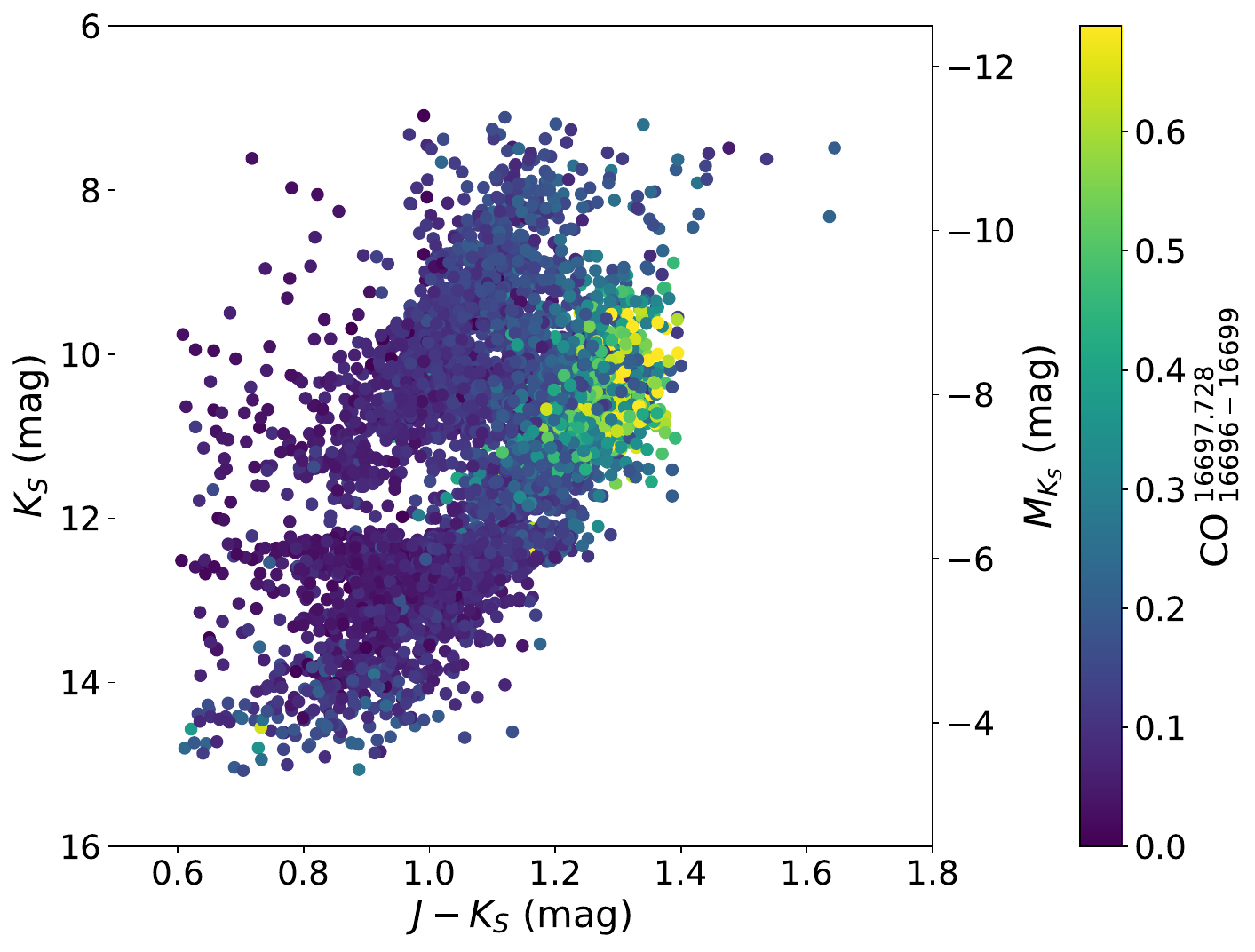}
\includegraphics[scale=0.24]{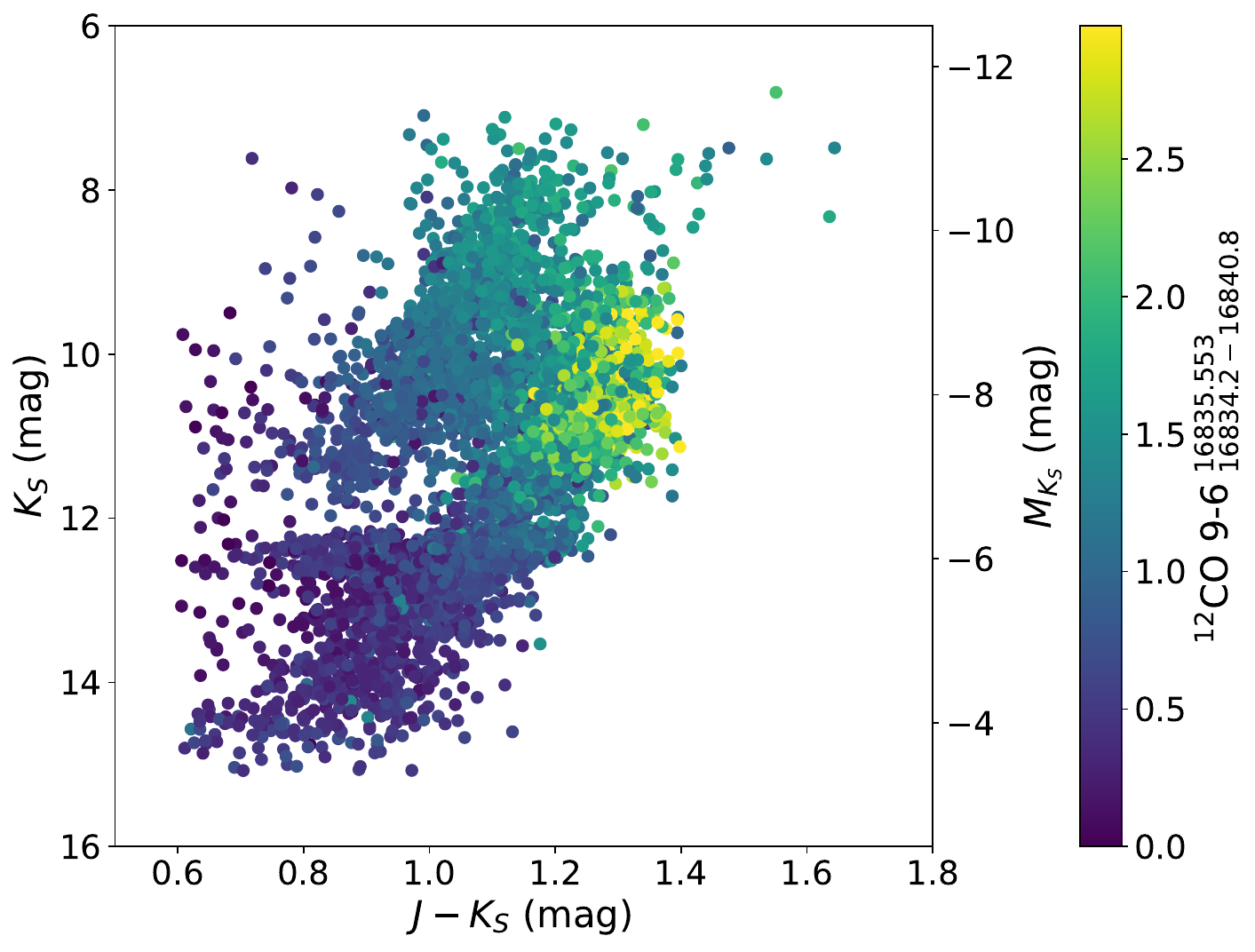}
\caption{Same as Figure~\ref{fig:ew}, but for the last 7 spectral lines or line complexes in Table~\ref{tbl:ewtable}. 
\label{fig:ew3}}
\end{figure*}

Figure~\ref{fig:ew1}, Figure~\ref{fig:ew2}, and Figure~\ref{fig:ew3} show the 2MASS CMDs color coded with the calculated EWs of different lines and line complexes. 

\end{CJK*}
\end{document}